# DDEC6: A Method for Computing Even-Tempered Net Atomic Charges in Periodic and Nonperiodic Materials

Thomas A. Manz* and Nidia Gabaldon Limas
Department of Chemical & Materials Engineering, New Mexico State University, Las Cruces, New Mexico, 88003-8001. *Corresponding author email: tmanz@nmsu.edu

**Abstract:**

Net atomic charges (NACs) are widely used in all chemical sciences to concisely summarize key information about the partitioning of electrons among atoms in materials. Although widely used, there is currently no atomic population analysis method suitable for being used as a default method in quantum chemistry programs. To address this challenge, we introduce a new atoms-in-materials method with the following nine properties: (1) exactly one electron distribution is assigned to each atom, (2) core electrons are assigned to the correct host atom, (3) NACs are formally independent of the basis set type because they are functionals of the total electron distribution, (4) the assigned atomic electron distributions give an efficiently converging polyatomic multipole expansion, (5) the assigned NACs usually follow Pauling scale electronegativity trends, (6) NACs for a particular element have good transferability among different conformations that are equivalently bonded, (7) the assigned NACs are chemically consistent with the assigned atomic spin moments, (8) the method has predictably rapid and robust convergence to a unique solution, and (9) the computational cost of charge partitioning scales linearly with increasing system size. Across a broad range of material types, the DDEC6 NACs reproduced electron transfer trends, core electron binding energy shift trends, and electrostatic potentials across multiple system conformations with excellent accuracy compared to other charge assignment methods. Due to non-nuclear attractors, Bader's quantum chemical topology could not assign NACs for some of these materials. The DDEC6 method alleviates the bifurcation or runaway charges problem exhibited by earlier DDEC variants and the Iterative Hirshfeld method. These characteristics make the DDEC6 method ideally suited for use as a default charge assignment method in quantum chemistry programs.

## 1. Introduction

**Net atomic charges (NACs)** are a ubiquitous concept in all chemical sciences. It is difficult to imagine chemistry being learned at either an introductory or advanced level without some reference to NACs.[1] For example, the pH scale measuring hydrogen ion activities in solution embodies the concept of hydrogen atoms carrying positive NACs. In biochemistry, many cellular functions depend on the transport of charged atoms such as $K^+$, $Na^+$, $Ca^{+2}$, and $Cl^-$ across cell membranes.[2] Experiments measuring the water molecule's dipole moment imply a negative NAC on its oxygen atom and a positive NAC on each of its two hydrogen atoms.[3] NACs also play an important role in solid state physics, where oxygen atoms in solid oxides carry negative NACs to enable oxygen ion transport.[4] Zwitterions, which are widely encountered in amino acids, illustrate that important chemical behaviors depend not only on the overall molecular charge but also on the net charges of local regions within a molecule.[5]

NACs concern more generally the question of how to partition the properties of a material among its constituent atoms. Most importantly, not all schemes proposed for such a partitioning are valid. According to the well-established and accepted scientific method, a concept can be scientifically valid only to the extent that it reproduces experimental observables. Schemes with no defined mathematical value as the complete basis limit is approached have no direct correspondence with experimental observables, because physical materials correspond to the complete basis set limit. In such explicitly basis-set-dependent schemes, the computed properties are highly dependent on the basis set choice so that the same electron density distribution yields differing results depending on the particular basis set representation. This adverse dependence on the basis set representation is completely unphysical and should be avoided. The Mulliken[6] and Davidson-Löwdin[7] methods lack a complete basis set limit, because in these methods the electrons are partitioned according to the atoms a basis function product is centered on rather than according to any physical principles. A complete basis set can be equivalently represented in terms of various basis function types. For example, a complete set of orthonormal spherical harmonics (multiplied by appropriate radial functions) centered on one atom in a material provides a complete basis set. Depending on whether we choose to center all of these basis functions on this or that atom, the Mulliken[6] and Davidson-Löwdin[7] methods would assign all electrons to either this or that atom in the system—a meaningless and arbitrary partition. Consequently, a quantum chemistry calculation performed at the complete basis set limit has ill-defined Mulliken[6] or Davidson-Löwdin[7] populations. Although this serious problem has been recognized for decades,[8-10] the computational chemistry community has thus far been slow to adapt, with the unphysical Mulliken or Löwdin atomic population analysis scheme still used as the default method in some popular quantum chemistry programs. While it has been proposed that the density matrix be projected onto a small basis set to make the Mulliken populations more consistent,[9] using small basis sets does not accurately represent the material's electron distribution. This emphasizes the urgent need to develop physically sound and computationally convenient alternatives such as the method described in this article.

Natural Population Analysis (NPA) assigns electrons to each atom in a material to maximize the weighted occupancy of orthonormal natural orbitals.[8] NPA exhibits much lower basis set sensitivity than Mulliken population analysis.[8] NPA and the related natural bond orbital (NBO) methods are useful for understanding the roles of natural atomic orbitals (NAOs) and their hybridization to produce chemical bonds.[8, 11, 12] However, these methods still fall short of the goal to represent NACs as a functional of the total electron distribution, $\rho(\vec{r})$.

Chemical systems are comprised of atomic nuclei surrounded by an electron cloud. This electron cloud can be computed using quantum chemistry calculations. Throughout this article we use the Born-Oppenheimer approximation, in which the electron cloud is assumed to equilibrate rapidly with respect to the nuclear motions. The NAC for atom A ($q_A$) equals its nuclear charge ($z_A$) minus the number of electrons assigned to it ($N_A$):

$$q_A = z_A - N_A . \qquad (1)$$

Herein we use the same notation as previously, except we use $\left(L_1, L_2, L_3\right)$ instead of $\left(k_1, k_2, k_3\right)$ to specify a translated image of atom A: "Following Manz and Sholl,[13] we begin by defining a material as a set of atoms $\{A\}$ located at positions $\{\vec{R}_A\}$, in a reference unit cell, **U**. For a nonperiodic system (e.g., a molecule), **U** is any parallelpiped enclosing the entire electron distribution. The reference unit cell has $L_1 = L_2 = L_3 = 0$, and summation over A means summation over all atoms in this unit cell. For a periodic direction, $L_i$ ranges over all integers with the associated lattice vector $\vec{v}_i$. For a nonperiodic direction, $L_i$=0 and $\vec{v}_i$ is the corresponding edge of **U**. Using this notation, the vector and distance relative to atom A are given by

$$\vec{r}_A = \vec{r} - L_1\vec{v}_1 - L_2\vec{v}_2 - L_3\vec{v}_3 - \vec{R}_A \qquad (2)$$

and $r_A = \left|\vec{r}_A\right|$."[14]

In this article, we are only interested in studying time-independent energy eigenstates of chemical systems. For such systems, a time-independent electron distribution,

$$\rho(\vec{r}) = \langle \Psi_{el} | \hat{\rho}(\vec{r}) | \Psi_{el} \rangle \qquad (3)$$

can be theoretically computed or experimentally measured[15, 16] where $\Psi_{el}$ is the system's multi-electronic wavefunction within the Born-Oppenheimer approximation. A well-posed atomic partitioning sums to the correct quantum operator

$$\hat{\rho}(\vec{r}) = \sum_{A,L} \hat{\rho}_A(\hat{r}) \qquad (4)$$

for the electron density operator $\hat{\rho}(\vec{r})$ and more generally

$$\hat{o} = \sum_{A,L} \hat{o}_A \qquad (5)$$

for any observable system property, $o_{sys}$, having the quantum operator $\hat{o}$

$$o_{sys} = \langle \Psi_{el} | \hat{o} | \Psi_{el} \rangle \qquad (6)$$

such that

$$\rho_A(\vec{r}) = \langle \Psi_{el} | \hat{\rho}_A(\vec{r}) | \Psi_{el} \rangle \qquad (7)$$

$$\Theta(\vec{r}) = \rho(\vec{r}) - \sum_{A,L} \rho_A(\vec{r}_A) \to 0 \qquad (8)$$

$$o_A = \langle \Psi_{el} | \hat{o}_A | \Psi_{el} \rangle \qquad (9)$$

$$o_{sys} = \sum_{A,L} o_A . \qquad (10)$$

Here $\sum_{A,L}$ means $\sum_{A}\sum_{L_1}\sum_{L_2}\sum_{L_3}$ denoting summation over all atoms in the material. For the reasons described above, $\hat{\rho}_A(\hat{r})$ and $\hat{o}_A$ must be constructed from observables with a well-defined complete basis set limit. $\{o_A\}$ are the assigned atomic properties that sum to the observable system property, $o_{sys}$. Examples of such atomic properties include atomic spin moments (ASMs), atomic masses, atomic energy partitions, etc.

Before considering specifically how to partition the electron density operator among atoms in a material at each spatial position, we first consider various schemes that partition only the integrated number of electrons. Electrostatic potential fitting (ESP,[17] Chelp,[18] Chelpg,[19] REPEAT[20]) methods optimize NACs by minimizing the root-mean-squared-error (RMSE) over a chosen set of grid points located outside the material's van der Waals surface. The atomic polar tensor (APT) charge quantifies the change in dipole moment due to the displacement of a nucleus.[21] APT and related dipole-change-derived NACs are useful for representing infrared (IR) spectra intensities.[22] In periodic materials, Born effective and related charge methods quantify the change in electric polarization due to the displacement of a nucleus and its periodic images.[23] A key limitation of electrostatic potential fitting, APT, and Born effective charges is that they are not designed to retain core electrons. For example, the Born effective charge of Ti in the cubic phases of $BaTiO_3$, $CaTiO_3$, $SrTiO_3$, and $PbTiO_3$ ranges from 6.7 to 7.2, which exceeds the nominal number of 4 valence electrons for a free Ti atom.[24] Charge Model 5 (CM5), which uses Hirshfeld (HD) charges as input, was parameterized to give NACs that approximately reproduce static molecular dipole moments.[25] The CM5 NACs do a much better (but not perfect) job of retaining core electrons on the host atom.[25, 26] The Voronoi deformation density (VDD) method assigns NACs according to the integral of the deformation density (i.e., the difference between $\rho(\vec{r})$ and a sum of spherically symmetric neutral reference atoms) over the Voronoi cell enclosing each atom.[27]

Atoms-in-materials (AIM) methods partition the electron density operator at each spatial position as described in Eqs. (4), (7), and (8) above subject to the constraint

$$\rho_A(\vec{r}) \geq 0. \qquad (11)$$

Because $\rho(\vec{r}) \geq 0$, constraint (10) allows $f_A(\vec{r}) = \rho_A(\vec{r})/\rho(\vec{r})$ to be interpreted as the probability of assigning an electron at position $\vec{r}$ to atom A. AIM methods include HD,[28] Iterative Hirshfeld (IH) and related charge partitioning methods,[29-33] Iterated Stockhold Atoms (ISA),[34, 35] Density Derived Electrostatic and Chemical (DDEC),[13, 14] radical Voronoi tessellation,[36, 37] etc. There has been some debate on how to best define the atomic probability factors, $\{f_A(\vec{r})\}$. The ISA method optimizes the set of atomic electron density distributions $\{\rho_A(\vec{r}_A)\}$ to resemble their spherical averages $\{\rho_A^{avg}(r_A)\}$.[34, 35] The HD[28] and IH[29] methods optimize $\{\rho_A(\vec{r}_A)\}$ to resemble a set of spherical reference atoms $\{\rho_A^{ref}(r_A)\}$. Nalewajski and Parr[38] and Parr et al.[39] argued for the HD definition based on information theory and philosophical considerations with the $\{f_A(\vec{r})\}$ considered as noumenons. Matta and Bader argued for a definition based on Virial compartments describing experimentally observed additive property relationships.[40] Bader's quantum chemical topology (QCT) partitions the electron cloud into non-overlapping compartments that satisfy the Virial theorem because they have zero-flux surfaces: $\nabla\rho \bullet d\hat{n} = 0$ where $d\hat{n}$ is the differential surface normal unit vector.[41-43] Because non-atomic Bader compartments exist in materials with non-nuclear attractors,[44] Bader's QCT is not strictly a partition into atomic electron distributions. However, Bader's QCT has historically been categorized with AIM methods, because it pioneered the theoretical development of the AIM concept.[41-43, 45] For the study of electrides, a non-nuclear attractor describing the electron ion would normally be considered an advantage.[46]

Ours is a scientific engineering design approach that resembles the process used to build airplanes. Similar to constructing an AIM partitioning, there is more than one conceivable way to build an airplane. One could make an airplane longer or shorter, for example. Yet, it is not accurate to say airplane design (or AIM partitioning design) is an arbitrary process, because the scientific method is utilized. Like airplanes, our AIM partitioning method has been scientifically engineered to meet chosen performance goals. This involved a process of constructing and testing prototypes to refine the design until all performance goals were achieved. When a new airplane is designed, prototypes are built and tested in wind tunnels to determine which shapes achieve appropriate lift, minimize drag, and respond favorably to air turbulence. Not only should an airplane fly, but it should take off and land smoothly, have good fuel efficiency, and so forth. All of these aspects are tested when developing a new airplane design. Our process for building an AIM partitioning method is similar. Specifically, we built and tested prototypes to improve the control, efficiency, accuracy, and robustness. As described in the Results and Discussion (Section 5), we made extensive comparisons to experimental data during this development process.

This scientific engineering design approach requires choosing performance goals. Table 1 lists nine desirable features we chose for assigning NACs. The first criterion is to assign exactly one electron distribution per atom in the material. This criterion is fulfilled by many but not all charge assignment methods. For example, Bader's QCT yields non-atomic electron distributions in materials with non-nuclear attractors.[44, 46, 47] The second criterion is to assign core electrons to their host atom. This criterion is not appropriate for APT and Born effective charges that quantify the system's response to nuclear displacements. Methods that directly fit the electrostatic potential without regard for atomic chemical states also do not satisfy this criterion. Since a goal of AIM methods is to describe atomic chemical states, they should preferably assign core electrons to the host atom. The third criterion is to assign NACs as functionals of $\{\rho(\vec{r})\}$. The main purposes of our NACs are to convey information about charge transfer between atoms and to approximately reproduce the electrostatic potential surrounding a material. Since charge transfer between atoms cannot occur without effecting $\rho(\vec{r})$ and $\rho(\vec{r})$ determines the electrostatic field surrounding the material, it makes sense to construct the NACs as functionals of $\{\rho(\vec{r})\}$. The fourth criterion is to assign atomic electron distributions to give an efficiently converging polyatomic multipole expansion. Polyatomic multipole expansions including multipolar and charge penetration terms of arbitrarily high order provide a formally exact representation of the electrostatic potential.[48-53] In practice, this expansion is normally truncated at some finite order; therefore, we wish to reproduce the electrostatic potential with good accuracy using the leading terms of the polyatomic multipole expansion. The fifth criterion is the assigned NACs should usually follow Pauling scale electronegativity trends. The Pauling scale electronegativity was parameterized to describe typical electron transfer directions in chemical bonds, where higher electronegativity elements typically take electrons from lower electronegativity elements.[54, 55] The sixth criterion is that NACs for a particular element have good transferability among different conformations that are equivalently bonded. We choose this criterion, because one of our goals is to assign NACs with good conformational transferability that are well-suited to construct flexible force-fields for classical atomistic simulations of materials. The seventh criterion is that the assigned NACs should be chemically consistent with the assigned atomic spin moments (ASMs). We will have more to say about this seventh criterion in Section 5.8.3. The eighth criterion is that the AIM distributions should have predictably rapid and robust convergence to a unique solution. The ninth criterion is that the computation cost of charge partitioning should ideally scale linearly with increasing system size. This criterion is desirable to have the method's computational cost remain competitive as the number of atoms in the unit cell increases.

Table 1: Nine desirable features (performance goals) we have chosen for assigning NACs

| |
|---|
| 1. exactly one assigned electron distribution per atom |
| 2. core electrons remain assigned to the host atom |
| 3. NACs are functionals of the total electron density distribution |
| 4. assigned atomic electron distributions give an efficiently converging polyatomic multipole expansion |
| 5. NACs usually follow Pauling scale electronegativity trends |
| 6. NACs for a particular element have good transferability among different conformations that are equivalently bonded |
| 7. the assigned NACs are chemically consistent with the assigned ASMs |
| 8. predictably rapid and robust convergence to a unique solution |
| 9. computational cost of charge partitioning scales linearly with increasing system size |

As a point of clarification, we intend that these criteria should be satisfied across a broad range of materials encompassing molecules, ions, nanostructures, solid surfaces, porous solids, nonporous solids, and other complex materials. Notably, developing a reliable method for charge partitioning in dense periodic solids is not simply the task of adding periodic boundary conditions to a charge partitioning method initially developed for small molecules. Small molecules are comprised mainly of surface atoms with few buried atoms. In contrast, dense solids are comprised mainly of buried atoms with few surface atoms. Therefore, charge assignment methods that work well for surface atoms but poorly for buried atoms are problematic for bulk solids. Currently, the most commonly used charge partitioning method for dense solids is Bader's QCT.[56] Because two charge partitioning methods that give practically equivalent results for molecules with lots of surface atoms sometimes produce spectacularly different results when applied to dense solids,[14] correlations between NAC methods for molecular test sets should not be extrapolated to dense solids. In summary, charge partitioning in dense solids is an intrinsically more difficult problem than charge partitioning in small molecules.

The remainder of this article is organized as following. Section 2 is a theoretical description of the DDEC6 method with emphasis on how it differs from the DDEC3 method. Section 3 describes the calculation and tabulation of reference ion densities and reference core densities. Section 4 summarizes computational parameters for the quantum chemistry calculations and Ewald summation. Section 5 contains the results and discussion for a variety of important chemical applications: 5.1 Compressed sodium chloride crystals with unusual stoichiometries, 5.2 Representing electron transfer between atoms in dense solids, 5.3 Comparison to spectroscopic results for various materials, 5.4 Reproducing the electrostatic potential in one system conformation, 5.5 Reproducing the electrostatic potential across multiple system conformations for constructing flexible force-fields, 5.6 Systems comprised almost entirely of surface atoms, 5.7 Solid surfaces, and 5.8 Collinear and non-collinear magnetic materials. Section 6 contains our conclusions.

A few remarks are appropriate pertaining to the charge assignment methods against which the new DDEC6 charge partitioning is compared. Because there are so many different charge assignment methods, it was impractical to compare all charge assignment methods for each material studied here. Therefore, we adopted the policy to compare against an appropriate subset of charge assignment methods for each material. Since DDEC6 is the successor to DDEC3, we compared DDEC6 to DDEC3 in most cases. In most cases, we included the charge assignment methods one would expect to perform the best for each kind of material. For example, electrostatic potential fitting (ESP or REPEAT) NACs were included in most comparisons based on the electrostatic potential RMSE and RRMSE. For dense solids, Bader's QCT was included, because it is currently the most widely used charge partitioning method for dense solids. We avoided Mulliken and Davidson-Löwdin charges, because these are extremely sensitive to the basis set choice.[8, 57] We included comparisons to the HD method in many cases, because it is easy to do even though the HD method usually underestimates NAC magnitudes.[14, 25, 29, 58] With the exception of Section 2.3, we restricted comparisons to IH and related methods to previously published results, because the several different variations of these methods and their various reference ion densities is beyond the scope of this article.[29-32, 59-61] (In Section 2.3, we present data for three systems proving for the first time that the IH optimization landscape is sometimes non-convex and converges to non-unique solutions.) Because several of the systems studied here were suggested in an article by Wang et al.[26] focusing on applications of CM5, we compared DDEC6 to CM5 results in those cases and a few others. For a few molecular systems, we also compared results to NPA and ISA charges. None of the dense materials included comparisons to the ISA charges, because these are known to perform poorly for dense solids.[13] We do not include comparisons to APT or Born effective charges, because DDEC6 charges quantify a system's static electron distribution while APT and Born effective charges quantify the system's response to a perturbation. As mentioned above, APT and Born effective charges are not designed to assign core electrons to the host atom.

## 2. Theory

### 2.1 Fundamentals of vectorized charge partitioning

We use the term *vectorized charge partitioning* to denote the class of AIM charge partitioning methods for which the relative probability of assigning electrons at position $\vec{r}_A$ to atom A can be represented in terms of some spherically symmetric atomic weighting factor, $w_A(r_A)$:

$$\rho_A(\vec{r}_A)/\rho(\vec{r}) = w_A(r_A)/W(\vec{r}) \qquad (12)$$

where

$$W(\vec{r}) = \sum_{A,L} w_A(r_A). \qquad (13)$$

We call this *vectorized charge partitioning,* because for each atom $w_A(r_A)$ forms a one-dimensional array of $w_A$ values corresponding to a series of $r_A$ values. The whole quest to define the charge partitioning method thus reduces the problem of finding a three-dimensional array of $f_A(\vec{r})$ values for each atom to that of finding a one-dimensional array of $w_A(r_A)$ values for each atom. This reduction in parameter space from three to one degrees of freedom per atom makes vectorized charge partitioning computationally efficient, because one-dimensional rather than three-dimensional arrays need to be computed and stored for each atom.

A key use of NACs is to construct point-charge models to regenerate the electrostatic potential in classical molecular dynamics and Monte Carlo simulations.[62] From Gauss's Law of Electrostatics it directly follows that the electrostatic potential exerted outside a spherically symmetric charge distribution is identical to an equivalent point charge placed at the sphere's center. Hence, it is wise to assign approximately spherically symmetric atomic electron distributions

$$\rho_A(\vec{r}_A) \approx \rho_A^{avg}(r_A) \qquad (14)$$

so that a point-charge model comprised of the NACs will approximately reproduce the electrostatic potential surrounding the material. This can be accomplished by making

$$W(\vec{r}) \approx \rho(\vec{r}) \qquad (15)$$

for then Eq. (12) simplifies to

$$\rho_A(\vec{r}_A) \approx w_A(r_A). \qquad (16)$$

If Eq. (15) is true everywhere in the system, then Eqs. (14) and (16) are also true everywhere in the system. This will make a point-charge model constructed from the NACs approximately reproduce the electrostatic potential surrounding the material. Thus, satisfying Eq. (15) is a key objective.

The differential path action dS allows us to study convergence properties of vectorized charge partitioning methods:[14]

$$dS = \sum_A \oint \delta\rho_A(\vec{r}_A) \ln\left(\zeta(\vec{r}_A)\right) d^3\vec{r}_A \qquad (17)$$

where

$$\zeta(\vec{r}_A) = \frac{\rho_A(\vec{r}_A) W(\vec{r})}{w_A(r_A)\rho(\vec{r})}. \qquad (18)$$

Stationary points of the path action $S = \int dS$ occur where

$$\frac{\delta S}{\delta\rho_A(\vec{r}_A)} = \ln\left(\zeta(\vec{r}_A)\right) = 0 \qquad (19)$$

for every atom, which yields Eq. (12) as the only solution(s).[14] Due to its path dependence, S is not a functional of $\{\rho_A(\vec{r}_A)\}$. The path action S is a kind of affine mapping in which we do not care which particular $\{\rho_A(\vec{r}_A)\}$ is chosen as the starting point (aka 'origin') for the integral, and we only care about its differential change, dS.

A second key purpose of NACs is to represent the chemical states of atoms in materials. This requires the assigned $\{\rho_A(\vec{r}_A)\}$ to have atomic-like properties. The spherically averaged electron distributions of isolated atoms decay approximately exponentially with increasing $r_A \geq 2\,\text{Å}$:

$$\frac{d^2 \ln\left(\rho_A^{avg}(r_A)\right)}{dr^2} \approx 0 \quad \text{for } r_A \geq 2\,\text{Å}. \qquad (20)$$

To maximize the transferability of atomic chemical properties between isolated atoms, molecules, porous solids, solid surfaces, non-porous solids, and nano-structures, the atomic electron distributions in materials should be assigned to approximately follow Eq. (20).

Another key consideration is the number of electrons assigned to each atom should resemble the number of electrons contained in the volume of space dominated by that atom. The volume of space dominated by atom A is defined as the spatial region for which $\rho_A(\vec{r}_A) > \rho_B(\vec{r}_B)$ for every B ≠ A. If the $w_A(r_A)$ for anions are too diffuse in ionic crystals, this might cause too many electrons in the volume of space dominated by the cations to be mistakenly assigned to the anions. This can lead to situations where the total number of electrons assigned to the cations is even lower than their number of core electrons. To avoid this mistake, some care should be given to quantify how many electrons are in the volume of space dominated by each atom. The number of electrons assigned to each atom should then be optimized to resemble this value, subject to additional optimization criteria. Also, we find (see Sections 5.2.1 and 5.3.2) that electron transfer trends and core-electron binding energy shifts and are more accurately described when the number of electrons assigned to each atom resembles that in the volume dominated by each atom.

To maximize chemical transferability, it is desirable to have each atom in a material resemble a reference ion of the same element having the same net charge. For example, a $Na^{+1}$ ion in a material should resemble a $Na^{+1}$ reference ion. Therefore, we use reference ions to construct part of the atomic weighting factors, $\{w_A(r_A)\}$. The HD method uses neutral atoms as the reference states:[28]

$$w_A^{HD}(r_A) = \rho_A^{ref}\left(r_A, q_A^{ref} = 0\right). \quad (21)$$

The extremely poor performance of the HD method can be explained by the fact that in partially or totally ionic systems $\rho(\vec{r})$ does not approximately equal the sum of neutral atom densities:

$$\rho(\vec{r}) \neq W^{HD}(\vec{r}) = \sum_{A,L} \rho_A^{ref}\left(r_A, q_A^{ref} = 0\right). \qquad (22)$$

Thus, Eqs. (14)–(16) are not consistently satisfied by the HD method. Consequently, the HD NACs usually give a poor representation of the electrostatic potential surrounding a material. The IH method improves upon the HD method by using self-consistently charged reference states:[29]

$$w_A^{IH}(r_A) = \rho_A^{ref}\left(r_A, q_A^{ref} = q_A\right). \quad (23)$$

While the IH method offers a clear improvement over the HD method, the performance of the IH method is still not optimal. Specifically, the IH method does not accurately account for the relative contraction or expansion of each ionic state due to its local environment. For example, an atomic anion in an ionic crystal is usually more contracted than the corresponding isolated atomic anion, because the cations in the ionic crystal provide charge balance and electrostatic screening that reduces electrostatic repulsion between excess electrons in the bound atomic anion. While it is possible to use charge-compensated reference ions in the IH method,[32, 63] the overall accuracy of constructing $W(\vec{r}) \approx \rho(\vec{r})$ is still limited in the IH method by using a single set of reference ions that do not respond to their local environment. This problem is overcome in the DDEC3 and DDEC6 methods by conditioning the reference ion densities to match the specific material. This conditioning describes the contraction or expansion of reference ions in response to their local environment while still only requiring a single reference ion library as input.[14]

Eq. (15) will be fulfilled across the widest variety of systems if $\{w_A(r_A)\}$ are themselves derived from partitions of $\rho(\vec{r})$. The ISA method is an early example of a charge partitioning scheme in which $\{w_A(r_A)\}$ are derived from a partition of $\rho(\vec{r})$.[34, 35] In the ISA method,

$$w_A^{ISA}(r_A) = \rho_A^{avg}(r_A). \quad (24)$$

Although the ISA method clearly fulfills Eqs. (14)–(16), the ISA NACs have poor conformational transferability and are chemically inaccurate for many materials (especially, non-porous materials).[13, 14, 64-66] In the next sections, we will show how to construct a new charge partitioning scheme, called DDEC6, that combines electron localization, reference ion weighting, and spherical averaging to create an accurate and robust charge partitioning method.

**2.2 The Density Derived Electrostatic and Chemical (DDEC) approach**

The DDEC approach optimizes $\{w_A(r_A)\}$ to simultaneously resemble reference ion states and $\{\rho_A^{avg}(r_A)\}$.[13, 14] Making $\{w_A(r_A)\}$ resemble reference ion states maximizes the chemical transferability of the assigned $\{\rho_A(\vec{r}_A)\}$ and NACs, while making $\{w_A(r_A)\}$ resemble $\{\rho_A^{avg}(r_A)\}$ causes the NACs to approximately reproduce the electrostatic potential surrounding the material.[13, 14]

Three variants of the DDEC method have been previously published.[13, 14] In the DDEC/c1 and DDEC/c2 methods, the atomic weighting factors are defined by

$$w_A^{DDEC/c1,c2}(r_A) = \left(\rho_A^{ref}\left(r_A, q_A^{ref} = q_A\right)\right)^{\chi}\left(\rho_A^{avg}(r_A)\right)^{1-\chi} \quad (25)$$

with $\chi^{DDEC/c1} = \chi^{DDEC/c2} = 1/10$.[13] During the iterative updates, the reference ion charges are updated to match the AIM charges, as done for the IH method. The reference ions, $\{\rho_A^{ref}(r_A, q_A^{ref})\}$, are computed using charge compensation and dielectric screening.[13] In the DDEC/c1 method, charge compensation and dielectric screening were modeled by computing the reference ion densities for atoms placed in a periodic array with a uniform compensating background charge.[13] In the DDEC/c2 method, charge compensation and dielectric screening were modeled by computing the reference ion densities for atoms enclosed by a spherical charge compensation shell.[13] For anions, the shell radius and compensating charge are carefully selected to minimize the system's total energy.[13] For cations of charge +q, the compensating charge is –q and the shell radius is the average radius of the outermost q occupied Kohn-Sham orbitals of the isolated neutral atom.[13]

In this article, the term *charge partitioning functional* means a functional F whose stationary points yield the corresponding atomic charge distributions. A point is a stationary point if and only if the full derivative of F is zero: $dF = 0$. This requires all of the first-order partial and variational derivatives of F with respect to the independent variables and functions, respectively, to be zero. Nalewajski and Parr showed the HD method minimizes the charge partitioning functional

$$F^{HD} = \sum_A \oint \rho_A(\vec{r}_A) \ln\left(\frac{\rho_A(\vec{r}_A)}{\rho_A^{ref}\left(r_A, q_A^{ref} = 0\right)}\right) d^3\vec{r}_A - \int_U \Gamma(\vec{r})\Theta(\vec{r}) d^3\vec{r} \quad (26)$$

where $\Gamma(\vec{r})$ is a Lagrange multiplier enforcing constraint (8).[38] Later, Bultinck et al. showed the ISA method minimizes a similar charge partitioning functional where $\rho_A^{avg}(r_A)$ appears instead of $\rho_A^{ref}(r_A, q_A^{ref}=0)$:[66]

$$F^{ISA} = \sum_A \oint \rho_A(\vec{r}_A) \ln\left(\frac{\rho_A(\vec{r}_A)}{\rho_A^{avg}(r_A)}\right) d^3\vec{r}_A - \int_U \Gamma(\vec{r})\Theta(\vec{r}) d^3\vec{r} \qquad (27)$$

A functional or affine mapping is convex if and only if its curvature is non-negative. Smooth convex functionals and affine mappings have a unique minimum. Both the ISA and non-iterative Hirshfeld charge partitioning functionals are convex and possess a single minimum leading to unique solution.[66]

The DDEC/c1,c2 methods minimize the partial derivative of

$$F = \sum_A \oint \rho_A(\vec{r}_A) \ln\left(\frac{\rho_A(\vec{r}_A)}{\left(\rho_A^{ref}(r_A, q_A^{ref})\right)^{\chi}\left(\rho_A^{avg}(r_A)\right)^{1-\chi}}\right) d^3\vec{r}_A - \int_U \Gamma(\vec{r})\Theta(\vec{r}) d^3\vec{r} \qquad (28)$$

with respect to $\{\rho_A(\vec{r}_A)\}$ while holding the $\{q_A^{ref}\}$ constant if $\{q_A^{ref}\}=\{q_A\}$.[13] Most importantly, Eq. (28) is emphatically ***not*** a charge partitioning functional for the DDEC/c1,c2 methods. Specifically, minimizing F does not automatically enforce the constraint $\{q_A^{ref}\}=\{q_A\}$. This constraint can be enforced by simply replacing $\{q_A^{ref}\}$ with $\{q_A\}$ in Eq. (28), but minimizing the resulting functional does ***not*** yield the weighting factors shown in Eq. (25). Instead of directly replacing $\{q_A^{ref}\}$ with $\{q_A\}$ in Eq. (28), the constraint $\{q_A^{ref}\}=\{q_A\}$ could be enforced using the method of Lagrange multipliers to yield a completely equivalent result that once again does not reproduce Eq. (25). Therefore, the object to be minimized for the DDEC/c1,c2 methods is not the functional shown in Eq. (28), but rather Manz'z path action S described in the previous section. This can be a potential source of confusion, because the first paper on the DDEC/c1,c2 method predated the introduction of the path action S and relied on partial (not proper) minimization of Eq. (28).[13] With the introduction of the path action S, this earlier approach that was not rigorous should be abandoned.

The same problem has also occurred in early literature on the IH method that predated Manz's path action S. Specifically, setting $\chi = 1$ in Eq. (28) does emphatically ***not*** yield a charge partitioning functional for the IH method, because it fails to properly impose the constraint $\{q_A^{ref}\}=\{q_A\}$ employed in the IH method. Enforcing $\{q_A^{ref}\}=\{q_A\}$ via a direct substitution in Eq. (28) or through the addition of Lagrange multipliers to Eq. (28) does ***not*** yield the IH weighting factors, but rather yields a new AIM method whose performance was not as good as the IH method.[33] The object to be minimized for the IH method is the path action S. Interestingly, a proposed proof[67] that the IH method always converges to a unique minimum (independent of the starting conditions) is invalid, because it was based on partial (not proper) minimization of Eq. (28) rather than using the correct object to be minimized. In the following section, we present specific examples of materials for which the converged IH NACs depend on the initial guess, thus proving the IH optimization landscape is sometimes non-convex.

The DDEC3 method was introduced to improve the accuracy of charge partitioning in dense solids containing short bond lengths.[14] For these materials, the DDEC/c1,c2 methods yielded extremely poor results.[14] The DDEC3 method improved the accuracy by introducing reference ion conditioning (see Section 2.4) and a constraint forcing the $w_A(r_A)$ tails of buried atoms to decay at least as fast as exp(-1.75 $r_A$).[14] The DDEC3 method starts with the same reference ion library as the DDEC/c2 method. DDEC3 smooths these reference ions to improve the optimization landscape curvature.[14] Finally, the reference ion weighting, $\chi^{\text{DDEC3}} = 3/14$ , was set to a theoretically derived value that achieves a balance for all materials.[14]

**2.3 The bifurcation or 'runaway charges' problem**

The 'runaway charges' problem was first noted for the DDEC/c2 method in the paper by Manz and Sholl introducing the DDEC3 method.[14] The DDEC3 method was introduced to correct this problem for dense materials by introducing reference ion conditioning and constraints preventing the tails of buried atoms from becoming too diffuse.[14] The DDEC3 method dramatically improved over DDEC/c2, but stopped short of a provably convex optimization functional.[14] Among the hundreds of materials studied to date, we have only noticed one system for which the 'runaway charges' problem still occurs for the DDEC3 method: the $H_2$ triplet state for a constrained bond length of 50 pm. The discovery of one system with this problem indicates other systems with this same problem probably exist. The two H atoms in $H_2$ are symmetrically equivalent in the CISD/aug-cc-pvqz wavefunction and electron distribution. As shown in Figure 1, during DDEC3 partitioning the NACs diverged from the initial values (HD charges) of $+3.3\times10^{-4}$ and $-3.3\times10^{-4}$ to final converged values of +0.50 and -0.50 after 37 iterations. This indicates the optimization landscape contains a bifurcation instability that leads to symmetry breaking. Only tiny initial differences in the input density grid files determine which of the two H atoms will head towards a NAC of +0.50 while the remaining one heads towards -0.50. The same type of bifurcation instability also occurs for the IH method in some dense solids. As shown in Figure 1, during IH partitioning the NACs of symmetry-equivalent atoms bifurcate from the initial values (HD charges) of $+4.8\times10^{-4}$ and $-4.8\times10^{-4}$ to +0.97 and -0.97 for the Mn crystal and from $+4.3\times10^{-3}$, $-5.4\times10^{-3}$, $-3.5\times10^{-3}$, and $+4.6\times10^{-3}$ to -0.36, -0.39, -0.38, and +1.1 for the Rh crystal before the VASP program gave up after 150 charge cycles. (This IH analysis was performed in VASP 5.3.5 using the PBE functional.) Again, tiny differences in the initial conditions determines which of the symmetry-equivalent atoms will head towards the large positive NACs and which will head towards the negative NACs.

To understand why previous DDEC and IH methods sometimes yield bifurcation, we now compute their optimization landscape curvature. The path action's curvature is given by[14]

$$d^2S = \sum_A \oint \delta\rho_A(\vec{r}_A)\left(\frac{\delta\rho_A(\vec{r}_A)}{\rho_A(\vec{r}_A)} - \frac{\delta w_A(r_A)}{w_A(r_A)} + \frac{\delta W(\vec{r})}{W(\vec{r})}\right)d^3\vec{r}_A \,. \quad (29)$$

"If the curvature is positive everywhere, i.e., $d^2S > 0$, then the action has only one stationary point, and this stationary point is its global minimum."[14] The solution lies within the search space satisfying constraint (8). Restricting $\{\rho_A(\vec{r}_A)\}$ to this valid search space ,

$$\sum_A \delta\rho_A(\vec{r}_A) = \sum_A \delta\rho(\vec{r}) = 0 \,. \quad (30)$$

Let $\bar{S}$ be the path action S restricted to paths lying inside this valid search space. Combining Eqs. (29) and (30) yields the optimization landscape curvature:

$$d^2\bar{S} = \sum_A \oint \delta\rho_A(\vec{r}_A)\left(\frac{\delta\rho_A(\vec{r}_A)}{\rho_A(\vec{r}_A)} - \frac{\delta w_A(r_A)}{w_A(r_A)}\right) d^3\vec{r}_A . \qquad (31)$$

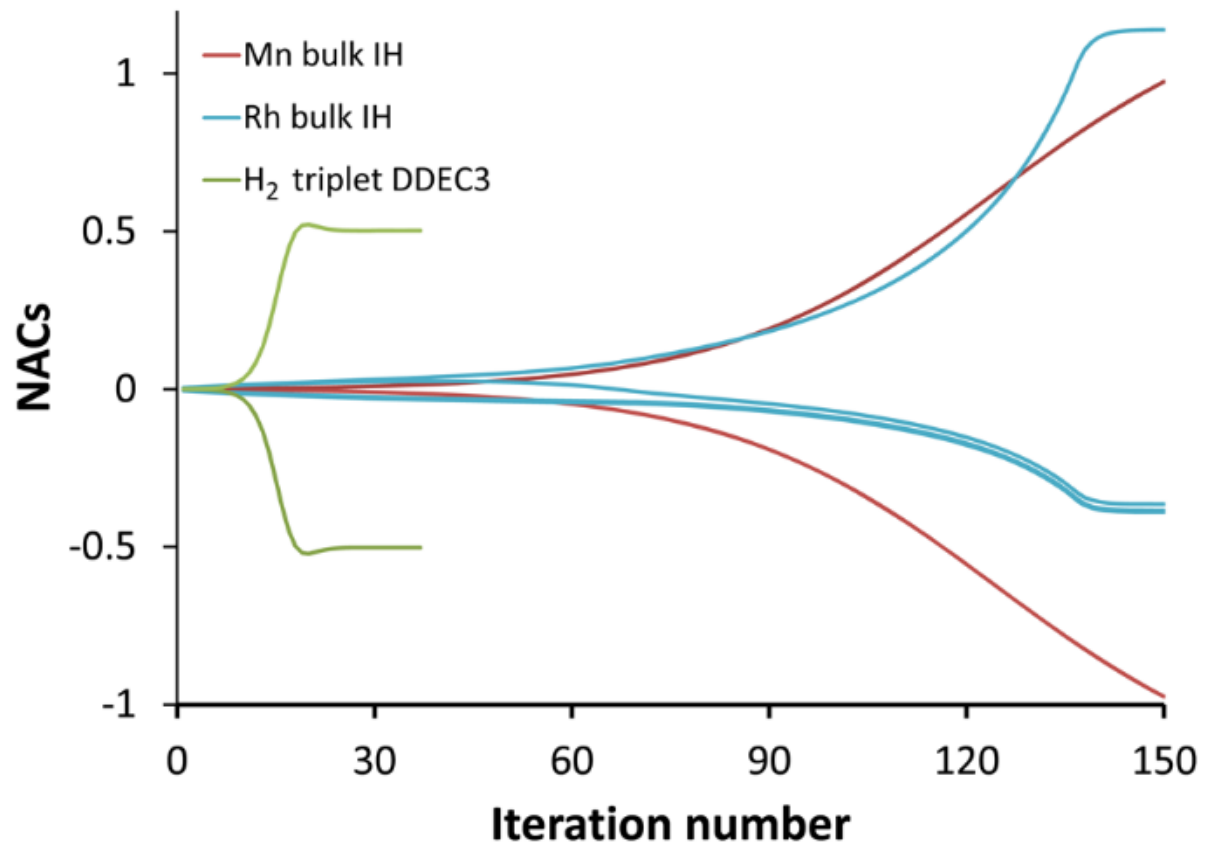

Figure 1. Bifurcation (i.e., spontaneous symmetry breaking) during DDEC3 and IH charge partitioning. This is called the 'runaway charges' problem.

Substituting $w_A^{model} = \left(\rho_A^{some_ref}(r_A, q_A^{ref})\right)^{\chi}\left(\rho_A^{avg}(r_A)\right)^{1-\chi}$ as a DDEC-style weighting factor yields the curvature

$$d^2\bar{S}^{model} = (1-\chi)d^2\bar{S}^{ISA} + \chi d^2\bar{S}^{ref} \qquad (32)$$

where

$$d^2\bar{S}^{ISA} = d^2F^{ISA} = \sum_A \oint \left(\frac{\left(\delta\rho_A(\vec{r}_A)\right)^2}{\rho_A(\vec{r}_A)} - \frac{\left(\delta\rho_A^{avg}(r_A)\right)^2}{\rho_A^{avg}(r_A)}\right) d^3\vec{r}_A \geq 0 \qquad (33)$$

$$d^2\bar{S}^{ref} = \sum_A \oint \left(\frac{\left(\delta\rho_A(\vec{r}_A)\right)^2}{\rho_A(\vec{r}_A)} + \frac{\delta\rho_A(\vec{r}_A)}{\rho_A^{some_ref}(r_A, q_A^{ref})} \frac{\partial\rho_A^{some_ref}(r_A, q_A^{ref})}{\partial q_A^{ref}} dq_A^{ref}\right) d^3\vec{r}_A . \qquad (34)$$

Bultinck et al. previously proved the ISA curvature (Eq. (33)) is non-negative.[66]

We now consider several special cases. Case 1: The ISA method, which corresponds to $\chi = 0$. This case gives $d^2\bar{S}^{ISA} \geq 0$ (i.e., positive semi-definite curvature) indicating a convex optimization landscape with a unique solution. However, the optimization landscape can be approximately flat ( $d^2\bar{S}^{ISA} \to 0$) in regions. Case 2: The non-iterative Hirshfeld method, which corresponds to $\chi = 1$ and $dq_A^{ref} = 0$. This case gives

$$d^2\bar{S}^{HD} = d^2F^{HD} = \sum_A \oint \left(\frac{\left(\delta\rho_A(\vec{r}_A)\right)^2}{\rho_A(\vec{r}_A)}\right) d^3\vec{r}_A > 0 \qquad (35)$$

for any $\left|\delta\rho_A(\vec{r}_A)\right| > 0$. This positive definite curvature indicates a convex optimization landscape with a unique solution. Because the curvature is positive definite, the optimization landscape is not approximately flat anywhere. Case 3: The IH method, which corresponds to $\chi = 1$ and

$$dq_A^{ref} = dq_A = -\oint \delta\rho_A(\vec{r}_A) d^3\vec{r}_A \,. \qquad (36)$$

Since $\{dq_A^{ref}\} \neq \{0\}$, the second term appearing in Eq. (34) is not necessarily non-negative. Consequently, we cannot guarantee that $d^2\bar{S}^{ref}$ is non-negative. From Eq. (34), the convexness or non-convexness of the IH method for a specific material depends on the particulars of the reference ion set used. Thus, under some conditions the IH method may have a negative optimization curvature. This yields the bifurcation or 'runaway charges' problem shown in Figure 1 and Table 2. Case 4: The DDEC/c2 method, which corresponds to $\chi = 1/10$ and $dq_A^{ref}$ as defined in Eq. (36). This case has similar characteristics to Case 3 and yields the bifurcation or 'runaway charges' problem for the same reason. Case 5: The DDEC3 method improved over the DDEC/c2 method by increasing the $dq_A^{ref}$ curvature term in Eq. (34),[14] but it stopped short of a provably convex optimization landscape. Thus, under rare circumstances (e.g., $H_2$ triplet with 50 pm constrained bond length in Figure 1) it leads to the bifurcation or 'runaway charges' problem.

In the following sections, we introduce a new charge partitioning method called DDEC6 that alleviates the 'runaway charges' problem. Examining Eq. (34), the iterative update of $q_A^{ref}$ by requiring the self-consistency condition $q_A^{ref} = q_A$ yields the possibility of negative optimization landscape curvature and hence bifurcation. The DDEC6 method replaces the self-consistency requirement $q_A^{ref} = q_A$ with a new strategy for computing $q_A^{ref}$. Within a magnified integration tolerance, the DDEC6 NACs were symmetrically equivalent: $\pm 2.1\times10^{-2}$ ($H_2$ triplet), $\pm 1.8\times10^{-5}$ (Mn solid), and $-2.8\times10^{-5}$, $+1.5\times10^{-4}$, $-7.4\times10^{-5}$, and $-4.7\times10^{-5}$ (Rh solid).

The DDEC6 method cannot lead to 'runaway charges' in any material, because it involves a prescribed sequence of exactly seven sequential charge partitioning steps. For purposes of illustration, assume the initial symmetry breaking present in the input grid files is some small positive amount $\varepsilon$. For purposes of illustration, let us further assume that during each subsequent charge partitioning step the amount of symmetry breaking is multiplied by some finite factor K. After X charge partitioning steps, the amount of symmetry breaking will thus be $K^X\varepsilon$. If X is small or if $|K| \leq 1$, the amount of symmetry breaking will be a modest multiple of $\varepsilon$. In the DDEC6 method, X = 7, so that even when $|K| > 1$ the magnitude of symmetry breaking can be contained as long as $\varepsilon$ is small. By making the input density grids more precise, the value of $\varepsilon$ and hence the final DDEC6 symmetry breaking ($\sim K^7\varepsilon$) can be made as small as desired. On the other hand, if X is large and $|K| > 1$, the symmetry breaking will be profoundly severe (i.e., 'runaway charges'). For the DDEC/c2, DDEC3, and IH methods, X may be arbitrarily large leading to 'runaway charges' when $|K| > 1$. Since X is arbitrarily large for these methods, the value of $K^X\varepsilon$ cannot be contained for $|K| > 1$ even if $\varepsilon$ is made a smaller positive number. Thus, improving the input density grid precision does not necessarily alleviate the 'runaway charges' problem for the DDEC/c2, DDEC3, and IH methods.

We also tested a second strategy that solves the bifurcation or 'runaway charges' problem by constructing the self-consistent convex charge partitioning functional:

$$F^{convex} = \sum_A \oint \rho_A(\vec{r}_A) \ln\left( \frac{\rho_A(\vec{r}_A)}{\left(\rho_A^{fixed_ref}(r_A)\right)^{\chi} \left(\rho_A^{avg}(r_A)\right)^{1-\chi}} \right) d^3\vec{r}_A - \int_U \Gamma(\vec{r})\Theta(\vec{r}) d^3\vec{r} - \sum_A \kappa_A N_A^{val} . \quad (37)$$

$\kappa_A$ is a Lagrange multiplier that enforces the constraint

$$N_A^{val} = \oint \rho_A(\vec{r}_A) - N_A^{core} \geq 0 . \quad (38)$$

$\left\{\rho_A^{fixed_ref}(r_A)\right\}$ is a set of fixed reference densities that is not updated during the self-consistent cycles. Minimizing $F^{convex}$

$$\frac{\delta F^{convex}}{\delta \rho_A(\vec{r}_A)} = \ln\left( \frac{\rho_A(\vec{r}_A)}{\left(\rho_A^{fixed_ref}(r_A)\right)^{\chi} \left(\rho_A^{avg}(r_A)\right)^{1-\chi}} \right) + \chi - \Gamma(\vec{r}) - \kappa_A \to 0 \quad (39)$$

gives the solution

$$w_A^{convex}(r_A) = e^{\kappa_A} \left(\rho_A^{fixed_ref}(r_A)\right)^{\chi} \left(\rho_A^{avg}(r_A)\right)^{1-\chi} . \quad (40)$$

The curvature is

$$d^2F^{convex} = \sum_A \oint \left( \frac{\left(\delta\rho_A(\vec{r}_A)\right)^2}{\rho_A(\vec{r}_A)} - (1-\chi) \frac{\left(\delta\rho_A^{avg}(r_A)\right)^2}{\rho_A^{avg}(r_A)} \right) d^3\vec{r}_A . \quad (41)$$

Inserting Eq. (33) into (41) yields

$$d^2F^{convex} \geq \chi \sum_A \oint \frac{\left(\delta\rho_A(\vec{r}_A)\right)^2}{\rho_A(\vec{r}_A)} d^3\vec{r}_A \quad (42)$$

which is positive definite if $\chi > 0$ . Thus, the functional is convex with a unique minimum. Through computational tests, we found nearly optimal results are obtained by setting $\chi^{convex} = 1/2$ . We computed $\left\{\rho_A^{fixed_ref}(r_A)\right\}$ by applying the upper and lower bound exponential decay constraints (see Section 2.7) while minimizing a distance measure between $\rho_A^{fixed_ref}(r_A)$ and $\rho_A^{cond}(r_A)\left\langle \rho(\vec{r})/\rho^{cond}(\vec{r}) \right\rangle_{r_A}$ . As shown by results in Section 2.9, the overall performance of this method is slightly inferior to the DDEC6 method.

We performed computational tests proving the bifurcation or ‘runaway charges’ problem is associated with non-unique minima on the optimization landscape. Table 2 summarizes computational results for the $H_2$ triplet system. For consistency, all four methods compared used the same density grids, same reference ion library (Section 3), and same integration method. (This integration method is explained in the ESI†.). The pair of numbers ($q_1$,$q_2$) denote the charges associated with the first and second H atoms, respectively. The initial guess for the IH and DDEC3 methods is the starting reference ion charge. For the DDEC3 method, the conditioned reference densities and the $w_A(r_A)$ were initialized to equal the starting reference ion densities. For the convex functional (Eq. (37)), $\rho_A^{avg}(r_A)$ was initialized to equal a reference ion density having the charge listed as the initial guess. The DDEC3 and IH methods exhibited pronounced bifurcation, with the converged NACs highly dependent on the initial guess. In contrast, the convex functional exhibited the unique solution (-0.0093,0.0093) independent of the initial guess. Because the DDEC6 NACs follow a fixed protocol of seven charge partitioning steps, they do not require any form of initial guess and converged to the unique solution (-0.0209,0.0209). These results prove the DDEC3 and IH methods do not have a convex optimization functional for some materials.

Table 2: Effect of initial guess on the converged NACs for $H_2$ triplet with a 50 pm constrained bond length (CISD/aug-cc-pvqz electron distribution). Because the DDEC3 and IH results depend on the initial guess, they do not have a convex optimization functional or a unique solution. Because the DDEC6 NACs follow a fixed protocol of seven charge partitioning steps, they do not require any form of initial guess and converged to the unique solution (-0.0209,0.0209). Also, the convex charge partitioning functional converges to a unique solution independent of the starting guess.

| **run #** | **initial guess** | **DDEC3 NACs** | **IH NACs** | **convex functional NACs** |
|---|---|---|---|---|
| 1 | (0,0) | (-0.5025,0.5025) | (-0.5978,0.5978) | (-0.0093,0.0093) |
| 2 | (0.5,-0.5) | (0.5014,-0.5014) | (0.5950,-0.5950) | (-0.0093,0.0093) |
| 3 | (-0.5,0.5) | (-0.5025,0.5025) | (-0.5978,0.5978) | (-0.0093,0.0093) |

### 2.4 Conditioning steps and the equivalent reference ion weighting $\chi_{\text{equiv}}$

The previous section demonstrated that using a fixed number of charge partitioning steps (e.g., X = 7) can alleviate the 'runaway charges' problem. In this section, we show how to construct schemes having a fixed number of charge partitioning steps that achieve the DDEC goal of simultaneously optimizing $\{\rho_A(\vec{r}_A)\}$ to resemble the reference ion densities $\{\bar{\rho}_A^{\text{ref}}(r_A, q_A^{\text{ref}})\}$ and the spherical average densities $\{\rho_A^{\text{avg}}(r_A)\}$.

First, we review the meaning of 'conditioning'. Conditioning refers to the process in which some set of weighting functions—let us call them by the generic name $\{\rho_A^{\text{generic}}(r_A)\}$—are used in a stockholder partitioning to obtain a new set of spherically averaged weighting factors, $\{\rho_A^{\text{conditioned}}(r_A)\}$:

$$\rho_A^{\text{conditioned}}(r_A) = \rho_A^{\text{generic}}(r_A)\left\langle \rho(\vec{r})/\rho^{\text{generic}}(\vec{r})\right\rangle_{r_A} \qquad (43)$$

$$\rho^{\text{generic}}(\vec{r}) = \sum_{A,L} \rho_A^{\text{generic}}(r_A) \qquad (44)$$

where $\langle\ \rangle_{r_A}$ denotes spherical averaging. We could perform another conditioning step by reinserting $\rho_A^{\text{generic}}(r_A) = \rho_A^{\text{conditioned}}(r_A)$ into the right-hand sides of Eq. (43)–(44). This process could be repeated until some desired number, c, of conditioning steps have been performed. Starting with the reference ions, $\bar{\rho}_A^{\text{ref}}(r_A)$, a single conditioning step produces

$$Y_A^{\text{avg}}(r_A) = \bar{\rho}_A^{\text{ref}}(r_A)\left\langle \rho(\vec{r})/\bar{\rho}^{\text{ref}}(\vec{r})\right\rangle_{r_A}. \qquad (45)$$

A second conditioning step produces

$$\rho_A^{\text{double_cond}}(r_A) = \bar{\rho}_A^{\text{ref}}(r_A)\left\langle \rho(\vec{r})/\bar{\rho}^{\text{ref}}(\vec{r})\right\rangle_{r_A}\left\langle \rho(\vec{r})/Y^{\text{avg}}(\vec{r})\right\rangle_{r_A}. \qquad (46)$$

After c conditioning steps,

$$\rho_A^{some_ref}(r_A) = \bar{\rho}_A^{ref}(r_A)\left(\overline{\left\langle \rho(\vec{r})/\rho^{generic}(\vec{r})\right\rangle_{r_A}}\right)^c \qquad (47)$$

where $\overline{\left\langle \rho(\vec{r})/\rho^{generic}(\vec{r})\right\rangle_{r_A}}$ is a geometric average of all of the $\left\langle \rho(\vec{r})/\rho^{generic}(\vec{r})\right\rangle_{r_A}$ style terms.

All of the previous DDEC schemes had

$$w_A(r_A) \approx \left(\rho_A^{some_ref}(r_A)\right)^{\chi}\left(\rho_A^{avg}(r_A)\right)^{1-\chi} . \qquad (48)$$

Inserting in Eq. (12) and rearranging yields

$$\rho_A^{avg}(r_A) \approx \rho_A^{some_ref}(r_A)\left\langle \rho(\vec{r})/W(\vec{r})\right\rangle_{r_A}^{1/\chi} . \qquad (49)$$

In general, $\left\{\rho_A^{some_ref}(r_A)\right\}$ may be produced by conditioning the reference ions $\bar{\rho}_A^{ref}(r_A)$ a total of c times. Substituting Eq. (47) into (49) yields

$$\rho_A^{avg}(r_A) \approx \bar{\rho}_A^{ref}(r_A)\overline{\left\langle \rho(\vec{r})/\rho^{generic}(\vec{r})\right\rangle_{r_A}}^{c+1/\chi} \qquad (50)$$

where the overbar denotes the (weighted) geometric average of $\left\langle \rho(\vec{r})/\rho^{generic}(\vec{r})\right\rangle_{r_A}$ style terms (such as those appearing in Eqs. (45), (46), and (49)). Comparing Eqs. (49) and (50), the equivalent amount of reference ion weighting on a conditioning adjusted basis is therefore

$$\frac{1}{\chi_{equiv}} = c + \frac{1}{\chi} . \qquad (51)$$

We now discuss specific examples. Case 1: DDEC/c1 and DDEC/c2 methods used no reference ion conditioning (i.e., c = 0) with $\chi = 1/10$ to give $\chi_{equiv}^{DDEC2} = 1/10$. Case 2: The HD and IH methods use no reference ion conditioning (i.e., c = 0) with $\chi = 1$ to give $\chi_{equiv}^{HD/IH} = 1$. Case 3: The DDEC3 method uses one conditioning step (i.e., c = 1) with $\chi = 3/14$ to give $\chi_{equiv}^{DDEC3} = 3/17$. Case 4: The ISA method has two completely equivalent representations: either an infinite number of conditioning steps (i.e., $c = \infty$ ) or $\chi = 0$ to yield $\chi_{equiv}^{ISA} = 0$. (Exactly the same $\left\{\rho_A^{ISA}(\vec{r}_A)\right\}$ are obtained in either representation. In practice, the two representations are not distinguishable and have the same computational algorithm.)

When developing the DDEC3 method, Manz and Sholl showed that one conditioning step (i.e., c = 1) combined with $\chi = 3/14$ yields an appropriate balance between reference ion weighting and spherical averaging independent of the material.[14] A scheme containing a fixed number of charge partitioning steps can be constructed to yield a similar $\chi_{equiv}$ value. Specifically, if $\left\{\bar{\rho}_A^{ref}(r_A)\right\}$ are conditioned four times to yield $\left\{w_A^{DDEC6}(r_A)\right\}$ (i.e., c = 4 and $\chi = 1$) and hence five times to yield the final DDEC6 $\left\{\rho_A^{avg}(r_A)\right\}$, this corresponds to $\chi_{equiv}^{DDEC6} = 1/5$. $\chi_{equiv}^{DDEC3} = 3/17$ lies between this value of 1/5 and the value of 1/6 that would be obtained from a total of six conditioning steps. We chose five conditioning steps, because this is more conservative. Increasing the number of conditioning steps leads to a slight decrease in the atomic multipoles at the expense of losing some of the chemical transferability. Yet even with five conditioning steps, we achieved DDEC6 atomic multipoles approximately 2–5% lower in magnitude on average (see Section 2.9) than the DDEC3 atomic multipoles. As explained in subsequent sections, this is due to using an accurate fixed reference ion charge and a weighted spherical average during some of the conditioning steps.

We now return to the bifurcation or 'runaway charges' problem demonstrated in the previous section. The first possible solution is to use a fixed number of charge partitioning steps. In other words, to use c = 4 and $\chi = 1$ for a total of five conditioning steps: $\chi_{\text{equiv}}^{\text{DDEC6}} = 1/5$. The second possible solution is to use $0 < \chi < 1$ where $\{\rho_A(\vec{r}_A)\}$ are recovered by minimizing a provably convex optimization functional or path action. For example, $F^{\text{convex}}$ shown in Eq. (37). Two conditioning steps (i.e., c = 2) are involved in computing $\{\rho_A^{\text{fixed_ref}}(r_A)\}$. Using $\chi^{\text{convex}} = 1/2$ yields $\chi_{\text{equiv}}^{\text{convex}} = 1/4$. We programmed both of these strategies and tested them for all systems described in this article. (Our computational method for computing the convex functional NACs is described in the ESI†.) Both strategies used identical $\{q_A^{\text{ref}}\}$ values. Both strategies alleviated the bifurcation or 'runaway charges' problem. While both approaches yielded reasonable results, the first approach proved superior to the second. The first approach allowed using a weighted spherical average in place of a simple spherical average during some of the conditioning steps. The second approach required a simple spherical average during the self-consistency iterations to prove convexness of the path action (see Eq. (33). The first approach was more computationally efficient and converged in 7 charge cycles compared to 11–14 charge cycles for the second approach. As shown in Section 2.9, the first approach yielded slightly lower atomic multipole moments on average, reproduced the electrostatic potential slightly more accurately on average, and described electron transfer trends in dense solids slightly more accurately on average. Therefore, in the end we decided to go with the fixed number of charge partitioning steps (approach 1: DDEC6) as opposed to minimizing $F^{\text{convex}}$ (approach 2).

Finally, we performed additional computational tests proving the equivalence relations described in this section. Specifically, we also analyzed nearly all systems in this article using c = 1 and $\chi = 1/3$ (yielding $\chi_{\text{equiv}} = 1/4$) and obtained similar (but not strictly identical) results to the c = 2 and $\chi = 1/2$ case discussed above. The computational cost was increased to ~24 charge cycles for convergence.

**2.5 Determining the DDEC6 reference ion charge value (charge cycles 1 and 2)**

The reference ion charge is the most significant difference between the DDEC6 and DDEC3 methods. In the DDEC3 method, the reference ion charge is the same as the AIM NAC: $q_A^{\text{ref}} = q_A$. While setting $q_A^{\text{ref}} = q_A$ has some theoretical appeal, it also comes with an important disadvantage. In dense materials, the diffuse nature of anions can cause the number of electrons assigned to an anion to be much greater than the number of electrons in the volume dominated by the anion. This causes several related problems. First, NACs assigned with $q_A^{\text{ref}} = q_A$ may fail to assign core electrons to the proper atom in some materials (see Section 5.1). Second, they do not properly describe electron transfer trends in some materials (see Section 5.2.1). Third, the correlation between NACs and core electron binding energy shifts will be weakened due to the overly delocalized assignment of electrons to the anions (see Ti-containing containing compounds in Section 5.3.2). Fourth, the accuracy of reproducing the electrostatic potential may be slightly degraded in some materials for which the anion charges are overestimated in magnitude (see Section 5.5.2).

There are two competing philosophies of electrons belonging to an atom: localized and stockholder atomic charges. For localized atomic charges, electrons in the volume dominated by an atom are assigned almost entirely to that atom. Non-overlapping compartments, such as those encountered in Bader's QCT and Voronoi cell partitioning, are the most extreme limit of localized charge partitioning.[27, 36, 37, 41, 42] Localized NACs (e.g., Bader NACs) convey useful information about charge transfer trends (see Section 5.2) and core-electron binding energy shifts (see Section 5.3.2), but they are not well-suited to reproducing the electrostatic potential surrounding a material (see Sections 5.2.3, 5.3.1). In stockholder partitioning schemes, the density assigned to each atom is proportional to the density of some spherical proto-atom, and this leads to overlapping atomic electron distributions.[28] DDEC3 NACs, which do not incorporate localized atomic charge information, are usually more accurate than Bader NACs at reproducing $V(\vec{r})$ but suffer the problems mentioned in the previous paragraph.

To achieve the best of both worlds, the DDEC6 method uses a fixed reference ion charge consisting of a weighted average of localized and stockholder charges. The ratio of localized to stockholder atomic charges used to compute these fixed reference ion charges was decided through a scientific engineering design approach in which we tested dozens of alternatives. (See ESI† for a summary of alternatives explored.) Here, we present the DDEC6 steps and equations in a way that is independent of the integration grids employed. Specific integration procedures are described in the ESI†.

Specifically, the DDEC6 reference ion charge value, $q_A^{ref}$, is set using the following scheme:

$$q_A^{ref} = q_A^{2,ref} \qquad (52)$$

$$q_A^{s,ref} = \frac{1}{3} q_A^{s,Stock} + \frac{2}{3} q_A^{s,Loc}, \; s = \{1, 2\} \qquad (53)$$

$$q_A^{i,Stock/Loc} = z_A - N_A^{i,Stock/Loc} \qquad (54)$$

$$N_A^{i,Stock/Loc} = \oint \frac{w_A^{i,Stock/Loc}(r_A)}{W^{i,Stock/Loc}(\vec{r})} \rho(\vec{r}) d^3\vec{r} \qquad (55)$$

$$W^{i,Stock/Loc}(\vec{r}) = \sum_{B,L} w_B^{i,Stock/Loc}(r_B) \qquad (56)$$

$$w_A^{1,Stock}(r_A) = \bar{\rho}_A^{ref}(r_A, 0) \qquad (57)$$

$$w_A^{1,Loc}(r_A) = \left(\bar{\rho}_A^{ref}(r_A, 0)\right)^4 \qquad (58)$$

$$w_A^{2,Stock}(r_A) = \bar{\rho}_A^{ref}\left(r_A, q_A^{1,ref}\right) \qquad (59)$$

$$w_A^{2,Loc}(r_A) = \left(\bar{\rho}_A^{ref}\left(r_A, q_A^{1,ref}\right)\right)^4. \qquad (60)$$

$N_A^{1,Loc}$ and $N_A^{2,Loc}$ are measures of the number of electrons in the volume dominated by each atom. $N_A^{1,Loc}$ is computed using the neutral atom reference densities (Eq. (58)). $N_A^{2,Loc}$ is computed using charged reference ions (Eq. (60)). The fourth power appearing in Eqs. (58) and (60) makes $w_A^{s,Loc}(r_A)$ $\{s = 1, 2\}$ change continuously while also becoming negligible in the volume of space for which $\bar{\rho}_A^{ref}(r_A) < \bar{\rho}_B^{ref}(r_B)$. When using an exponent of 4, the ratio $\bar{\rho}_A^{ref}(r_A) / \bar{\rho}_B^{ref}(r_B) = 2$ corresponds to assigning 16 times higher density to atom A compared to atom B at this grid point. This provides an appropriate balance between transition sharpness and smoothness. Using an exponent less (more) than 4 would cause the density transition between atoms to be more gradual (sharper). An arbitrarily high exponent would correspond to the limit of non-overlapping atoms. Eq. (53) updates the reference ion charges by adding 2/3 of this localized charge to 1/3 of the stockholder charge based on minimizing the information distance to the (prior) reference ion densities. The final reference ion charge (Eq. (52)) determined in this manner is a compromise between counting electrons in the volume dominated by each atom and counting electrons in proportion to the (prior) reference ion densities.

The superscript numerals 1 and 2 refer to the charge cycle in which that quantity is computed. The weighing factors for the first charge cycle are computed using neutral reference atoms (Eqs. (57)–(58)). The weighting factors for the second charge cycle are computed using Eqs. (59)–(60) based on the results of the first charge cycle. Each of the first two charge cycles first computes $W^{i,Stock}(\vec{r})$ and $W^{i,Loc}(\vec{r})$ by a loop over atoms and grid points to compute the sum in Eq. (56). Each of the first two charge cycles then performs another loop over atoms and grid points to compute $N_A^{i,Stock}$ and $N_A^{i,Loc}$ via Eq. (55). The stockholder, localized, and updated reference charges are then computed via Eqs. (53)–(54). The first charge cycle yields the HD NACs: $q_A^{Hirshfeld} = q_A^{1,Stockholder}$ . (Even though the CM5 NACs are not used in DDEC6 charge partitioning, they were computed at this stage using the HD NACs and the CM5 definition of Marenich et al.[25]) The second charge cycle yields the DDEC6 reference ion charges (Eq. (52)).

**2.6 Computing the conditioned reference ion density (charge cycle 3)**

The next step (i.e., third charge cycle) is to compute the conditioned reference ion densities, $\rho_A^{cond}(r_A)$. This conditioning matches the reference ion densities to the specific material of interest. First, a loop of over grid points and atoms is performed to accumulate the following sum

$$\bar{\rho}^{ref}(\vec{r}) = \sum_{B,L} \bar{\rho}_B^{ref}\left(r_B, q_B^{ref}\right). \qquad (61)$$

Then another loop over grid points and atoms is performed to compute the following spherical average:

$$Y_A^{avg}(r_A) = \bar{\rho}_A^{ref}\left(r_A, q_A^{ref}\right)\left\langle \rho(\vec{r}) / \bar{\rho}^{ref}(\vec{r}) \right\rangle_{r_A}. \qquad (62)$$

<u>Constraints applied to each conditioned reference ion density</u>: In the DDEC6 method, each conditioned reference density, $\rho_A^{cond}(r_A)$, is constrained to monotonically decrease with increasing $r_A$

$$\phi^I(r_A) = \frac{d\rho_A^{cond}(r_A)}{dr_A} \le 0 \qquad (63)$$

and to integrate to the number of electrons in the reference ion:

$$\varphi_A^I = \int_0^{r_{cutoff}} \rho_A^{cond}(r_A) 4\pi (r_A)^2 dr_A - z_A + q_A^{ref} \to 0 \qquad (64)$$

These constraints were introduced for theoretical appeal to ensure expected behavior. In tests we performed, these constraints had only a small effect on the NACs. They are not present in the DDEC3 method.

$\{\rho_A^{cond}(r_A)\}$ is found by minimizing the functional

$$h^{I}\left(\rho_{A}^{cond}\left(r_{A}\right)\right)=\int_{0}^{\infty}\left[\frac{\left(\rho_{A}^{cond}\left(r_{A}\right)-Y_{A}^{avg}\left(r_{A}\right)\right)^{2}}{2\sqrt{Y_{A}^{avg}\left(r_{A}\right)}}+\Gamma_{A}^{I}\left(r_{A}\right)\phi^{I}\left(r_{A}\right)\right]4\pi\left(r_{A}\right)^{2}dr_{A}-\Phi_{A}^{I}\varphi_{A}^{I} \qquad (65)$$

where $\Gamma_{A}^{I}\left(r_{A}\right)$ and $\Phi_{A}^{I}$ are Lagrange multipliers enforcing constraints (63) and (64), respectively. The minimum of $h^I$ is found by setting

$$\frac{\delta h^{I}}{\delta\rho_{A}^{cond}\left(r_{A}\right)}=0\,. \qquad (66)$$

Inserting eq. (65) into (66) and using integration by parts to simplify gives the solution

$$\rho_{A}^{cond}\left(r_{A}\right)=Y_{A}^{avg}\left(r_{A}\right)+\sqrt{Y_{A}^{avg}\left(r_{A}\right)}\left(\Phi_{A}^{I}+\frac{d\Gamma_{A}^{I}\left(r_{A}\right)}{dr_{A}}+\frac{2\Gamma_{A}^{I}\left(r_{A}\right)}{r_{A}}\right). \qquad (67)$$

Because

$$\frac{\delta^{2}h^{I}}{\delta\rho_{A}^{cond}\left(r_{A}\right)\partial\rho_{A}^{cond}\left(r_{A}'\right)}=\frac{4\pi\left(r_{A}\right)^{2}\delta^{dirac}\left(r_{A}-r_{A}'\right)}{\sqrt{Y_{A}^{avg}\left(r_{A}\right)}}\geq 0 \qquad (68)$$

it directly follows that $h^{I}\left(\rho_{A}^{cond}\left(r_{A}\right)\right)$ is a convex functional. Therefore, $\left\{\rho_{A}^{cond}\left(r_{A}\right)\right\}$ is uniquely determined for a given $\left\{Y_{A}^{avg}\left(r_{A}\right)\right\}$ input.

A robust and rapidly converging iterative algorithm was used to compute $\left\{\rho_{A}^{cond}\left(r_{A}\right)\right\}$. In each iteration, an estimate of $\Phi_{A}^{I}$ is used to compute the estimate

$$\rho_{A}^{cond}\left(r_{A}\right)=Y_{A}^{avg}\left(r_{A}\right)+\Phi_{A}^{I}\sqrt{Y_{A}^{avg}\left(r_{A}\right)}\,. \qquad (69)$$

Then, constraint (63) is enforced by recursively setting

$$\rho_{A}^{cond}\left(r_{A}\right)=\min\left(\rho_{A}^{cond}\left(r_{A}\right),\rho_{A}^{cond}\left(r_{A}-\Delta r_{A}\right)\right) \qquad (70)$$

beginning with the second radial shell and continuing outward until the last radial shell. To complete one iteration, $\varphi_{A}^{I}$ is computed via Eq. (64). A scheme is required to generate a new $\Phi_{A}^{I}$ estimate for the next iteration. Since $\varphi_{A}^{I}$ is a monotonically increasing function of $\Phi_{A}^{I}$, $\Phi_{A}^{I}$ should be increased (decreased) when $\varphi_{A}^{I}<0$ ($\varphi_{A}^{I}>0$). Convergence is achieved when $\left|\varphi_{A}^{I}\right|$ is less than some zero tolerance (e.g., $10^{-10}$ electrons).

The update scheme we used contained the following sequence of steps. In the first iteration, we set $\Phi_{A}^{I\,(1)}=0$. Then we set

$$\Phi_{A}^{I\,(i+1)}=2\Phi_{A}^{I\,(i)}-\frac{\varphi_{A}^{I\,(i)}}{\int_{0}^{r_{cutoff}}\sqrt{Y_{A}^{avg}\left(r_{A}\right)}4\pi\left(r_{A}\right)^{2}dr_{A}} \qquad (71)$$

where $\varphi_{A}^{I\,(i)}$ is the result of using $\Phi_{A}^{I\,(i)}$. Because $\varphi_{A}^{I\,(1)}<0$ if constraint (63) is binding, repetitively applying Eq. (71) increases $\Phi_{A}^{I\,(i+1)}$ until $\varphi_{A}^{I\,(i)}>0$. Since $\Phi_{A}^{I\,(i+1)}$ more than doubles between successive iterations, the number of iterations required to reach this upper bound is small. At this point, we stop using

Eq. (71) and set $\Phi_A^{I\,(\text{upper})}$ as the smallest known value yielding $\varphi_A^{I\,(\text{upper})} > 0$. We also set $\Phi_A^{I\,(\text{lower})}$ as the largest known value yielding $\varphi_A^{I\,(\text{lower})} < 0$. The remainder of the steps are simply aimed at squeezing the upper and lower bounds as quickly as possible. First, we try the midpoint $\Phi_A^{I\,(\text{mid})} = \left(\Phi_A^{I\,(\text{upper})} + \Phi_A^{I\,(\text{lower})}\right)/2$ to get $\varphi_A^{I\,(\text{mid})}$. Then, we fit the triple of points $\left(\Phi_A^{I\,(\text{lower})}, \varphi_A^{I\,(\text{lower})}\right)$, $\left(\Phi_A^{I\,(\text{mid})}, \varphi_A^{I\,(\text{mid})}\right)$, $\left(\Phi_A^{I\,(\text{upper})}, \varphi_A^{I\,(\text{upper})}\right)$ to a parabola. Then, we set $\Phi_A^{I\,(\text{parabolic})}$ equal to the root of the parabola where $\varphi_A^{I\,(\text{parabolic})}$ is predicted to be zero. After computing the actual value of $\varphi_A^{I\,(\text{parabolic})}$ via Eq. (64), we identify which points among $\varphi_A^{I\,(\text{lower})}$, $\varphi_A^{I\,(\text{mid})}$, $\varphi_A^{I\,(\text{upper})}$, and $\varphi_A^{I\,(\text{parabolic})}$ are the closest to zero from above (i.e., $\varphi_A^{I\,(\text{above})}$) and below (i.e., $\varphi_A^{I\,(\text{below})}$). A linear interpolation between these two closest-to-zero points is performed to identify $\Phi_A^{I\,(\text{corrector})}$ as the point where $\varphi_A^{I\,(\text{corrector})}$ is predicted to be: (a) $-\varphi_A^{I\,(\text{above})}$ if $3\left|\varphi_A^{I\,(\text{above})}\right| < \left|\varphi_A^{I\,(\text{below})}\right|$, (b) $-\varphi_A^{I\,(\text{below})}$ if $3\left|\varphi_A^{I\,(\text{below})}\right| < \left|\varphi_A^{I\,(\text{above})}\right|$, and (c) $\left(\varphi_A^{I\,(\text{above})} + \varphi_A^{I\,(\text{below})}\right)/2$ otherwise. This procedure places $\varphi_A^{I} = 0$ approximately half-way between $\varphi_A^{I\,(\text{corrector})}$ and $\varphi_A^{I\,(\text{above})}$ or $\varphi_A^{I\,(\text{below})}$ (whichever is smaller in magnitude), subject to the constraint that the resulting interval size is no more than $\left(\varphi_A^{I\,(\text{above})} + \varphi_A^{I\,(\text{below})}\right)/2$. At the close of this iteration, we identify the new lower (upper) bound as the largest (smallest) $\Phi_A^{I}$ among $\Phi_A^{I\,(\text{lower})}$, $\Phi_A^{I\,(\text{mid})}$, $\Phi_A^{I\,(\text{upper})}$, $\Phi_A^{I\,(\text{parabolic})}$, and $\Phi_A^{I\,(\text{corrector})}$ yielding a corresponding $\varphi_A^{I} < 0$ ($\varphi_A^{I} > 0$). Having refined the lower and upper bounds, we repeat the bisection, parabolic fitting, and linear interpolation steps in the next iteration to reach a tighter yet lower and upper bound. This process is repeated until convergence. In practice, we found this process converges magnificently, with one or two cycles of parabolic fitting and linear interpolation sufficient to achieve a precision of $10^{-10}$ electrons. Because this algorithm cuts the size of the search domain by better than half in each iteration, it is mathematically guaranteed to always converge in a few iterations.

After computing $\left\{\rho_A^{\text{cond}}\left(r_A\right)\right\}$, a loop over grid points and atoms is performed to compute

$$\rho^{\text{cond}}\left(\vec{r}\right) = \sum_{B,L} \rho_B^{\text{cond}}\left(r_B\right). \qquad (72)$$

Then another loop over grid points and atoms is performed to compute

$$\tau_A\left(r_A\right) = \left\langle \frac{\rho_A^{\text{cond}}\left(r_A\right)}{\sqrt{\rho^{\text{cond}}\left(\vec{r}\right)}} \right\rangle_{r_A} \left(\left\langle \sqrt{\rho^{\text{cond}}\left(\vec{r}\right)} \right\rangle_{r_A}\right)^{-1}. \qquad (73)$$

### 2.7 Updating $\left\{w_A\left(r_A\right)\right\}$ (charge partitioning steps 4 to 7)

The fourth (i.e. i = 4) charge cycle uses:

$$w_A\left(r_A\right) = \rho_A^{\text{cond}}\left(r_A\right). \quad (74)$$

The logical variable update_kappa is initialized to FALSE, and the integer completed_steps is initialized to 3. The fourth and later charge cycles use the following sequence of steps:

1. In the first loop over grid points and atoms, the sum in Eq. (13) is computed at each grid point.
2. In the second loop over grid points and atoms, the following quantities are computed:

$$\rho_A\left(\vec{r}_A\right) = w_A\left(r_A\right)\rho\left(\vec{r}\right)/W\left(\vec{r}\right) \qquad (75)$$

$$\rho_A^{avg}(r_A) = \langle \rho_A(\vec{r}_A) \rangle_{r_A} \quad (76)$$

$$\theta(r_A) = \left\langle \left(1 - \frac{w_A(r_A)}{W(\vec{r})}\right) \rho_A(\vec{r}_A) \right\rangle_{r_A} \quad (77)$$

$$\left\langle \frac{w_A(r_A)}{W(\vec{r})} \right\rangle_{r_A} \quad (78)$$

$$N_A = \oint \rho_A(\vec{r}_A) d^3\vec{r} \quad (79)$$

All of these quantities except $\{\rho_A(\vec{r}_A)\}$ are stored.

**3.** A weighted spherical average density, $\rho_A^{wavg}(r_A)$, is then computed:

$$\rho_A^{wavg}(r_A) = \frac{\theta(r_A) + \frac{\rho_A^{avg}(r_A)}{5} \left\langle \frac{w_A(r_A)}{W(\vec{r})} \right\rangle_{r_A}}{1 - \frac{4}{5} \left\langle \frac{w_A(r_A)}{W(\vec{r})} \right\rangle_{r_A}} \quad (80)$$

Eq. (80) has the form of a weighted spherical average density

$$\rho_A^{wavg}(r_A) = \frac{\langle \omega_A(\vec{r}_A) \rho_A(\vec{r}_A) \rangle_{r_A}}{\langle \omega_A(\vec{r}_A) \rangle_{r_A}}. \quad (81)$$

where $\rho_A(\vec{r}_A)$ is weighted at each grid point proportional to

$$\omega_A(\vec{r}_A) = \left(1 - \frac{w_A(r_A)}{W(\vec{r})}\right) + \frac{1}{5} \left\langle \frac{w_A(r_A)}{W(\vec{r})} \right\rangle_{r_A}. \quad (82)$$

Examining eq. (82), the relative weight assigned to each point is bounded by

$$0.1 \lesssim \omega_A(\vec{r}_A) < 1.2. \quad (83)$$

The lower bound of ~0.1 occurs, because it is not possible to have $\langle w_A(r_A)/W(\vec{r}) \rangle_{r_A} = 0$ if $w_A(r_A)/W(\vec{r}) \approx 1$ for any grid point on the same radial shell. For a specific $r_A$, positions with larger $w_A(r_A)/W(\vec{r})$ receive smaller $\omega_A(\vec{r}_A)$, and positions with smaller $w_A(r_A)/W(\vec{r})$ receive larger $\omega_A(\vec{r}_A)$. Thus, $\rho_A^{wavg}(r_A)$ weights portions of $\rho_A(\vec{r}_A)$ that overlap other atoms more heavily than those portions that do not overlap other atoms. At this stage, we also compute:

$$u_A = \frac{\partial N_A}{\partial \kappa_A} = \int_0^{r_{cutoff}} 4\pi (r_A)^2 \theta(r_A) dr_A \quad (84)$$

**4.** On the fourth charge cycle, update_kappa is not altered. On the fifth and subsequent charge cycles, update_kappa is set to TRUE if $N_A^{val} < -10^{-5}$ electrons for any atom. If update_kappa is TRUE and the $N_A$ and $\{\rho_A^{wavg}(r_A)\}$ changes between successive charge cycles were less than $10^{-5}$e and $10^{-5}$e/bohr$^3$, respectively, for each atom two consecutive times in a row then update_kappa is reset to FALSE and $\{\kappa_A\}$ is reset to $\{0\}$. Otherwise, the current value of update_kappa is not altered. On charge cycles *changing* update_kappa *from FALSE to TRUE*, the the current value of $\{w_A(r_A)\}$ is placed into $\{w_A^{fixed}(r_A)\}$.

**5.** If the value of update_kappa is FALSE, then completed_steps is incremented by +1. If completed_steps = 7, the iterative charge partitioning cycles are finished and exit at this point.

**6a.** If the value of update_kappa is FALSE, then $\{w_A(r_A)\}$ is updated to impose the constraints

$$\frac{-2.5\,\text{bohr}^{-1}}{\left(1-\left[\tau_A(r_A)\right]^2\right)} \le \frac{d\left(\ln\left(w_A(r_A)\right)\right)}{dr_A} \le \left(-1.75\,\text{bohr}^{-1}\right)\left(1-\left[\tau_A(r_A)\right]^2\right) \quad (85)$$

which are illustrated in Figure 2.

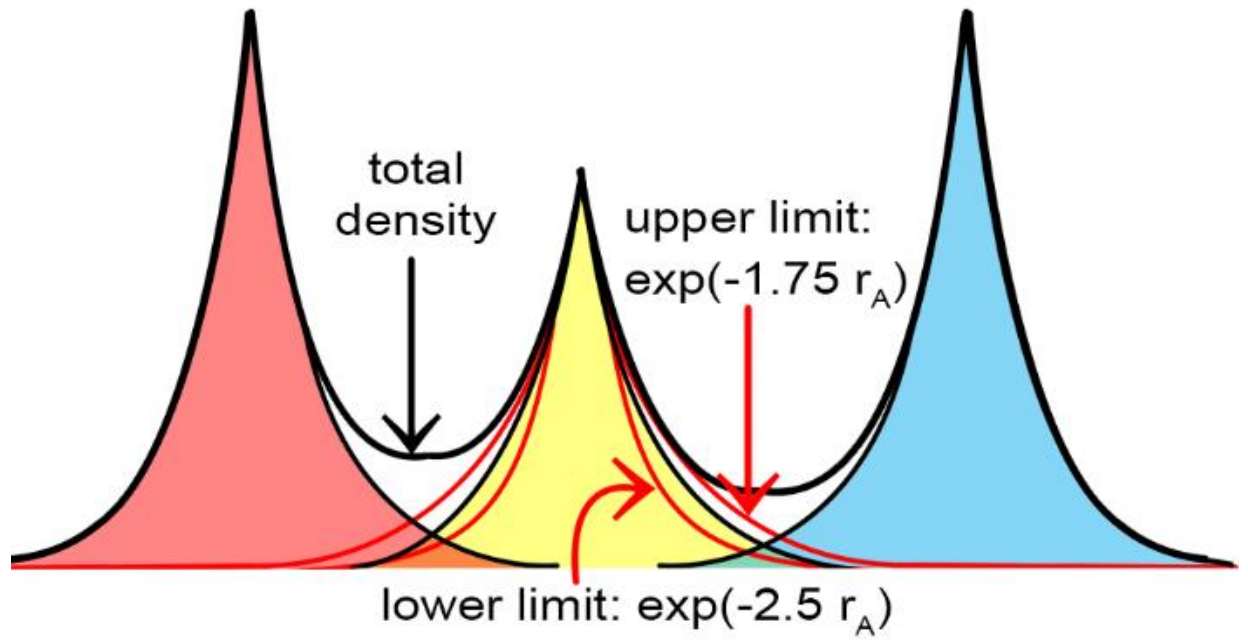

Figure 2. Illustration of exponential decay constraints applied to $w_A(r_A)$ in the buried atom tails. The red curves are not drawn to scale.

i) <u>Constraint preventing buried tails from becoming too diffuse</u>: Analogous to the DDEC3 method,[14] $G_A(r_A)$ is computed to make sure the tails of buried atoms do not become too diffuse. In the DDEC6 method, we set

$$\sigma_A(r_A) = \rho_A^{wavg}(r_A) \quad (86)$$

instead of the DDEC3 expression for $\sigma_A(r_A)$. The constraints

$$\phi^{II}(r_A) = \frac{dG_A(r_A)}{dr_A} + \eta_A^{lower}(r_A)G_A(r_A) \le 0 \quad (87)$$

$$\eta_A^{lower}(r_A) = \left(1.75\,\text{bohr}^{-1}\right)\left(1-\left(\tau_A(r_A)\right)^2\right) \quad (88)$$

$$\varphi_A^{II} = \int_0^{r_{cutoff}} \left(G_A(r_A)-\sigma_A(r_A)\right)4\pi(r_A)^2\,dr_A \to 0 \quad (89)$$

are imposed by minimizing the following optimization functional

$$h^{II}\left(G_A(r_A)\right) = \int_0^{\infty}\left[\frac{\left(G_A(r_A)-\sigma_A(r_A)\right)^2}{2\sqrt{\sigma_A(r_A)}} + \Gamma_A^{II}(r_A)\phi^{II}(r_A)\right]4\pi(r_A)^2\,dr_A - \Phi_A^{II}\varphi_A^{II} \quad (90)$$

where $\Gamma_A^{II}(r_A)$ and $\Phi_A^{II}$ are Lagrange multipliers enforcing constraints (87) and (89), respectively.[14] $h^{II}\left(G_A(r_A)\right)$ is a convex functional with the unique minimum:[14]

$$G_A(r_A) = \sigma_A(r_A) + \sqrt{\sigma_A(r_A)}\left(\Phi_A^{II} + \frac{d\Gamma_A^{II}(r_A)}{dr_A} + \frac{2\Gamma_A^{II}(r_A)}{r_A} - \Gamma_A^{II}(r_A)\eta_A^{lower}(r_A)\right). \quad (91)$$

A robust and rapidly converging iterative algorithm was used to compute $\{G_A(r_A)\}$. In each iteration, an estimate of $\Phi_A^{II}$ is used to compute the estimate

$$G_A(r_A) = \rho_A^{wavg}(r_A) + \Phi_A^{II}\sqrt{\sigma_A(r_A)}. \quad (92)$$

Then, constraint (87) is enforced by recursively setting

$$G_A(r_A) = \min\left(G_A(r_A), G_A(r_A - \Delta r_A)\exp\left(-\eta_A^{lower}(r_A)\Delta r_A\right)\right). \quad (93)$$

beginning with the second radial shell and continuing outward until the last radial shell. To complete one iteration, $\varphi_A^{II}$ is computed via Eq. (89). A scheme is required to generate a new $\Phi_A^{II}$ estimate for the next iteration. Since $\varphi_A^{II}$ is a monotonically increasing function of $\Phi_A^{II}$, $\Phi_A^{II}$ should be increased (decreased) when $\varphi_A^{II} < 0$ ($\varphi_A^{II} > 0$). Convergence is achieved when $\left|\varphi_A^{II}\right|$ is less than some zero tolerance (e.g., $10^{-10}$ electrons). The update scheme we used contained the same sequence of steps described in Section 2.6, except $\Phi_A^{II}$, $\varphi_A^{II}$, and $\sigma_A(r_A)$ replace $\Phi_A^{I}$, $\varphi_A^{I}$, and $Y_A^{avg}(r_A)$. Specifically, this update scheme determines upper and lower bound values, evaluates the midpoint, fits these three points to a parabola whose root is used as the fourth point, and linearly interpolates to find a corrector point. Then, the new upper and lower bound values are identified, and a new iteration is performed until convergence. Because this algorithm cuts the size of the search domain by better than half in each iteration, it is mathematically guaranteed to always converge in a few iterations.

ii) Constraint preventing buried tails from becoming too contracted: $H_A(r_A)$ is then computed to make sure the tails of buried atoms do not become too contracted. Specifically, the constraints

$$\phi^{III}(r_A) = \frac{dH_A(r_A)}{dr_A} + \eta_A^{upper}(r_A)H_A(r_A) \geq 0 \quad (94)$$

$$\eta_A^{upper}(r_A) = \frac{2.5\ \text{bohr}^{-1}}{\left(1 - \left(\tau_A(r_A)\right)^2\right)} \quad (95)$$

$$\varphi_A^{III} = \int_0^{r_{cutoff}} \left(H_A(r_A) - G_A(r_A)\right)4\pi(r_A)^2\, dr_A \to 0 \quad (96)$$

are imposed by minimizing the following optimization functional

$$h^{III}\left(H_A(r_A)\right) = \int_0^{\infty}\left[\frac{\left(H_A(r_A) - G_A(r_A)\right)^2}{2G_A(r_A)} + \Gamma_A^{III}(r_A)\phi^{III}(r_A)\right]4\pi(r_A)^2\, dr_A - \Phi_A^{III}\varphi_A^{III} \quad (97)$$

where $\Gamma_A^{III}(r_A)$ and $\Phi_A^{III}$ are Lagrange multipliers enforcing constraints (94) and (96), respectively. $h^{III}\left(H_A(r_A)\right)$ is a convex functional:

$$\frac{\delta^2 h^{III}}{\delta H_A(r_A)\delta H_A(r_A')} = \frac{4\pi(r_A)^2 \delta^{dirac}(r_A - r_A')}{G_A(r_A)} \geq 0. \quad (98)$$

The unique minimum is

$$H_A(r_A) = G_A(r_A)\left(1 + \Phi_A^{III} + \frac{d\Gamma_A^{III}(r_A)}{dr_A} + \frac{2\Gamma_A^{III}(r_A)}{r_A} - \Gamma_A^{III}(r_A)\eta_A^{upper}(r_A)\right). \quad (99)$$

In practice, $H_A(r_A)$ can be easily found by starting with the initial estimate $H_A^{est}(r_A) = G_A(r_A)$ and enforcing constraint (94) by recursively setting

$$H_A^{est}(r_A) = \max\left(H_A^{est}(r_A), H_A^{est}(r_A - \Delta r_A)\exp\left(-\eta_A^{upper}(r_A)\Delta r_A\right)\right) \quad (100)$$

beginning with the second radial shell and continuing outward until the last radial shell. Finally, $H_A(r_A)$ is normalized to satisfy constraint (96):

$$H_A(r_A) = H_A^{est}(r_A)\frac{\int_0^{r_{cutoff}} G_A(r_A)4\pi(r_A)^2 dr_A}{\int_0^{r_{cutoff}} H_A^{est}(r_A)4\pi(r_A)^2 dr_A}. \quad (101)$$

The updated atomic weighting factor is

$$w_A(r_A) = H_A(r_A). \quad (102)$$

Comments on these exponential tail constraints: Constraint (87) preventing the tails of buried atoms from becoming too diffuse is the same for the DDEC3 and DDEC6 methods. The DDEC6 method adds constraint (94) to prevent the tails of buried atoms from becoming too contracted. The limiting exponent of 2.5 $bohr^{-1}$ corresponds to $w_A(r_A)$ in the buried tail being cut in half for an $r_A$ increase of 0.277 bohr (0.147 Å). There is no reason for $w_A(r_A)$ to decrease more rapidly than this in the buried tail. The integrals of the optimization functionals $h^{I,II,III}$ in Eqs. (65) , (90), and (97) have similar forms, except that a square root appears in the denominator of the first integral in $h^{I,II}$ but not in $h^{III}$. This exponent in the denominator of the first integral of the optimization functional is called the reshaping exponent.[14] A reshaping exponent $\xi = 1/2$ is used in the optimization functional $h^{II}$ (Eq. (90)) that enforces constraint (87) preventing tails of buried atoms from becoming too diffuse.[14] A reshaping exponent $\xi = 1/2$ is also used in the optimization functional $h^{I}$ (Eq. (65)) that reshapes the conditioned reference ion densities. The value $\xi = 1/2$ is appropriate for these cases, because it shifts electron density from the tail region into the intermediate region where atoms interface.[14] A reshaping exponent $\xi = 1$ is used in the optimization functional $h^{III}$ (Eq. (97)) that enforces constraint (94) preventing tails of buried atoms from becoming too contracted. Eq. (100) implementing constraint (94) adds electron density to the tail region thereby requiring density to be removed during renormalization. Removing electron density during renormalization with $\xi < 1$ is ill-behaved, because this will completely deplete the electron density in the buried tail region due to $w_A(r_A)^{\xi} / w_A(r_A) \gg 1$ when $w_A(r_A) \ll 1$. Using $\xi = 1$ avoids this problem.

**6b.** If the value of update_kappa is TRUE, then $\{\kappa_A\}$ and $\{w_A(r_A)\}$ are updated by setting

$$\kappa_A = \max\left(0, \left(\kappa_A^{old} - N_A^{val}/u_A\right)\right) \quad (103)$$

$$w_A(r_A) = e^{\kappa_A(r_A)} w_A^{fixed}(r_A). \quad (104)$$

where $\kappa_A^{old}$ is the value of $\kappa_A$ before the update is applied. This process corresponds to minimizing the functional

$$F^{DDEC6} = \sum_A \oint \rho_A(\vec{r}_A) \ln\left(\frac{\rho_A(\vec{r}_A)}{w_A^{fixed}(r_A)}\right) d^3\vec{r}_A - \int_U \Gamma(\vec{r})\Theta(\vec{r}) d^3\vec{r} - \sum_A \kappa_A N_A^{val} \quad (105)$$

where $\Gamma(\vec{r})$ and $\{\kappa_A\}$ are Lagrange multipliers enforcing constraints (8) and (38), respectively. During this process, the values of $\{w_A^{fixed}(r_A)\}$ do not change. Consequently, the curvature

$$d^2F^{DDEC6} = \sum_A \oint \left(\frac{\left(\delta\rho_A(\vec{r}_A)\right)^2}{\rho_A(\vec{r}_A)}\right) d^3\vec{r}_A > 0 \quad (106)$$

for any $|\delta\rho_A(\vec{r}_A)| > 0$. This positive definite curvature indicates a convex optimization landscape with a unique solution. For an atom without any overlaps (i.e., isolated atomic ion limit), $u_A = 0$ and the converged $\rho_A(\vec{r}_A)$ is independent of $\kappa_A$. Therefore, when $u_A$ is negligible (e.g., $u_A < 10^{-7}$ ) we set $\kappa_A = 0$ to avoid division by zero in Eq. (103).

**7.** The charge cycle number if incremented by +1, and the calculation returns to # 1 above to start the next charge cycle.

## 2.8 Computational Speed and Convergence Robustness

DDEC charge and spin partitioning use a cutoff radius (e.g., 5 Å) to achieve a computational cost that scales linearly with increasing number of atoms in the unit cell after the initial electron and spin density grids have been generated.[14, 68] When combined with a linearly scaling quantum chemistry program (e.g., ONETEP), this provides computationally efficient charge analysis even for systems containing thousands of atoms in the unit cell.[69, 70] Figure 3 plots the wall time for computing DDEC3 and DDEC6 NACs, atomic multipoles, and electron cloud decay exponents for the NaCl crystal (ambient pressure) as a function of the number of atoms in the unit cell. The unit cells containing 16 and 54 atoms were constructed by forming 2×2×2 and 3×3×3 supercells, respectively, that were used as input for computing the density grid files in VASP. This calculation utilized a volume of $2\times10^{-3}$ $bohr^3$ per grid point. The wall time in Figure 3 begins when the CHARGEMOL program is first entered prior to reading the VASP density grid files and continues until the moment after the computed NACs, atomic multipoles, and electron cloud decay exponents have been written to the net_atomic_charges.xyz file. As expected, Figure 3 shows the required wall time depends linearly on the number of atoms in the unit cell. Even though the computation was run on a single processor core, only six minutes were required for DDEC6 charge analysis of the unit cell containing 54 atoms. This was only one-fifth of the time required for DDEC3 charge analysis of the same material. Much larger times are required for DDEC3 or DDEC6 calculations reading in Gaussian basis set coefficients (e.g., GAUSSIAN 09 generated .wfx files), because in such cases the density grids must be explicitly computed within the CHARGEMOL program.

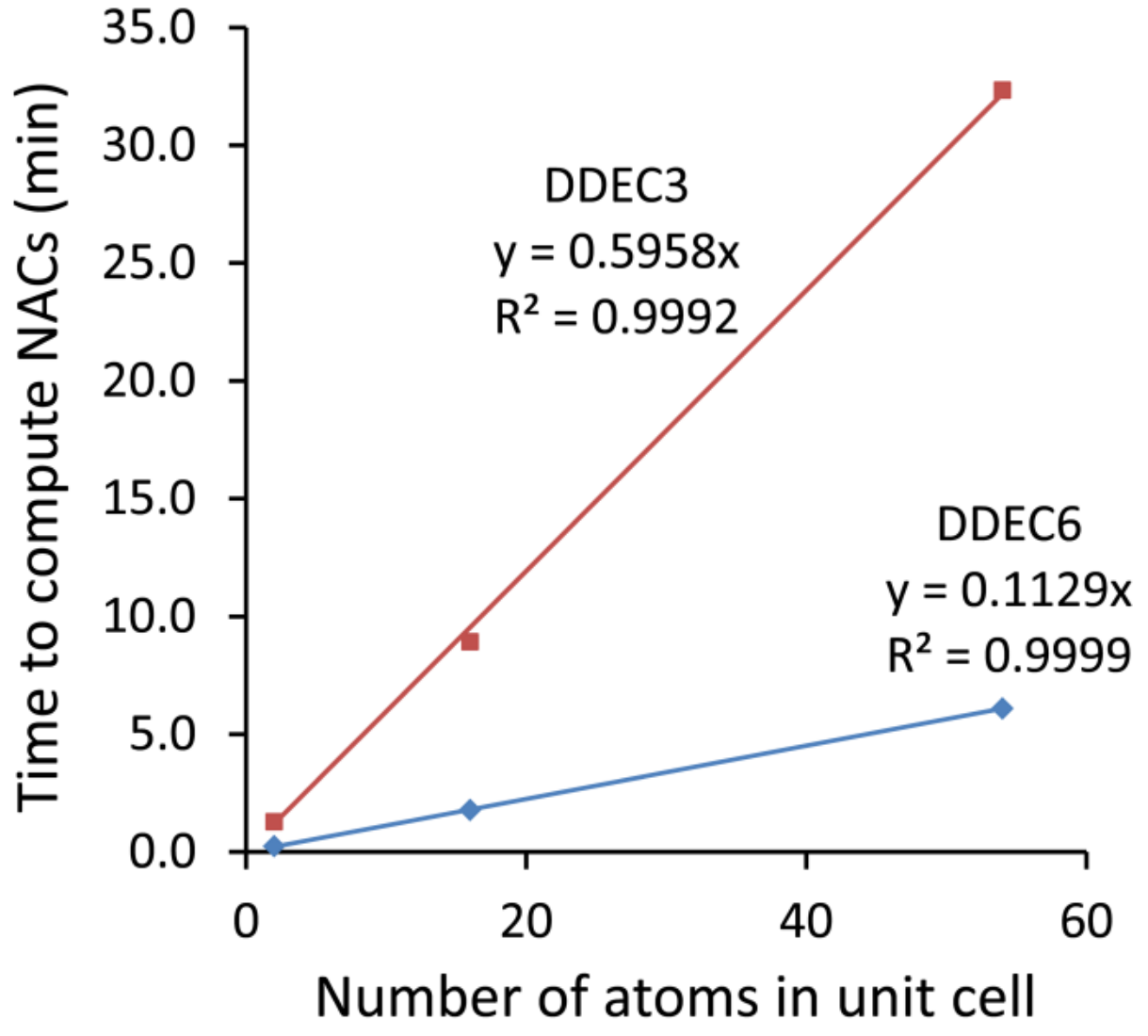

Figure 3: The computational cost of DDEC3 and DDEC6 charge partitioning scales linearly with increasing system size. Computation for NaCl crystal (ambient pressure) performed with serial Fortran CHARGEMOL program executed on a single processor core in Intel Xeon E5-2680v3 on the Comet supercomputing cluster at the San Diego Supercomputing Center.

Table 3: Number of charge cycles to convergence for selected systems

| | **DDEC3** | **DDEC6** |
|---|---|---|
| NaCl crystal (ambient pressure) | 107–109[a] | 7[a] |
| TiO solid | 130 | 7 |
| $Mo_2C$ slab with K adatom | 87[b] | 7 |
| $Fe_2O_3$ solid | 173[b] | 7 |
| Zn nicotinate MOF | 75 | 7 |
| water molecule | 37 | 7 |
| DNA decamer | 83 | 7 |
| linear $Li_2O$ molecule | 52 | 7 |
| $Na_3Cl$ (P4mmm crystal) | 62[c], 75[d] | 7[c,d] |
| $Fe_4O_{12}N_4C_{40}H_{52}$ noncollinear single molecule magnet | 90[a] | 7 |
| $Cs@C_{60}$ | 51[e] | 7[e] (18[f]) |

[a] For unit cells containing 2, 16, and 54 atoms. [b] From reference [14]. [c] Using 10 frozen Na core electrons. [d] Using 2 frozen Na core electrons. [e] Using 46 frozen Cs core electrons. [f] Using 54 simulated frozen Cs core electrons by treating 8 of the 9 PAW valence electrons as core.

The main difference in computational cost between DDEC3 and DDEC6 arises from the number of charge cycles required for convergence. In fact, the electron partitioning scheme is the only computational difference between DDEC3 and DDEC6. As shown in Table 3, more charge cycles are required for DDEC3 convergence than for DDEC6 convergence. For all materials we studied, fewer than 200 DDEC3 charge cycles were required.[14] For all materials, seven DDEC6 charge partitioning steps are required. More than one DDEC6 charge cycle per charge partitioning step is required only when the $N_A \geq N_A^{core}$ constraint is binding, because in this case $\{\kappa_A\}$ must be iterative computed. For Cs@$C_{60}$ with 54 simulated frozen core electrons, 18 DDEC6 charge cycles were required to complete the seven charge partitioning steps. All other materials we studied converged in seven DDEC6 charge cycles. It is gratifying that the extra accuracy of DDEC6 compared to DDEC3 comes with a reduced computational cost.

Our overall objective is to develop an atomic population analysis method with characteristics suitable for use as a default method in popular quantum chemistry programs. This requires the method to maximize broad applicability and minimize failure. The algorithm should converge rapidly and reliably to a unique solution with low dependence on the basis set choice. The assigned NACs, ASMs, and other AIM properties should be chemically consistent, accurately describe electron transfer directions, and correlate to experimental data for reference materials.

Existing atomic population analysis methods are not well-suited for use as a default method in quantum chemistry programs: (1) Mulliken[6] and Davidson-Löwdin[7] population analyses do not have any mathematical limits as the basis set is systematically improved toward completeness, and they are not directly applicable to plane-wave basis sets.[8] (2) Bader's quantum chemical topology has many theoretically desirable properties, but it can lead to non-nuclear attractors that produce undefined NACs.[44] (3) Electrostatic potential fitting methods (e.g., ESP,[17] Chelp,[18] Chelpg,[19] REPEAT[20]) do not have good conformational transferability and assign unreasonable charge values to some buried atoms.[20, 71-74] Including constraints (e.g., RESP[20, 71] methods) improves this, but the form of these constraints is flexible leading to numerous possible charge values. Simultaneous fitting across multiple conformations is an another possible solution, but this requires computing electron distributions for many different system geometries.[74] Also, nonporous systems do not possess a surface outside which to fit the electrostatic potential. (4) The original HD[28] method, which is based on neutral reference atom densities, usually underestimates NAC magnitudes.[14, 25, 29, 58] While the IH method improves the NAC magnitudes,[29] it exhibits the bifurcation problem shown in Figure 1. (5) The CM5 method adds an empirical correction to the HD NACs, but it does not self-consistently update the HD ASMs.[25, 26] As shown in Section 5.8 below, this causes the CM5 NACs and ASMs to be inconsistent with each other in some materials. (6) Because APT[21] and Born effective charges[23] require computing system response properties (via perturbation theory or atomic displacements), they are not well-suited for use as a default atomic population analysis method. (7) Among existing methods, NPA is probably the closest to a reasonable and broadly applicable atomic population analysis method that could be used as a default in quantum chemistry programs.[8] Recently, the NPA and related Natural Bond Orbital (NBO) and Adaptive Natural Density Partitioning (ANDP) methods have been extended to periodic materials.[11, 75] Although NPA is a dramatic improvement over the Mulliken method, NPA is not strictly a functional of $\rho(\vec{r})$ and retains some explicit basis set dependence.[8] In plane-wave calculations, the natural atomic orbitals are found by constructing an auxiliary localized basis set onto which the plane-waves are projected.[11] This creates a small charge spillage if the plane-wave and localized basis sets do not span exactly the same function spaces.[11]

The DDEC6 method is an appealing approach, because it is an explicit functional of the electron and spin distributions with no explicit basis set dependence. There is, therefore, no need to create an auxiliary localized basis set when computing DDEC6 charges in plane-wave calculations. As shown by the extensive tests presented in this article, the DDEC6 method achieves an extremely broad range of applicability across diverse material classes. It converges rapidly with a computational cost scaling linearly with increasing number of atoms in the unit cell. Using five fixed conditioning steps ensures robust convergence to a unique solution. As shown in Section 2.3 above, the DDEC6 method alleviates the bifurcation or ‘runaway charges’ problem exhibited by earlier DDEC and IH methods. As shown in Section 5.8.3 below, the DDEC6 NACs and ASMs achieve good chemical consistency. Therefore, DDEC6 is well-suited for use as a default charge assignment method in popular quantum chemistry programs. Actually incorporating DDEC6 into popular quantum chemistry programs will require additional work. For example, it might be desirable to implement DDEC6 on the same integration grid already used in the respective quantum chemistry program.

**2.9 Summary of changes between DDEC3 and DDEC6 methods**

Table 4 lists the six differences between the DDEC3 and DDEC6 methods. First, the DDEC6 method uses fixed reference ion charge values rather than self-consistently updating them as in the DDEC3 method. Second, the DDEC6 method uses c=4 and $\chi = 1$ to yield a fixed number of charge partitioning steps with $\chi_{\text{equiv}}^{\text{DDEC6}} = 1/5$. In contrast, the DDEC3 method uses a self-consistent iterative scheme with c=1 and $\chi^{\text{DDEC3}} = 3/14$ to yield $\chi_{\text{equiv}}^{\text{DDEC3}} = 3/17$. Third, the DDEC6 method uses a weighted spherical average in place of the simple spherical average used in the DDEC3 method. Fourth, the DDEC6 method incorporates constraints to make the conditioned reference ion densities monotonically decreasing and to integrate to $N_A^{\text{ref}}$. Fifth, the DDEC6 method adds a constraint to ensure the buried tails of $w_A(r_A)$ do not decay too quickly. Both the DDEC3 and DDEC6 methods include the same constraint to ensure the buried tails of $w_A(r_A)$ do not decay too slowly. These constraints make the buried tail of $w_A(r_A)$ decay exponentially with increasing $r_A$. Sixth, we have now computed a complete reference ion library for elements 1 to 109 (Section 3).

Figure 4 compares DDEC3 to DDEC6 atomic dipole magnitudes in atomic units. The left panel contains a set of materials comprised almost entirely of surface atoms. This is the same set of materials for which DDEC3 and DDEC6 NACs are compared in Section 5.6. The right panel contains the following dense materials: $TiCl_4$ crystal, $SrTiO_3$ surface slab, Pnma $NaCl_3$ crystal (2 frozen Na core electrons), P4mmm $Na_2Cl$ crystal (2 frozen Na core electrons), natrolite, NaF surface slab, $Mo_2C$ surface with K adatom, Cmmm $Na_2Cl$ crystal (2 frozen Na core electrons), Pd crystal with interstitial H atom, $Pd_3Hf$ crystal with interstitial H atom, $Pd_3In$ crystal with interstitial H atom, $Pd_3V$ crystal with interstitial H atom. (These structures are described in more detail in Section 5.) The slopes of the best fit lines constrained to have an intercept of (0,0) were 1.0462 (1.0222) with R-squared correlation coefficient = 0.9263 (0.9496) for the surface atom materials (dense materials). This shows the atomic dipole magnitudes are about 2–5% larger in magnitude for DDEC3 compared to DDEC6. These small atomic multipoles allow DDEC6 NACs to approximately reproduce $V(\vec{r})$ surrounding a material.

Table 4. List of differences between DDEC3 and DDEC6 methods

| | | **DDEC3** | **DDEC6** |
|---|---|---|---|
| **1** | **reference ion charge** | iteratively set to AIM charge: $q_A^{ref} = q_A$ | non-iteratively set based on a special partitioning: $q_A^{ref} = q_A^{2,ref}$ |
| **2** | **how spherical averaging is incorporated** | c=1, $\chi = 3/14$, $\chi_{equiv}^{DDEC3} = 3/17$ | c=4, $\chi = 1$, $\chi_{equiv}^{DDEC6} = 1/5$ |
| **3** | **type of spherical average used** | simple | weighted |
| **4** | **constraints applied to each conditioned reference ion density** | — | monotonically decreasing with increasing $r_A$ and integrates to $N_A^{ref} = z_A - q_A^{ref}$ |
| **5** | **tail constraints applied to $w_A(r_A)$ during 4th and later charge cycles** | tails of buried atoms decay no slower than exp(-1.75 $r_A$) | tails of buried atoms decay no slower than exp(-1.75 $r_A$) and no faster than exp(-2.5 $r_A$) |
| **6** | **reference ion library** | originally incomplete, now a completed set[a] | completed set |

[a] As originally published, the DDEC3 method required explicit computation of each reference ion which led to termination of calculations where a required reference ion was not precomputed. We recommend using the new equations (131)–(132) together with the explicitly computed reference ion ranges in Table 8 to complete this library for the DDEC3 method. Accordingly, the DDEC3 and DDEC6 methods use the same reference ion library.

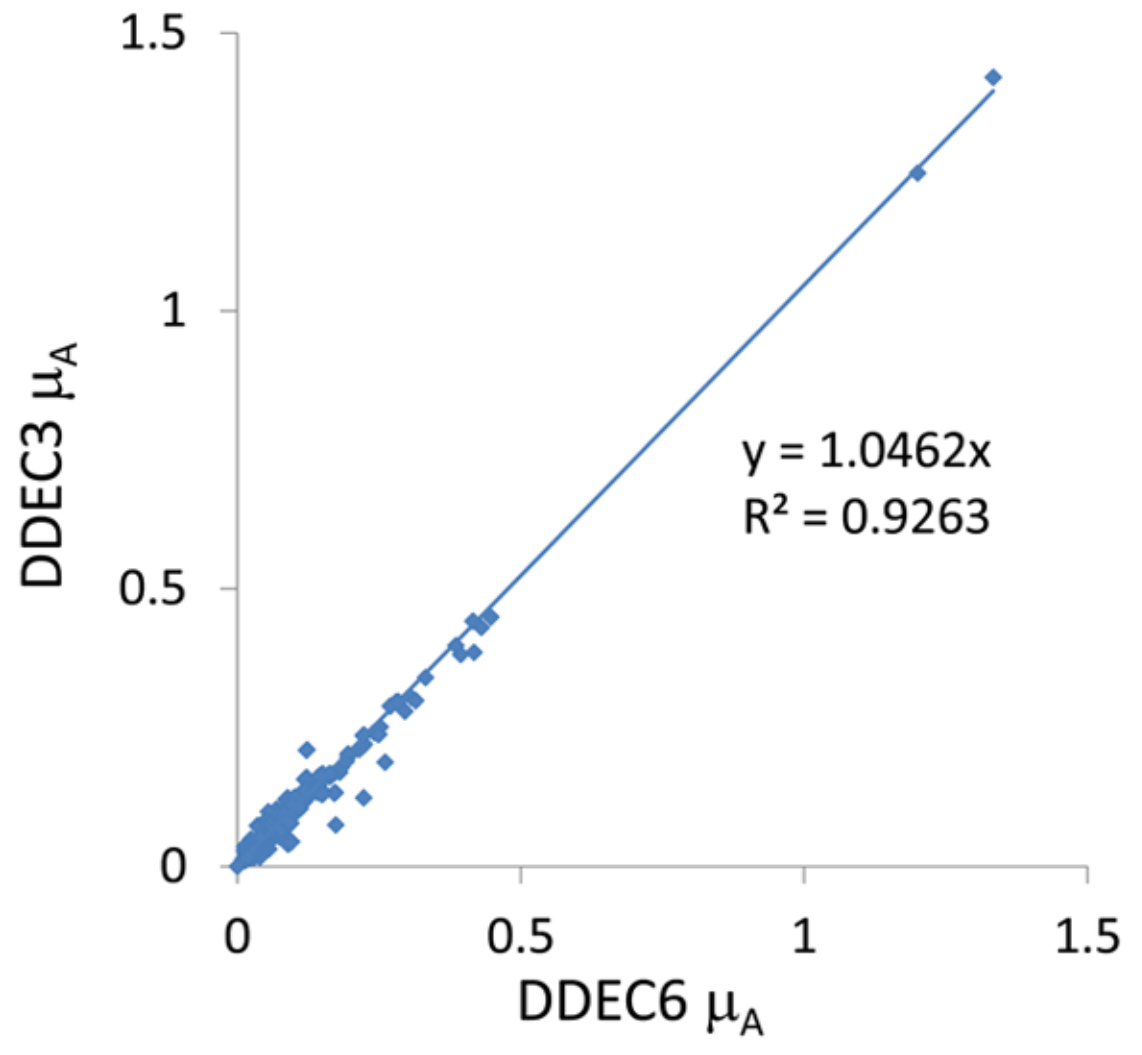

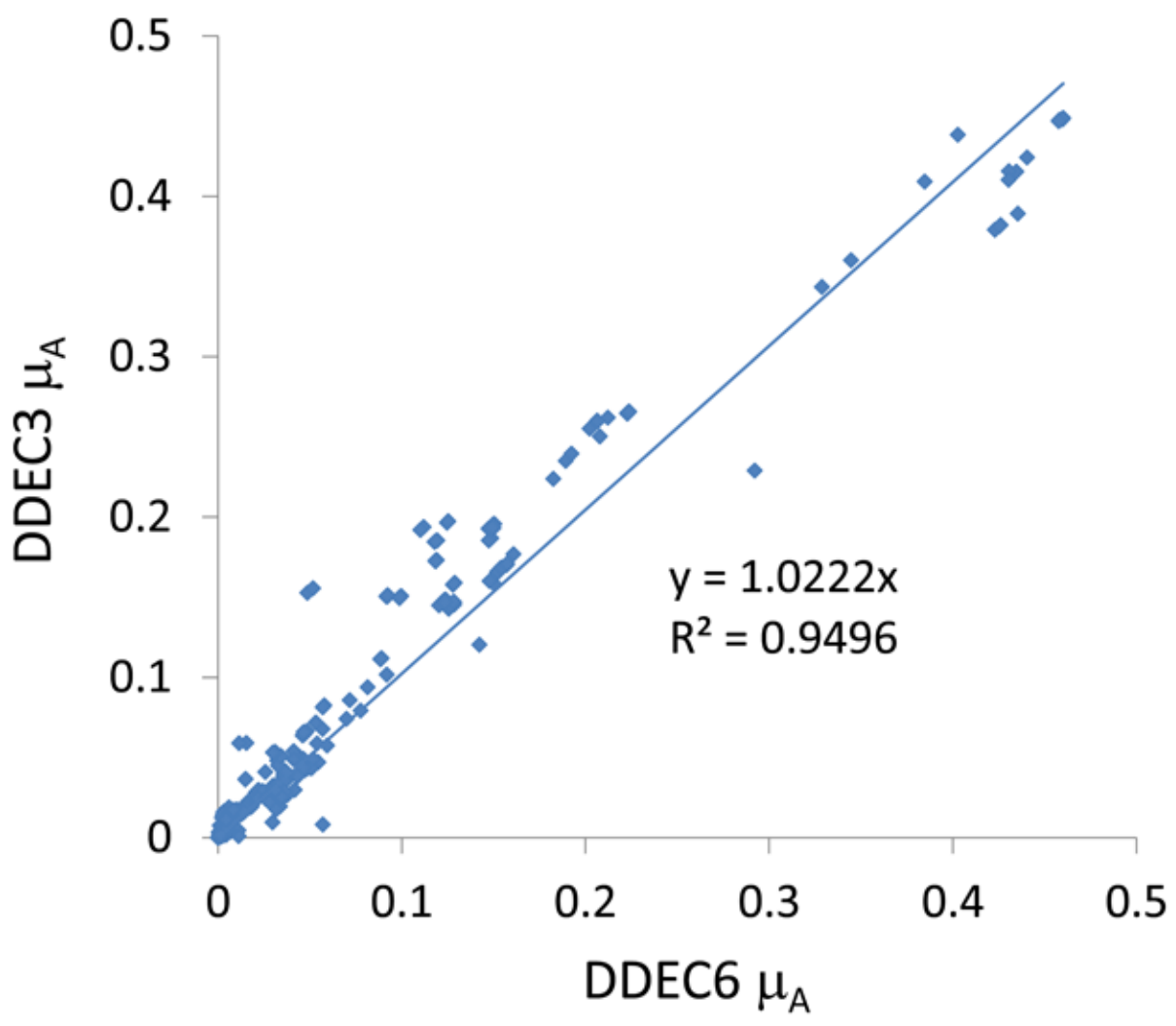

Figure 4: Comparison of DDEC3 to DDEC6 atomic dipole magnitudes. *Left:* Materials comprised almost entirely of surface atoms. *Right*: Dense materials comprised mainly of buried atoms.

Why are the DDEC6 atomic dipole magnitudes slightly smaller on average than those for the DDEC3 method? The atomic dipole magnitude, $\mu_A$, can be written as

$$\mu_A = \left| \oint \left( w_A(r_A)\hat{r}_A \left( \frac{\rho(\vec{r})}{W(\vec{r})} - 1 \right) \right) (r_A) d^3\vec{r}_A \right|. \quad (107)$$

One strategy to slightly reduce $\mu_A$ is to make the assigned $\{\rho_A(\vec{r}_A)\}$ slightly less diffuse without significantly altering $\rho(\vec{r})/W(\vec{r})$. Because the integral contributions in Eq. (107) are proportional to $r_A$, minimizing the contributions for large $r_A$ values will decrease $\mu_A$. The DDEC6 anions are typically less diffuse than the DDEC3 anions, because DDEC6 uses a partially localized reference ion charge ( $q_A^{ref} = q_A^{2,ref}$ ) instead of the AIM charge used as reference ( $q_A^{ref} = q_A$ ) in DDEC3. A second strategy for minimizing $\mu_A$ is to make $\rho(\vec{r})/W(\vec{r})$ as close to 1 as feasible in the bonding regions where atoms overlap. The integral contributions in Eq. (107) do not depend on $w_A(r_A)$ in regions where atoms do not overlap significantly, because $w_A(r_A) \approx W(\vec{r})$ in those regions. The weighted spherical average, $\rho_A^{wavg}(r_A)$, weights more heavily the regions where atom overlaps are substantial. Thus, $\rho_A^{wavg}(r_A)$ makes $\rho(\vec{r})/W(\vec{r})$ as close to 1 as feasible specifically within the regions where atom overlaps are substantial. This is precisely those regions where integral contributions to $\mu_A$ can be suppressed. For this reason, using $\rho_A^{wavg}(r_A)$ substantially outperforms the simple spherical average, $\rho_A^{avg}(r_A)$, for the purpose of minimizing $\mu_A$. This reduction in $\mu_A$ also causes the NACs to more accurately reproduce $V(\vec{r})$ surrounding the material.

Table 5 summarizes computational tests for six different DDEC algorithms. The column labeled 'Simple spherical average' uses an algorithm identical to the DDEC6 method, except $\rho_A^{avg}(r_A)$ is used in place of $\rho_A^{wavg}(r_A)$. The performance of this algorithm is substantially worse than DDEC6 but still an improvement over DDEC3. The column labeled 'Convex functional' uses the charge partitioning functional of Eq. (37). The Convex functional also did not perform as well on average as DDEC6. The columns labeled '4 charge partitioning steps' and '10 charge partitioning steps' use an algorithm identical to the DDEC6 method but stop after 4 and 10 charge partitioning steps, respectively, instead of the 7 charge partitioning steps used in the DDEC6 method. The overall performance of the '4 charge partitioning steps' and '10 charge partitioning steps' algorithms were similar to that of the DDEC6 method. Specifically, results for about half of the materials are improved upon going from 7 to 4 charge partitioning steps, while results for the other half of the materials are improved upon going from 7 to 10 charge partitioning steps. We chose 7 charge partitioning steps for the DDEC6 method, because it offers a suitable compromise.

The overall performance of each method was scored based on twelve computational tests:

1. Squared correlation coefficient ($R^2$) between computed NACs and core-electron binding energy shifts for the Ti-containing solids described in Section 5.3.2.

2. The slope for a plot analogous to Figure 4 (left panel). A smaller slope (considered better) indicates smaller $\mu_A$ for the test set containing mostly surface atoms.
3. The slope for a plot analogous to Figure 4 (right panel). A smaller slope (considered better) indicates smaller $\mu_A$ for the test set comprising dense materials.
4. The NAC of Mg in crystalline MgO. A NAC closer to ~1.7 was considered better.
5. The NAC of Ru in crystalline $Li_3RuO_2$. A NAC closer to ~0.1 was considered better.
6. The NAC of P in the $H_2PO_4^-$ molecular ion. A NAC closer to ~1.5 is considered better.
7. The NAC of Li in linear $Li_2O$. A NAC closer to ~0.87 is considered better.
8. The NAC of H in the $H_2O$ molecule. A NAC closer to ~0.37 is considered better.
9. The NAC of Cl in the high-pressure Imma $Na_2Cl$ crystal. A NAC closer to -1.35 is considered better.
10. The NAC of H in the $H_2$ molecule (triplet state, 50 pm constrained bond length, CISD/aug-cc-pvqz). A NAC closer to zero is better.
11. The $Li_3$ molecule dipole error (atomic units) quantified as $\mu_A$ for the NAC model minus $\mu_A$ for the DFT-computed electron distribution (PBE+D3[76]/aug-cc-pvtz optimized geometry and electron distribution). An error closer to zero is better.
12. The time in minutes required to perform charge analysis on the NaCl crystal supercell containing 54 atoms running on a single processor core in Intel Xeon E5-2680v3 on the Comet supercomputing cluster at SDSC. This is the total wall time from program start to program finish, including the input file reading, core electron partitioning, valence electron partitioning, computation of multipole moments, output file printing, etc.

The target values for tests 4 to 9 were chosen based on comparisons to Bader charges (dense solids) and charges that would more closely reproduce the electrostatic potential (molecules).

NACs are often used to construct force-fields for classical molecular dynamics and Monte Carlo simulations. For these applications, the NACs should approximately reproduce the molecular electrostatic potential (MEP). Conformational transferability is important for the construction of flexible force-fields. Table 6 compares the accuracy of the DDEC6 and Convex functional methods for reproducing the electrostatic potential of various conformations of carboxylic acids, $Li_2O$ molecule, and Zn-nicotinate metal-organic framework. For these tests, we used the same set of molecular conformations as described in Section 5.5. Results listed under the 'carboxylic acids' column in Table 6 are the averages over five carboxylic acids. When NACs were optimized specifically for each molecular conformation, $V(\vec{r})$ was described most accurately using the Convex functional for the carboxylic acids and most accurately by the DDEC6 method for the $Li_2O$ molecule and Zn-nicotinate MOF. When the conformation averaged and low-energy conformation NACs were used to construct point-charge models for every system conformation, the DDEC6 method reproduced $V(\vec{r})$ most accurately for all three types of materials. This shows the DDEC6 NACs are more accurate than the Convex functional NACs for reproducing $V(\vec{r})$ across multiple system conformations. In summary, the DDEC6 NACs have better overall performance than the Convex functional NACs.

Table 5. Performance of different DDEC algorithms compared to DDEC6. Bold numbers indicate a result better than DDEC6. Italic numbers indicate a result worse than DDEC6. Numbers neither bold nor italic are within ±0.01 with respect to DDEC6 and are considered to be the same as the DDEC6 values.

| | DDEC6 (7 steps) | DDEC3 | 4 charge partitioning steps | 10 charge partitioning steps | Simple spherical average | Convex functional |
|---|---|---|---|---|---|---|
| $R^2$ for Ti solids | 0.704 | *0.360* | **0.739** | *0.671* | 0.704 | **0.718** |
| slope [$R^2$] for atomic dipoles (surface atoms) | 1.000 [1.000] | *1.046* [0.926] | *1.119* [0.825] | **0.956** [0.980] | *1.041* [0.955] | *1.061* [0.944] |
| slope [$R^2$] for atomic dipoles (solids) | 1.000 [1.000] | *1.022* [0.950] | *1.122* [0.950] | **0.943** [0.987] | 0.995 [0.995] | *1.011* [0.987] |
| Mg NAC (for MgO) | 1.465 | *2.012* | *1.312* | **1.508** | 1.473 | *1.436* |
| Ru NAC (for $Li_3RuO_2$) | -0.083 | *-0.172* | **0.193** | *-0.149* | *-0.195* | -0.088 |
| P NAC (for $H_2PO_4^-$) | 1.622 | *1.800* | *1.656* | **1.586** | *1.682* | *1.673* |
| Li NAC (for $Li_2O$) | 0.902 | *0.984* | **0.864** | *0.914* | *0.927* | 0.912 |
| H NAC (for water) | 0.395 | *0.417* | *0.418* | **0.381** | *0.410* | *0.414* |
| Cl NAC (Imma $Na_2Cl$) | -1.628 | *-2.439* | *-1.972* | **-1.492** | *-1.679* | *-1.809* |
| H NAC ($H_2$ triplet) | ±0.021 | *±0.503* | **±0.004** | *±0.040* | ±0.019 | **±0.009** |
| $Li_3$ dipole error | 0.645 | *0.900* | **0.420** | *0.666* | *0.771* | 0.635 |
| minutes (NaCl 54 atoms) | 6.6 | *34.6* | **6.0** | *7.6* | 6.6 | *7.8* |
| Score for the method | 0 | -12 | 0 | 0 | -7 | -5 |

Table 6: Comparison of accuracy for representing the electrostatic potential across multiple conformations of carboxylic acids, $Li_2O$ molecule, and Zn-nicotinate MOF. The first column lists the source of the NACs used to model the electrostatic potential across the various conformations. The reported values are the RMSE in kcal/mol averaged across all conformations. For each comparison, the best value is shown in boldface type.

| **NACs** | **carboxylic acids** | | **$Li_2O$ molecule** | | **Zn-nicotinate MOF** | |
|---|---|---|---|---|---|---|
| | **DDEC6** | **Convex functional** | **DDEC6** | **Convex functional** | **DDEC6** | **Convex functional** |
| conformation specific | 1.06 | **0.98** | **5.76** | 5.80 | **2.99** | 3.31 |
| conformation averaged | **1.32** | 1.38 | **6.48** | 6.56 | **3.13** | 3.49 |
| low energy conformation | **1.42** | 1.52 | **7.17** | 7.36 | **3.39** | 3.84 |

## 2.10 Exact for isolated atomic ion limit

The isolated atomic ion limit corresponds to spatially separated and negligibly overlapping atomic ions, such as occurs when each atomic ion is ~10 Å or more from all other atoms. It is the only situation for which an exact $\{w_A(r_A)\}$ is uniquely defined. Specifically, in the isolated atomic ion limit, the atomic weighting factors should equal the spherical average of each isolated density: $w_A(r_A) = \rho_A^{avg}(r_A)$.

Consider the specific example of a periodic cubic array with Na and F ions at alternating vertices. By symmetry, the atomic dipole, quadrupole, and octapole moments are zero. Since atomic hexadecapole moments are the leading non-zero atomic multipoles, the atomic electron distributions are approximately (but not exactly) spherically symmetric. Thus, in this idealized example, the total electron density can be approximated as the sum of individual spherical ion densities:

$$\rho(\vec{r})\Big|_{\text{NaF array}} \approx \sum_{A,L} \rho_A^{avg}(r_A). \qquad (108)$$

We consider two limiting cases: (a) a periodic array having a 20 Å distance between nearest Na atoms and (b) the PBE-optimized low energy crystal structure having 2.27 Å between nearest Na atoms.

There are two possible strategies for the IH method: (i) use reference ions for the isolated ions without charge compensation (as done when the IH method was introduced[29]) or (ii) use reference ions that mimic the ion shapes in condensed crystals by including charge compensation effects (as done in later modifications of the IH method[32, 59, 61, 63, 77]). If choice (i) is made, the reference ion shapes will match the ions in example (a). If choice (ii) is made, the reference ion shapes will match the ions in example (b). Due to charge compensation and electrostatic screening effects, anions in the condensed phase are more contracted than their isolated gas-phase counterparts. Therefore, the IH method must choose which of these two limits to reproduce. Moreover, to yield $w_A(r_A) = \rho_A^{avg}(r_A)$ the IH reference ions would also need to be computed using a similar exchange-correlation theory and basis set as used to study the material of interest.

The DDEC6 method accurately reproduces both limits, because the reference ion densities are conditioned to the material of interest. In the isolated atomic ion limit (structure (a)), $w_A^{DDEC6}(r_A) = \rho_A^{cond}(r_A) = \rho_A^{ref}(r_A, q_A^{ref})\left\langle \rho(\vec{r})/\rho^{ref}(\vec{r})\right\rangle_{r_A} = \rho_A^{avg}(r_A)$ and additional conditioning steps do not alter $w_A^{DDEC6}(r_A)$. Because the DDEC6 method derives $\{w_A^{DDEC6}(r_A)\}$ from partitions of $\rho(\vec{r})$ (i.e., a conditioning process), the DDEC6 method returns $w_A(r_A) = \rho_A^{avg}(r_A)$ in the isolated atomic ion limit *even when the DDEC6 reference ions are computed using a different exchange-correlation theory, different basis sets, and different local chemical environment than used in the system of interest!* In the optimized crystal geometry (structure (b)), the symmetry makes the atomic dipole, quadrupole, and octapole moments zero. Consequently, $\rho_A(\vec{r}_A) \approx \rho_A^{avg}(r_A)$ which implies $\left\langle \left(\rho(\vec{r})/W(\vec{r})\right)^2 \right\rangle_{r_A} - \left\langle \rho(\vec{r})/W(\vec{r})\right\rangle_{r_A}^2 \approx 0$. Since $\{w_A^{DDEC6}(r_A)\}$ are computed via conditioning steps that make $\left\langle \rho(\vec{r})/W(\vec{r})\right\rangle_{r_A} \approx 1$, this means the conditioning process combined with the crystal symmetry makes $\rho(\vec{r})/W(\vec{r}) \approx 1$. This gives $\rho_A(\vec{r}_A) \approx w_A^{DDEC6}(r_A) \approx \rho_A^{avg}(r_A)$. Thus, the DDEC6 method accurately recovers the nearly exact limit for both structures (a) and (b).

Now consider the NACs for structures (a) and (b). Since the atoms are fully separated in structure (a), the computed Bader, DDEC6, HD, and ISA NACs were the same within an integration tolerance. In this fully separated atomic ion limit, the results depend only on the exchange-correlation theory used to generate the electron distribution. For structure (a), the atomic charge magnitudes were 1.00 (Hartree-Fock method), 0.56 (HSE06[78] functional), and 0.58 (PBE functional). Now consider the PBE-optimized low-energy crystal structure having 2.27 Å between nearest Na atoms. Analysis of experimental data shows the NaF crystal is mostly ionic.[79] The computed HD (0.28) and ISA (0.48) atomic charge magnitudes are too small chemically. The Bader (0.86) and DDEC6 (0.85) atomic charge magnitudes are more reasonable. Thus, even when considering these simple NaF structures, some key advantages of DDEC6 over the HD and ISA methods are apparent.

**2.11 Electrostatic potential expansion: Atomic multipole moments and charge penetration terms**

In this section, we review the basic principles of expanding the electrostatic potential, $V(\vec{r})$, into atomic contributions. AIM methods provide a formally exact expansion of $V(\vec{r})$ by partitioning the total electron distribution $\rho(\vec{r})$ into atomic electron distributions, $\{\rho_A(\vec{r}_A)\}$:

$$V(\vec{r}) = \sum_{A,L} V_A(\vec{r}_A) \qquad (109)$$

$$V_A(\vec{r}_A) = \frac{z_A}{r_A} - \oint \frac{\rho_A(\vec{r}_A')}{|\vec{r}_A - \vec{r}_A'|} d^3\vec{r}_A' = \frac{q_A}{r_A} + B_A(\vec{r}_A) + C_A(\vec{r}_A) \qquad (110)$$

where $B_A(\vec{r}_A)$ and $C_A(\vec{r}_A)$ are terms due to atomic multipoles (AMs) and penetration of the atom's electron cloud, respectively.[13, 51, 80, 81] Outside $\rho_A(\vec{r}_A)$, the charge penetration term $C_A(\vec{r}_A)$ vanishes reducing $V_A(\vec{r}_A)$ to a multipole expansion.[82] Although Eq. (110) is formally exact for all AIM methods, in practice $B_A(\vec{r}_A)$ and $C_A(\vec{r}_A)$ are truncated at some finite orders leading to approximate $V(\vec{r})$ expansions.[80, 81, 83] Therefore, AIM methods providing more rapidly converging $V(\vec{r})$ expansions are more convenient for constructing force-fields. For constructing flexible force-fields, the AIM NACs should preferably have good conformational transferability. Alternative expansions of $V(\vec{r})$ and system multipole moments based on distributed multipole analysis and Gaussian density functions are given in the related literature.[48-50, 52, 53]

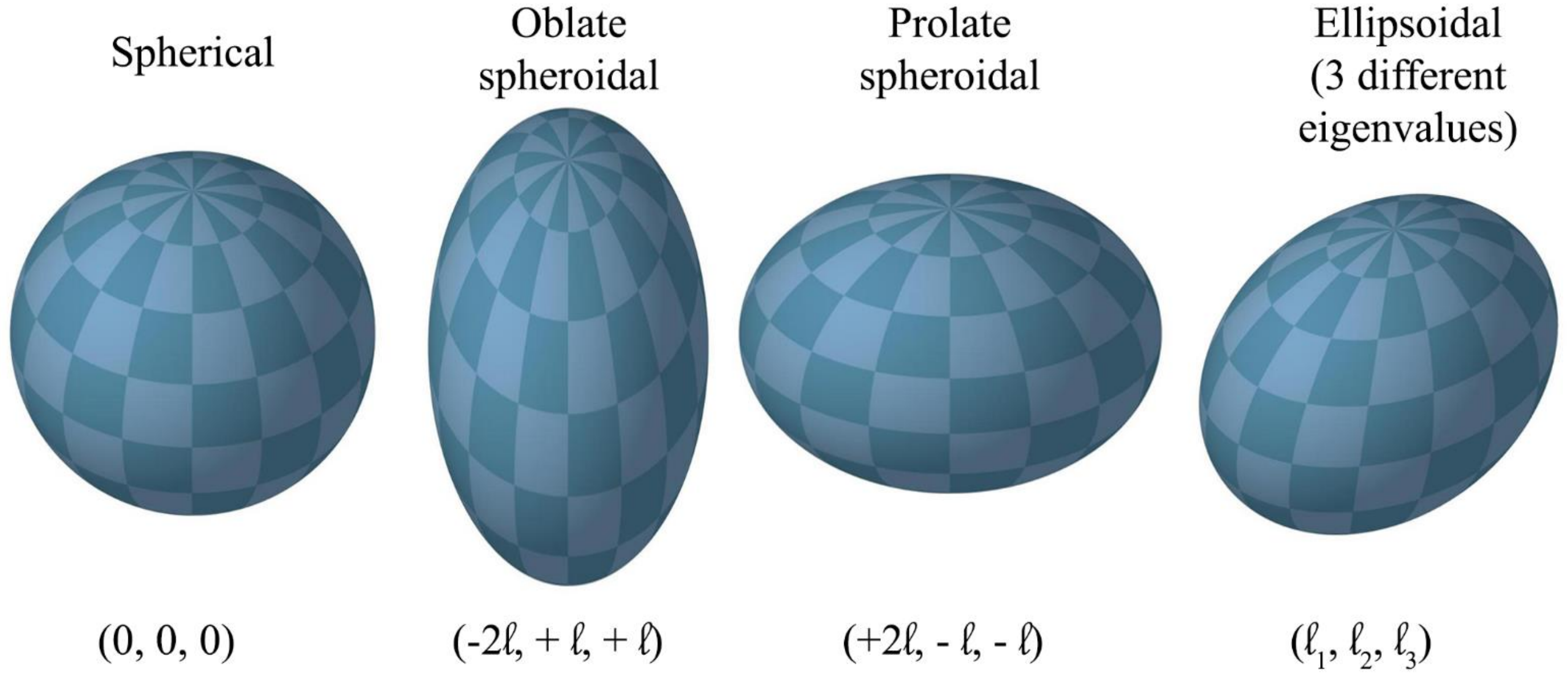

Figure 5: Relationship between atomic shape and traceless atomic quadrupole eigenvalues. (This image uses a modification of a free image of a sphere from http://www.kidsmathgamesonline.com/images/pictures/shapes/sphere.jpg.)

For each system, we computed atomic multipoles up to quadrupole order using well-known formulas. Atomic dipoles,

$$\vec{\mu}_A = -\oint \vec{r}_A \rho_A(\vec{r}_A) d^3\vec{r}_A \qquad (111)$$

$$\mu_A = |\vec{\mu}_A| \qquad (112)$$

are the leading component of $B_A(\vec{r}_A)$. Atomic quadrupoles have the form

$$Q_T = -\oint T_A \rho_A(\vec{r}_A) d^3\vec{r}_A \qquad (113)$$

where T is a second degree polynomial. The five linearly independent quadrupole components can be expressed as (i) $T_A = (X - X_A)(Y - Y_A)$ for $Q_{xy}$, (ii) $T_A = (X - X_A)(Z - Z_A)$ for $Q_{xz}$ , (iii) $T_A = (Y - Y_A)(Z - Z_A)$ for $Q_{yz}$,, (iv) $T_A = (X - X_A)^2$ for $Q_{x^2}$, (v) $T_A = (Y - Y_A)^2$ for $Q_{y^2}$, and (v) $T_A = 3(Z - Z_A)^2 - (r_A)^2$ for $Q_{3z^2-r^2}$, where $\vec{R}_A = (X_A, Y_A, Z_A)$ is the nuclear position. The traceless atomic quadrupole tensor, $\vec{\vec{Q}}_A$ , is defined by $\vec{\vec{T}}_A = \vec{r}_A \vec{r}_A - (r_A)^2 \vec{\vec{\delta}}/3$ , where $\vec{\vec{\delta}}$ is the identity tensor. The three eigenvalues of $\vec{\vec{Q}}_A$ are independent of molecular orientation and coordinate system, and they sum to zero. Figure 5 illustrates the relationship between atomic shape and traceless atomic quadrupole eigenvalues. For a spherical atom, all three quadrupole eigenvalues are zero. However, three zero quadrupole eigenvalues does not mean the atom is necessarily spherical, because such an atom could have non-zero dipole or higher order multipole (e.g., octapole) moments that indicate deviation from spherical symmetry. An oblate spheroidal density has one negative and two equal positive quadrupole eigenvalues. A prolate spheroidal density has one positive and two equal negative quadrupole eigenvalues. An ellipsoidal density can have three unequal quadrupole eigenvalues.

For finite clusters, molecules, and ions, the total dipole moment, $\mu = |\vec{\mu}|$ , is given by

$$\vec{\mu} = \sum_{A} q_A \vec{R}_A + \sum_{A} \vec{\mu}_A . \qquad (114)$$

Therefore, a charge model including both NACs and atomic dipoles reproduces $\vec{\mu}$ exactly to within a grid integration toleration. A point-charge only model includes the first term in Eq. (114) but neglects the atomic dipole terms. Unless a point-charge only model is explicitly defined to reproduce $\vec{\mu}$ (or higher order multipole) exactly, it will generally reproduce $\vec{\mu}$ (or higher order multipole) only approximately except in cases where $\vec{\mu}$ (or higher order multipole) is zero by symmetry.[84] (The general idea to constrain atom-centered point charges to exactly reproduce the molecular dipole moment is impossible for planar molecules placed in an electric field perpendicular to the molecule's plane. For this reason, we abandon the idea to constrain atom-centered point charges to exactly reproduce the system's dipole moment.) The total $p^{th}$ order multipole of a finite cluster, molecule, or ion can be expressed as a sum over NACs and atomic multipoles up to $p^{th}$ order.[80, 85]

Spherical cloud penetration, $C_A^{avg}(r_A)$, is the leading term in $C_A(\vec{r}_A)$:

$$C_A(\vec{r}_A) = C_A^{avg}(r_A) + C_A^{non-spherical}(\vec{r}_A) \quad (115)$$

$$C_A^{avg}(r_A) = \int_{r_A}^{\infty} \rho_A^{avg}(r'_A) \left( \frac{1}{r_A} - \frac{1}{r'_A} \right) 4\pi (r'_A)^2 \, dr'_A . \qquad (116)$$

Eq. (116) is the well-known result arising from basic electrostatic principles. The tail of $\rho_A^{avg}(r_A)$ decays approximately exponentially with increasing $r_A$ :

$$\rho_A^{avg}(r_A) \approx e^{a - b r_A} \text{ for } r_{min_fit} \le r_A \le r_{cutoff} \qquad (117)$$

Inserting (117) into (116) yields,

$$C_A^{avg}(r_A) \cong \frac{4\pi}{b^2} e^{a - b r_A} \left( 1 + \frac{2}{b r_A} \right). \qquad (118)$$

To determine the parameters $a$ and $b$ for each atom, the CHARGEMOL program performed a linear least squares fit of $a - b r_A$ to $\ln(\rho_A^{avg}(r_A))$ over the range $r_{min_fit}$ = 2 Å to $r_{cutoff}$ = 5 Å.[14] The R-squared correlation coefficient for this linear regression was usually > 0.99 indicating a nearly exact fit. Previous studies have used different variations of exponential or Gaussian decaying densities (sometimes multiplied by polynomials in $r_A$) or exponential damping of the 1/r or multipole potential to provide approximations of charge penetration energies.[53, 86-90]

A charge model's accuracy for reproducing $V(\vec{r})$ can be quantified by the root mean-squared error (RMSE) in the electrostatic potential quantified over a chosen a set of grid points.[14, 18-20, 72, 91, 92] The specific method we used to compute RMSE is detailed in previous work and includes a constant potential adjustment to equalize the average electrostatic potentials (over the chosen set of grid points) of the charge model and full electron distribution.[14, 20] The charge model may include point charges, dipoles and/or higher order multipoles, spherical cloud penetration, and/or aspherical cloud penetration terms. The relative root mean squared error (RRMSE) is the RMSE for the charge model divided by the RMSE of a null charge model having $\{V_A(\vec{r}_A)\}=0$.[17, 20, 72, 73, 93] The RMSE and RRMSE were computed over a uniform grid of points lying between surfaces defined by $\gamma_{inner}$ and $\gamma_{outer}$ times the van der Waals (vdW) radii. We set $(\gamma_{inner},\gamma_{outer})=(1.4,2.0)$ for nonperiodic materials and $(\gamma_{inner},\gamma_{outer})=(1.3,20.0)$ for periodic materials.[72, 73, 93] For three-dimensional materials containing only nanopores (e.g., MOFs, zeolites), $\gamma_{outer}=20.0$ is sufficiently large that the RMSE grid points span the entire pores. We used the same UFF vdW radii as Campana et al.[20] (REPEAT program) which are listed in the Supporting Information of Watanabe et al.[72]

## 3. Calculation and tabulation of reference ion densities

### 3.1 DFT calculation of the reference ions

Table 7. Basis function composition of universal Gaussian basis set (UGBS).

| $\alpha$ | Type of shell | $\alpha$ | Type of shell | $\alpha$ | Type of shell |
|---|---|---|---|---|---|
| 0.021494 | SPDF | 3.320117 | SPDF | 512.858511 | SPDF |
| 0.044157 | SPDF | 6.820958 | SPDF | 1053.633557 | SPDF |
| 0.090718 | SPDF | 14.013204 | SPDF | 2164.619772 | SPDF |
| 0.186374 | SPDF | 28.789191 | SPDF | 4447.066748 | SPD |
| 0.382893 | SPDF | 59.145470 | SPDF | 9136.201616 | SPD |
| 0.786628 | SPDF | 121.510418 | SPDF | 18769.716020 | SP |
| 1.616074 | SPDF | 249.635037 | SPDF | | |

The DDEC6 method utilizes the same library of reference ions as the DDEC3 method. As explained in earlier publications, these reference ions are computed using the PW91[94] functional near the complete basis set limit using a fourth-order Douglas-Kroll-Hess relativistic Hamiltonian with spin-orbit coupling and a spherical shell of compensating charge for the charged oxidation states.[13, 14] The uncharged oxidation state (i.e., neutral atom) for each element did not employ any compensating charge. These calculations use the Gaussian nuclear model of Visscher and Dyall.[95] Fully relaxed all-electron calculations were performed. Table 7 lists the universal Gaussian basis set we used for all elements atomic number 1 to 109, which is an even-tempered basis set with a constant ratio of 2.054... between adjacent exponents in this series.[13] For anions, the radius of the spherical shell of compensating charge and the value of the compensating charge are simultaneously optimized to minimize the anion's total energy.[13] The anion calculations were performed in GAUSSIAN 09 using the conductor-like polarizable continuum model (CPCM)[96]. For cations of charge +q, the radius of the spherical shell of compensating charge is fixed at the average radius of the q outermost occupied Kohn-Sham orbitals of the neutral atom, and the compensating charge value is fixed at -q.[13] For each reference ion, a series of calculations for different spin states were run and the lowest energy spin state was selected. For anions, this required independent optimization of the charge compensation radius and compensating charge value for each spin state. Wavefunction stability analysis was performed on each reference ion (GAUSSIAN 09 keyword stability = opt) to make sure the lowest energy ground state was obtained. An ultrafine pruned grid with 99 radial shells or finer grid (e.g., unpruned grid with 300 radial shells) with 590 angular points per radial shell was used for integrating the exchange-correlation potential.

Table 8 lists explicitly computed reference ions for each element. Some of these reference ions were computed by Manz and Sholl.[13, 14] We have extended and refined the list of explicitly computed reference ions to compile a complete set for all elements atomic number 1 to 109. In Table 8, the lower and upper bounds were set to include oxidation states -2 to +2 for every element, plus all oxidation states from the minimum to the maximum known oxidation states for each element, plus one oxidation state more positive than the maximum known oxidation state for each element if the maximum known oxidation state was ≤ +8, plus one oxidation state more negative than the minimum known oxidation state for each element. The list of known oxidation states for each element was taken from Wikipedia.org on 28 October 2014,[97] which includes numerous additions and corrections to the table of Greenwood and Earnshaw.[98] Since no data was available for element 109, we set its range of explicitly computed reference ions (-2 to 9) similar to element 108. The lower ( $q_A^{lower}$ ) and upper ( $q_A^{upper}$ ) bounds for each of these elements will not be modified upon the discovery of new oxidation states for any of these elements, because the set of explicitly computed reference ions provided here already covers a wide range and NACs are typically less extreme than oxidation states. In Section 3.3 below, we provide an algorithm for implicitly extending this library to all possible charge states without requiring explicit computation of any more reference ions.

Table 8. Explicitly computed reference ions.

| Atomic number | Atomic symbol | Lower bound | Upper bound | Atomic number | Atomic symbol | Lower bound | Upper bound | Atomic number | Atomic symbol | Lower bound | Upper bound |
|---|---|---|---|---|---|---|---|---|---|---|---|
| 1 | H | -2 | 1 | 38 | Sr | -2 | 3 | 75 | Re | -4 | 8 |
| 2 | He | -2 | 2 | 39 | Y | -2 | 4 | 76 | Os | -3 | 9 |
| 3 | Li | -2 | 2 | 40 | Zr | -2 | 5 | 77 | Ir | -4 | 9 |
| 4 | Be | -2 | 3 | 41 | Nb | -2 | 6 | 78 | Pt | -3 | 7 |
| 5 | B | -2 | 4 | 42 | Mo | -3 | 7 | 79 | Au | -2 | 6 |
| 6 | C | -5 | 5 | 43 | Tc | -4 | 8 | 80 | Hg | -2 | 5 |
| 7 | N | -4 | 6 | 44 | Ru | -3 | 9 | 81 | Tl | -2 | 4 |
| 8 | O | -3 | 3 | 45 | Rh | -2 | 7 | 82 | Pb | -5 | 5 |
| 9 | F | -2 | 2 | 46 | Pd | -2 | 7 | 83 | Bi | -4 | 6 |
| 10 | Ne | -2 | 2 | 47 | Ag | -2 | 5 | 84 | Po | -3 | 7 |
| 11 | Na | -2 | 2 | 48 | Cd | -2 | 3 | 85 | At | -2 | 8 |
| 12 | Mg | -2 | 3 | 49 | In | -2 | 4 | 86 | Rn | -2 | 7 |
| 13 | Al | -2 | 4 | 50 | Sn | -5 | 5 | 87 | Fr | -2 | 2 |
| 14 | Si | -5 | 5 | 51 | Sb | -4 | 6 | 88 | Ra | -2 | 3 |
| 15 | P | -4 | 6 | 52 | Te | -3 | 7 | 89 | Ac | -2 | 4 |
| 16 | S | -3 | 7 | 53 | I | -2 | 8 | 90 | Th | -2 | 5 |
| 17 | Cl | -2 | 8 | 54 | Xe | -2 | 9 | 91 | Pa | -2 | 6 |
| 18 | Ar | -2 | 2 | 55 | Cs | -2 | 2 | 92 | U | -2 | 7 |
| 19 | K | -2 | 2 | 56 | Ba | -2 | 3 | 93 | Np | -2 | 8 |
| 20 | Ca | -2 | 3 | 57 | La | -2 | 4 | 94 | Pu | -2 | 9 |
| 21 | Sc | -2 | 4 | 58 | Ce | -2 | 5 | 95 | Am | -2 | 8 |
| 22 | Ti | -2 | 5 | 59 | Pr | -2 | 5 | 96 | Cm | -2 | 9 |
| 23 | V | -2 | 6 | 60 | Nd | -2 | 5 | 97 | Bk | -2 | 5 |
| 24 | Cr | -3 | 7 | 61 | Pm | -2 | 4 | 98 | Cf | -2 | 5 |
| 25 | Mn | -4 | 8 | 62 | Sm | -2 | 4 | 99 | Es | -2 | 5 |
| 26 | Fe | -3 | 7 | 63 | Eu | -2 | 4 | 100 | Fm | -2 | 4 |
| 27 | Co | -2 | 6 | 64 | Gd | -2 | 4 | 101 | Md | -2 | 4 |
| 28 | Ni | -2 | 5 | 65 | Tb | -2 | 5 | 102 | No | -2 | 4 |
| 29 | Cu | -2 | 5 | 66 | Dy | -2 | 5 | 103 | Lr | -2 | 4 |
| 30 | Zn | -2 | 3 | 67 | Ho | -2 | 4 | 104 | Rf | -2 | 5 |
| 31 | Ga | -2 | 4 | 68 | Er | -2 | 4 | 105 | Db | -2 | 6 |
| 32 | Ge | -5 | 5 | 69 | Tm | -2 | 4 | 106 | Sg | -2 | 7 |
| 33 | As | -4 | 6 | 70 | Yb | -2 | 4 | 107 | Bh | -2 | 8 |
| 34 | Se | -3 | 7 | 71 | Lu | -2 | 4 | 108 | Hs | -2 | 9 |
| 35 | Br | -2 | 8 | 72 | Hf | -2 | 5 | 109 | Mt | -2 | 9 |
| 36 | Kr | -2 | 3 | 73 | Ta | -2 | 6 | | | | |
| 37 | Rb | -2 | 2 | 74 | W | -3 | 7 | | | | |

We emphasize that this fixed reference ion library forms part of the definition of the DDEC6 method, and we intend that the DDEC6 method always be used with this fixed reference ion library. This definition is motivated by practical considerations. First, requiring the reference ions to be converged using the same basis set family as employed in the quantum mechanical calculation of the system's total electron density distribution is not optimal. For example, an atom-centered basis set for a large cluster of water molecules (e.g., aug-cc-pvtz) is more complete than the same family of atom-centered basis set (e.g., aug-cc-pvtz) applied to a single isolated atom. Consequently, requiring the aug-cc-pvtz basis set to be used for each isolated reference ion would be requiring the reference ions to be converged using a less complete basis set than that used for the polyatomic system. Therefore, we believe the most appropriate approach is to compute the reference ions near the complete basis set limit.

Second, it is impractical to require the reference ions to be converged using the same exchange-correlation (XC) theory as used in the quantum mechanical calculation of the system's total electron density distribution. Converging the reference ions is a tricky and time-consuming process that is best completed off-line rather than on-the-fly at the beginning of a charge partitioning calculation. Aside from the nearly impossible task of optimizing the reference ions separately for each of the countless different XC theories, there are theoretical motivations for using a single fixed reference ion library. Consider a situation in which the same material M is studied with two different XC theories called XC1 and XC2. Suppose we construct some quantitative measure of similarity between two electron distributions and use it to quantify (a) how similar the electron distribution of the material M computed with XC1 is to the electron distribution of the material M computed with XC2 and (b) how similar a reference ion library (containing the chemical elements in M) computed with XC1 is to a reference ion library computed with XC2. There are three possibilities: (i) XC1 and XC2 give much more similar electron distributions for the reference ions than they do for the material M, (ii) XC1 and XC2 give much more similar electron distributions for the material M than they do for the reference ions, and (iii) XC1 and XC2 give electron distributions of comparable similarity for the material M and the reference ions. In case (i), XC1 and XC2 give similar reference ion densities compared to their different electron distributions for the material M, so in this case we can confidently use the XC1 reference ions to analyze the electron distribution of material M computed via either XC1 or XC2. In case (ii), XC1 and XC2 give similar electron distributions for the material M compared to their different reference ion densities, so in this case the use of two different reference ion sets (i.e., XC1 and XC2) would introduce an artificial difference in NACs for the material that reflects more the change in reference ion sets than the change in the material's electron distribution. Accordingly, in case (ii), we would be better off to choose one of the reference ion sets (e.g., XC1) and use it consistently to analyze the material's electron distribution computed via either XC1 or XC2. In case (iii), where the changes in reference ion densities between XC1 and XC2 are of comparable magnitude to the changes in the material's electron density, it cannot be determined a priori whether using a fixed reference ion library (i.e., using XC1 reference ions to analyze the material's electron distribution computed with either XC1 or XC2) or a variable reference ion library (i.e., using the same XC functional to compute the reference ions as was used to compute the material's electron distribution) is better. While cases (i) and (iii) are not decisive, case (ii) clearly favors using a fixed reference ion library computed with one XC functional irrespective of the XC functional used to compute the material's electron distribution. Therefore, we believe the most appropriate approach is to define the reference ions using a specific XC theory (i.e., PW91) irrespective of the XC theory employed in the quantum mechanical calculation of the system's total electron density distribution.

Table S1 compares the calculated to the experimental ground spin states for the neutral atoms. With seven exceptions (Ti, V, Zr, Ce, W, Pt, and Cm) in the transition metals having closely spaced energy levels, the PW91 method reproduces the correct ground spin states. (A similar comparison is not possible for the charged ions. Because the computed reference ions include charge compensation effects, they cannot be compared to experiments on isolated ions.) During DDEC6 charge and spin partitioning, the spin densities of the reference ions is not utilized as input. Therefore, the particular spin state of each reference ion matters only to the extent that it effects the spherically averaged electron distribution of each reference ion. Table S2 compares the second radial moment of the reference ion electron distribution (aka 'electronic spatial extent') for each of these seven elements in the +1, 0, and -1 charge states. For the neutral atoms (i.e., 0 charge state), the electronic spatial extents are listed both for the experimental and PW91 low energy spin states. All calculations were performed using the PW91 method near the complete basis set limit as described above. The +1 and -1 charge states used the charge compensation scheme described above. Examining Table S2, the electronic spatial extents for the experimental and PW91 spin states of each element were more similar to each other than to either the +1 or -1 charge states. Because the difference in electronic spatial extent between the experimental and PW91 spin states was comparatively small, the reference ions computed using the PW91 low energy spin states are a reasonable approximation.

As demonstrated by the extensive results in this article, the overall performance of the PW91 reference ions is superb within the DDEC6 method. However, we do sympathize with the desire to remove the density functional approximation and compute the reference ions exactly. We do not believe it would be a useful activity to repeat the calculation of the reference ions using other density functional approximations. Rather, we believe the best path forward is to use the PW91 reference ions until it is possible to circumvent density functional approximations altogether and compute the reference ions nearly exactly using high-level all-electron relativistic (including spin-orbit coupling) multi-reference configuration interaction (or similar) methods near the complete basis set limit using a similar charge-compensation model with wave-function stability analysis.

### 3.2 Core reference densities

DDEC analysis is always performed using an effective all-electron density. When the quantum mechanical computation of a material's electron distribution is done using effective core potentials (pseudo-potentials) in place of some core electrons, reference core densities are used to add these missing core electrons back into the system at the start of DDEC analysis. For each atom in the material, the reference core density with the same number of core electrons substituted by the pseudo-potential is added to the pseudo-valence density output from the quantum chemistry calculation to account for all electrons in each atom. For calculations using frozen core electrons rather than pseudo-potentials, both the frozen core electrons and the valence electrons are always included during DDEC analysis. The projector augmented wave (PAW) method is an example of an all-electron frozen core method.[99, 100] Finally, DDEC analysis can be performed on all-electron densities in which all electrons (core and valence) were fully relaxed during the quantum chemistry calculation.

In cases where the frozen or missing core electron densities are available from the quantum chemistry program (e.g., PAW calculations from VASP or GAUSSIAN 09 generated wfx files), these are used to construct the total electron distribution, $\rho(\vec{r})$. In other cases, the reference core densities described below are used.

Table 9 lists the reference core densities available for each element. These were extracted from the corresponding Kohn-Sham orbitals of each element's neutral reference atom. The same neutral reference atom was used here as described in Section 3.1 above. For each element, the average radius

$$\langle r \rangle_i = \oint \varphi_i{}^*(\vec{r}) \varphi_i(\vec{r}) r d^3\vec{r} \qquad (119)$$

of each Kohn-Sham orbital in the neutral reference atom was computed. The sets of core reference densities for each element were then assembled by starting with the closed subshell having smallest $\langle r \rangle_i$ (i.e., the 1s subshell) and successively adding closed subshells having the next larger $\langle r \rangle_i$. In this manner, the Kohn-Sham orbitals not included in the reference core density always had larger $\langle r \rangle_i$ than those included in the reference core density. This process was used to generate the full set of reference core densities for each element. For example, the Pt element contains the follow sets of core electrons: 2, 4 , 10 , 12 , 18 , 28 , 30 , 36 , 46 , 60 , 62 , and 68 which are built up by including successive subshells in the series $1s^2 2s^2 2p^6 3s^2 3p^6 3d^{10} 4s^2 4p^6 4d^{10} 4f^{14} 5s^2 5p^6$. Interestingly, the core electron filling order is not identical for some elements. For example, Cs and Ba (which have no 4f electrons) fill the $5s^2$ core electrons immediately after the $4d^{10}$ electrons, while elements with atomic number ≥ 70 (which have a filled $4f^{14}$ subshell) fill the $4f^{14}$ core electrons immediately after the $4d^{10}$ electrons and before the $5s^2$ core electrons. This is because the $4d^{10}$ electrons have a smaller average radius than the $5s^2$ electrons. Partially filled subshells were always treated as valence electrons. Thus, for atomic numbers 57–69, neither any of the 4f electrons nor the $5s^2$ or $5p^6$ subshells were included as core electrons, because the 4f subshell was only partially filled and the $5s^2$ and $5p^6$ orbitals had a larger average radius than the 4f valence orbitals. A similar situation occurs for elements 89–101, which have a partially filled 5f subshell.

Table 9. List of reference core densities available (954 total).

| Atomic number | Atomic symbol | Core electrons |
|---|---|---|
| 1–2 | H,He | – |
| 3–10 | Li,Be,B,C,N,O,F,Ne | 2 |
| 11–18 | Na,Mg,Al,Si,P,S,Cl,Ar | 2,4,10 |
| 19–30 | K,Ca,Sc,Ti,V,Cr,Mn,Fe,Co,Ni,Cu,Zn | 2,4,10,12,18 |
| 31–36 | Ga,Ge,As,Se,Br,Kr | 2,4,10,12,18,28 |
| 37–48 | Rb,Sr,Y,Zr,Nb,Mo,Tc,Ru,Rh,Pd,Ag,Cd | 2,4,10,12,18,28,30,36 |
| 49–54 | In,Sn,Sb,Te,I,Xe | 2,4,10,12,18,28,30,36,46 |
| 55–56 | Cs,Ba | 2,4,10,12,18,28,30,36,46,48,54 |
| 57–69 | La,Ce,Pr,Nd,Pm,Sm,Eu,Gd,Tb,Dy,Ho,Er,Tm | 2,4,10,12,18,28,30,36,46 |
| 70–80 | Yb,Lu,Hf,Ta,W,Re,Os,Ir,Pt,Au,Hg | 2,4,10,12,18,28,30,36,46,60,62,68 |
| 81–86 | Tl,Pb,Bi,Po,At,Rn | 2,4,10,12,18,28,30,36,46,60,62,68,78 |
| 87–88 | Fr,Ra | 2,4,10,12,18,28,30,36,46,60,62,68,78,80,86 |
| 89–101 | Ac,Th,Pa,U,Np,Pu,Am,Cm,Bk,Cf,Es,Fm,Md | 2,4,10,12,18,28,30,36,46,60,62,68,78 |
| 102–109 | No,Lr,Rf,Db,Sg,Bh,Hs,Mt | 2,4,10,12,18,28,30,36,46,60,62,68,78,92,94,100 |

Only the spherically averaged reference ion and reference core densities are utilized in DDEC analysis. Therefore, these reference densities can be expressed in the form:

$$\rho_A^{ref}(r_A) = \sum_{p=0}^{3}\sum_{i=1}^{210} C_{p,i} {r_A}^{2p} e^{-\alpha_i^{pair} {r_A}^2} \qquad (120)$$

The summation from i = 1 to 210 represents the number of distinct overlap pairs that can be constructed from the 20 different α values listed in Table 7:

$$\text{number of } \alpha^{pair} = \frac{(20)(20+1)}{2} = 210 \qquad (121)$$

The summation from p = 0 to 3 occurs, because the basis set contains s to f functions (Table 7). The product of two s basis functions generates a p = 0 term, and the product of two f basis functions generates a p = 3 term. Only the even (i.e., 2p) powers of $r_A$ occur in Eq. (120), because the product of two basis functions generating an odd power (e.g., product of s and p basis functions) always spherically averages to zero. The $\{C_{p,i}\}$ values for each reference ion and reference core density were computed using an in-house

Fortran program that analyzed the first-order density matrices of the reference ions generated by GAUSSIAN 09.

Tests were performed to make sure these reference core densities were computed correctly. We checked that:

a) Each reference core density integrated to the correct number of core electrons:

$$n_A^{ref} = \int_0^{\infty} \rho_A^{ref}(r_A) 4\pi (r_A)^2 dr_A \qquad (122)$$

b) The average radius

$$r_m^{avg} = \sqrt[m]{\frac{\langle r^m \rangle}{n^{ref}}} \qquad (123)$$

based on the m$^{th}$ radial moment

$$\langle r^m \rangle = \int_0^{\infty} \rho_A^{ref}(r_A) 4\pi (r_A)^{2+m} dr_A \qquad (124)$$

increased with increasing m = -1, 1, 2, 3.

c) For each element, the root-mean-squared radius $r_{rms}^{core} = r_2^{avg}$ of the core reference density increased with increasing number of core electrons.

**3.3 Constraints applied to the reference ion densities**

We wrote a program to check whether the DFT-computed reference ion densities are monotonically decreasing with increasing radius. The program compared the spherically averaged reference ion densities on 100 radial shells equally spaced between 0 and 5 Å. All of the 944 reference ion densities were already monotonically decreasing for density values ≥ $10^{-3}$ e/bohr$^3$. However, a few exceptions occurred for smaller density values. Five reference ions ($He^{-2}$, $He^{-1}$, $Ne^{-2}$, $Ne^{-1}$, $Ar^{-2}$) had monotonicity exceptions for densities ≥ $10^{-4}$ e/bohr$^3$; these are light noble gases that do not want to be in anionic states. No change in the number of exceptions occurred for densities ≥ $10^{-6}$ e/bohr$^3$. Eight additional reference ions ($Cl^{+6}$, $Ru^{+8}$, $Te^{+4}$, $Te^{+5}$, $I^{+5}$, $I^{+6}$, $Os^{+8}$, and $U^{+5}$) had monotonicity exceptions for densities ≥ $10^{-7}$ e/bohr$^3$; these are all highly charged cations. The number of exceptions increased for smaller density values, with a total of 109 individual reference ions exhibiting monotonicity exceptions for densities ≥ $10^{-10}$ e/bohr$^3$; all of these exceptions were for anions and cations. None of the neutral atoms exhibited any monotonicity exceptions at all for density values all the way out to the 5 Å cutoff radius. Our results are consistent with the general belief that isolated neutral atoms have a monotonically decreasing spherically averaged electron density near the complete basis set limit.[101, 102]

A second property of reference ions deserves special consideration. At first one might expect the electron density will not increase anywhere if electrons are removed from the system. However, multi-reference single and double-excitation configuration interaction calculations with a nearly complete basis set showed the $Ne^{+1}$ atom has a slightly higher electron density near its nucleus than the neutral Ne atom does.[103] Our PW91 calculations with a charge-compensated $Ne^{+1}$ ion show the same trend of a slightly higher electron density near its nucleus than for the neutral Ne atom. Similar situations occur for many of the other reference ions. To understand how this unusual feature of reference ions could affect results, we examine the optimization landscape curvature for iterative Hirshfeld-like partitioning (Eq. (34)). When adding an electron to the reference ion leads to a *decrease* in the electron density near its nucleus, this decrease in electron count near the nucleus will have to be offset by a corresponding increase in electron count farther from the nucleus where the electron density is smaller. For example, if adding an electron to the reference ion decreases the electron count near the nucleus by $\varepsilon$ electrons, then the number of electrons farther from the nucleus will have to increase by $1+\varepsilon$. Because the second term in Eq. (34) is proportional to $-\delta\rho_A^{ref}\left(r_A, q_A^{ref}\right)/\rho_A^{ref}\left(r_A, q_A^{ref}\right)$, moving added electrons away from the nucleus makes this contribution to the curvature more negative. In extreme cases, this could potentially make the optimization landscape curvature nearly zero (i.e., nearly flat optimization landscape) or even negative.

A positive optimization landscape curvature is desirable to facilitate convergence. The optimization landscape curvature can be improved by smoothing the reference ion densities. Here, $\rho_A^{ref}\left(r_A, q_A^{ref}\right)$ represents the quantum-mechanical computed (i.e., pre-smoothed) reference ion density, and $\bar{\rho}_A^{ref}\left(r_A, q_A^{ref}\right)$ represents the smoothed reference ion density. The reference ions are smoothed by applying the following series of constraints.[14] Both the pre-smoothed and smoothed reference ion densities integrate to the correct number of electrons:

$$n_A^{ref} = z_A - q_A^{ref} = \int_0^{r_{cutoff}} \bar{\rho}_A^{ref}\left(r_A, q_A^{ref}\right) 4\pi\left(r_A\right)^2 dr_A . \qquad (125)$$

The smoothed reference ion density decreases monotonically with increasing $r_A$

$$\frac{\partial \bar{\rho}_A^{ref}\left(r_A, q_A^{ref}\right)}{\partial r_A} \leq 0 \qquad (126)$$

where $q_A^{ref}$ is the reference ion charge. The smoothed reference ion density increases as electrons are added to the reference ion:

$$\frac{\partial \bar{\rho}_A^{ref}\left(r_A, q_A^{ref}\right)}{\partial q_A^{ref}} \leq 0 \qquad (127)$$

The electron density added to each smoothed reference ion with decreasing $q_A^{ref}$ decreases monotonically with increasing $r_A$:

$$\frac{\partial^2 \bar{\rho}_A^{ref}\left(r_A, q_A^{ref}\right)}{\partial q_A^{ref} \partial r_A} \geq 0 \qquad (128)$$

Without constraint (128), it would be possible to add electrons in the function $\bar{\rho}_A^{ref}\left(r_A, q_A^{ref}\right)$ only to large $r_A$ values as $q_A^{ref}$ decreases, which might correspond to the buried tail regions of atoms in dense materials. As discussed above, this would add large negative terms to the optimization landscape curvature that would be detrimental to convergence stability. Constraint (128) ensures that smoothed reference ion density changes also occur at smaller $r_A$ values as $q_A^{ref}$ varies, and this ensures the smoothed reference ion changes also affect regions near the atomic nuclei. Moving some of the added electrons nearer the nucleus where $\bar{\rho}_A^{ref}\left(r_A, q_A^{ref}\right)$ is larger increases the optimization landscape curvature and hence the convergence stability. These constraints are analogous to those used in the DDEC3 method,[14] except the notation has been changed here to reflect the fact that the reference ion charge ( $q_A^{ref}$ ) is not the same as the AIM charge ( $q_A$ ) in the DDEC6 method. To avoid division by zero errors, we also constrained

$$\bar{\rho}_A^{ref}\left(r_A, q_A^{ref}\right) \geq 10^{-16} \text{ atomic units} \qquad (129)$$

for $r_A \leq r_{cutoff}$ .

These constraints were enforced as following. First, we corrected the electron density of $\rho_A^{ref}\left(r_A, q_A^{ref}\right)$ in the first radial shell (i.e., the radial shell with smallest $r_A$) to give the correct total number of electrons, where the volume of each radial shell was computed analytically. Then, we initialized the smoothed reference ion density with the estimate $\bar{\rho}_A^{ref}\left(r_A, q_A^{ref}\right) = \rho_A^{ref}\left(r_A, q_A^{ref}\right)$. Then, we enforced constraint (126) by recursively setting

$$\bar{\rho}_A^{ref}\left(r_A, q_A^{ref}\right) = \min\left(\bar{\rho}_A^{ref}\left(r_A - \Delta r_A, q_A^{ref}\right), \bar{\rho}_A^{ref}\left(r_A, q_A^{ref}\right)\right) \qquad (130)$$

starting with the second radial shell and continuing outward to the last radial shell. (In general, the distance between adjacent radial shells, $\Delta r_A > 0$, could be different for different radial shells, but for simplicity we utilized a constant $\Delta r_A$ .) Then $\bar{\rho}_A^{ref}\left(r_A, q_A^{ref}\right)$ was normalized to ensure it integrated to the correct total number of electrons, $n_A^{ref} = z_A - q_A^{ref}$ , subject to constraint (129). Constraints (127) and (128) were then enforced according to a previously published procedure: "The density of the neutral atom is unchanged during this process. (i) First, density is added to the -1 anion or removed from the +1 cation as necessary to make the density difference with respect to the neutral atom monotonically decreasing with increasing $r_A$. (ii) Then, the [smoothed] reference ion densities are normalized .... Steps (i) and (ii) are repeated until [the smoothed reference density] converges for the -1 and +1 ions. After the smoothed -1 anion and +1 cation reference densities are determined, a similar process is applied to the -2 and +2 ions. Specifically, density is added to (removed from) to the -2 anion (+2 cation) to make the density difference with respect to the -1 anion (+1 cation) monotonically decreasing with increasing $r_A$. The -2 and +2 ion densities are then normalized .... These last two steps are repeated until [the smoothed reference density] converges for the -2 and +2 ions. Next, the process is applied to the -3 and +3 ions, and so forth until all of the reference densities have been smoothed ...." (Manz and Sholl,[14] page S3)

In the rare event $q_A^{ref}$ exceeds the range of explicitly computed reference ions listed in Table 8 for that element, the smoothed reference ion density is computed by scaling proportional to the number of electrons

$$\bar{\rho}_A^{ref}\left(r_A, q_A^{ref} > q_A^{upper}\right) = \left(\frac{z_A - q_A^{ref}}{z_A - q_A^{upper}}\right)\bar{\rho}_A^{ref}\left(r_A, q_A^{upper}\right) \qquad (131)$$

$$\bar{\rho}_A^{ref}\left(r_A, q_A^{ref} < q_A^{lower}\right) = \left(\frac{z_A - q_A^{ref}}{z_A - q_A^{lower}}\right)\bar{\rho}_A^{ref}\left(r_A, q_A^{lower}\right). \qquad (132)$$

Eqs. (131)–(132) ensure the reference ion library is complete in the sense that all needed reference ions are available within the library. There is, therefore, no reason for any DDEC6 charge partitioning calculation to terminate for lack of reference ion availability. The three justifications for Eqs. (131)–(132) are: (A) Cations more positively net charged than $q_A^{upper}$ are likely to be found only as isolated gas-phase atomic ions, and for isolated atoms the assigned AIM electron distribution is independent of the reference ion density. (B) Anions more negatively net charged than $q_A^{lower}$ are unlikely to be found, because for free atomic ions these electrons would be unbound and in compounds these electrons would tend to be unbound or captured by other atoms in the material. (C) Highly positively charged and highly negatively charged buried atoms are likely to approach the limits of the tail constraint (Eq. (85)) that reduces the dependence of $w_A(r_A)$ on the precise form of the reference ion densities. Eqs. (131)–(132) automatically satisfy constraints (125)–(128).

In practice, the above constraints are first applied to integer values of $q_A^{ref}$. For non-integer $q_A^{ref}$, linear interpolation between the two nearest integers is used:

$$\bar{\rho}_A^{ref}\left(r_A, q_A^{ref}\right) = \left(1 + q_A^{floor} - q_A^{ref}\right)\bar{\rho}_A^{ref}\left(r_A, q_A^{floor}\right) + \left(q_A^{ref} - q_A^{floor}\right)\bar{\rho}_A^{ref}\left(r_A, q_A^{floor} + 1\right) \quad (133)$$

$$q_A^{floor} = \text{floor}\left(q_A^{ref}\right) \quad (134)$$

This linear interpolation ensures that constraints (125)–(129), (131), and (132) also hold for non-integer $q_A^{ref}$.

## 4. Computational details

### 4.1 Quantum chemistry calculations

We performed periodic quantum chemistry calculations using VASP[104, 105] software. Our VASP calculations used the projector augmented wave (PAW) method[99, 100] to perform all-electron frozen-core calculations including scalar relativistic effects with a plane-wave basis set cutoff energy of 400 eV. Calculations specifying “2 frozen Na core electrons” or “10 frozen Na core electrons” used PAWs for the Na atom including 2 or 10 frozen core electrons, respectively. For all systems, the number of k-points times the unit cell volume exceeded 4000 $Å^3$. This is enough k-points to converge relevant properties including geometries and AIM properties (NACs, ASMs, etc.). Except for the solid surfaces in Section 5.7, geometry optimizations relaxed both the unit cell vectors and ionic positions. The solid surface calculations used the DFT-optimized bulk lattice vectors and relaxed the ionic positions. Where noted, experimental crystal structures or other geometries from the published literature were used. A Prec=Accurate (~0.14 bohr) electron density grid spacing was used. Bader NACs were computed using the program of Henkelman and coworkers.[56]

We performed non-periodic quantum chemistry calculations using GAUSSIAN 09[106] software. ESP NACs were computed in GAUSSIAN 09 using the Merz-Singh-Kollman scheme.[17, 93]

### 4.2 Ewald summation

In the periodic materials, the Ewald summation method of Smith, including NACs and (optionally) atomic dipoles, was used to compute electrostatic potentials for RMSE calculations.[107] This Ewald summation separates the Coulomb potential into a short-range portion summed in real space and a long-range portion summed in reciprocal space:

$$V(\vec{r}) = V^{short-range}(\vec{r}) + V^{long-range}(\vec{r}) \quad (135)$$

$$V^{\text{short-range}}(\vec{r}) = \sum_{A} \sum_{L_1=-L_1^{max}}^{L_1^{max}} \sum_{L_2=-L_2^{max}}^{L_2^{max}} \sum_{L_3=-L_3^{max}}^{L_3^{max}} \left[ \frac{q_A \operatorname{erfc}(\alpha_E r_A)}{r_A} + (\vec{\mu}_A \bullet \vec{r}_A)\left( \frac{\operatorname{erfc}(\alpha_E r_A)}{(r_A)^3} + \frac{2\alpha_E \exp\left(-(\alpha_E r_A)^2\right)}{\sqrt{\pi}(r_A)^2} \right) \right]$$

(136)

We set the Ewald summation convergence parameter to $\alpha_E = \sqrt{\pi}/(10\,\text{Å})$. Enough real space replications of the unit cell were included such that every point in the reference unit cell was surrounded by at least $3/\alpha_E$ real space distance in all directions:

$$L_i^{max} = \operatorname{ceil}\left( \frac{3}{\alpha_E} \max_{h \neq i} \left( \sqrt{\frac{(\vec{v}_h \bullet \vec{v}_h)}{(\vec{v}_i \bullet \vec{v}_i)(\vec{v}_h \bullet \vec{v}_h) - (\vec{v}_i \bullet \vec{v}_h)^2}} \right) \right). \quad (137)$$

This corresponds to an $\operatorname{erfc}(\alpha_E r_A)/r_A$ cutoff of $(\alpha_E/3)\operatorname{erfc}(3) = 6.9\times10^{-7}$ bohr$^{-1}$. The reciprocal lattice vectors are defined by

$$\vec{u}_i = 2\pi \frac{(\vec{v}_h \times \vec{v}_j)}{\vec{v}_i \bullet (\vec{v}_h \times \vec{v}_j)}, \; h \neq i \neq j\;. \quad (138)$$

The reciprocal space summation encompassed integer multiples $b_i$ of the corresponding reciprocal lattice vectors

$$\vec{k} = b_1\vec{u}_1 + b_2\vec{u}_2 + b_3\vec{u}_3 \quad (139)$$

$$k^2 = \vec{k} \bullet \vec{k} \quad (140)$$

to yield the long-range portion of the electrostatic potential

$$V^{\text{long-range}}(\vec{r}) = \sum_{A} \sum_{b_1=-b_1^{max}}^{b_1^{max}} \sum_{b_2=-b_2^{max}}^{b_2^{max}} \sum_{b_3=-b_3^{max}}^{b_3^{max}} \left[ 4\pi \frac{\exp\left(\frac{-k^2}{4\alpha_E^2}\right)}{V_{unit_cell} k^2} \left( q_A \cos(\vec{k} \bullet \vec{r}_A) + (\vec{\mu}_A \bullet \vec{k}) \sin(\vec{k} \bullet \vec{r}_A) \right) \right].$$

(141)

where $\vec{r}_A$ in Eq. (141) is computed using $(L_1, L_2, L_3) = (0,0,0)$. The term $(b_1, b_2, b_3) = (0,0,0)$ is excluded from the sum in Eq. (141). $V_{unit_cell}$ is the unit cell volume. Our reciprocal space cutoff

$$b_i^{max} = \operatorname{ceil}\left( 4\sqrt{\pi}\alpha_E \max_{h \neq i} \left( \sqrt{\frac{(\vec{u}_h \bullet \vec{u}_h)}{(\vec{u}_i \bullet \vec{u}_i)(\vec{u}_h \bullet \vec{u}_h) - (\vec{u}_i \bullet \vec{u}_h)^2}} \right) \right) \quad (142)$$

includes at least all reciprocal space vectors having $0 < |\vec{k}| \leq 4\sqrt{\pi}\alpha_E$. Noting that each term in the reciprocal space term includes $\exp\left(-k^2/(4\alpha_E^2)\right)$ as a multiplier, our reciprocal space cutoff corresponds to $\exp\left(-k^2/(4\alpha_E^2)\right) \approx \exp(-4\pi) \approx 3.5\times10^{-6}$. Because it is a short-range effect, spherical charge penetration can be included entirely in the real space summation using the analytic potential of Eq. (118) . While the spherical cloud penetration effect is small over grid points used to compute RMSE, it becomes increasingly important for smaller $r_A$ values.

## 5. Results and discussion

A diverse materials set was carefully selected to evaluate the accuracy of our new charge partitioning method. To test whether the DDEC6 method consistently performs better than the DDEC3 method, we included many systems for which the DDEC3 method was originally tested.[14] In addition, we study many new materials carefully selected for their ability to make falsifiable tests of a charge assignment method's ability to describe electron transfer: (a) compressed sodium chloride crystals of unusual stoichiometries, (b) Li-containing and dilithiated transition metal oxide and sulfide crystals,[26] (c) endohedral fullerenes, (d) Ti-containing solids, and (e) electrostatic potential comparisons across a wider range of small molecules, a large biomolecule, and porous solids. One of the most frequent concerns about charge assignment methods is that it is difficult to compare them directly to experimental data. Therefore, we included many materials having strong experimental data. These comparisons to experimental data allow our Results and Discussion to be viewed not only as applications of the DDEC6 method but also as performance tests.

### 5.1 Compressed sodium chloride crystals with unusual stoichiometries

We were first motivated to improve upon the DDEC3 method by a series of calculations on sodium chloride crystals having unusual stoichiometries. Specifically, we computed NACs for the ten high-pressure crystal structures reported by Zhang et al.[108] and the ambient-pressure NaCl structure shown in Figure 6. We generated the electron density in VASP using the PBE[109] functional and (a) the experimental x-ray diffraction geometries[108] for the ten high-pressure crystals and (b) the PBE-optimized geometry for the ambient-pressure Fm3m-NaCl. As shown in Table 10 and Figure **7**, the DDEC3 NAC for at least one Na atom is larger than +1.0 for the following cases: (a) 1.311 for Na(3) atoms in Cmmm-$Na_2Cl$ crystal at 180 GPa, (b) 1.235 for Na(1) and 1.295 for Na(2) atoms in Cmmm-$Na_3Cl_2$ crystal at 280 GPa, (c) 1.219 for Na atoms in Imma-$Na_2Cl$ crystal at 300 GPa, (d) 1.035 for Na(1) and 1.108 for Na(2) atoms in P4/m-$Na_3Cl_2$ crystal at 140 GPa, (e) 1.063 for Na(1) atoms in P4/mmm-$Na_2Cl$ crystal at 120 GPa, (f) 1.078 for Na(1) atoms in Pm3-$NaCl_7$ crystal at 200 GPa, (g) 1.136 for Na atoms in Pm3m-NaCl crystal at 140 GPa, (h) 1.140 for Na atoms in Pm3n-$NaCl_3$ crystal at 200 GPa, and (i) 1.011 for Na atoms in Pnma-$NaCl_3$ crystal at 40 GPa. Because a neutral sodium atom has one electron in its valence shell, an AIM-based NAC for a sodium atom in sodium-containing solids should ideally be ≤ +1.0. (Non-AIM-based NACs such as APT, Born effective, and ESP charges are not expected to have this property.) A Na NAC greater than +1.0 would indicate that some electrons from the closed [Ne] core are donated to other atoms, but such a donation should be energetically unfavorable under chemically relevant conditions due to the high ionization energy of closed shell configurations. (For comparison, the ionization energy of a Ne atom is 21.56 eV.[110]) Based on these results, we concluded that the DDEC3 method overestimates atomic charge magnitudes in some materials. If ten Na core electrons are frozen, the DDEC3 NACs for the Na atoms are constrained to be ≤ +1.0 as shown in Table 10, but this is not a satisfactory solution because we want NACs to be approximately independent of the number of frozen core electrons.

This observation led us to explore numerous potential modifications to the DDEC method, which after testing dozens of potential modifications culminated in the DDEC6 method. As shown in Table 10 and Figure 7, the DDEC6 NACs have the expected behavior being ≤ +1.0 for each of the Na atoms. Moreover, the DDEC6 NACs were nearly insensitive to whether 2 or 10 frozen Na core electrons were used.

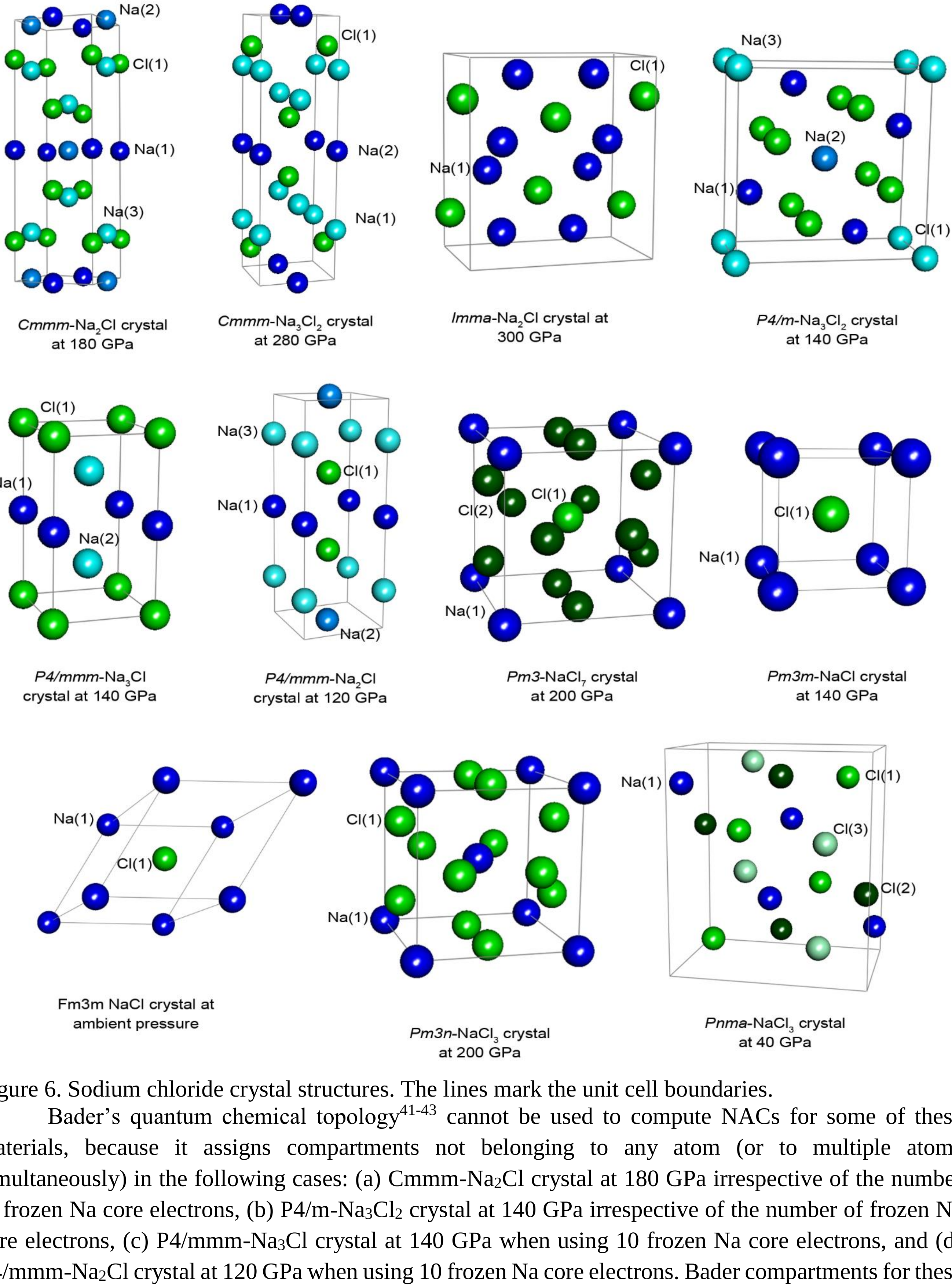

Figure 6. Sodium chloride crystal structures. The lines mark the unit cell boundaries.

Bader's quantum chemical topology[41-43] cannot be used to compute NACs for some of these materials, because it assigns compartments not belonging to any atom (or to multiple atoms simultaneously) in the following cases: (a) Cmmm-$Na_2Cl$ crystal at 180 GPa irrespective of the number of frozen Na core electrons, (b) P4/m-$Na_3Cl_2$ crystal at 140 GPa irrespective of the number of frozen Na core electrons, (c) P4/mmm-$Na_3Cl$ crystal at 140 GPa when using 10 frozen Na core electrons, and (d) P4/mmm-$Na_2Cl$ crystal at 120 GPa when using 10 frozen Na core electrons. Bader compartments for these four materials are detailed in Table 11. As it should be, the assignment of these Bader compartments was based on the full (i.e., valence + (frozen) core) electron density, not simply the valence density or the valence pseudodensity. At first, one might propose each non-nuclear attractor could be assigned to one of the nearby atoms, but this is not satisfactory because in some cases such an assignment cannot be made without destroying the crystalline symmetry. For example, the *P4/mmm*-$Na_2Cl$ crystal at 120 GPa (modeled with 10 frozen Na core electrons) contains one non-nuclear attractor whose closest atoms are the two equivalent Na(3) atoms; therefore, it is impossible to assign this non-nuclear attractor to one of the closest atoms without breaking the crystal symmetry. Alternatively, one could propose to divide the electron density and/or volume of each non-nuclear attractor amongst the nearby atoms in a way that preserves the system's symmetry, but it is not presently clear whether this could be done in a way that preserves most of the important properties of the Virial compartments. Specifically, each of the Bader compartments satisfies the Virial theorem and behaves as an open quantum system, but divided pieces of such compartments may not.[41, 43] It might be possible that divisions of a non-nuclear attractor could be made that satisfy the net zero flux condition (and Virial theorem) over each division volume but not the local zero flux condition in the bounding surfaces, but it is not presently clear whether such a partitioning would always have a unique definition even if constrained to preserve the system's symmetry.

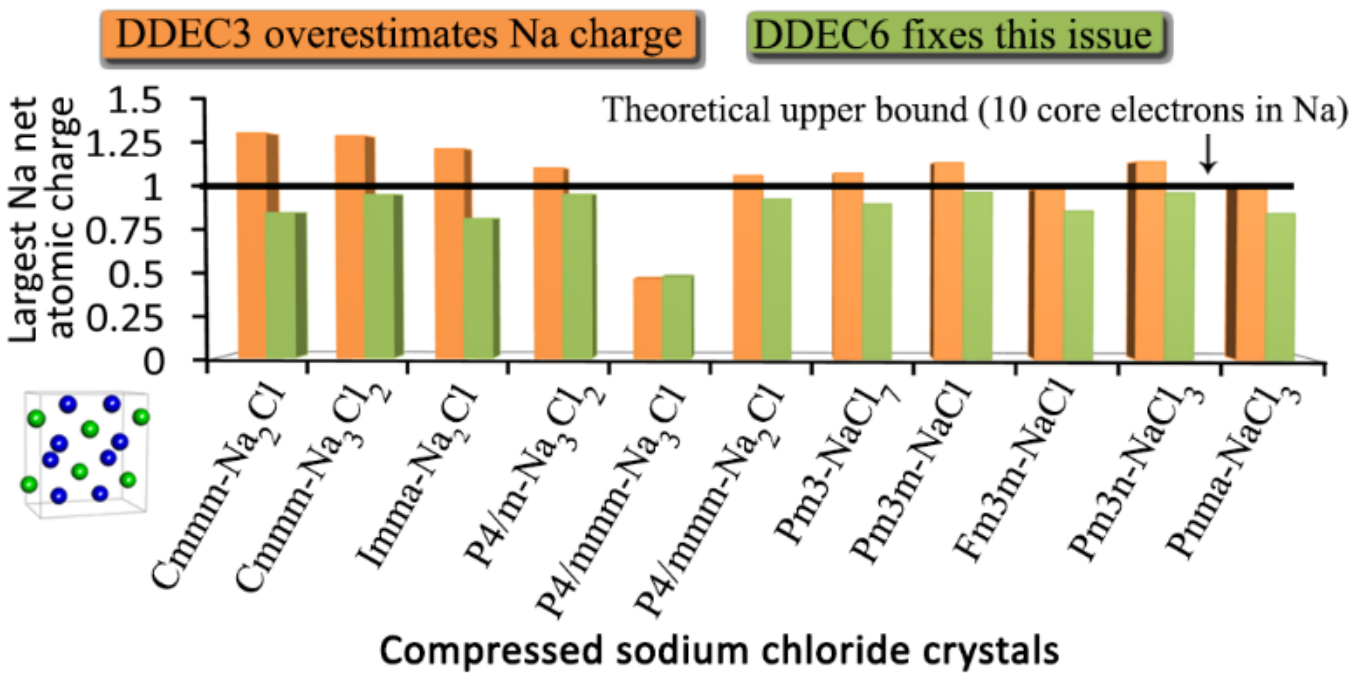

Figure 7: Largest magnitude Na atomic charges in compressed sodium chloride crystals. These were computed using 2 frozen Na core electrons. Based on chemical arguments, at least 10 electrons should be assigned to each Na atom. The DDEC3 method gives many Na atom charges > 1, which indicates some electrons are not assigned to the correct atom. The DDEC6 method fixes this problem.

In materials for which there is a one-to-one correspondence between Bader compartments and atoms (i.e., no non-nuclear attractors), the Bader NACs are computed by integrating the number of electrons over each compartment. In such cases, the Bader method often yields reasonable NACs for dense ionic solids. Examining Table 10, the DDEC6 and Bader NACs using 2 frozen Na core electrons exhibited similar trends for all of the sodium chloride crystals where the Bader NACs were defined. Of particular interest, the Cl NAC was significantly more negative than -1.0 for some of the materials. The Bader NACs were more sensitive than the DDEC6 NACs to whether 2 or 10 frozen Na core electrons were used. For example, in Imma-$Na_2Cl$ crystal at 300 GPa the DDEC6 NAC for the Cl atom was -1.628 (2 frozen Na core electrons) and -1.570 (10 frozen Na core electrons) compared to the Bader Cl NAC of -1.351 (2 frozen Na core electrons) and -0.633 (10 frozen Na core electrons). The reason for this larger sensitivity of the Bader NACs on the number of frozen core electrons is that according to an integration routine now used in popular Bader analysis programs the frozen core electrons are assigned wholly to the host atom while non-frozen electrons crossing into neighboring Bader compartments are divided amongst several atoms.[56] This artifact could be removed by partitioning all electrons (i.e., both frozen and non-frozen) according to their density in each of the Bader compartments, yet even so the sensitivity of the number of Bader compartments on the number of frozen core electrons (e.g., P4/mmm-$Na_3Cl$ crystal at 140 GPa and P4/mmm-$Na_2Cl$ crystal at 120 GPa) would persist. Alternatively, one could choose a small number of frozen core electrons to ensure the amount of frozen core electron density spilling into neighboring compartments is negligible. Consequently, Bader NACs with 2 frozen Na core electrons are more reliable than those with 10 frozen Na core electrons.

Table 10. DDEC and Bader net atomic charges of sodium chloride crystals.

| Atom type | Number of atoms | DDEC3 [a] | DDEC6 [a] | Bader [a] |
|---|---|---|---|---|
| | | Cmmm-$Na_2Cl$ crystal at 180 GPa | | |
| Na(1) | 2 | 0.392 (0.319) | 0.316 (0.334) | [d] |
| Na(2) | 2 | 0.566 (0.592) | 0.547 (0.563) | [d] |
| Na(3) | 4 | 1.311 (1.000) | 0.849 (0.842) | [d] |
| Cl(1) | 4 | -1.790 (-1.455) | -1.281 (-1.290) | [d] |
| | | Cmmm-$Na_3Cl_2$ crystal at 280 GPa | | |
| Na(1) | 2 | 1.235 (1.000) | 0.954 (0.891) | 0.780 (0.560) |
| Na(2) | 4 | 1.295 (1.000) | 0.871 (0.866) | 0.643 (0.291) |
| Cl(1) | 4 | -1.912 (-1.500) | -1.348 (-1.311) | -1.033 (-0.571) |
| | | Imma-$Na_2Cl$ crystal at 300 GPa | | |
| Na(1) | 8 | 1.219 (1.000) | 0.814 (0.785) | 0.676 (0.317) |
| Cl(1) | 4 | -2.439 (-2.000) | -1.628 (-1.570) | -1.351 (-0.633) |
| | | P4/m-$Na_3Cl_2$ crystal at 140 GPa | | |
| Na(1) | 4 | 1.035 (1.000) | 0.808 (0.777) | [d] |
| Na(2) | 1 | 1.108 (1.000) | 0.956 (0.902) | [d] |
| Na(3) | 1 | -0.461 (-0.396) | -0.310 (-0.226) | [d] |
| Cl(1) | 4 | -1.197 (-1.151) | -0.969 (-0.946) | [d] |
| | | P4/mmm-$Na_3Cl$ crystal at 140 GPa | | |
| Na(1) | 1 | -0.237 (0.184) | -0.246 (-0.202) | 0.06 ([d]) |
| Na(2) | 2 | 0.465 (0.466) | 0.477 (0.480) | 0.531 ([d]) |
| Cl(1) | 1 | -0.693 (-0.749) | -0.709 (-0.758) | -1.122 ([d]) |
| | | P4/mmm-$Na_2Cl$ crystal at 120 GPa | | |
| Na(1) | 1 | 1.063 (1.000) | 0.927 (0.889) | 0.756 ([d]) |
| Na(2) | 1 | -0.259 (-0.187) | -0.242 (-0.201) | 0.041 ([d]) |
| Na(3) | 2 | 0.541 (0.503) | 0.487 (0.486) | 0.511 ([d]) |
| Cl(1) | 2 | -0.943 (-0.910) | -0.830 (-0.830) | -0.909 ([d]) |
| | | Pm3-$NaCl_7$ crystal at 200 GPa[b] | | |
| Na(1) | 1 | 1.078 (1.000) | 0.899 (0.874) | 0.883 (0.652) |
| Cl(1) | 1 | 0.297 (0.260) | 0.202 (0.196) | 0.090 (0.088) |
| Cl(2) | 6 | -0.229 (-0.210) | -0.184 (-0.178) | -0.162 (-0.123) |
| | | Pm3m-NaCl crystal at 140 GPa | | |
| Na(1) | 1 | 1.136 (1.000) | 0.966 (0.916) | 0.862 (0.673) |
| Cl(1) | 1 | -1.136 (-1.000) | -0.966 (-0.916) | -0.862 (-0.673) |
| | | NaCl crystal at ambient pressure[c] | | |
| Na(1) | 1 | 0.981 (0.978) | 0.859 (0.848) | 0.840 (0.829) |
| Cl(1) | 1 | -0.981 (-0.978) | -0.859 (-0.848) | -0.840 (-0.829) |
| | | Pm3n-$NaCl_3$ crystal at 200 GPa[b] | | |
| Na(1) | 2 | 1.140 (1.000) | 0.962 (0.909) | 0.913 (0.653) |
| Cl(1) | 6 | -0.380 (-0.333) | -0.321 (-0.303) | -0.304 (-0.218) |
| | | Pnma-$NaCl_3$ crystal at 40 GPa [b] | | |
| Na(1) | 4 | 1.011 (1.000) | 0.842 (0.853) | 0.815 (0.770) |
| Cl(1) | 4 | -0.718 (-0.709) | -0.590 (-0.597) | -0.530 (-0.501) |
| Cl(2) | 4 | 0.105 (0.101) | 0.054 (0.055) | -0.030 (-0.028) |
| Cl(3) | 4 | -0.398 (-0.392) | -0.307 (-0.311) | -0.255 (-0.242) |

[a] Values listed for 2 frozen Na core electrons; values in parentheses for 10 frozen Na core electrons. [b] Similar Bader NACs were reported previously in reference [108]. [c] Na charge of 1.05 computed with IH/R3 all-electron reported previously in reference [63]. [d] Bader NACs cannot be reported because Bader analysis yields more compartments than atoms (see Table 11)

Table 11. Bader compartment populations for crystals with non-nuclear attractors

| Compartment type | Number of compartments | Enclosed atom | Number of electrons [a] |
|---|---|---|---|
| Population for *Cmmm*-$Na_2Cl$ crystal at 180 GPa | | | |
| 1 | 2 | Na(1) | 10.416 (10.425) |
| 2 | 2 | Na(2) | 10.326 (10.437) |
| 3 | 4 | Na(3) | 10.260 (10.637) |
| 4 | 4 | Cl(1) | 18.116 (17.623) |
| 5 | 2 | none | 0.253 (0.308) |
| 6 | 2 | none | 0.252 (0.308) |
| population for *P4/m*-$Na_3Cl_2$ crystal at 140 GPa | | | |
| 1 | 4 | Na(1) | 10.222 (10.392) |
| 2 | 1 | Na(2) | 10.197 (10.340) |
| 3 | 1 | Na(3) | 10.616 (10.543) |
| 4 | 4 | Cl(1) | 17.955 (17.752) |
| 5 | 1 | none | 0.456 (0.544) |
| population for *P4/mmm*-$Na_3Cl$ crystal at 140 GPa | | | |
| 1 | 1 | Na(1) | 10.940 (10.655) |
| 2 | 2 | Na(2) | 10.469 (10.507) |
| 3 | 1 | Cl(1) | 18.122 (17.871) |
| 4 | 4 | none | (0.115) |
| population for *P4/mmm*-$Na_2Cl$ crystal at 120 GPa | | | |
| 1 | 1 | Na(1) | 10.244 (10.326) |
| 2 | 1 | Na(2) | 10.959 (10.655) |
| 3 | 2 | Na(3) | 10.489 (10.486) |
| 4 | 2 | Cl(1) | 17.909 (17.776) |
| 5 | 1 | none | (0.495) |

[a] Values listed for 2 frozen Na core electrons; values in parentheses for 10 frozen Na core electrons.

## 5.2 Representing electron transfer between atoms in dense solids

### 5.2.1 Metal oxides and sulfides

We now study electron transfer between atoms in the dense solids shown in Figure 8. While we were testing modifications of the DDEC method for the sodium chloride crystals, Wang et al.[26] pointed out a related problem with the DDEC3 method. Specifically, when DDEC3 NACs are compared for a series of transition metal oxide solids with and without Li atoms, the DDEC3 NACs on the transition metal atoms exhibit a trend that does not match chemical expectations. As shown in Table 12, the DDEC3 NAC on the Co atom is lower in crystalline $CoO_2$ than in $LiCoO_2$. In contrast, the Bader, CM5, and HD NACs on the Co atom are higher for crystalline $CoO_2$ than for $LiCoO_2$. To assess which trend is correct, Wang et al. plotted isosurfaces of the electron density difference between $CoO_2$ and $LiCoO_2$ using the M06L[111] functional and found a slight increase in electron density around the Co and O atoms upon Li addition to $CoO_2$ to create $LiCoO_2$.[26] Thus, the Bader, CM5, and HD methods predict the correct charge transfer direction between these two materials, but the DDEC3 method predicts the wrong charge transfer direction between these two materials.[26] For reasons clearly explained in prior publications, charge transfer magnitudes predicted by the HD method are usually much too small.[14, 25, 29, 58] In Table 12, we compare NACs computed using the PBE optimized geometries and electron distributions. For DDEC3, the previously reported M06L results are also listed for comparison.[26] As shown in Table 12 (PBE results) and Wang et al.[26] (M06L results), the CM5 and Bader methods predict a decrease of the transition metal NAC upon lithiation for the solids $TiS_2 \rightarrow LiTiS_2$, $LiTi_2O_4 \rightarrow Li_2Ti_2O_4$, $Mn_2O_4 \rightarrow LiMn_2O_4$, while the DDEC3 method predicts an increase for all except $LiTi_2O_4 \rightarrow Li_2Ti_2O_4$. Assuming these materials behavior similar to the $CoO_2$ material, a decrease in the transition metal NAC upon lithiation is chemically expected. Thus, we employed these materials as a test set to evaluate the performance of potential modifications to the DDEC method when developing the DDEC6 method. In addition, we studied the $Li_3RuO_2$ crystal suggested to us by Ayorinde Hassan. Charge partitioning for the $Li_3RuO_2$ crystal is challenging due to the large proportion of Li atoms and the nearly neutral Ru atoms, because the neutral Li and Ru reference atoms are much more diffuse than the cationic ones leading to large sensitivity of the reference ion densities on the reference ion charges. As shown in Table 12, the DDEC6 algorithm yields reasonable NACs for all of these materials. Only for $TiS_2 \rightarrow LiTiS_2$ is there a small increase from 1.32 to 1.38 in the transition metal DDEC6 NAC upon lithiation. For all of these materials, the Bader and DDEC6 methods give similar Li NACs, while the CM5 and HD methods gave substantially smaller Li NACs.

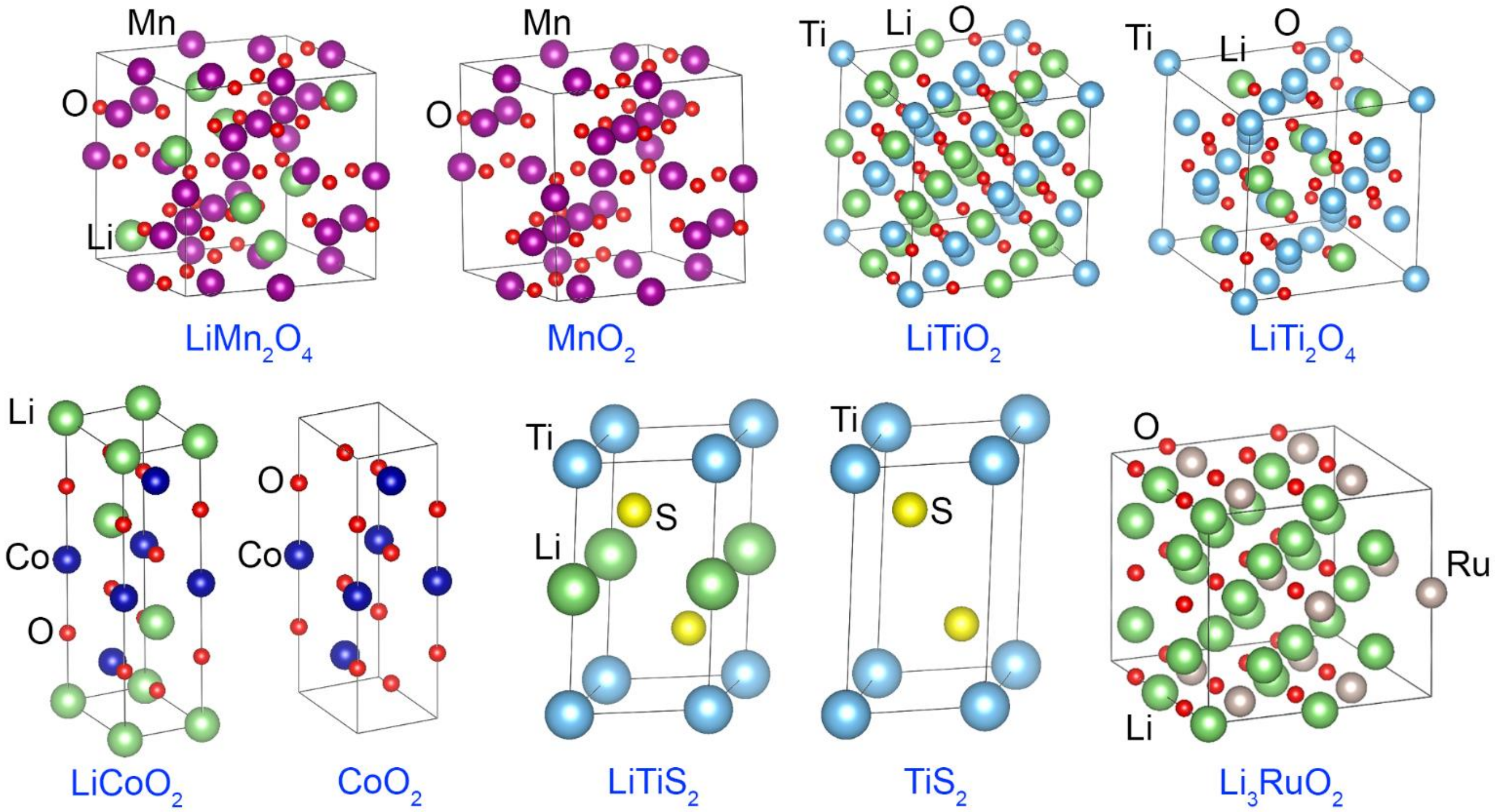

Figure 8: Unit cells used to model metal oxide and sulfide solids. The lines mark the unit cell boundaries. Atoms are colored by element: Li (green), O (red), S (yellow), Ti (light blue), Co (dark blue), Mn (magenta), Ru (beige).

Table 12. Average HD, CM5, DDEC3, DDEC6, and Bader charges of Li, transition metal (TM), and nonmetal atoms. NACs shown are for the PBE optimized geometries and electron densities.

| | HD | | | CM5 | | | DDEC3 | | | DDEC6 | | | Bader | | |
|---|---|---|---|---|---|---|---|---|---|---|---|---|---|---|---|
| crystal | Li | TM | anion | Li | TM | anion | Li | TM | anion | Li | TM | anion | Li | TM | anion |
| $LiCoO_2$ | 0.11 | 0.34 | -0.23 | 0.49 | 0.73 | -0.61 | 1.03 (1.00[a]) | 1.47 (1.45[a]) | -1.25 (-1.23[a]) | 0.87 | 1.07 | -0.97 | 0.88 | 1.22 | -1.05 |
| $CoO_2$ | — | 0.35 | -0.18 | — | 0.80 | -0.40 | — | 1.14 (1.23[a]) | -0.57 (-0.62[a]) | — | 1.12 | -0.56 | — | 1.39 | -0.69 |
| $LiTiS_2$ | 0.07 | 0.40 | -0.23 | 0.27 | 0.79 | -0.53 | 0.98 (0.97[a]) | 1.67 (1.48[a]) | -1.33 (-1.23[a]) | 0.86 | 1.38 | -1.12 | 0.89 | 1.48 | -1.18 |
| $TiS_2$ | — | 0.43 | -0.21 | — | 0.86 | -0.43 | — | 1.06 (1.06[a]) | -0.53 (-0.53[a]) | — | 1.32 | -0.66 | — | 1.61 | -0.80 |
| $Li_2Ti_2O_4$ | 0.11 | 0.56 | -0.34 | 0.46 | 1.16 | -0.81 | 1.05 (1.00[a]) | 2.17 (2.10[a]) | -1.61 (-1.55[a]) | 0.89 | 1.65 | -1.27 | 0.89 | 1.57 | -1.23 |
| $LiTi_2O_4$ | 0.16 | 0.64 | -0.36 | 0.48 | 1.31 | -0.78 | 1.03 (1.00[a]) | 2.33 (2.32[a]) | -1.42 (-1.41[a]) | 0.90 | 1.94 | -1.19 | 0.91 | 1.84 | -1.15 |
| $LiMn_2O_4$ [b] | 0.17 | 0.34 | -0.21 | 0.53 | 0.84 | -0.55 | 0.99 (1.00[a]) | 1.56 (1.95[a]) | -1.03 (-1.23[a]) | 0.86 | 1.23 | -0.83 | 0.89 | 1.59 | -1.02 |
| $Mn_2O_4$ | — | 0.36 | -0.18 | — | 0.88 | -0.44 | — | 1.24 (1.47[a]) | -0.62 (-0.73[a]) | — | 1.25 | -0.63 | — | 1.69 | -0.85 |
| $Li_3RuO_2$ | 0.11 | 0.31 | -0.32 | 0.33 | 0.58 | -0.79 | 0.83 | -0.18 | -1.15 | 0.72 | -0.08 | -1.04 | 0.82 | 0.12 | -1.30 |

[a] NACs from reference [26] using M06L optimized geometries and electron distributions. [b] $LiMn_2O_4$ has a spinel structure that undergoes a charge-ordering transition as shown in experiments;[112-114] the PBE functional shows charge disproportionation between the Mn sites (i.e., a charge-ordered phase) while the M06L functional gives equal NACs on all Mn sites[26] (i.e., a high-temperature phase without charge ordering).

### 5.2.2 Palladium-containing crystals

As additional examples of charge transfer in solids, we studied an interstitial H atom in Pd, $Pd_3V$, $Pd_3In$, and $Pd_3Hf$ crystals, plus $Pd_3V$ with no interstitial H atom. Manz and Sholl previously studied these materials with the DDEC2, DDEC3, and Bader methods,[14] and we used their geometries and PW91 electron densities to now compute the HD, CM5, and DDEC6 NACs. (These geometries are representative local energy minima, not necessarily global energy minima.[14]) Interestingly, the HD and CM5 NACs are negative for V, In, and Hf atoms and positive for Pd atoms, even though the Pauling scale electronegativity[54] of Pd (2.20) is greater than V (1.63), In (1.78), and Hf (1.3). The Bader, DDEC3, and DDEC6 NACs followed the Pauling scale electronegativity trends with a negative average Pd NAC and positive X NACs following the expected trend Hf > V > In. All methods gave slightly negative to nearly neutral H NACs within the range -0.32 and +0.02.

Table 13. Average NACs for Interstitial H in Ordered $Pd_3X$ Alloys.

| Material | H charge[a] | | | | | Pd charge[a] | | | | | X charge[a] | | | | |
|---|---|---|---|---|---|---|---|---|---|---|---|---|---|---|---|
| | Bader | DDEC3 | DDEC6 | HD | CM5 | Bader | DDEC3 | DDEC6 | HD | CM5 | Bader | DDEC3 | DDEC6 | HD | CM5 |
| H in Pd | -0.04 | -0.25 | -0.05 | -0.13 | -0.12 | 0.00 | 0.01 | 0.00 | 0.004 | 0.004 | n.a. | n.a. | n.a. | n.a. | n.a. |
| $Pd_3V$ | n.a. | n.a. | n.a. | n.a. | n.a. | -0.35 | -0.10 | -0.15 | 0.02 | 0.02 | 1.04 | 0.31 | 0.44 | -0.06 | -0.06 |
| H in $Pd_3V$ | -0.22 | -0.32 | -0.14 | -0.15 | -0.13 | -0.34 | -0.09 | -0.14 | 0.03 | 0.02 | 1.04 | 0.32 | 0.44 | -0.06 | -0.06 |
| H in $Pd_3In$ | -0.05 | -0.18 | -0.05 | -0.11 | -0.09 | -0.21 | -0.08 | -0.07 | 0.06 | 0.16 | 0.64 | 0.27 | 0.22 | -0.16 | -0.47 |
| H in $Pd_3Hf$ | -0.05 | 0.02 | -0.01 | -0.10 | -0.09 | -0.53 | -0.31 | -0.23 | 0.08 | 0.08 | 1.58 | 0.92 | 0.69 | -0.22 | -0.22 |

[a] Bader and DDEC3 NACs are from reference [14].

Why did the HD and CM5 methods yield negative NACs for the V, In, and Hf atoms? It is well-known that isolated neutral atoms usually become more contracted upon going from left to right within the same subshell of a periodic table row due to the increasing nuclear charge that contracts the subshell. (Deviations from this trend can occur where electron configurations deviate from the Aufbau principle, such as Pd through Cd.) Moreover, atoms usually become slightly more diffuse or remain about the same size down a periodic table column. Accordingly, an isolated neutral Hf atom is more diffuse than an isolated neutral Pd atom. The Pauling scale electronegativity will usually follow the opposite trend, with the Pauling scale electronegativity increasing left to right within the same subshell of a periodic table row and decreasing down a periodic column except where electron configurations deviate from the Aufbau principle. Because the neutral Hf reference atom is more diffuse than the neutral Pd reference atom, during HD partitioning the Hf atoms steal electrons from the more electronegative Pd atoms. Thus, in this case, the HD method predicts the wrong charge transfer direction. The CM5 method adds a correction to the HD NACs, but this correction is zero between two transition metal atoms.[25] Consequently, the HD and CM5 NACs are identical for $Pd_3V$. In the other materials, there is a non-zero CM5 correction between the main-group elements H and In and the other elements, which causes the CM5 NACs to slightly differ from the HD NACs.

To avoid this problem, the DDEC3 and DDEC6 methods include a constraint that forces $w_A(r_A)$ for tails of buried atoms to decay at least as fast as $\exp(-1.75 r_A/\text{bohr})$.[14] Second, the DDEC6 method sets the reference ion charge for each atom in the material to a weighted average of a stockholder type charge partitioning and a smoothed localized charge partitioning. This ensures the reference ion charge resembles the charge in the local vicinity of the atom in order to prevent atoms from becoming too diffuse or too contracted. This makes DDEC6 NACs more accurately describe the true charge transfer direction.

### 5.2.3 Magnesium oxide

Table 14 compares six different charge assignment methods for $(MgO)_n$ molecules (n = 1 to 6) and crystalline MgO. Geometries of the $(MgO)_n$ molecules and their HD, CM5, and DDEC3 NACs and dipole moments were taken from Wang et al.[26] These geometries (Figure 9) were built by removing Mg and O atoms from a rigid $(MgO)_6$ cluster, rather than optimizing the geometries with DFT.[26] Following Wang et al.,[26] we computed electron distributions for the $(MgO)_n$ clusters in GAUSSIAN 09 using the M06L functional and def2-TZVP[115] basis set. The geometry and electron distribution of crystalline MgO were optimized in VASP using the PBE functional.

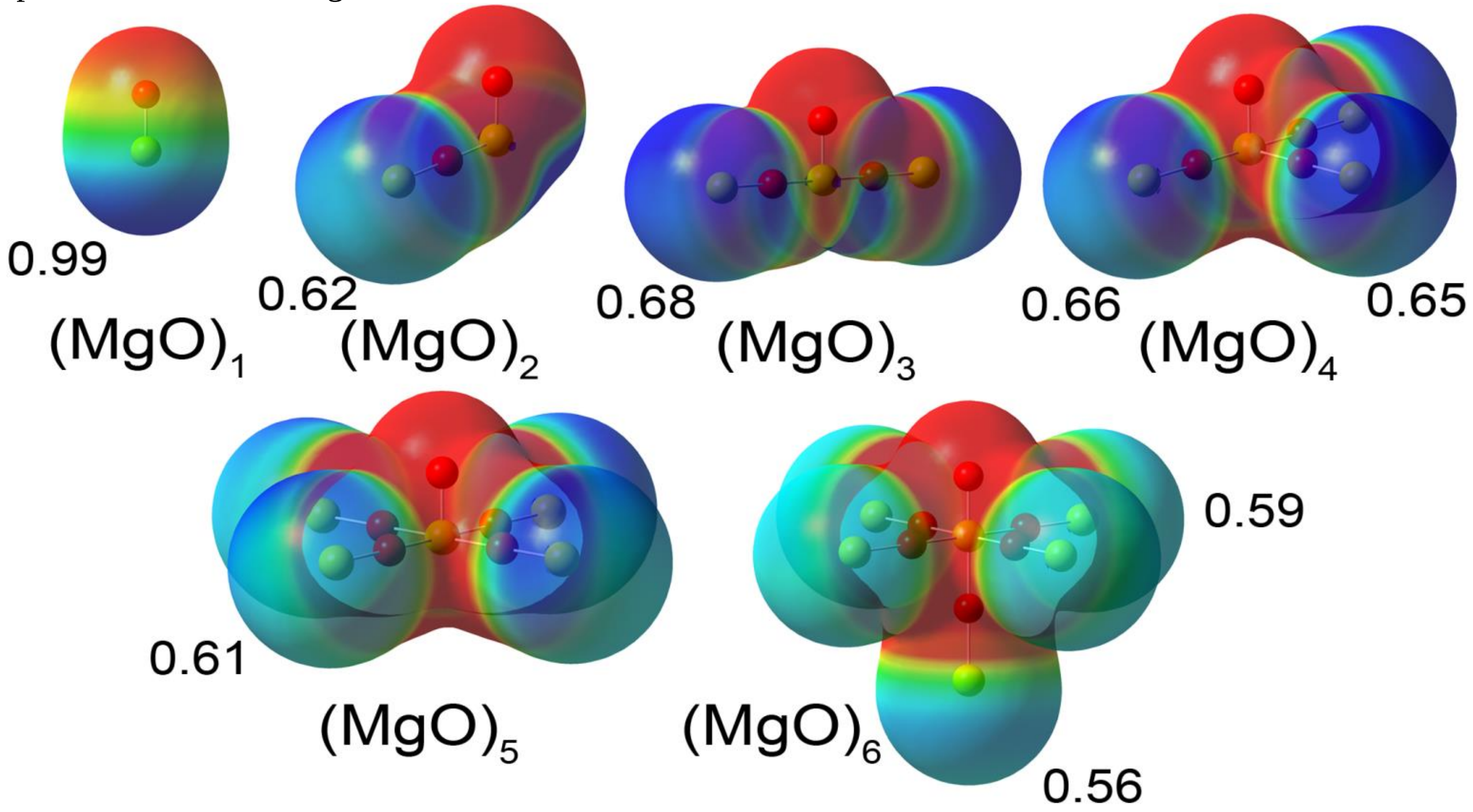

Figure 9. Molecular electrostatic potential (MEP) of the six $(MgO)_n$ molecules (n=1 to 6) studied. The MEP is shown on the 0.0004 electrons/bohr$^3$ density contour with a MEP scale ranging from -0.78 volts (red) to 0.78 volts (blue). The numbers appearing beside the terminal Mg atoms are their DDEC6 NACs.

Based on the much lower Pauling scale electronegativity of Mg (1.31) than O (3.44), a substantial transfer of electrons from Mg to O is expected. A simple chemical argument suggests that as the central Mg atom is surrounded by more oxygen anions, electrostatic stabilization of the central Mg cation by the oxygen anions should increase, thereby stabilizing more electron transfer from the central Mg atom to the adjacent oxygen atoms. This simple chemical argument predicts an increase in the central Mg atom NAC as the number of adjacent O atoms increases. Examining Table 14, only the DDEC3 method consistently followed this trend. The trend for terminal Mg NACs can be inferred from the electrostatic potential values. As shown in Figure 9, the electrostatic potential and DDEC6 NACs are most positive near the terminal Mg atoms following the trend MgO > $(MgO)_3$ > $(MgO)_4$ > $(MgO)_2$ > $(MgO)_5$ > $(MgO)_6$.

As shown in Table 14, the Mg NAC in bulk MgO followed the trend DDEC3 (2.01) > Bader (1.70) > DDEC6 (1.47) > CM5 (0.77) > HD (0.33). The DDEC3 NAC of 2.01 for bulk MgO is similar to some recent high-resolution diffraction experiments and their interpretations in terms of fully ionized $Mg^{+2}$ and $O^{2-}$ ions.[116, 117] However, the situation is not as straightforward as it first appears, because (i) charge partitioning in the experimentally measured electron distribution depends on model definitions used to assign NACs and (ii) the low-order structure factors in simple cubic crystals (e.g., MgO and NaCl) have low sensitivity to the amount of charge transfer.[26, 63, 116, 117] Zuo et al. used a convergent beam electron diffraction technique to improve the resolution of the low-order structure factors and concluded the crystal's electron distribution is consistent with fully ionized $Mg^{2+}$ and $O^{2-}$ ions,[116] but this does not rule out other interpretations.

Table 14. Comparison of different charge assignment methods for $(MgO)_n$ molecules and crystalline MgO. The NAC methods are listed from smallest to largest NAC magnitudes in bulk MgO. For the DDEC6 method, M represents NACs only, D represents the inclusion of atomic dipoles, and SCP represents the inclusion of the spherical charge penetration term.

| Method | Molecule[a] | | | | | | Bulk MgO |
|---|---|---|---|---|---|---|---|
| | MgO | $(MgO)_2$ | $(MgO)_3$ | $(MgO)_4$ | $(MgO)_5$ | $(MgO)_6$ | |
| | | | NAC of central Mg | | | | |
| HD | 0.59 | 0.69 | 0.56 | 0.42 | 0.28 | 0.15 | 0.33 |
| CM5 | 0.79 | 0.97 | 0.93 | 0.88 | 0.82 | 0.77 | 0.77 |
| DDEC6 | 0.99 | 1.45 | 1.57 | 1.60 | 1.61 | 1.61 | 1.47 |
| Bader | 1.18 | 1.54 | 1.65 | 1.62 | 1.57 | 1.48 | 1.70 |
| DDEC3 | 1.00 | 1.55 | 1.72 | 1.76 | 1.79 | 1.84 | 2.01 |
| ESP | 0.89 | 1.16 | 1.04 | 0.96 | 0.58 | -1.72 | [b] |
| | | | Dipole moment in a.u. | | | | MAE |
| full density | 2.71 | 2.32 | 1.39 | 1.70 | 0.48 | 1.97 | 0.00 |
| HD | 1.87 | 1.97 | 1.28 | 2.08 | 0.88 | 1.98 | 0.35 |
| ESP | 2.82 | 2.74 | 1.87 | 2.13 | 1.12 | 2.43 | 0.42 |
| CM5 | 2.51 | 2.54 | 1.91 | 2.47 | 1.51 | 2.61 | 0.56 |
| DDEC6 | | | | | | | |
| M | 3.14 | 3.18 | 2.37 | 2.49 | 1.68 | 2.91 | 0.87 |
| D | 2.71 | 2.32 | 1.39 | 1.70 | 0.48 | 1.97 | 0.00 |
| DDEC3 | 3.17 | 3.31 | 2.48 | 2.57 | 1.79 | 3.08 | 0.97 |
| Bader | 3.71 | 3.39 | 2.46 | 3.09 | 1.63 | 3.77 | 1.25 |
| | | RMSE in kcal/mol (RRMSE) | | | | | Average RMSE |
| ESP | 2.81(0.10) | 5.40 (0.23) | 4.59 (0.24) | 5.16 (0.27) | 4.96 (0.27) | 3.79 (0.21) | 4.45 |
| CM5 | 3.98 (0.15) | 6.50 (0.28) | 4.77 (0.25) | 5.90 (0.31) | 6.12 (0.33) | 5.28 (0.29) | 5.43 |
| HD | 9.27 (0.35) | 10.05 (0.43) | 6.06 (0.32) | 5.99 (0.31) | 4.90 (0.26) | 4.01 (0.22) | 6.71 |
| DDEC6 | | | | | | | |
| M | 4.25 (0.16) | 7.24 (0.31) | 6.75 (0.35) | 7.70 (0.40) | 7.77 (0.42) | 6.58 (0.36 ) | 6.71 |
| M + SCP | 4.63 (0.17) | 7.40 (0.32) | 6.98 (0.36) | 7.82 (0.41) | 7.98 (0.43) | 6.78 (0.37) | 6.93 |
| D | 1.31 (0.05) | 2.60 (0.11) | 2.87 (0.15) | 2.86 (0.15) | 3.09 (0.17) | 3.03 (0.16) | 2.63 |
| D + SCP | 0.86 (0.03) | 1.55 (0.07) | 1.67 (0.09) | 1.61 (0.08) | 1.74 (0.09) | 1.70 (0.09) | 1.52 |
| DDEC3 | 4.44 (0.17) | 8.48 (0.36) | 7.52 (0.39) | 8.48 (0.44) | 8.42 (0.45) | 7.08 (0.39) | 7.40 |
| Bader | 9.10 (0.34) | 9.26 (0.40) | 10.81 (0.56) | 12.87 (0.67) | 14.20 (0.76) | 14.11 (0.77) | 11.73 |

[a] NACs for the HD, CM5, and DDEC3 methods for the $(MgO)_n$ molecules are from reference [26]. [b] ESP NAC cannot be reported for bulk MgO, because there are no surface atoms.

These tests on $(MgO)_n$ and bulk MgO illustrate some possible compromises between matching the chemical state trends on the one hand and the electrostatic potential trends on the other hand. Dipole MAE followed the trend HD (0.35) < ESP (0.42) < CM5 (0.56) < DDEC6 (0.87) < DDEC3 (0.97) < Bader (1.25). Electrostatic potential RMSE (kcal/mol) followed the trend ESP (4.45) < CM5 (5.43) < HD,

DDEC6 (6.71) < DDEC3 (7.40) < Bader (11.73). Although the ESP method gave low dipole MAE and electrostatic potential RMSE, we do not recommend the ESP method for assigning NACs, because the ESP NACs of the central Mg atom fluctuated erratically from 1.16 for $(MgO)_2$ to -1.72 for $(MgO)_6$. Because the Bader point charges had the highest dipole moment MAE and the highest average electrostatic potential RMSE, we do not recommend Bader NACs for use in force-field point charge models for classical atomistic simulations. Choosing between the remaining four point charge methods (i.e., HD, CM5, DDEC3, and DDEC6) is complicated by the fact that dipole MAE and electrostatic potential RMSE followed a different trend than the central Mg atom NAC. On the basis of the central Mg atom NAC increasing monotonically from MgO molecule to $(MgO)_6$ molecule to bulk MgO, the DDEC3 method would be preferable, but the DDEC3 method gave the largest dipole MAE and average electrostatic potential RMSE among these four charge assignment methods. The HD and CM5 methods had comparatively low dipole MAE and average electrostatic potential RMSE, but yielded low values of 0.33 (HD) and 0.77 (CM5) for the Mg NAC in bulk MgO. Results for the DDEC6 method were intermediate for central Mg NAC, dipole MAE, and average electrostatic potential RMSE.

Finally, Table 14 investigates effects of atomic dipoles and spherical charge penetration. Including atomic dipoles for any AIM method (e.g., HD, DDEC6, Bader, DDEC3), eliminates the dipole prediction error to within a grid integration tolerance (e.g., ~0.01). Because the dipole moment of a spherical charge distribution is zero, the spherical charge penetration term has no effect on the computed dipoles. Including DDEC6 atomic dipoles decreased the average RMSE from 6.71 to 2.63 kcal/mol. Although the spherical charge penetration term slightly increased the average RMSE at the DDEC6 (M+SCP) level, it dramatically reduced the average RMSE to 1.52 kcal/mol at the DDEC6 (D+SCP) level. Notably, the DDEC6 (D+SCP) average RMSE was ~3 times lower than any of the point charge models.

### 5.2.4 Materials containing four or more different elements

Table 15 lists the Spearman rank coefficient between DDEC6 NACs and Pauling scale electronegativity for the materials in this paper containing four or more different elements, except the substituted carboxylic acids studied in Section 5.5.1. (For the substituted carboxylic acids, a separate comparison based on the σ' substituent constants is presented in Section 5.5.1 to verify the chemical meaning of the DDEC6 NACs.) For each material, the average DDEC6 NAC was computed for each element. Nine of the 14 materials had a Spearman rank coefficient of 1.00. A Spearman rank coefficient of 1.00 indicates the average DDEC6 NACs followed exactly the same order as the element electronegativities. The remaining five materials had Spearman rank coefficients between 0.60 and 0.94, indicating the average DDEC6 NACs followed approximately but not exactly the same order as the element electronegativities. These results show DDEC6 NACs usually (but not always) follow Pauling scale electronegativity trends. The exceptions are not to be regarded as a deficiency of either the DDEC6 NACs or the Pauling scale electronegativities, because element electronegativities can only describe the usual direction of electron transfer. The specific direction of electron transfer is affected by the chemical environment. For example, while electrons are usually transferred from carbon to the more electronegative oxygen, experiments show carbon monoxide is an exception with electron transfer from oxygen towards carbon.[118] Boron monofluoride is another exception with electrons transferred from fluorine towards boron.[119] Furthermore, multivalent cations can sometimes acquire a positive NAC greater than that of less electronegative monovalent cations, because the multivalent cations may acquire a NAC greater than +1.

For example, the multivalent P, Al, and Si atoms in B-DNA and natrolite acquired higher NACs than the monovalent Na atoms.

Table 15. Spearman rank coefficient quantifying the ordering relationship between average DDEC6 NACs for each element and the Pauling scale electronegativities.

| Material | Elements | Spearman rank coefficient |
|---|---|---|
| B-DNA | C, H, N, Na, O, P | 0.94 |
| $Cu_2$ pyridine complex | C, Cu, H, N | 1.00 |
| CuBTC | C, Cu, H, O | 0.80 |
| $Fe_4O_{12}N_4C_40H_{52}$ noncollinear SMM | C, Fe, H, N, O | 1.00 |
| Formamide[a] | C, H, N, O | 0.60 |
| IRMOF[b] | C, H, O, Zn | 1.00 |
| lp-MIL-53 | Al, C, H, O | 1.00 |
| $Mn_{12}$-acetate SMM[c] | C, H, Mn, O | 1.00 |
| Natrolite | Al, H, Na, O, Si | 0.60 |
| ZIF8 | C, H, N, Zn | 1.00 |
| ZIF90 | C, H, N, O, Zn | 0.90 |
| Zn nicotinate[d] | C, H, N, O, Zn | 1.00 |
| Zr bisperoxy complex | C, H, N, O, Zr | 1.00 |
| Zr puckered bare complex | C, H, N, Zr | 1.00 |

[a] Formamide geometry optimized with B3LYP/6-311++G**. [b] IRMOF x-ray structure. [c] $Mn_{12}$-acetate single molecule magnet geometry optimized with PBE/LANL2DZ. [d] Zn nicotinate geometry optimized with PBE/planewave.

### 5.3 Comparison to spectroscopic results for various materials

#### 5.3.1 Net atomic charges extracted from high resolution diffraction experiments

Extracting NACs from high-resolution diffraction data is not straightforward, but it can be done using approximations and models. In 'Kappa refinement', the high-resolution diffraction data is first fit to a multipolar model[117, 120, 121] to determine atomic coordinates, thermal parameters, and an electron density map and then refit to a spherical pseudoatom model[122, 123] to determine the NACs. In Kappa refinement, the spherical pseudoatoms have the form $\rho_{at}(r_A) = \rho_A^{core}(r_A) + n_A^{val}\kappa^3 p_A^{val}(\kappa r_A)$, where $p_A^{val}(r_A)$ is the normalized shape function of the valence density of the neutral reference atom.[122] The two primary limitations of Kappa refinement are that the pseudoatom densities do not necessarily sum to the correct total density $\rho(\vec{r})$ and the shape functions for the charged atoms are represented as expanded or contracted versions of the neutral atoms.[14] Here, we revisit two examples for which DDEC3 and experimentally extracted NACs were compared in reference [14]: the formamide and natrolite structures shown Figure 10. We refer the reader to the earlier publications for a discussion of the experimental details and analysis.[14, 15, 123, 124] The same geometries and electron distributions are used in this work as in reference [14].

As shown in Table 16, both the DDEC3 and DDEC6 NACs follow a trend similar to the experimentally extracted NACs for natrolite, except the DDEC3 NACs on all atoms except Na are significantly higher in magnitude than the experimentally extracted ones. For all atoms, the DDEC6 NACs are slightly lower in magnitude than the DDEC3 NACs, leading to an overall better agreement between the DDEC6 and experimentally extracted NACs. Only for the Na atom, which was fixed to a value of 1.00 in the experimental analysis,[124] is the experimentally extracted NAC closer to the DDEC3 value than the DDEC6 value.

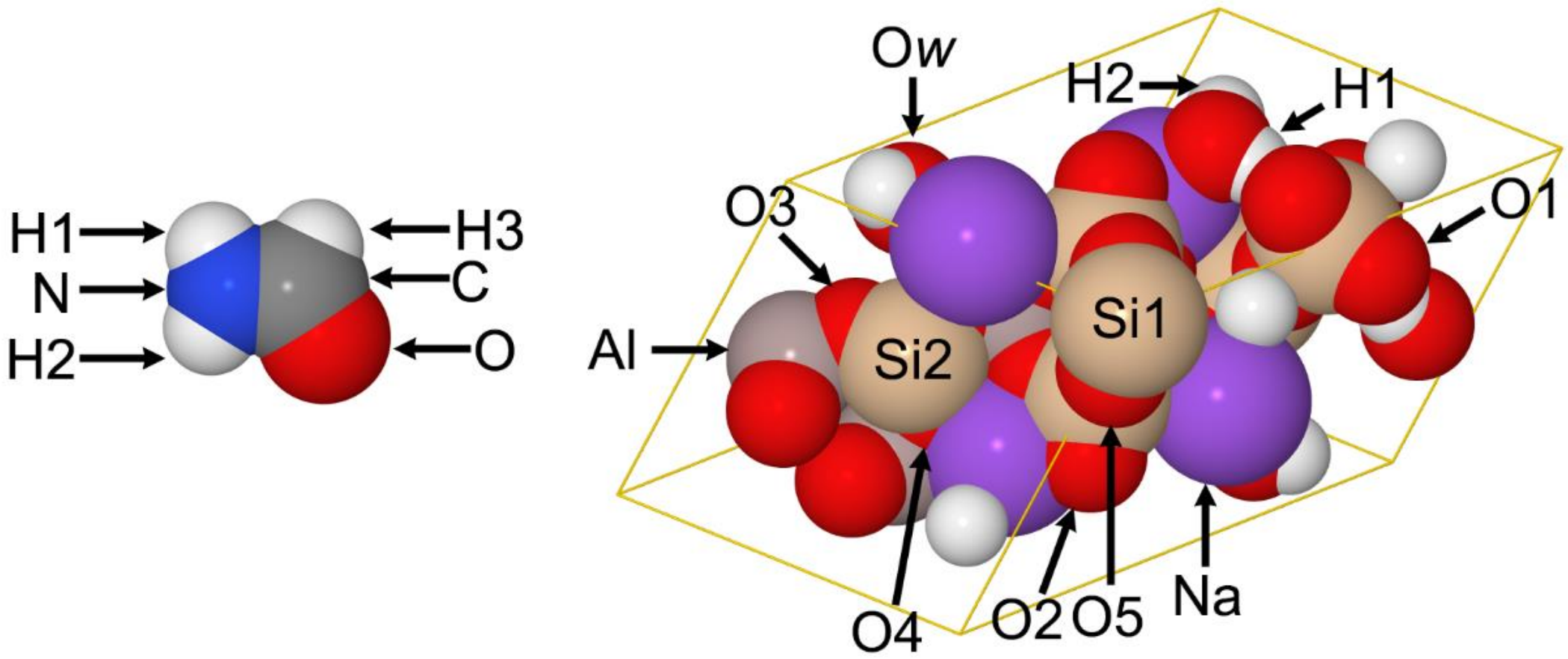

Figure 10. Formamide and natrolite structures. The lines in natrolite indicate the unit cell boundaries. Figure reproduced with permission from reference [14]. © ACS 2012.

Table 16. Experimental and theoretical natrolite NACs. DDEC3 and DDEC6 results computed using the PBE-optimized geometries.

| | High res. XRD[a] | DDEC3[b] | DDEC6 |
|---|---|---|---|
| Si1 | 1.84 ± 0.12 | 2.172 | 1.772 |
| Si2 | 1.65 ± 0.10 | 2.207 | 1.760 |
| Al | 1.51 ± 0.11 | 2.067 | 1.762 |
| O1 | -0.90 ± 0.05 | -1.227 | -1.036 |
| O2 | -1.21 ± 0.05 | -1.318 | -1.103 |
| O3 | -1.03 ± 0.05 | -1.337 | -1.094 |
| O4 | -1.07 ± 0.05 | -1.320 | -1.110 |
| O5 | -0.87 ± 0.05 | -1.113 | -0.913 |
| Na | 1.00 | 1.000 | 0.896 |
| O*w* | -0.59 ± 0.03 | -0.926 | -0.862 |
| H1 | 0.24 ± 0.03 | 0.446 | 0.408 |
| H2 | 0.36 ± 0.03 | 0.435 | 0.405 |

[a] High resolution XRD data from reference [124]. [b] DDEC3 NACs from reference [14].

Table 17 summarizes experimentally extracted and computed NACs for formamide. Theoretical charges were computed using the B3LYP[125, 126] functional with aug-cc-pvtz[127] basis set. The high-resolution x-ray diffraction results were extracted using fully optimized radial factors for all atoms.[123] Maximum absolute differences from the experimentally extracted NACs are 0.07 (NPA), 0.11 (DDEC3), 0.12 (DDEC6), 0.13 (IH), 0.17 (ESP), 0.22 (ISA), 0.64 (HD), and 0.96 (Bader). The DDEC6, ESP,

DDEC3, and ISA point charge dipoles were within ±5% of the full wavefunction value of 1.55. Errors for the other point charge dipoles were -6% (IH), +23% (NPA), and +66% (Bader). Of course, when atomic dipoles are included, all of the AIM methods (Bader, DDEC3, DDEC6, HD, IH, and ISA) yield the exact dipole moment to the integration grid precision. The accuracy of the point charge models for reproducing the electrostatic potential followed the trend ESP > DDEC6 > DDEC3 > ISA > IH > NPA > HD > Bader. When atomic dipoles were included, the RMSE for the DDEC6 method decreased from 0.74 to 0.49 kcal/mol. When the spherical charge penetration term was included, the RMSE values for the DDEC6 method were unchanged (within a computational tolerance of 0.01 kcal/mol) at 0.74 (M + SCP) and 0.49 (D + SCP), indicating a negligible impact of spherical charge penetration over the RMSE grid points.

Table 17. Experimental and theoretical formamide NACs. Dipoles in atomic units. RMSE in kcal/mol.

| | High res. XRD[a] | Bader[b] | DDEC3[b] | DDEC6 M (D)[d] | ESP[b] | HD[b] | IH[b] | ISA[b] | NPA[b] |
|---|---|---|---|---|---|---|---|---|---|
| O | -0.55 ± 0.04 | -1.149 | -0.557 | -0.506 | -0.562 | -0.304 | -0.537 | -0.593 | -0.605 |
| N | -0.78 ± 0.07 | -1.183 | -0.788 | -0.662 | -0.923 | -0.136 | -0.862 | -0.911 | -0.808 |
| C | 0.51 ± 0.08 | 1.469 | 0.624 | 0.519 | 0.680 | 0.139 | 0.644 | 0.726 | 0.534 |
| H1 | 0.39 ± 0.03 | 0.411 | 0.352 | 0.329 | 0.389 | 0.128 | 0.360 | 0.389 | 0.394 |
| H2 | 0.40 ± 0.03 | 0.426 | 0.369 | 0.313 | 0.429 | 0.133 | 0.377 | 0.407 | 0.388 |
| H3 | 0.03 ±0.03 | 0.026 | 0.000 | 0.007 | -0.012 | 0.040 | 0.018 | -0.019 | 0.096 |
| | Dipole moment[c] | 2.57 | 1.59 | 1.53 (1.55) | 1.57 | 1.13 | 1.46 | 1.62 | 1.91 |
| | RMSE | 9.85 | 0.85 | 0.74 (0.49) | 0.58 | 3.32 | 1.43 | 0.99 | 3.13 |
| | RRMSE | 0.89 | 0.08 | 0.07 (0.04) | 0.05 | 0.30 | 0.13 | 0.09 | 0.28 |

[a] From reference [123]. [b] Bader, DDEC3, ESP, HD, IH, ISA, and NPA NACs are from reference [14]. [c] Dipole moment of the B3LYP/aug-cc-pvtz wavefunction was 1.55. [d] M denotes point charge (monopole) model; D denotes the inclusion of atomic dipoles.

### 5.3.2 Correlations between NACs and spectroscopically measured core electron binding energy shifts

The core electron binding energy shift is defined as the binding energy of a particular core orbital level for an atom-in-a-material compared to the same core orbital level for an atom of the same element in a reference compound.[128-133] Core electron binding energies can be measured using x-ray photoelectron spectroscopy (XPS) or x-ray absorption near edge structure (XANES). Several key factors affect core electron binding energy shifts.[128-133] First, a change in the valence electron population of this atom affects its core electron binding energy, because more valence electrons cause electrostatic shielding of the nuclear charge and a decrease in the core electron binding energy.[128-132] Second, the core electron binding energy is directly affected by the electrostatic potential exerted on this atom by the other atoms in the material: lots of anions nearby will decrease the core electron binding energy and lots of cations nearby will increase the core electron binding energy.[128-132] Third, the core electron binding energy is affected by relaxation in which the electrons rearrange to partially fill the hole left by the ejected photoelectron.[129, 131-133] Various simple model equations have been developed to correlate core electron binding energy shifts to easily computed chemical descriptors such as (a) the NAC of the atom emitting the photoelectron, (b) the electrostatic potential exerted on the atom emitting the photoelectron by all the other atoms in the material (as computed using electron distributions or simple point charge models), (c) quantum mechanically computed electrostatic potential near the nucleus of the atom emitting the photoelectron, (d) orbital eigenvalues (aka ‘orbital energies’) computed using the Hartree-Fock or other quantum chemistry methods for chemical models of the initial and final states, and (e) two-electron integrals describing exchange and electrostatic interactions between valence and core orbitals.[128-132]

Table 18. R-squared correlation coefficients between NACs and spectroscopically measured core electron binding energies. NAC methods ordered from highest to lowest average R-squared correlation coefficient.

| | Ti compounds | Mo compounds | Fe compounds |
|---|---|---|---|
| HD | 0.795 | 0.987 | 0.819 |
| DDEC6 | 0.704 | 0.978 | 0.868 |
| Bader | 0.727 | 0.911 | 0.817 |
| DDEC3 | 0.360 | 0.977 | 0.905 |
| CM5 | 0.345 | 0.898 | 0.747 |

Table 19. Experimental Ti $2p_{3/2}$ core electron binding energies (eV) measured by x-ray photoelectron spectroscopy and theoretically computed NACs. Compounds ordered from lowest to highest average binding energy.

| solid | $2p_{3/2}$ binding energy (eV)[a] lower limit | $2p_{3/2}$ binding energy (eV)[a] upper limit | ICSD code | Ti net atomic charge: Bader | CM5 | DDEC3 | DDEC6 | HD |
|---|---|---|---|---|---|---|---|---|
| Ti | 453.66 | 454.14 | 44872 | 0.00 | 0.00 | 0.00 | 0.00 | 0.00 |
| $TiB_2$ | 454.14 | 454.50 | 78848 | 1.30 | 1.53 | 1.74 | 1.36 | 0.44 |
| TiO | 454.88 | 455.33 | 56612 | 1.50 | 0.87 | 2.16 | 1.29 | 0.35 |
| TiN | 455.66 | 456.00 | 105128 | 1.59 | 1.20 | 2.59 | 1.61 | 0.42 |
| $BaTiO_3$ | 458.28 | 458.71 | 99737 | 2.13 | 1.42 | 2.55 | 2.20 | 0.72 |
| $TiCl_4$ | 458.36 | 458.71 | 280981 | 1.90 | 0.95 | 1.27 | 1.45 | 0.56 |
| $PbTiO_3$ | 458.43 | 458.78 | 165498 | 2.12 | 1.39 | 2.48 | 2.16 | 0.63 |
| $CaTiO_3$ | 458.57 | 459.00 | 163662 | 2.01 | 1.48 | 2.61 | 2.26 | 0.75 |
| $SrTiO_3$ | 458.57 | 459.00 | [b] | 2.09 | 1.45 | 2.64 | 2.26 | 0.72 |
| $TiO_2$ | 458.64 | 459.33 | 39166 | 2.08 | 1.47 | 2.53 | 2.28 | 0.72 |

[a] XPS values from reference [133]. [b] $SrTiO_3$ was geometry optimized using PBE.

Here, we are most interested in correlations between core electron binding energy shifts and NACs that occur for some crystalline materials.[134-138] We now consider a series of Ti, Mo, and Fe compounds as examples. Table 18 summarizes linear correlations between core electron binding energies and NACs. The HD, DDEC6, and Bader methods gave reasonable performance (i.e., R-squared ≥ 0.704) for all three elements, while the DDEC3 and CM5 methods performed poorly (i.e., R-squared ≤ 0.360) for the Ti compounds. Overall, the strength of the correlation between NACs and core electron binding energies followed the trend HD > DDEC6 > Bader > DDEC3 > CM5. Table 19 summarizes details for the Ti-containing solids. NACs were computed using the PBE electron distributions for the experimental geometries defined by the ICSD codes in Table 19, except for $SrTiO_3$ which was geometry optimized. The poor correlation of the DDEC3 method for the Ti-containing solids was primarily due to high NACs for TiO, TiN, and $TiB_2$ and a lower NAC for $TiCl_4$ than for TiO.

Table 20 summarizes results for the Mo-containing solids. NACs were computed using the PBE electron distributions for the experimental geometries defined by the ICSD codes in Table 20. For structures listing two ISCD codes, both crystal structures were included in the correlation to the experimental K-edge energy. Li et al. measured these K-edge energies using XANES.[138] The K-edge energy is correlated to the binding energy of the K-shell (i.e., 1s) core electrons.[131, 139] Our analysis and correlation for these Mo-containing solids is identical to that of Li et al.[138] using DDEC3 and Bader NACs, except we have extended it to DDEC6, CM5, and HD NACs.

Table 20. Experimental K-edge energy (eV) of molybdenum-containing compounds and average Mo NAC computed by various charge assignment methods. Compounds ordered from lowest to highest K-edge energy.

| solid | K-edge energy(eV)[a] | ICSD code | average Mo net atomic charge | | | | |
|---|---|---|---|---|---|---|---|
| | | | Bader[a] | CM5 | DDEC3[a] | DDEC6 | HD |
| Mo | 20005.3 | 41513 (*Fm-3m*) | 0 | 0 | 0 | 0 | 0 |
| $MoS_2$ | 20006.5 | 43560(*R*3*mH*),<br>95570(*P*63/*mmc*) | 1.09,<br>1.09 | 0.75,<br>0.71 | 0.23,<br>0.21 | 0.58,<br>0.53 | 0.26,<br>0.25 |
| $Mo_2C$ | 20006.9 | 43322(*Pbcn*) | 0.66 | 0.39 | 0.57 | 0.44 | 0.19 |
| $MoO_2$ | 20011.0 | 152316(*P*121/*c*1) | 1.88 | 1.21 | 1.78 | 1.65 | 0.58 |
| $Rb_2MoO_4$ | 20013.5 | 24904(*C*12/*m*1) | 2.13 | 1.36 | 2.12 | 1.92 | 0.80 |
| $MoO_3$ | 20013.7 | 151751(*Pnma*),<br>152312(*Pbnm*) | 2.36,<br>2.29 | 1.62,<br>1.63 | 2.38,<br>2.36 | 2.27,<br>2.23 | 0.82,<br>0.80 |

[a] From reference [138].

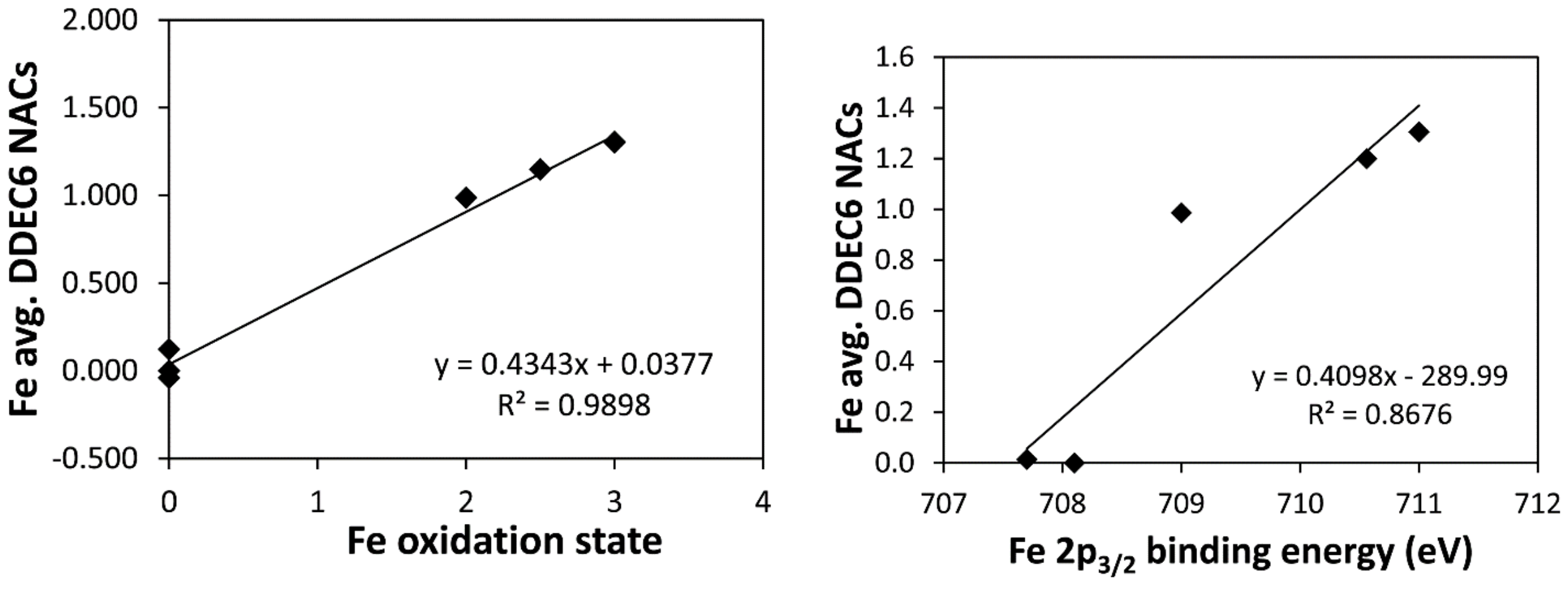

Figure 11. *Left*: Correlation between Fe oxidation state and average Fe DDEC6 NAC for Fe, $Fe_2SiO_4$, $Fe_2O_3$, $Fe_3O_4$, and $Fe_3Si$ solids. *Right*: Correlation between $2p_{3/2}$ core electron binding energy (as measured using XPS) and average Fe DDEC6 NAC for these materials.

Figure 11 shows linear regression plots between the average Fe DDEC6 NACs and the oxidation state (left panel) and the $2p_{3/2}$ core electron binding energy (right panel). NACs were computed using the PBE electron distributions based on the following geometries: Fe (NAC is zero due to symmetry), $Fe_2SiO_4$ (PBE-optimized geometry of anti-ferromagnetic spinel phase[14]), $Fe_2O_3$ (PBE-optimized geometry of anti-ferromagnetic phase[14]), $Fe_3O_4$ (PBE-optimized geometry of anti-ferrimagnetic phase[14]), and $Fe_3Si$ (experimental crystal structure[140]). Our analysis for Fe-containing solids is similar to that of Manz and Sholl[14] using DDEC3 NACs, except we have extended it to DDEC6, CM5, HD, and Bader NACs.

### 5.4 Reproducing the electrostatic potential in one system conformation

As embodied in Eqs. (109)–(110), all AIM methods yield a formally exact representation of the electrostatic potential in the form of a polyatomic multipole expansion with charge-penetration terms. For conciseness, it is desirable to have this polyatomic multipole expansion converge rapidly with most of the electrostatic potential described by the leading-order terms. Many force-fields used in classical molecular dynamics and Monte Carlo simulations use point-charge models to estimate the electrostatic interaction energies between chemical species.[62, 141, 142] These types of force-fields can be parameterized using NACs and optionally atomic multipoles computed via quantum chemistry calculations. To be suitable for this purpose, we desire the DDEC6 NACs to approximately reproduce the electrostatic potential surrounding a material.

Table 21. Accuracy of fitting the electrostatic potential. (Values in parentheses include spherical cloud penetration.) The best values for a point charge model are shown in boldface type. Values at M+SCP, D, or D+SCP are shown in boldface type if they are equal to or better than the best point charge model value.

| | | | | RMSE (kcal/mol) | | | | RRMSE | | | |
|---|---|---|---|---|---|---|---|---|---|---|---|
| material | geom | XC | basis set | DDEC3[a] | | DDEC6 | | DDEC3[a] | | DDEC6 | |
| | | | | M | D | M (M+SCP) | D (D+SCP) | M | D | M (M+SCP) | D (D+SCP) |
| $B_4N_4$ | DFT[b] | PW91 | 6-311+G* | 0.26 | 0.33 | **0.17** (0.19) | 0.35 (0.35) | 0.08 | 0.10 | **0.05** (**0.05**) | 0.10 (0.10) |
| BN tube | DFT[b] | PW91 | planewave | 8.81 | **2.40** | **5.91** (5.94) | **1.11** (**1.09**) | 2.13 | **0.58** | **1.43** (1.44) | **0.27** (**0.26**) |
| BN sheet | DFT | PBE | planewave | **0.07** | **0.07** | **0.07** (**0.07**) | **0.07** (**0.07**) | **0.64** | **0.64** | **0.64** (**0.61**) | **0.64** (**0.61**) |
| formamide | DFT[c] | B3LYP | aug-cc-pvtz | 0.85 | **0.40** | **0.74** (**0.74**) | **0.49** (**0.49**) | 0.08 | **0.04** | **0.07** (**0.07**) | **0.04** (**0.04**) |
| 1p-MIL-53(Al) | XRD[d] | PW91 | planewave | 1.57 | **0.59** | **1.46** (1.47) | **0.60** (**0.58**) | 0.80 | **0.30** | **0.74** (0.75) | **0.30** (**0.30**) |
| IRMOF-1 | XRD[e] | PW91 | planewave | **0.83** | **0.44** | 0.86 (0.86) | **0.26** (**0.24**) | **0.39** | **0.20** | 0.40 (0.40) | **0.12** (**0.11**) |
| IRMOF-1 | DFT[f] | PW91 | planewave | **0.65** | **0.58** | 0.82 (0.81) | **0.28** (**0.27**) | **0.27** | **0.24** | 0.33 (0.33) | **0.12** (**0.11**) |
| ZIF-8 | DFT[f] | PW91 | planewave | 0.88 | **0.72** | **0.85** (**0.81**) | **0.79** (**0.76**) | 0.57 | **0.47** | **0.56** (**0.53**) | **0.52** (**0.50**) |
| ZIF-90 | DFT[f] | PW91 | planewave | **0.81** | 0.84 | 1.03 (0.97) | 0.93 (0.88) | **0.12** | **0.12** | 0.15 (0.14) | 0.14 (0.13) |
| Zn-nicotinate | DFT | PBE | planewave | **0.82** | **0.44** | 0.90 (0.89) | **0.41** (**0.40**) | **0.46** | **0.25** | 0.51 (0.51) | **0.23** (**0.23**) |
| water | DFT | B3LYP | 6-311++G** | 1.31 | **0.80** | **1.16** (**1.16**) | **0.88** (**0.88**) | 0.14 | **0.08** | **0.12** (**0.12**) | **0.09** (**0.09**) |
| $H_2PO_4^-$ | DFT | M06L | aug-cc-pvtz | 2.16 | **0.41** | **1.65** (**1.65**) | **0.49** (**0.49**) | 0.17 | **0.03** | **0.13** (**0.13**) | **0.04** (**0.04**) |
| DNA | DFT | PBE | planewave | 13.91 | 13.79 | **12.67** (12.68) | 12.77 (12.77) | 0.59 | 0.58 | **0.54** (**0.54**) | **0.54** (**0.54**) |

[a] DDEC3 data (except BN sheet, formamide, Zn-nicotinate, water, $H_2PO_4^-$, and DNA) is from reference [14]. [b]From reference [13]. [c]From reference [14]. [d]From reference [143]. [e]From reference [144]. [f]From reference [72].

In this section, we compare the accuracy of the DDEC3 and DDEC6 NACs for reproducing the electrostatic potential surrounding a single geometric conformation. Table 21 lists 13 materials including small molecules and ion, a large biomolecule, several metal-organic frameworks, a nanosheet, and a nanotube. This represents several kinds of materials often encountered in classical molecular dynamics or Monte Carlo simulations.[62, 71, 141, 145] For each material, the same electrostatic potential grid point files were used to compute the DDEC3 and DDEC6 RMSE values. The DDEC6 NACs reproduced the electrostatic potential better than the DDEC3 NACs in 8 of these systems. This shows the DDEC6 NACs are a slight improvement compared to the DDEC3 NACs for reproducing the electrostatic potential surrounding a material. Including DDEC6 atomic dipoles improved the RSME by > 0.4 kcal/mol for the BN nanotube, lp-MIL-53(Al), IRMOF-1 (XRD and DFT geometries), Zn-nicotinate, and $H_2PO_4^-$. This shows that overall including atomic dipoles produces a modest improvement in the RMSE accuracy. Adding spherical charge penetration at the monopole or dipole levels (i.e., M+SCP and D+SCP) had no significant effect for these materials.

Table 22. Commonly used 3-site water models listed in alphabetical order.

| model | O-H distance (Å) | H-O-H angle (°) | H NAC | O NAC |
|---|---|---|---|---|
| SPC[a] | 1.00 | 109.47 | 0.41 | -0.82 |
| SPC/E[b] | 1.00 | 109.47 | 0.4238 | -0.8476 |
| TIP3P[c] | 0.9572 | 104.52 | 0.417 | -0.834 |
| TIPS[d] | 0.9572 | 104.52 | 0.40 | -0.80 |

[a] From reference [146]. [b] From reference [147]. [c] From reference [148]. [d] From reference [149].

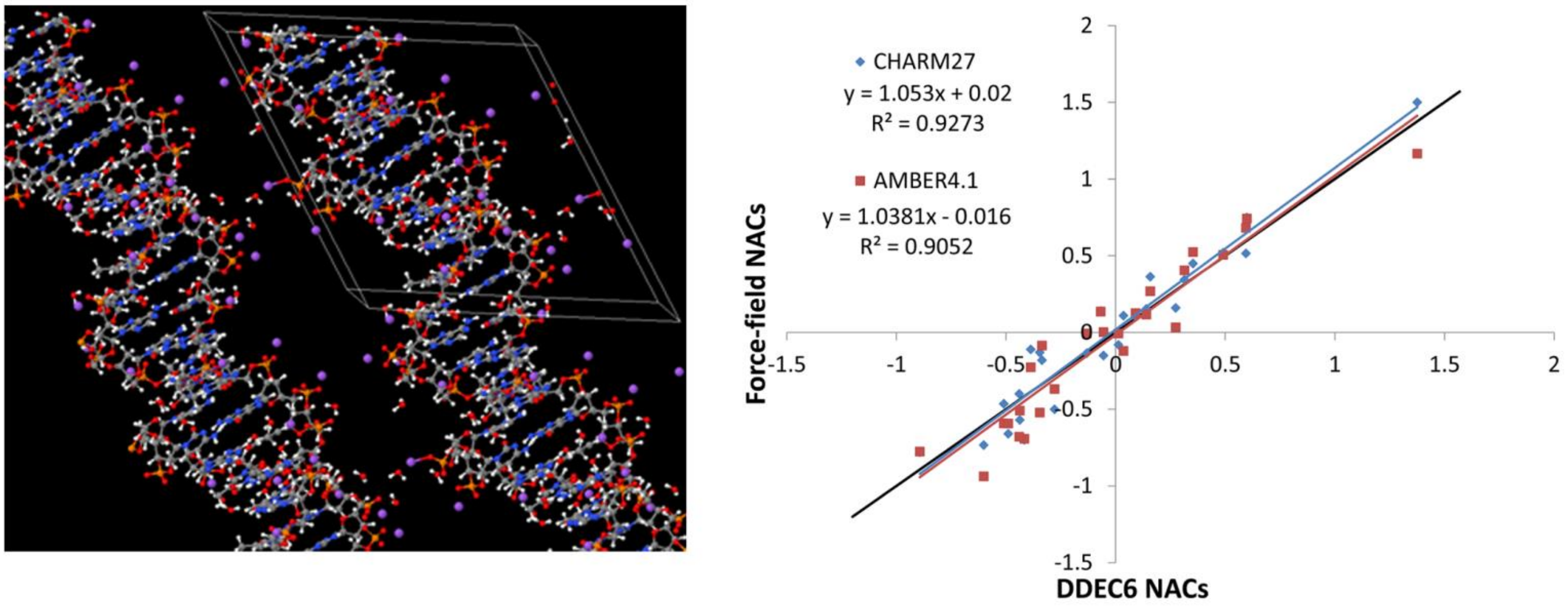

Figure 12. *Left*: B-DNA decamer (CCATTAATGG)$_2$, the lines mark the unit cell boundaries. *Right*: Correlation between DDEC6, CHARMM, and AMBER force-field NACs for all atoms excluding the bound water molecules and added $Na^+$ atoms. The black line has a slope of 1 and intercept of 0. CHARMM27 NACs from Foloppe and Mackerell.[150] AMBER4.1 NACs from Cornell et al.[151]

Water is the most abundant solvent in biology and chemical processing. Because water is vital to life on earth, it plays a key role in nearly all health applications. Water also plays a key role in environmental, weather, and climate change processes. Consequently, water is the most important molecule for molecular modeling in general. Because many classical atomistic molecular dynamics and Monte Carlo simulations will use DDEC6 NACs for non-water molecules combined with a well-established commonly used water model for the water solvent, it is desirable for the DDEC6 NACs for the water molecule to be approximately consistent with those of commonly used water models. Lee et al. computed NACs for large unit cells of simulated bulk water (~2500 atoms with PBE functional and large psinc basis sets) and showed the DDEC/cc2 ($q_H$ = 0.3915, $q_O$ = -0.783) and DDEC3 ($q_H$ = 0.402, $q_O$ = -0.804) results are similar to common 3-site water models.[69, 70] Table 22 lists commonly used 3-site water models that have been optimized to reproduce various properties of bulk water in classical atomistic simulations.[146-149] For comparison, DDEC6 results for the isolated water molecule with B3LYP/6-311++G** optimized geometry and electron distribution are $q_H$ = 0.3953, $q_O$ = -0.7906. A recent study by Farmahini et al. computed DDEC3 NACs to study changes in the hydrophobicity/hydrophilicity of nanoporous silicon carbide-derived carbon upon fluorine doping.[152] These results show the DDEC methods are well-suited for studying water molecules.

The B-DNA decamer (CCATTAATGG) structure was obtained from a neutron diffraction experiment performed by Arai et al. (PDB ID: 1WQZ).[153] The 25 $H_2O$ molecules in the crystal structure are from the solvent and are hydrogen-bonded to the B-DNA as shown in Figure 12. We added a $Na^+$ ion next to each phosphate group, following previous studies to simulate the B-DNA being in a real solution.[154] We optimized the positions of the $Na^+$ ions in VASP while keeping the experimental B-DNA structure fixed. We used the PBE functional with a planewave cutoff energy of 400 eV. The left panel of Figure 12 shows the optimized B-DNA decamer including the $Na^+$ ions and hydrogen-bonded water molecules. The right panel of Figure 12 compares the DDEC6 NACs to the CHARMM27 and AMBER4.1 forcefield NACs for DNA. There is some scatter in the data, but the overall correlation between DDEC6 and force-field NACs is good with R-squared correlation coefficients of 0.93 (CHARMM27) and 0.91 (AMBER4.1). The phosphorus NAC was 1.5[150] for CHARMM compared to 1.166[151] for AMBER, with the DDEC6 value of 1.38 in-between. Recent articles by Lee et al. studied applications of DDEC NACs to atomistic simulations of large biomolecules including a comparison of force-fields based on AMBER and DDEC NACs for several large proteins.[69, 70]

### 5.5 Reproducing the electrostatic potential across multiple system conformations for constructing flexible force-fields

For some applications, the preferred strategy is to use NACs from quantum chemistry calculations to build an electrostatic model in flexible force-fields for classical molecular dynamics and Monte Carlo simulations. Gas adsorption and diffusion in porous crystalline materials is a common example.[62] The simulations of large biomolecules is another common example.[150] For these applications, flexibility of the material may play a key role.[155] Thus, it is important for the NACs to simultaneously have good conformational transferability and approximately reproduce the electrostatic potential around the material. This is a challenging criterion, because NACs directly fit to the electrostatic potential (without additional fitting criteria) often have poor conformational transferability.[71, 74]

#### 5.5.1 Carboxylic acids

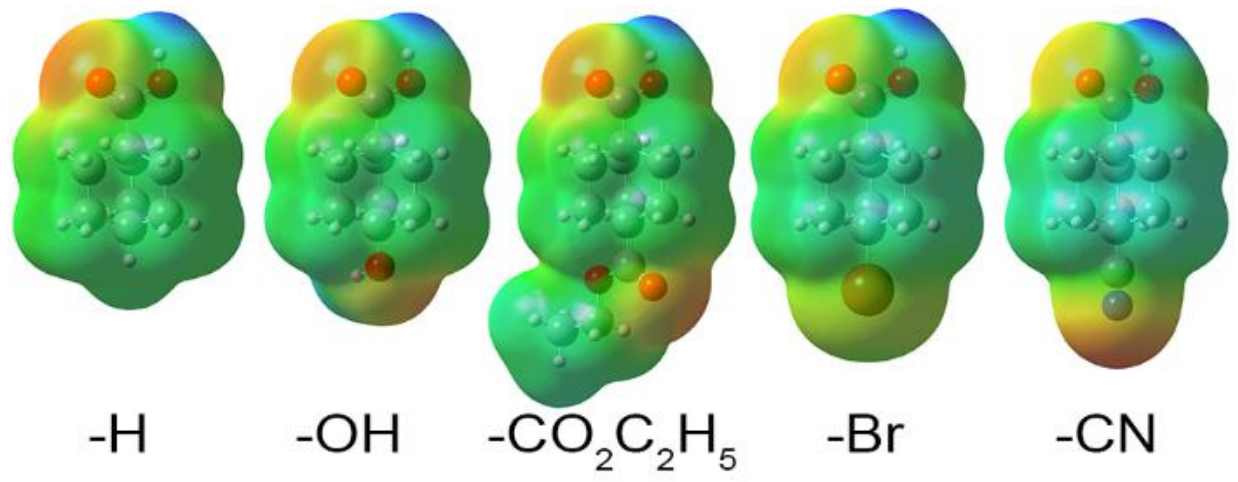

Figure 13. Structures and molecular electrostatic potentials (MEPs) of the low energy conformations (B3LYP/6-311++G** level of theory) of 4-X-substituted bicyclo[2,2,2]octane-1-carboxylic acids: X = (a) –H, (b) –OH, (c) $–CO_2C_2H_5$, (d) –Br, and (e) –CN. The MEP is shown on the 0.0004 electrons/bohr$^3$ density contour with a MEP scale ranging from -1.6 volts (red) to 1.6 volts (blue). The electrostatic potential is negative near the oxygen, bromine, and nitrogen atoms and postive near the proton of the carboxylate group.

In a previous publication, Manz and Sholl studied the accuracy of HD, DDEC3, ISA, IH, NPA, and ESP NACs for reproducing the electrostatic potential across various conformations of the five 4-X-substituted bicyclo[2,2,2]octane-1-carboxylic acids shown in Figure 13.[14] They found the ESP NACs reproduce the electrostatic potential as accurately as possible when optimized individually for each conformation, but have low conformational transferability.[14] When using a conformationally averaged set of NACs to reproduce the electrostatic potential across the various conformations of each molecule, the ESP NACs performed slightly better than the DDEC3 NACs.[14] When using NACs from the low energy conformation to reproduce the electrostatic potential across the various conformations of each molecule, the DDEC3 NACs performed slightly better than the ESP NACs.[14] The DDEC3 NACs also had excellent conformational transferability.[14]

We now show these desirable properties are further improved by the DDEC6 NACs. The B3LYP/6-311++G** optimized geometries and electron distributions from reference [14] are used. Table 23 summarizes the fragment charges for each of these charge assignment methods, where the weighted sum as defined by Manz and Sholl[14] is:

$$q_{frag} = \sum_{A} (0.75)^{N_{bonds}} q_A \qquad (143)$$

where $q_A$ is the NAC for atom A and $N_{bonds}$ is the number of bonds in the shortest chain connecting the atom to the substituent group. The purpose of this weighted sum is to smooth out the effects of the NACs, where all of the atoms in the substituent group are weighted by $q_A$ and those not in the substituent group receive a diminished weight that tends towards zero as the atom is far removed from the substituent group. Roberts and Moreland determined σ' substituent constants using experimentally measured acid dissociation constants.[156] As shown in Table 23, the HD NACs were most closely correlated to the σ' values, where the R-squared correlation coefficient is that for linear regression: $q_{frag} = a_0 + a_1\sigma'$.[14] The DDEC6 NACs showed the second strongest correlation to the σ' values, with an R-squared correlation coefficient of 0.90 for the weighted sum in Eq. (143). This shows the DDEC6 NACs captured the important chemical trend among the substituent groups.

Table 23. Fragment charges for the low energy conformation. NAC methods ordered from highest to lowest R-squared correlation coefficient for weighted sum.

| | | substituent net charge[b] | | | | | | | weighted sum of eq (143)[b] | | | | | | |
|---|---|---|---|---|---|---|---|---|---|---|---|---|---|---|---|
| X | σ'[a] | HD | DDEC6 | DDEC3 | ISA | IH | NPA | ESP | HD | DDEC6 | DDEC3 | ISA | IH | NPA | ESP |
| H | 0.000 | 0.03 | 0.05 | 0.04 | -0.01 | 0.07 | 0.21 | -0.02 | -0.01 | -0.04 | -0.03 | 0.02 | -0.06 | -0.20 | -0.01 |
| OH | 0.283 | -0.08 | -0.22 | -0.25 | -0.31 | -0.25 | -0.29 | -0.33 | -0.05 | -0.12 | -0.12 | -0.09 | -0.15 | -0.34 | -0.10 |
| $CO_2C_2H_5$ | 0.297 | -0.03 | -0.02 | -0.02 | -0.07 | 0.06 | 0.02 | -0.04 | -0.04 | -0.10 | -0.08 | -0.04 | -0.09 | -0.29 | -0.02 |
| Br | 0.454 | -0.1 | -0.23 | -0.25 | -0.29 | -0.11 | -0.02 | -0.19 | -0.07 | -0.17 | -0.16 | -0.15 | -0.13 | -0.30 | -0.18 |
| CN | 0.579 | -0.16 | -0.17 | -0.16 | -0.25 | -0.12 | -0.02 | -0.31 | -0.11 | -0.16 | -0.14 | -0.10 | -0.16 | -0.32 | -0.10 |
| $R^2$ corr. coef. | | 0.92 | 0.54 | 0.44 | 0.51 | 0.26 | 0.17 | 0.44 | 0.93 | 0.90 | 0.81 | 0.71 | 0.69 | 0.59 | 0.47 |

[a] From reference [156]. [b] DDEC3, ESP, HD, IH, ISA, and NPA NACs are from reference [14].

We now consider accuracy of these charge assignment methods for reproducing the electrostatic potential across various system conformations. As shown in Table 24, the conformational transferability of the charge assignment methods from best to worst ordered IH > NPA > DDEC6 > DDEC3 > ISA > HD. The excellent conformational transferability of IH charges and poor conformational transferability of ISA charges have also been shown in prior work.[14, 64, 65] Table 25 compares the electrostatic potential RMSE and RRMSE values averaged across all molecular conformations for each of the point charge models using (a) NACs optimized individually for each conformation, (b) conformation averaged NACs, and (c) the low energy conformation NACs. DDEC6 values at the M+SCP (individually optimized for each conformation and using the low energy conformation), D (individually optimized for each conformation), and D+SCP (individually optimized for each conformation) levels are also shown for comparison. As expected, the ESP NACs reproduced the electrostatic potential most accurately among all point charge models optimized individually for each conformation. Although the DDEC6 NACs gave significantly higher RMSE and RRMSE values than the ESP NACs, the DDEC6 RMSE and RRMSE values including atomic dipoles (e.g., D and D+SCP) were approximately the same as those for ESP NACs optimized individually for each conformation. When using the conformation averaged NACs, ESP NACs still yielded the best overall results with the DDEC3 and DDEC6 NACs not far behind. When using NACs from the low energy conformation, the DDEC6 method provided the best overall results.

Table 24. Assessment of the conformational transferability of different charge assignment methods. NAC methods ordered from highest to lowest conformational transferability.

| | Mean unsigned deviation of NACs[a] | | | | |
|---|---|---|---|---|---|
| Substituent: | H | Br | CN | OH | Ester |
| Conformations: | 4 | 4 | 4 | 8 | 16 |
| IH | 0.002 | 0.002 | 0.002 | 0.003 | 0.002 |
| NPA | 0.004 | 0.004 | 0.004 | 0.006 | 0.006 |
| DDEC6 | 0.005 | 0.005 | 0.005 | 0.007 | 0.006 |
| DDEC3 | 0.007 | 0.007 | 0.007 | 0.010 | 0.008 |
| ISA | 0.015 | 0.015 | 0.016 | 0.025 | 0.016 |
| HD | 0.038 | 0.057 | 0.051 | 0.045 | 0.041 |

[a] Mean unsigned deviations of NACs for the DDEC3, HD, IH, ISA, and NPA are from reference [14].

Finally, we considered the 25 conformations of the –OH substituted carboxylic acid generated by the ab initio molecular dynamics (AIMD) calculations of Manz and Sholl at 300 K (Nosé thermostat).[14] Following AIMD calculations in VASP using the PW91 functional with D2 dispersion corrections, Manz and Sholl computed the electron distributions and electrostatic potentials in GAUSSIAN 09 using the B3LYP/6-311++G** level of theory for each geometry.[14] We use these same geometries, electron distributions, and electrostatic potentials here. Our purpose here is to see how the DDEC6 NACs perform compared to the previously reported results[14] for the DDEC3, ESP, HD, IH, ISA, and NPA methods. As shown in Table 26, the DDEC6 NACs had lower electrostatic potential RMSE and RRMSE values across these AIMD conformations than any of the other six charge assignment methods when using either the low energy NACs or the conformation averaged NACs from Table 25. In summary, all of these tests for substituted carboxylic acids show the DDEC6 NACs have desirable properties for constructing flexible force-fields: (a) reproduce chemical trends, (b) good conformational transferability, and (b) reasonable accuracy for reproducing the electrostatic potential across various system conformations.

Table 25. Average RMSE and RRMSE values for charge assignment methods. NAC methods listed in alphabetical order. The best values for a point charge model are shown in boldface type. Values at M+SCP, D, or D+SCP are shown in boldface type if they are equal to or better than the best point charge model.

| | Avg. RMSE (kcal/mol)[a] | | | | | Avg. RRMSE[a] | | | | |
|---|---|---|---|---|---|---|---|---|---|---|
| Substituent → | H | Br | CN | OH | Ester | H | Br | CN | OH | Ester |
| NACs optimized separately for each conformation | | | | | | | | | | |
| DDEC3 | 0.81 | 1.15 | 0.87 | 0.89 | 0.77 | 0.13 | 0.18 | 0.10 | 0.13 | 0.10 |
| DDEC6 | | | | | | | | | | |
| M | 0.90 | 1.17 | 1.04 | 1.16 | 1.02 | 0.14 | 0.18 | 0.12 | 0.16 | 0.13 |
| M + SCP | 0.89 | 1.16 | 1.03 | 1.15 | 1.01 | 0.14 | 0.18 | 0.11 | 0.16 | 0.13 |
| D | **0.33** | **0.87** | **0.37** | **0.47** | **0.38** | **0.05** | **0.13** | **0.04** | **0.07** | 0.05 |
| D + SCP | **0.32** | **0.88** | **0.37** | **0.48** | **0.38** | **0.05** | **0.13** | **0.04** | **0.07** | 0.05 |
| ESP | **0.49** | **0.93** | **0.38** | **0.48** | **0.42** | **0.07** | **0.14** | **0.04** | **0.07** | **0.04** |
| HD | 2.85 | 3.26 | 3.67 | 3.73 | 3.27 | 0.41 | 0.48 | 0.40 | 0.51 | 0.41 |
| IH | 1.12 | 2.49 | 1.60 | 1.35 | 1.05 | 0.18 | 0.38 | 0.18 | 0.20 | 0.14 |
| ISA | 0.73 | 1.45 | 0.74 | 0.71 | 0.70 | 0.11 | 0.22 | 0.08 | 0.10 | 0.09 |
| NPA | 1.71 | 3.23 | 2.56 | 1.94 | 2.98 | 0.25 | 0.49 | 0.29 | 0.27 | 0.31 |
| Conformation averaged NACs | | | | | | | | | | |
| DDEC3 | 1.27 | 1.48 | 1.29 | 1.40 | **1.25** | 0.18 | 0.22 | 0.14 | **0.19** | 0.16 |
| DDEC6 | 1.21 | 1.41 | 1.31 | 1.38 | 1.27 | 0.18 | 0.21 | 0.14 | **0.19** | 0.16 |
| ESP | **1.10** | **1.36** | **1.00** | **1.37** | 1.38 | **0.15** | **0.20** | **0.11** | **0.19** | **0.14** |
| HD | 2.88 | 3.31 | 3.70 | 3.71 | 3.31 | 0.41 | 0.50 | 0.41 | 0.50 | 0.42 |
| IH | 1.48 | 2.65 | 1.84 | 1.75 | 1.39 | 0.23 | 0.41 | 0.21 | 0.25 | 0.18 |
| ISA | 1.33 | 1.81 | 1.31 | 1.57 | 1.43 | 0.18 | 0.26 | 0.14 | 0.21 | 0.18 |
| NPA | 2.12 | 3.47 | 2.86 | 2.42 | 3.71 | 0.30 | 0.52 | 0.32 | 0.33 | 0.38 |
| All conformations use NACs from low energy conformation | | | | | | | | | | |
| DDEC3 | 1.39 | 1.61 | 1.38 | 1.73 | 1.44 | 0.19 | 0.23 | 0.15 | 0.24 | 0.18 |
| DDEC6 | | | | | | | | | | |
| M | **1.31** | **1.49** | 1.35 | **1.63** | **1.34** | **0.18** | **0.22** | 0.15 | **0.22** | **0.17** |
| M + SCP | **1.30** | **1.49** | 1.44 | **1.62** | **1.33** | **0.18** | **0.22** | 0.15 | **0.22** | **0.17** |
| ESP | 1.49 | 1.73 | **1.26** | 2.12 | 1.91 | 0.19 | 0.24 | **0.13** | 0.28 | 0.20 |
| HD | 2.97 | 3.24 | 3.68 | 3.74 | 3.31 | 0.42 | 0.48 | 0.41 | 0.51 | 0.42 |
| IH | 1.55 | 2.61 | 1.85 | 1.98 | 1.48 | 0.23 | 0.40 | 0.21 | 0.28 | 0.19 |
| ISA | 1.46 | 1.97 | 1.41 | 1.98 | 1.74 | **0.19** | 0.29 | 0.15 | 0.27 | 0.21 |
| NPA[a] | 2.23 | 3.50 | 2.93 | 2.57 | 4.05 | 0.31 | 0.52 | 0.32 | 0.35 | 0.42 |

[a] RMSE and RRMSE for the DDEC3, ESP, HD, IH, ISA, and NPA are from reference [14].

Table 26. Average RMSE (kcal/mol) and RRMSE values for geometries of the –OH substituted carboxylic acid generated using ab initio molecular dynamics. NAC methods listed in alphabetical order. The best values are shown in boldface type.

| | DDEC3[a] | DDEC6 | ESP[a] | HD[a] | IH[a] | ISA[a] | NPA[a] |
|---|---|---|---|---|---|---|---|
| Using the low energy conformation NACs of Table 25 | | | | | | | |
| RMSE | 2.51 | **2.17** | 3.23 | 4.38 | 2.55 | 3.04 | 3.65 |
| RRMSE | 0.27 | **0.23** | 0.35 | 0.47 | 0.27 | 0.33 | 0.39 |
| Using the conformation averaged NACs of Table 25 | | | | | | | |
| RMSE | 2.08 | **1.88** | 2.11 | 4.31 | 2.11 | 2.44 | 3.39 |
| RRMSE | 0.23 | **0.20** | 0.23 | 0.47 | 0.23 | 0.26 | 0.37 |

[a] From reference [14].

### 5.5.2 $Li_2O$ molecule

Wang et al. compared several charge assignment methods for the $Li_2O$ molecule constrained to bent angles of 90, 100, … 170° with bond lengths and electron distributions at each of these angles optimized using the M06L functional and def2-TZVP basis set.[26] They found the CM5 NACs closely reproduced the $Li_2O$ dipole moment while the DDEC3 NACs significantly overestimated the $Li_2O$ dipole moment.[26] In symmetric non-linear conformations of $Li_2O$, the NACs that exactly reproduce the molecular dipole moment are uniquely defined (aka 'Dipole charge').[26] Here, we revisit this example to study in greater depth relationships between NACs, molecular dipole moments, electrostatic potential RMSE and RRMSE, and atomic dipole moments.

Table 27 summarizes computed NACs for each of the geometries studied by Wang et al.[26] plus the global low energy conformation using the M06L/def2-TZVP level of theory. The global low energy conformation is a linear molecule corresponding to a 180° angle. The Dipole charge cannot be computed for this low energy conformation, because its molecular dipole is zero irrespective of the NAC. For all of the charge assignment methods except the Bader method, the NAC increased monotonically as the angle increased. (For the Bader method, the increase was almost monotonic.) In order of smallest to largest Li NACs, the charge assignment methods were HD < CM5 < Dipole charge < NPA, ESP < DDEC6, Bader < DDEC3.

Table 27. Li NAC for different Li–O–Li angles in singlet $Li_2O$ molecules using various charge models. NAC methods listed in alphabetical order.

| | Li net atomic charge[a] | | | | | | | | | (geom opt) |
|---|---|---|---|---|---|---|---|---|---|---|
| Angle → | 90 | 100 | 110 | 120 | 130 | 140 | 150 | 160 | 170 | 180 |
| Bader | 0.83 | 0.85 | 0.85 | 0.86 | 0.86 | 0.86 | 0.86 | 0.87 | 0.86 | 0.87 |
| CM5 | 0.55 | 0.56 | 0.57 | 0.58 | 0.59 | 0.60 | 0.60 | 0.61 | 0.61 | 0.62 |
| DDEC3 | 0.83 | 0.88 | 0.92 | 0.94 | 0.96 | 0.97 | 0.98 | 0.98 | 0.98 | 0.98 |
| DDEC6 | 0.77 | 0.80 | 0.83 | 0.85 | 0.87 | 0.88 | 0.89 | 0.90 | 0.90 | 0.90 |
| Dipole charge | 0.59 | 0.59 | 0.60 | 0.60 | 0.60 | 0.61 | 0.62 | 0.64 | 0.65 | [b] |
| ESP | 0.61 | 0.63 | 0.65 | 0.67 | 0.70 | 0.74 | 0.79 | 0.83 | 0.86 | 0.87 |
| HD | 0.39 | 0.40 | 0.40 | 0.41 | 0.41 | 0.42 | 0.42 | 0.43 | 0.43 | 0.44 |
| NPA | 0.78 | 0.79 | 0.80 | 0.81 | 0.83 | 0.84 | 0.85 | 0.85 | 0.86 | 0.86 |

[a] Except for the linear molecule, NACs for the HD, CM5, DDEC3, ESP, NPA, and Dipole charge methods are from reference [26]. [b] Cannot be determined because the dipole moment is zero and the molecule is linear and symmetric.

To further understand these trends, Table 28 summarizes the electrostatic potential RMSE and RRMSE, dipole moment mean absolute error (MAE), and the mean unsigned deviation (MUD) from the conformation averaged NAC. From best to worst conformational transferability, the charge assignment methods ordered Bader > Dipole charge, HD > CM5, NPA > DDEC6 , DDEC3 > ESP. From best to worst accuracy in reproducing the dipole moment, the methods ordered Dipole charge > CM5 > ESP > HD > NPA > DDEC6 > Bader > DDEC3. For the RMSE and RRMSE, HD performed the worst of all the charge assignment methods, and DDEC3 performed the second worst. Among the different point charge models, the ESP NACs provided the lowest RMSE and RRMSE when the NACs were optimized separately for each molecular conformation.

Table 28. Average electrostatic potential RMSE (kcal/mol), RRMSE, dipole moment MAE in atomic units, and conformational transferability of $Li_2O$ for various charge assigment methods. Charge assignment methods listed in alphabetical order.

| | conformation averaged NACs | | NACs optimized separately for each conformation | | NACs from the lowest energy conformation | | dipole moment MAE | NAC conformational transferability (MUD) |
|---|---|---|---|---|---|---|---|---|
| | RMSE | RRMSE | RMSE | RRMSE | RMSE | RRMSE | | |
| Bader | 6.46 | 0.25 | 6.30 | 0.25 | 7.18 | 0.28 | 0.63 | 0.01 |
| CM5 | 6.92 | 0.29 | 6.90 | 0.29 | 6.31 | 0.26 | 0.07 | 0.03 |
| DDEC3 | 8.01 | 0.31 | 7.55 | 0.30 | 9.09 | 0.36 | 0.80 | 0.05 |
| DDEC6 | | | | | | | | |
| M | 6.48 | 0.26 | 5.76 | 0.23 | 7.17 | 0.28 | 0.59 | 0.05 |
| M + SCP | [a] | [a] | 5.85 | 0.23 | 7.19 | 0.28 | 0.59 | 0.05 |
| D | [b] | [b] | 1.43 | 0.06 | [b] | [b] | 0.00 | 0.05 |
| D + SCP | [b] | [b] | 1.20 | 0.05 | [b] | [b] | 0.00 | 0.05 |
| Dipole charge[d] | 6.68 | 0.29 | 6.62 | 0.28 | [c] | [c] | 0.00 | 0.02 |
| ESP | 5.55 | 0.23 | 4.40 | 0.18 | 7.27 | 0.28 | 0.19 | 0.11 |
| HD | 11.40 | 0.48 | 11.49 | 0.48 | 10.72 | 0.44 | 0.50 | 0.02 |
| NPA | 6.10 | 0.24 | 5.56 | 0.22 | 7.03 | 0.28 | 0.53 | 0.03 |

[a] Not computed. [b] Not computed, because the variation in the molecular conformation affects the orientation of the atomic dipoles. [c] Dipole charges cannot be determined, because the dipole moment of the lowest energy (i.e., linear) conformation is zero irrespective of the NAC values. [d] Since no Dipole charges were available for the linear molecule, these represent values for the nine non-linear conformations.

Across all of the accuracy measures listed in Table 28, the following overall trends were observed: (a) the NPA, DDEC6, CM5, and Bader NACs performed better than the DDEC3 NACs, (b) the NPA NACs performed better than the DDEC6 NACs, (c) across the subset of accuracy measures where the Dipole charges were defined, they performed as good as or better than the CM5 and DDEC3 NACs, and (d) all other comparisons between NAC methods yielded mixed results, with better performance for at least one accuracy measure and worse performance for at least one accuracy measure.

For comparison, Table 28 also lists DDEC6 results including spherical charge penetration and atomic dipoles. Adding spherical charge penetration to the point charges had negligible effect on the results. However, adding spherical charge penetration to the DDEC6 point charges plus atomic dipoles decreased the conformation specific RMSE to 1.20 kcal/mol, which was dramatically better than any of the point charge only models.

What conclusions can be drawn from these results? We can definitely say the HD NACs were too small in magnitude and the DDEC3 NACs were too large in magnitude for this material.[26] We can also say the Dipole charge is a limited concept, because it cannot be computed for some molecular conformations. Overall, this example illustrates some of the compromises involved in designing a general-purpose charge assignment method: (a) The molecular dipole moment can be reproduced exactly using a point charge plus atomic dipole model, but this makes the model more complicated than a point charge only model. (b) Fitting the electrostatic potential directly to a point charge model for each conformation leads to comparatively low RMSE values, but this degrades the conformational transferability as demonstrated by the ESP results. (c) The electrostatic potential can be more accurately reproduced by a (truncated) multipole model with charge penetration terms (e.g., D+SCP), but this results in more complicated force-field terms.

### 5.5.3 Zn-nicotinate metal-organic framework

Figure 14 shows the Zn-nicotinate MOF. This structure is comprised of one-dimensional pore channels having an approximately square cross-section. The electrostatic potential is most positive near the atomic nuclei and becomes most negative near the pore centers. In Figure 14, a contour of electrostatic potential isovalue is displayed as a green surface.

To assess the effects of framework flexibility on the Zn-nicotinate MOF, we performed an AIMD calculation in VASP. This AIMD simulation used a planewave cutoff energy of 400 eV, the PAW method, and PBE functional for 1200 fs with a time step of 1 fs. (Electronic energies were converged to $10^{-4}$ eV and a PREC = Normal grid was used.) A canonical ensemble at T = 300 K was simulated using a Nosé thermostat[157]. An initial run using a Nosé-mass setting SMASS = 0.03 exhibited unreasonably large temperature fluctuations (~1000K maximum temperature fluctuation), so the Nosé mass was set using SMASS = 0.005 and exhibited reasonable temperature fluctuations that preserved the MOF's chemical integrity. A period of ~250 fs was allowed for thermal equilibration. Twenty-one conformations were used for the subsequent charge analysis: the DFT-optimized minimum energy geometry and 20 AIMD conformations corresponding to time steps 250, 300, 350, ... 1200. For each of these conformations, the valence and total all-electron densities and electrostatic potential were generated in VASP using single-point (fixed-geometry) calculations with a PREC = Accurate grid.

Electrostatic potential fitting NACs were calculated using the REPEAT method and associated software code by Campaña, Mussard, and Woo.[20] For the REPEAT method, NACs were fit outside surfaces defined by $\gamma_R$=1.0 and 1.3 times the atomic vdW radii. As previously noted, REPEAT NACs are highly sensitive to the particular value of this vdW multiplier $\gamma_R$.[20, 72, 73] Campaña et al.[20] recommended the value $\gamma_R$ =1.0. Chen et al. recommended the value $\gamma_R$ =1.3.[73]

Table 29 summarizes electrostatic potential RMSE and RRMSE values averaged over all 21 system conformations. These were computed on a uniform grid defined by an inner vdW multiplier of 1.3 and an outer vdW multiplier limited only by the pore size.[72] When using the conformation averaged NACs and the conformation specific NACs, the REPEAT method produced a more accurate representation of the electrostatic potential than the DDEC6 NACs. By definition, electrostatic potential fitting methods (such as REPEAT) should produce a more accurate representation of the electrostatic potential than other types of atom-centered point charge models when using the conformation specific NACs. Including atomic dipoles in the conformation specific NACs dramatically lowered the DDEC6 RMSE from 2.99 to 0.55 kcal/mol, which was even better than the REPEAT values for both $\gamma_R$ =1.0 and 1.3. This means that for reproducing the electrostatic potential of a rigid framework, the DDEC6 method including atomic dipoles sometimes outperforms the REPEAT NACs. As shown in Table 29, including the spherical charge penetration term had negligible effect. When using the low energy conformation NACs, REPEAT with $\gamma_R$ =1.0 yielded the lowest RMSE and RRMSE values. For the low energy conformation NACs, the DDEC6 RMSE and RRMSE values were between the REPEAT values using $\gamma_R$ = 1.0 and 1.3.

Table 29. Average RMSE (kcal/mol) and RRMSE of Zn-nicotinate metal-organic framework at 21 different structural conformations. The values in parentheses include spherical charge penetration effects.

| | DDEC6 | | REPEAT | REPEAT |
|---|---|---|---|---|
| | M (M+SCP) | D (D+SCP) | ($\gamma_R$=1.0) | ($\gamma_R$=1.3) |
| | Using the conformation averaged NACs | | | |
| RMSE | 3.13 | [a] | 1.05 | 1.01 |
| RRMSE | 0.49 | [a] | 0.22 | 0.20 |
| | Using the conformation specific NACs | | | |
| RMSE | 2.99 (3.00) | 0.55 (0.47) | 0.65 | 0.60 |
| RRMSE | 0.47 (0.47) | 0.10 (0.09) | 0.11 | 0.10 |
| | Using the low energy conformation NACs | | | |
| RMSE | 3.39 (3.38) | [a] | 2.34 | 3.56 |
| RRMSE | 0.53 (0.53) | [a] | 0.40 | 0.66 |

[a] Not computed, because the variation in the molecular conformation affects the orientation of the atomic dipoles.

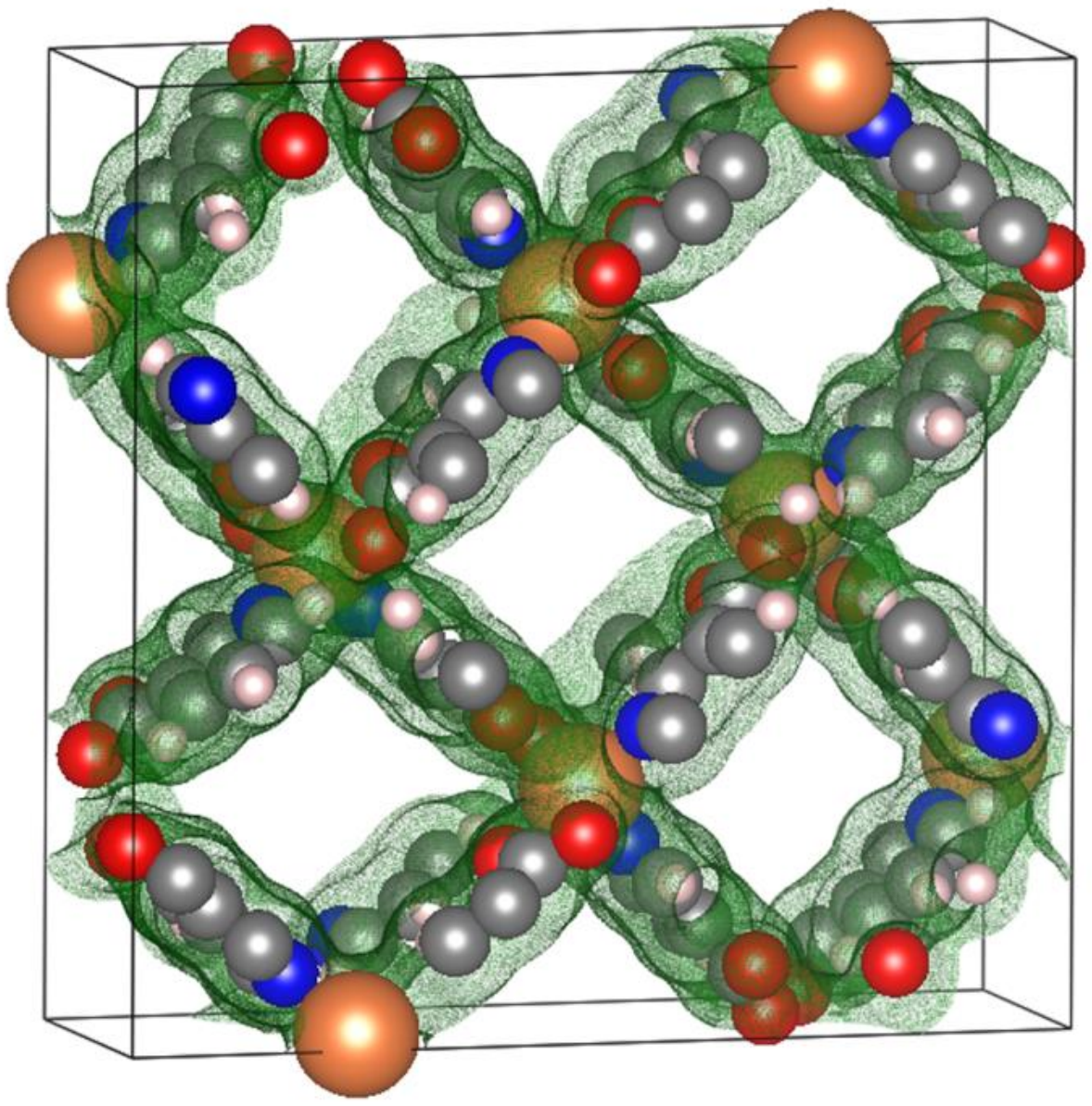

Figure 14. The geometry-optimized Zn-nicotinate MOF with one-dimensional pore channels. The lines mark the unit cell boundaries. The pore cross-sections are approximately square. The green surface corresponds to an electrostatic potential isovalue. The electrostatic potential becomes more negative closer to the pore centers and more positive closer to the atomic nuclei. Atom colors: C (gray), N (blue), O (red), Zn (orange), H (pink).

Table 30 summarizes information about the conformational transferability of the NACs. The DDEC6 NACs had excellent conformational transferability with a MUD ≤ ~0.01 for each atom type. Moreover, the max and min DDEC6 NACs for each atom type differed by < 0.1e. The REPEAT method had better conformational transferability with $\gamma_R$ =1.0 than with $\gamma_R$ =1.3. For $\gamma_R$ =1.0, three of the atom types exhibited fluctuations > 0.5 e as measured by the difference between max and min NACs. For $\gamma_R$ =1.3, the Zn NAC varied from -0.49 to 1.38 across the different conformations, and two of the other atom types also exhibited fluctuations > 1e. With the exception of atom type O(1), all of the min and max values of the DDEC6 NACs were between the min and max values of the REPEAT NACs using $\gamma_R$ =1.3. With the exception of atom types C(6), O(1), and O(2), all of the min and max values of the DDEC6 NACs were between the min and max values of the REPEAT NACs using $\gamma_R$ =1.0.

Table 30. Average, maximum, minimum, and mean unsigned deviation of NACs for each atom type in Zn-nicotinate using DDEC6 and REPEAT methods.

| atom type | DDEC6 | | | | REPEAT ($\gamma_R$ =1.0) | | | | REPEAT ($\gamma_R$ =1.3) | | | |
|---|---|---|---|---|---|---|---|---|---|---|---|---|
| | avg | max | min | MUD | avg | max | min | MUD | avg | max | min | MUD |
| C(1) | -0.12 | -0.10 | -0.14 | 0.01 | -0.16 | 0.08 | -0.36 | 0.09 | -0.14 | 0.34 | -0.50 | 0.18 |
| C(2) | -0.06 | -0.05 | -0.08 | 0.01 | -0.06 | 0.17 | -0.23 | 0.07 | 0.03 | 0.56 | -0.20 | 0.11 |
| C(3) | -0.02 | 0.00 | -0.04 | 0.01 | 0.05 | 0.25 | -0.14 | 0.07 | -0.04 | 0.32 | -0.42 | 0.13 |
| C(4) | 0.08 | 0.10 | 0.05 | 0.01 | 0.14 | 0.33 | -0.19 | 0.09 | 0.03 | 0.40 | -0.42 | 0.17 |
| C(5) | 0.09 | 0.14 | 0.06 | 0.01 | 0.19 | 0.50 | -0.05 | 0.11 | 0.20 | 0.77 | -0.33 | 0.21 |
| C(6) | 0.56 | 0.58 | 0.52 | 0.01 | 0.37 | 0.51 | 0.20 | 0.06 | 0.26 | 0.59 | 0.04 | 0.13 |
| H(1) | 0.09 | 0.10 | 0.07 | 0.01 | 0.04 | 0.16 | -0.03 | 0.03 | 0.10 | 0.26 | -0.04 | 0.07 |
| H(2) | 0.10 | 0.11 | 0.08 | 0.00 | 0.03 | 0.11 | -0.03 | 0.03 | 0.00 | 0.22 | -0.37 | 0.08 |
| H(3) | 0.12 | 0.13 | 0.11 | 0.01 | 0.09 | 0.20 | -0.04 | 0.04 | 0.05 | 0.28 | -0.18 | 0.09 |
| H(4) | 0.12 | 0.14 | 0.11 | 0.00 | 0.08 | 0.18 | 0.00 | 0.03 | 0.12 | 0.30 | -0.07 | 0.07 |
| N | -0.23 | -0.21 | -0.25 | 0.01 | -0.37 | -0.10 | -0.63 | 0.11 | -0.23 | 0.47 | -0.68 | 0.24 |
| O(1) | -0.56 | -0.53 | -0.58 | 0.01 | -0.39 | -0.28 | -0.49 | 0.04 | -0.23 | 0.05 | -0.51 | 0.13 |
| O(2) | -0.53 | -0.50 | -0.56 | 0.01 | -0.41 | -0.32 | -0.51 | 0.03 | -0.37 | -0.21 | -0.65 | 0.07 |
| Zn | 0.74 | 0.76 | 0.70 | 0.01 | 0.80 | 0.91 | 0.54 | 0.08 | 0.47 | 1.38 | -0.49 | 0.33 |

What are the implications of these results for developing force-fields to reproduce the electrostatic potential surrounding materials? If the goal is to reproduce the electrostatic potential as accurately as possible surrounding a rigid material using an atom-centered point-charge model without regard for the chemical meaning of those NACs, then methods such as ESP,[17] Chelp,[18] or Chelpg[19] for molecular systems or REPEAT[20] for periodic materials or the Wolf-summation technique of Chen et al.[73] are preferable, because these methods minimize RMSE without regard for the chemical meaning of the NACs. If the goal is to produce chemically meaningful NACs that reproduce the electrostatic potential as accurately as possible surrounding a rigid material, the DDEC6 method is preferable with or without including atomic dipoles, because this method assigns atomic electron distributions to resemble real atoms and reproduce the electrostatic potential. For constructing flexible, non-reactive force-fields, NACs based on the low-energy structure or an average across multiple system conformations can be used. Depending on the material and computational details, either the DDEC6, REPEAT[20] (or its extension to simultaneously fit multiple conformations[74]), ESP,[17] Chelp,[18] Chelpg,[19] or Wolf-summation technique[73] may yield the more accurate flexible force-field NACs.

### 5.6 Systems comprised almost entirely of surface atoms

Figure 15 compares DDEC6 to DDEC3 NACs for the same materials comprised almost entirely of surface atoms that were previously used by Manz and Sholl to prepare a similar plot comparing DDEC/c2 to DDEC3 NACs.[14] We used the same geometries and electron density files as reference [14]. These materials were: (a) $B_4N_4$ cluster, (b) BN nanotube, (c) h-BN sheet, (d) formamide (PW91 exchange-correlation functional with 6-311++G** and planewave basis sets), (e) the metal-organic frameworks IRMOF-1 (DFT-optimized and x-ray diffraction geometries), MIL-53(Al), ZIF-90, ZIF-8, Zn-nicotinate (PW91 optimized geometry), and CuBTC, (f) $ZrN_4C_{52}H_{72}$ organometallic complex, (g) $ZrO_4N_4C_{52}H_{72}$ organometallic complex, (h) $[GdI]^{+2}$ (SDD and planewave basis sets), (i) the MgI, MoI, SnI, TeI, and TiI molecules using both SDD and planewave basis sets, (j) $[Cr(CN)_6]^{3-}$, (k) the ozone singlet and triplet spin states using the PW91, B3LYP, CCSD, SAC-CI, and CAS-SCF exchange-correlation theories, (l) ozone +1 cation doublet (PW91, B3LYP, and CCSD methods), (m) the $Fe_4O_{12}N_4C_{40}H_{52}$ noncollinear single molecule magnet, and (n) $[Cu_2N_{10}C_{36}H_{52}]^{2+}$ spin triplet. As shown in Figure 15, the DDEC6 NACs follow a trend similar to the DDEC3 NACs for materials comprised almost entirely of surface atoms, but the two charge measures provide statistically significant differences.

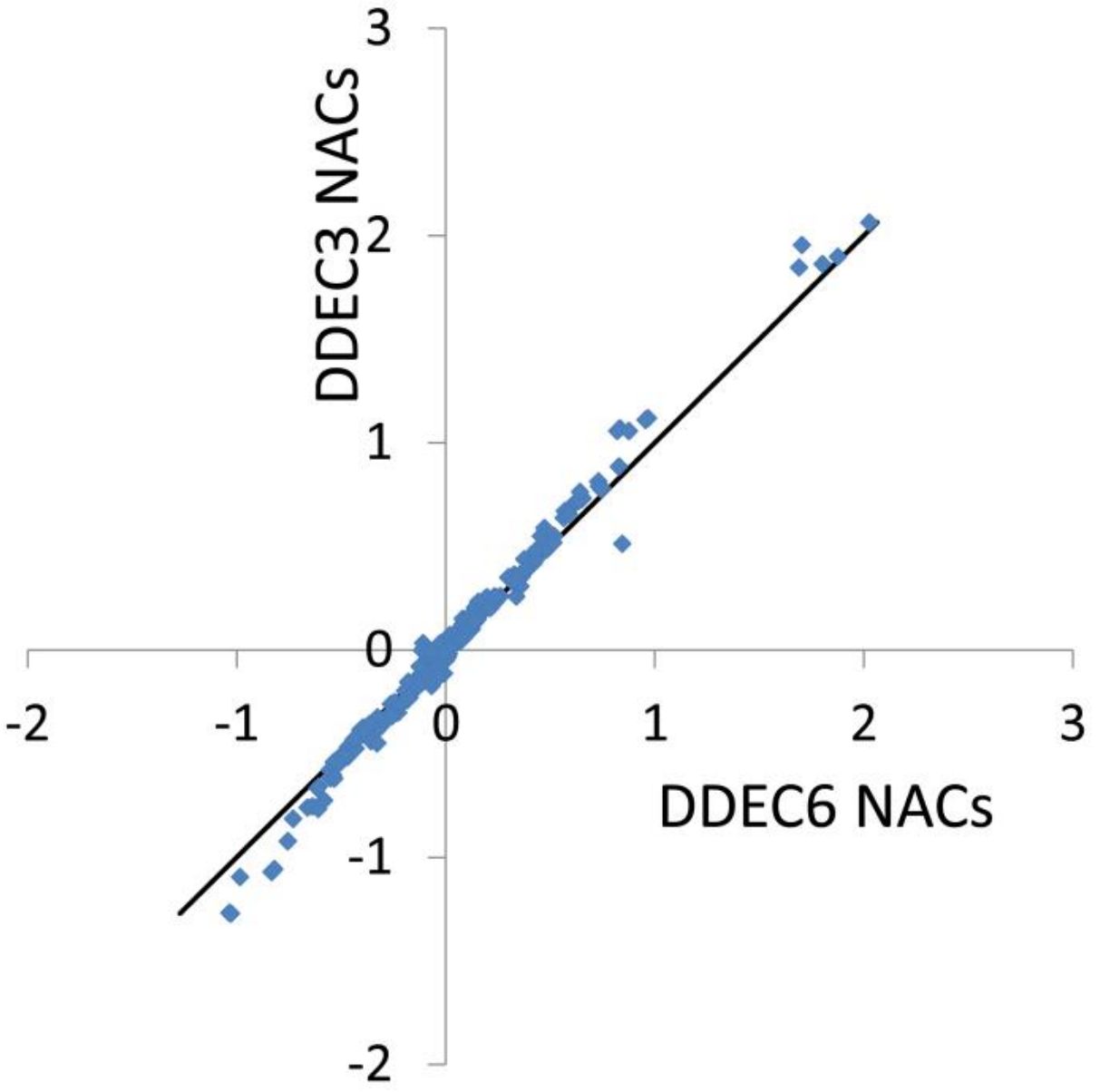

Figure 15. Comparison of DDEC3 and DDEC6 NACs for systems comprised almost entirely of surface atoms. The black line has a slope of 1 and an intercept of 0.

### 5.7 Solid surfaces

A general purpose method for assigning NACs should yield reasonable results for both surface and buried atoms. This is significant, because some methods for assigning NACs such as DDEC/c1, DDEC/c2, ISA, REPEAT, ESP, etc. do not work well for buried atoms.[13, 14, 66, 71, 72] Here, we show the DDEC6 method continues to give reasonable results for three solid surfaces already studied with the DDEC3 method: a K adatom on a $Mo_2C$ (110) surface, the NaF(001) surface, and the $SrTiO_3$(100) surface.[14] The K adatom on a $Mo_2C$ (110) surface slab geometry is from Han et al.,[158] and for this material we used the same PBE-generated electron distribution and DDEC3 NACs as Manz and Sholl[14]. We optimized the NaF(001) and $SrTiO_3$(100) surface slab geometries and electron distributions using the PBE functional at the PBE-optimized NaF and $SrTiO_3$ bulk lattice constants and used these to compute DDEC3 and DDEC6 NACs.

Figure 16 shows the geometries of these three surface slabs and compares the DDEC6 and DDEC3 NACs. For all three materials, the general trends displayed by the DDEC6 and DDEC3 NACs were similar, except the DDEC3 NACs were larger in magnitude than the DDEC6 NACs. For all three of these solid surfaces, the DDEC6 method gave similar (but not identical) NACs for the surface and buried atoms of the same element. This shows the DDEC6 method analyzes surface and buried atoms on a consistent basis.

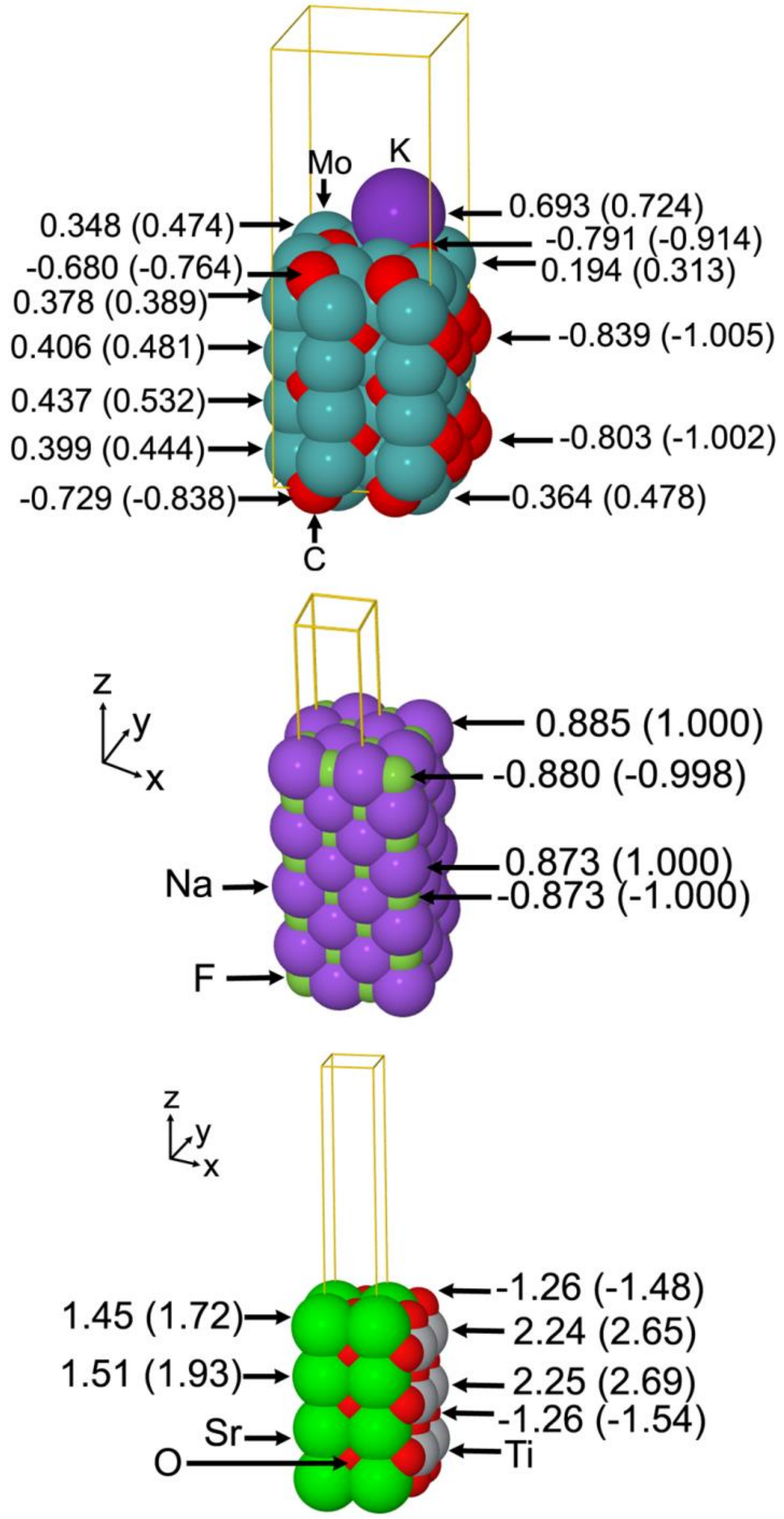

Figure 16: DDEC6 (and DDEC3 in parentheses) NACs of *top:* K adatom on a $Mo_2C$ (110) surface; *center:* NaF(001) slab; *bottom:* $SrTiO_3$(100) slab.

Table 31 compares the NACs and atomic multipoles for the $SrTiO_3$ slab to those of the bulk material. $SrTiO_3$ is comprised of alternating SrO and $TiO_2$ layers. The trends in atomic dipoles were similar for the DDEC3 and DDEC6 methods. All of the atomic dipoles in the bulk material were zero due to symmetry. In the surface slab, each atom had a non-zero atomic dipole parallel to the direction of the surface plane. In both the bulk and slab materials, the $Q_{x^2-y^2}$ quadrupolar component was zero for all atoms except the O atoms in the $TiO_2$ planes. Half the oxygen atoms in each $TiO_2$ plane had a positive value for the $Q_{x^2-y^2}$ quadrupolar component and the other half had a negative value. Because the surface breaks the crystal symmetry along the z-axis, the Sr and Ti atoms, which had $Q_{3z^2-r^2}=0$ in the bulk material, acquired non-zero $Q_{3z^2-r^2}$ moments in the surface slab.

Table 31. DDEC6 NACs and multipole moments (in a.u.) for $SrTiO_3$(100) and Bulk $SrTiO_3$[a]. DDEC3 NACs and multipole moments shown in parentheses.

| Layer | Atom | NAC | $\mu_z$ | $Q_{x^2-y^2}$ | $Q_{3z^2-r^2}$ |
|---|---|---|---|---|---|
| 1 | O | -1.260 (-1.479) | 0.111 (0.193) | 0.000 (0.000) | -0.302 (-0.083) |
| 1 | Sr | 1.443 (1.722) | -0.050 (-0.154) | 0.000 (0.000) | 0.349 (0.003) |
| 2 | O | -1.266 (-1.534) | -0.036 (-0.039) | ±0.145 (±0.254) | 0.107 (0.183) |
| 2 | Ti | 2.236 (2.648) | -0.007 (-0.010) | 0.000 (0.000) | -0.065 (-0.075) |
| 3 | O | -1.253 (-1.534) | 0.020 (0.027) | 0.000 (0.000) | -0.265 (-0.435) |
| 3 | Sr | 1.505 (1.932) | 0.005 (-0.018) | 0.000 (0.000) | 0.018 (0.009) |
| 4 | O | -1.265 (-1.562) | 0.002 (0.001) | ±0.143 (±0.245) | 0.128 (0.213) |
| 4 | Ti | 2.253 (2.686) | 0.001 (-0.002) | 0.000 (0.000) | -0.021 (-0.022) |
| Bulk | O (Sr) | -1.250 (-1.541) | 0.000 (0.000) | 0.000 (0.000) | -0.287 (-0.474) |
| Bulk | O (Ti) | -1.250 (-1.541) | 0.000 (0.000) | ±0.144 (±0.237) | 0.144 (0.237) |
| Bulk | Sr | 1.488 (1.927) | 0.000 (0.000) | 0.000 (0.000) | 0.000 (0.000) |
| Bulk | Ti | 2.262 (2.696) | 0.000 (0.000) | 0.000 (0.000) | 0.000 (0.000) |

[a] For all atoms, $\mu_x = \mu_y = Q_{xy} = Q_{xz} = Q_{yz} = 0.000$.

Table 32: DDEC and Bader NACs for $SrTiO_3$ crystal at ambient pressure. Results shown using the following number of frozen core electrons: (12) Ti, (28) Sr, and (2) O.

| Atom type | DDEC3 | DDEC6 | Bader | IH/R3[a] |
|---|---|---|---|---|
| O | -1.541 | -1.250 | -1.162 | -1.43 |
| Ti | 2.696 | 2.262 | 2.091 | 2.69 |
| Sr | 1.927 | 1.488 | 1.394 | 1.62 |

[a] IH results from reference [63] using the R3 reference ions.

Table 33: DDEC and Bader NACs for NaF crystal at ambient pressure.

| Atom type | DDEC3[a] | DDEC6[a] | Bader[a] | IH/R3[b] |
|---|---|---|---|---|
| Na | 1.014 (1.000) | 0.853 (0.877) | 0.855 (0.798) | 1.05 |
| F | -1.014 (-1.000) | -0.853 (-0.877) | -0.855 (-0.798) | -1.05 |

[a] Values listed for 2 frozen Na core electrons; values in parenthesis for 10 frozen Na core electrons. [b] IH results from reference [63] using the R3 reference ions.

Table 32 and Table 33 compare DDEC3, DDEC6, Bader, and IH/R3 NACs for bulk NaF and $SrTiO_3$, respectively. We used PBE optimized geometries and electron distributions to compute the DDEC3, DDEC6, and Bader results. For these materials, the NAC magnitudes followed the trend DDEC3, IH/R3 > DDEC6, Bader.

### 5.8 Collinear and non-collinear magnetic materials

For both the collinear and non-collinear magnetic systems described below, DDEC6 ASMs were computed with the method of Manz and Sholl[68] using the DDEC6 atomic electron distributions and the recommended value $\chi_{spin} = 0.5$. Integrations were performed over 100 uniformly spaced radial shells up to a cutoff radius of 5 Å.

#### 5.8.1 Collinear magnetism

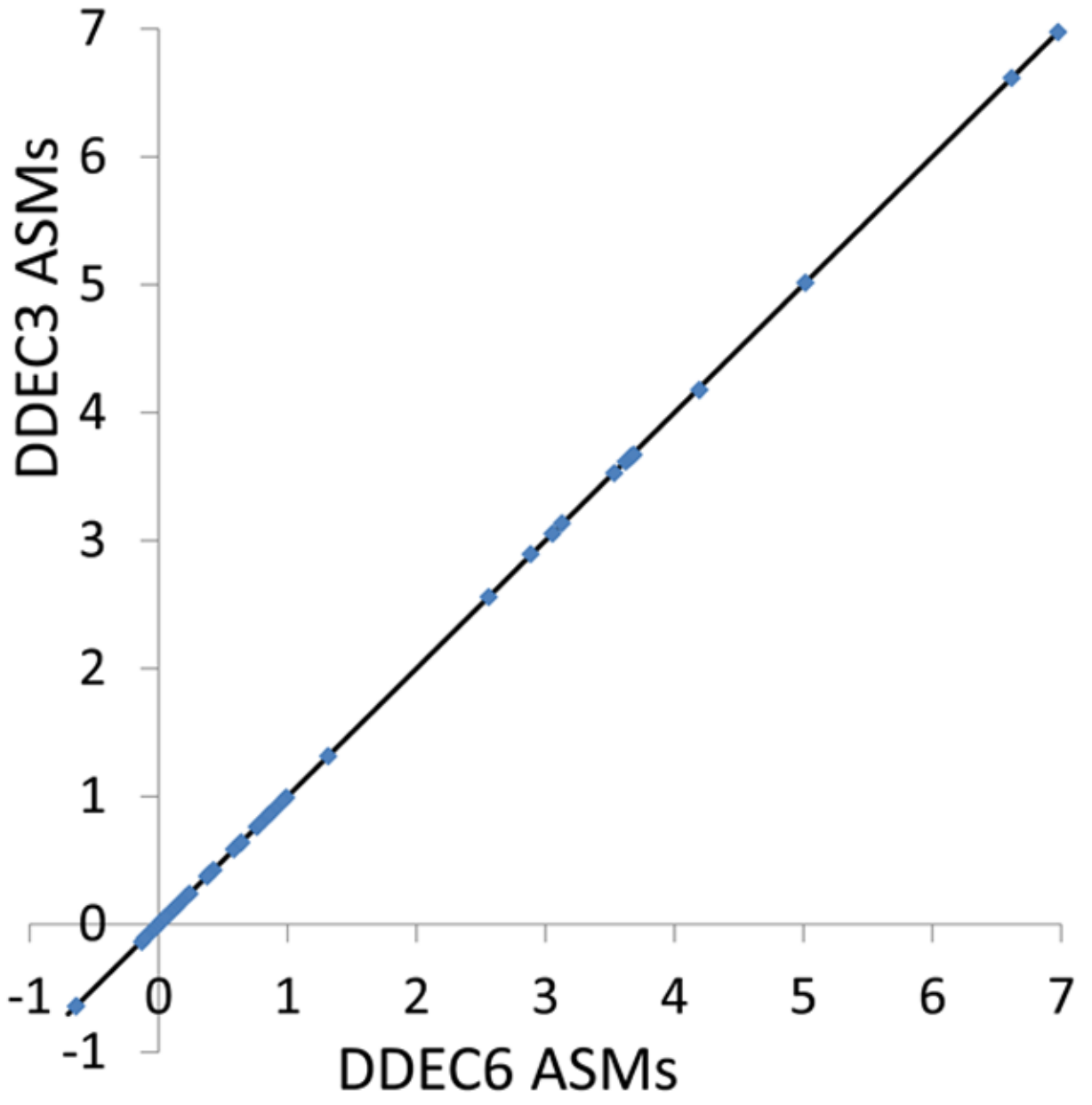

Figure 17. Comparison of DDEC3 and DDEC6 atomic spin moments for systems with collinear magnetism. The black line has a slope of 1 and an intercept of 0.

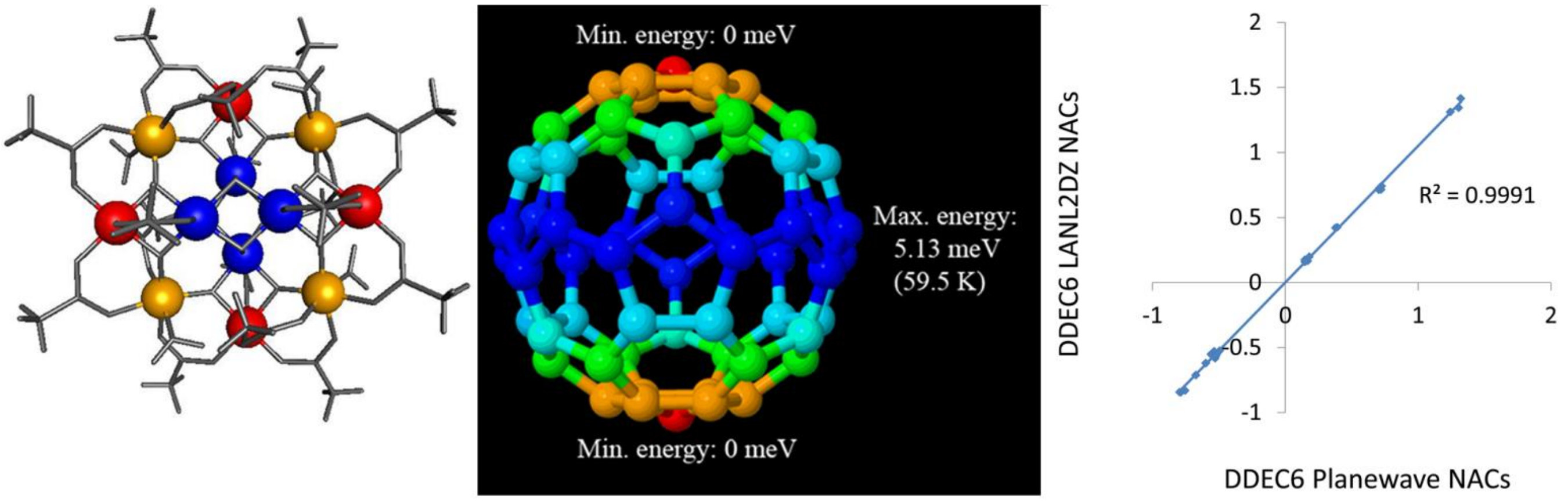

Figure 18. *Left:* Atomic structure of $Mn_{12}$-acetate single molecule magnet. Mn type 1 (blue), Mn type 2 (red), Mn type 3 (yellow). In the minimum energy conformations, the Mn ASM vectors are perpendicular to the plane of the page. *Middle:* Computed spin-orbit coupling potential energy surface. *Right:* Comparison of DDEC6 NACs computed with LANL2DZ and planewave basis sets.

Figure 17 compares DDEC6 to DDEC3 ASMs for all of the collinear magnetic materials studied in the article by Manz and Sholl[14] that introduced the DDEC3 method: $[Cr(CN)_6]^{3-}$ spin quartet, $[Cu_2N_{10}C_{36}H_{52}]^{2+}$ spin triplet, anti-ferromagnetic CuBTC metal-organic framework, anti-ferromagnetic $Fe_2O_3$ crystal, anti-ferromagnetic $Fe_2SiO_4$ crystal, anti-ferrimagnetic $Fe_3O_4$ crystal (PBE functional and PBE+$U_{eff}$ ($U_{eff}$ = 4.0 eV) functionals), $Fe_3Si$ crystal, $[GdI]^{+2}$ using both SDD and planewave basis sets, the MgI, MoI, SnI, TeI, and TiI molecules using both SDD and planewave basis sets, and the ozone triplet spin state and the ozone +1 cation doublet spin state using the PW91, B3LYP, and CCSD exchange-correlation theories. The same geometries, electron distributions, and spin distributions were used as input for DDEC6 analysis as were previously used for DDEC3 analysis[14]. As shown in Figure 17, the DDEC6 and DDEC3 ASMs are essentially identical. This follows the observation that ASMs are usually less sensitive than NACs to the choice of atomic population analysis method.[14, 68, 159]

Table 34.Comparison of DDEC6 ASMs for Mn atoms in the $Mn_{12}$-acetate single molecule magnet to prior experiments and computations. Our computed magnetic anisotropy barrier is also compared to prior experiments and computations.

| Atom Type | DDEC6 PBE Planewave | DDEC6 PBE LANL2DZ | experiments | Pederson Khanna PBE[b] |
|---|---|---|---|---|
| | | *Atomic Spin Moment* | | |
| Mn type 1 | -2.80 | -2.56 | -2.34±0.13[a] | -2.6 |
| Mn type 2 | 3.82 | 3.63 | 3.79±0.12[a] | 3.6 |
| Mn type 3 | 3.81 | 3.57 | 3.69±0.14[a] | 3.6 |
| | | *Magnetic Anisotropy Barrier* | | |
| | 59.5 | | 60–62[c] | 55.6–55.8 |

[a] Polarized neutron diffraction experiments of Robinson et al.[160] [b]Pederson and Khanna using integration of the spin density over spheres of 2.5 bohr radius to compute the ASMs.[161] [c]Fort et al.[162]

As an additional example, we consider the $Mn_{12}$-acetate (formula unit $Mn_{12}C_{32}H_{56}O_{48}$) single molecule magnet illustrated in Figure 18 (left panel). $Mn_{12}$-acetate is one of the most widely studied of all single molecule magnets since its synthesis and discovery by Lis.[163-165] We performed calculations in GAUSSIAN 09 using the PBE functional with LANL2DZ[166] basis sets and in VASP using the PBE/planewave method. Experiments support a conceptual model with an S = 10 and $S_Z$ = 10 ground state.[167] Accordingly, we set $S_Z$=10 as a constraint on the GAUSSIAN 09 and VASP electron and spin distributions we computed. In VASP, we optimized the atomic positions and used the experimental lattice parameters of Farrell et al.[168] (Cambridge Structural Database ID: BESXAA). In GAUSSIAN 09, we used an isolated molecule and optimized the atomic positions. As shown in Table 34, the DDEC6 ASMs computed using the PBE functional with both LANL2DZ and planewave basis sets were in good agreement with Robinson et al.'s[160] polarized neutron diffraction experiments and Pederson and Khanna's[161] PBE computations. ASMs on all atoms except Mn atoms were almost negligible in magnitude (i.e., ≤ 0.033 (planewave) and ≤ 0.077 (LANL2DZ)), which agrees with the experimental finding that "there is no evidence for net [magnetic] moments on the oxygen atoms"[160]. The magnetic anisotropy barrier of a single molecule magnet is the energy required to flip the magnetic moment orientation relative to the molecular structure.[164] We computed this barrier by performing 62 single-point spin-orbit coupling calculations in VASP, where the electron and spin distributions were kept constant while the magnetic direction was rotated (by varying the SAXIS parameter in VASP). A 1×1×2 Monkhorst-Pack k-point mesh was used with Fermi smearing (smearing width = 0.05 eV) and a spherical gradient field cutoff (i.e., GGA_COMPAT = .FALSE.) and one-center PAW charge densities stored using LMAXMIX = 6. As shown in Figure 18 (middle panel), the spin-orbit coupling potential energy surface had global energy minima at the poles and a global energy maximum at the equator with no other local energy minima or maxima. This yielded a magnetic anisotropy barrier of 5.13 meV (59.5 K), which is in good agreement with Fort et al.'s[162] experimental value of 60–62 K and Pederson and Khanna's computed value of 55.6–55.8 K[161]. As shown in Figure 18 (right panel), DDEC6 NACs computed with the LANL2DZ basis set are nearly identical to those computed using the planewave basis set. This shows the DDEC6 NACs are not overly sensitive to the basis set choice.

### 5.8.2 Noncollinear magnetism

The left panel of Figure 19 shows the globally minimized geometry and non-collinear magnetic structure of the $Fe_4O_{12}N_4C_{40}H_{52}$ noncollinear single molecule magnet. We computed DDEC6 NACs and the atomic spin magnetization vectors $\{\vec{M}_A\}$ for this material using the same electron and spin magnetization density files as reference [68]. As shown in the center panel, the DDEC6 NACs followed a similar trend and magnitude as the DDEC3 NACs reported in reference [14]. The atomic spin magnitudes are the magnitudes of the atomic spin magnetization vectors: $M_A = |\vec{M}_A|$. As shown in the right panel, the DDEC6 atomic spin magnitudes were virtually identical to the DDEC3 values. The total wall time from CHARGEMOL program start (before input file reading) to end (after output printing finished) was 16.3 minutes for this calculation run on a single processor core in Intel Xeon E5-2680v3 at the Comet supercomputing cluster. This works out to 8.7 seconds per atom. This calculation utilized a volume of $2.9\times10^{-3}$ $bohr^3$ per grid point. These results demonstrate that DDEC6 is well-suited for quantifying NACs and atomic spins in non-collinear magnets.

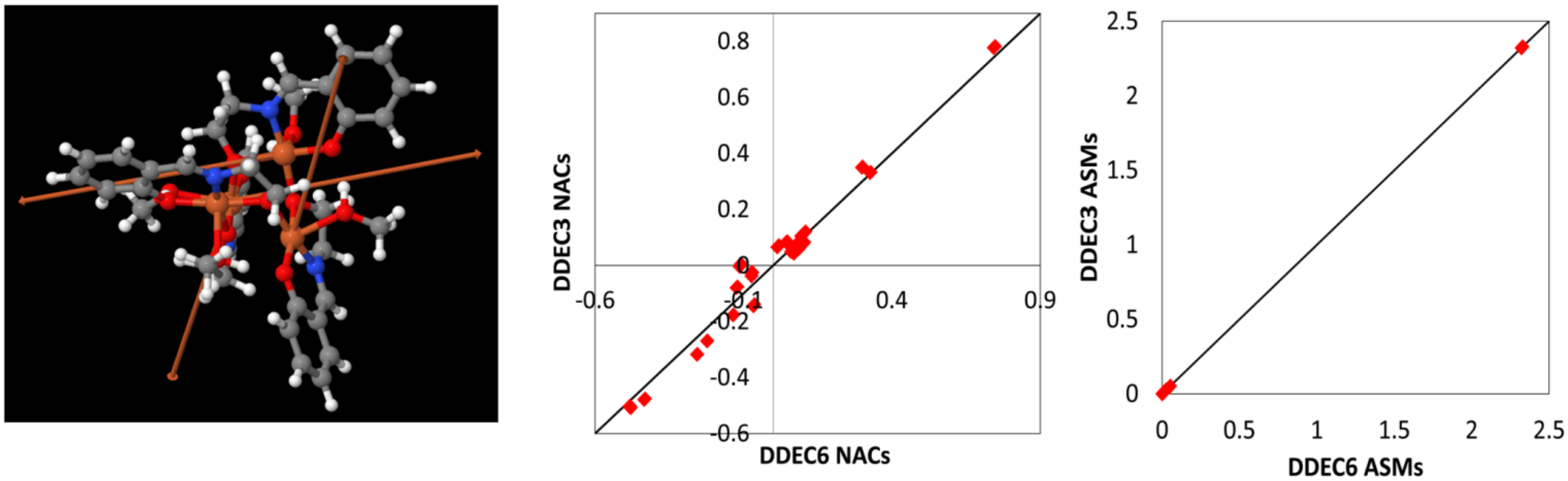

Figure 19. *Left*: $Fe_4O_{12}N_4C_{40}H_{52}$ noncollinear single molecule magnet structure reproduced with permission from reference [68] (© ACS 2011). The arrows show the magnitude and direction of the atomic spin magnetization vectors on each atom. The atomic spin magnitudes are small on all atoms except the four iron atoms. *Center*: Comparison of DDEC3 and DDEC6 net atomic charges. *Right*: Comparison of DDEC3 and DDEC6 atomic spin magnitudes. The black lines have a slope of 1 and an intercept of 0.

### 5.8.3 Quantifying the consistency between assigned NACs and ASMs

An AIM method should preferably yield chemically consistent NACs and ASMs. For a special type of system, the consistency between assigned NACs and ASMs can be quantitatively measured.

Consider a single neutral atom or a +1 atomic cation having only one easily removable electron. For convenience, we refer to these atoms or atomic ions as containing only one labile electron. Next, consider an uncharged host system containing only deeply bound electrons that are paired. If we combine the atom or atomic ion having one labile electron with the host system having paired electrons to form a weakly or ionically bound endohedral complex, a portion of the labile electron's density may be transferred to the host system's atoms. Since there is only one labile electron in a background of strongly held effectively paired electrons (in the endohedral and host system atoms) and (optionally) strongly held like-spin unpaired electrons (in the endohedral atom), the labile electron's spin cannot be locally cancelled by any other electrons in the system. In this case, the amount of electron density transferred from the endohedral atom to the host system should equal the amount of spin magnetization density transferred from the endohedral atom to the host system. This leads to the following quantification of consistency

$$\Delta = \underbrace{\left(q_{\text{total}} - \text{NAC}_{\text{endohedral}}\right)}_{\text{charge transferred to host}} + \underbrace{\left(M_{\text{total}} - \text{ASM}_{\text{endohedral}}\right)}_{\text{spin magnetization transferred to host}} \quad (144)$$

where $q_{\text{total}} \geq 0$ is the total system charge (in atomic units), $M_{\text{total}} \geq 0$ is the system's total spin magnetic moment (in atomic units), and $\text{NAC}_{\text{endohedral}}$ and $\text{ASM}_{\text{endohedral}}$ are the assigned NAC and ASM of the endohedral atom in the endohedral complex. In the ideal case, $\Delta \to 0$, because the single labile electron should transfer equal amounts of spin magnetization and negative charge to the host. This condition should also be fulfilled in situations where a weakly bound endohedral atom has approximately zero labile electrons. It is not necessarily fulfilled in cases where the number of labile electrons exceeds one, because in such cases the multiple labile electrons might be transferred into orbitals of opposing spins leading to physically different amounts of transferred spin magnetization and transferred negative charge. Nor should it be fulfilled in cases where a strong covalent bond forms between the endohedral atom and the host.

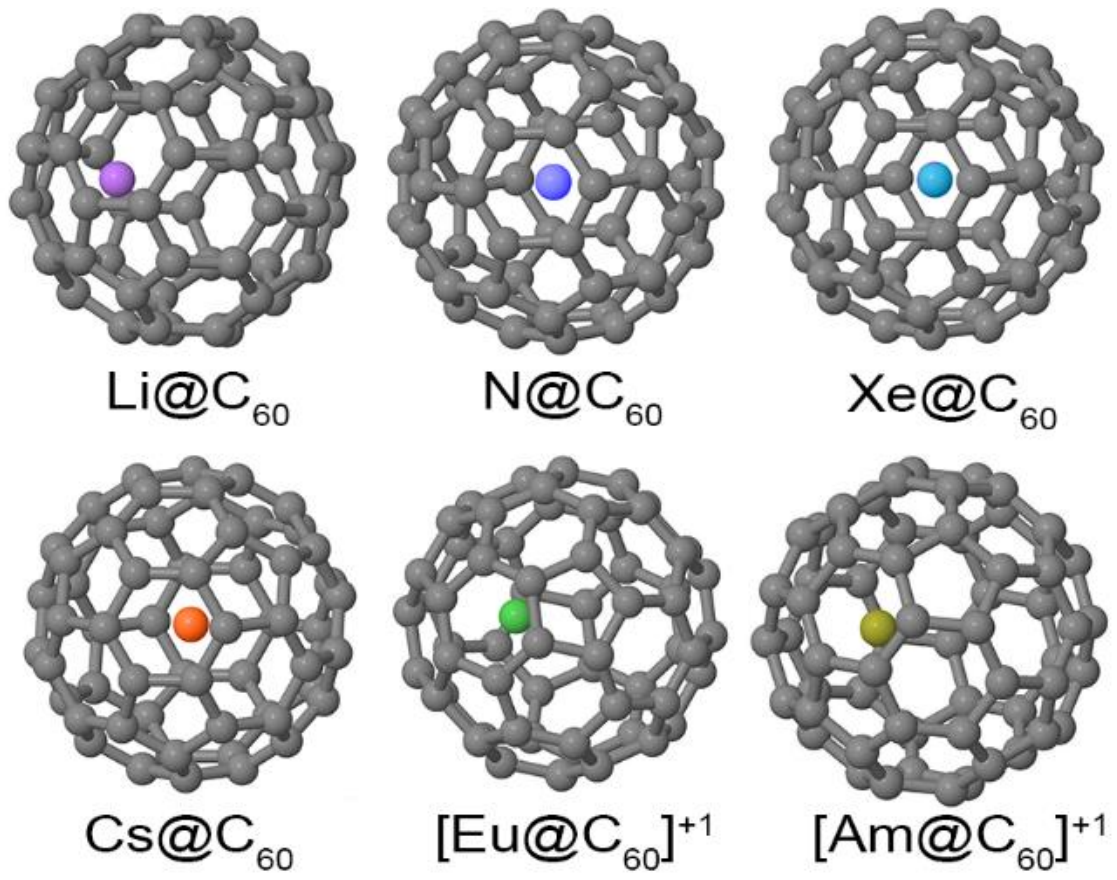

Figure 20. Endohedral complexes used to test the charge and spin transfer consistency. The N, Xe, and Cs atoms are approximately centered in the $C_{60}$ cage. The Li, Eu, and Am atoms attract to one side of the $C_{60}$ cage. The sizes of the endohedral atoms are not drawn to scale.

As specific examples, we consider the Li@$C_{60}$, N@$C_{60}$, Cs@$C_{60}$, Xe@$C_{60}$, $[\text{Eu@C}_{60}]^{+1}$, and $[\text{Am@C}_{60}]^{+1}$ endohedral fullerenes. The $C_{60}$ host has only deeply held paired electrons.[169] (Experiments show $C_{60}$ has a first ionization energy of 6.4–7.9 eV, an electron affinity of approx. 2.6–2.8 eV, and a first optical transition of approx. 3.2 eV.[170-173]) These complexes were chosen as examples, because they span a wide range from light to heavy elements having zero to one labile electrons. The Li and Cs elements have a nominal $s^1$ valence configuration; this outer s-electron is donated to the Li@$C_{60}$ and Cs@$C_{60}$

systems as the labile electron. The $Eu^{+1}$ and $Am^{+1}$ elements have a nominal $f^7s^1$ valence configuration; the half-filled f-shell is tightly held while the outer s-electron is donated to the $[Eu@C_{60}]^{+1}$ and $[Am@C_{60}]^{+1}$ systems as the labile electron. (Spectroscopic experiments show Eu in $Eu@C_{60}$ is in the +II oxidation state, meaning the seven f electrons remain bound to the Eu atom.[174, 175]) Because Xe is a noble gas element, the $Xe@C_{60}$ system has no labile electrons. Spectroscopic experiments show that in $N@C_{60}$, the endohedral N atom has a quartet spin state analogous to the isolated N atom.[176] This can be explained by the high electronegativity of the N atom, which retains its three unpaired electrons. Thus, we consider none of the electrons in $N@C_{60}$ to be labile.

We optimized the geometries and electron distributions of these endohedral fullerenes in VASP using the PBE functional with the PAW method and a 400 eV plane-wave cutoff. A 20Å × 20Å × 20Å cubic unit cell was used. The positions of all atoms in the system were optimized until the forces on every atom were negligible. Due to the almost spherical nature of the $C_{60}$ enclosure, only the equilibrium displacement of the endehedral atom from the cage's center should be considered significant. Therefore, we did not attempt multiple initial geometries with different angular variations in the endohedral atom's position. Figure 20 displays the optimized geometries. In Table 35, the optimized offset is the distance of the endohedral atom from the center of the $C_{60}$ group. (The center of the $C_{60}$ group was computed by averaging the (x, y, z) coordinates of the carbon atoms.) The N, Xe, and Cs atoms were located at the center of the $C_{60}$ group (within a computational tolerance). The central position of the N atom and quartet spin state agree with electron paramagnetic resonance (EPR) and electron-nuclear double resonance (ENDOR) experiments.[176] As reviewed by Popov et al., a wide variety of spectroscopic experiments show the noble gas atom in $Ng@C_{60}$ complexes (Ng = He, Ne, Ar, Kr, or Xe) resides at the cage's central position.[177] The $Cs@C_{60}$ geometry optimization converged to an energy minimum having a centrally located Cs atom, even though the calculation was started using a non-zero offset of 0.46 Å. This shows the central Cs position is at least a local (and perhaps global) energy minimum. In the optimized structures, the Li, Eu, and Am ions were displaced by > 1 Å from the center. Our calculated offset for $Li@C_{60}$ is in good agreement with previous computational studies.[178, 179] Prior calculations on neutral $Eu@C_{60}$ and $Am@C_{60}$ complexes also indicate an off-center position.[180-182]

Table 35: Computed NACs and ASMs for the enclosed atom in endohedral doped bucky-balls

| system | optimized offset (Å) | total unpaired electrons | HD | | CM5 | Bader | | DDEC6 | |
|---|---|---|---|---|---|---|---|---|---|
| | | | NAC | ASM | NAC | NAC | ASM | NAC | ASM |
| $Li@C_{60}$ | 1.52 | 1 | 0.323 | 0.002 | 0.566 | 0.899 | 0.000 | 0.903 | -0.001 |
| $N@C_{60}$ | 0.06 | 3 | 0.137 | 2.816 | 0.118 | 0.013 | 2.886 | 0.142 | 2.854 |
| $Xe@C_{60}$ | 0.01 | 0 | 0.293 | 0.000 | 0.304 | 0.092 | 0.000 | 0.316 | 0.000 |
| $Cs@C_{60}$ | 0.02 | 1 | 0.404 | 0.002 | 1.468 | 0.917 | -0.001 | 1.057[a] (1.000[b]) | -0.002[a] (-0.002[b]) |
| $[Eu@C_{60}]^{+1}$ | 1.10 | 8 | 0.542 | 7.501 | 1.050 | 1.566 | 6.879 | 1.368 | 7.515 |
| $[Am@C_{60}]^{+1}$ | 1.19 | 8 | 0.608 | 7.355 | 1.049 | 1.579 | 6.545 | 1.318 | 7.390 |
| **mean absolute inconsistency:** | | | 0.601 | | 0.386 | 0.302 | | 0.147 | |

[a] Using 46 frozen Cs core electrons. [b] Using 54 simulated frozen Cs core electrons by treating 8 of the 9 PAW valence electrons as core.

In Table 35, the mean absolute inconsistency between the assigned NACs and ASMs is quantified as the average of the absolute value of Δ for the six materials. Because CM5 is a correction to the HD NACs, the CM5 method utilized the HD ASMs. Among the four methods, the HD NACs and ASMs were the most inconsistent with an average inconsistency of 0.6 electrons. The DDEC6 NACs and ASMs were the most consistent with an average inconsistency of 0.15 electrons. The CM5 and Bader methods had intermediate performance. Because the HD and DDEC6 ASMs were nearly the same, the poor performance of the HD method must have been due to its inaccurate NACs.

Examining the DDEC6 results in Table 35, ~1 electron was transferred from the Li and Cs atoms, leaving a $Li^{+1}$ or $Cs^{+1}$ cation in the center having negligible unpaired spin. The CM5 NAC of 1.468 for the Cs atom seems too high, because this implies removal of some of its outer core electrons. The HD, CM5, and DDEC6 methods all gave ~0.3 electrons transferred in the Xe system, but the Bader method gave ~0.1 transferred electrons in this material. All four methods gave the least amount of electron transfer for the N system compared to other systems. In the Eu and Am cationic systems, ~0.5 electrons were transferred from the endohedral metal atom to the $C_{60}$ host to give a NAC of ~1.5 and an ASM of ~7.5 for the endohedral metal atom. Overall, these results demonstrate reasonable consistency between the assigned DDEC6 NACs and ASMs.

## 6. Conclusions

The main utility of net atomic charges (NACs) is they concisely convey important information about the electron distribution in materials. Due to the continuous nature of the electron cloud in a material, there is some flexibility in how to partition the total electron distribution among atoms-in-materials. In this article, we introduced a new and improved method, called DDEC6, for defining atoms-in-materials and computing NACs in periodic and non-periodic materials. Our method can be applied with equal validity to small and large molecules, ions, porous and non-porous solids, solid surfaces, nanostructures, and magnetic and non-magnetic materials irrespective of the basis set type used. This broad applicability makes it ideally suited for use as a default atomic population analysis method in quantum chemistry programs. The DDEC6 NACs are well-suited both for understanding charge-transfer in materials and for constructing flexible force-fields for classical atomistic simulations of materials.

A key advantage of our approach is that it includes a complete set of charge-compenstated reference ions for elements atomic number 1 to 109. To the best of our knowledge, this is the first complete reference ion library that has been computed for these elements. This is important, because approaches that compute the reference ions on-the-fly at the beginning of charge partitioning sometimes terminate with the error message that one or more reference ions could not be converged. Our fully computed reference ion library eliminates this problem.

We used a scientific engineering design approach to achieve nine performance goals: (1) the total electron distribution is partitioned among the atoms by assigning exactly one electron distribution to each atom, (2) core electrons remain assigned to the host atom, (3) NACs are formally independent of the basis set type because they are functionals of the total electron distribution, (4) the assigned atomic electron distributions give an efficiently converging polyatomic multipole expansion, (5) the assigned NACs usually follow Pauling scale electronegativity trends, (6) NACs for a particular element have good transferability among different conformations that are equivalently bonded, (7) the assigned NACs are chemically consistent with the assigned atomic spin moments, (8) the method has predictably rapid and robust convergence to a unique solution, and (9) the computational cost of charge partitioning scales linearly with increasing system size.

The DDEC6 method clearly improves over the DDEC3 method. We recommend the DDEC3 method be completely replaced with the DDEC6 method. The DDEC6 method alleviates the bifurcation or 'runaway charges' problem that occurs for some materials with earlier DDEC and Iterative Hirshfeld methods. The earlier DDEC and Iterative Hirshfeld methods sometimes have non-convex optimization functionals leading to non-unique solutions that depend on the starting conditions. For the $H_2$ triplet molecule with a constrained bond length of 50 pm, the DDEC3 method yielded NACs of +0.5 and -0.5. These NACs are unphysical, because they break the molecular electron density's symmetry. For this molecule, the DDEC6 NACs were +0.02 and -0.02, which reflects a magnification of integration error. For some materials, the DDEC3 NACs were too large in magnitude leading to core electrons being assigned to the wrong atom. For example, in compressed sodium chloride crystals some of the Na NACs were > +1. The DDEC6 method fixes this problem by making the following modifications relative to the DDEC3 method: (a) computing the electron population in a localized compartment surrounding each atom and using this localized population as one of the factors to determine a non-iterative reference ion charge for charge partitioning, (b) using a weighted spherical average to improve the effect of spherical averaging during charge partitioning, (c) constraining the atomic weighting factor $w_A(r_A)$ to decay no faster than $\exp(-2.5r_A)$ in an atom's buried tail, and (d) using five conditioning steps instead of $\left(\rho_A^{some_ref}\right)^{\chi}\left(\rho_A^{avg}\right)^{1-\chi}$ when constructing $w_A\left(r_A\right)$. For some materials, these improvements led to better correlations to: (i) core electron binding energy shifts (e.g., the studied Ti-containing compounds), (ii) electrostatic potentials (e.g., the $Li_2O$ molecule), and (iii) charge-transfer properties (e.g., the studied metal oxide and sulfide solids). Finally, the DDEC6 method is more computationally efficient than the DDEC3 method. For all materials, the DDEC6 method converges in exactly seven charge partitioning steps. For materials where the constraint $N_A \geq N_A^{core}$ is not binding, only one DDEC6 charge cycle per charge partitioning step is required.

We now summarize the main results of our computational tests. For a series of high-pressure sodium chloride crystals with unusual stoichiometries, we found the DDEC3 method sometimes gives NACs in excess of +1.0 for the Na atoms and Bader's quantum chemical topology sometimes yields non-nuclear attractors while the DDEC6 method exhibits neither of these problems. As pointed out by Wang et al.[26], the DDEC3 method predicts the incorrect electron transfer sign for the transition metal atom for the delithiation of solid $LiCoO_2$ to $CoO_2$. The DDEC6 method fixes this problem. For several Pd-containing alloys, we compared the electron transfer direction predicted by element electronegativities to computed NACs: the Bader, DDEC3, and DDEC6 NACs followed the Pauling scale electronegativity trends while the HD and CM5 NACs did not. For $(MgO)_n$ (n = 1 to 6) clusters, we found the DDEC6 method exhibits overall better performance than DDEC3 for reproducing the electrostatic potential and dipole moments. For natrolite, the DDEC6 NACs were smaller in magnitude than the DDEC3 NACs. For this material, the DDEC6 NACs were closer to NACs extracted from high-resolution diffraction data using Kappa refinement (with the exception of the Na atom which was not refined). DDEC3 and DDEC6 were both in excellent agreement with formamide NACs extracted from high-resolution diffraction data using spherical atom refinement. For a series of Ti-containing compounds, core-electron binding energy shifts were approximately linearly correlated to the DDEC6, HD, and Bader NACs but not to the DDEC3 and CM5 NACs. All five charge assignment methods gave reasonably good correlations between core electron binding shifts and computed NACs for the Mo-containing and Fe-containing compounds. For 13 materials studied at the low energy conformation, the DDEC6 NACs reproduced the electrostatic potential slightly better than the DDEC3 NACs in 8 of the 13 materials. A detailed study across various conformations of $Li_2O$, five carboxylic acids, and the Zn-nicotinate MOF showed the DDEC6 NACs have excellent conformational transferability and are ideally suited for constructing flexible force-fields to approximately reproduce the electrostatic potential across various system conformations. For a series of systems comprised almost entirely of surface atoms, the DDEC6 and DDEC3 NACs exhibited similar trends with some statistically significant differences in NAC values. Tests of three solid surfaces (K adatom on a $Mo_2C$ (110) surface, NaF(001) slab, and $SrTiO_3$(100) slab), showed the DDEC6 method maintains a consistent treatment of surface and buried atoms. Finally, we examined materials with collinear and non-collinear magnetism and found the DDEC6 atomic spin moments (ASMs) are essentially identical to the DDEC3 ASMs. For the $Mn_{12}$-acetate single molecule magnet, the computed DDEC6 ASMs were in excellent agreement with previous experiments[160] and computations[161]. We computed the spin-orbit coupling potential energy surface for this material and found the resulting magnetic anisotropy barrier (5.13 meV) to be in excellent agreement with previous experiments[162] and computations[161]. For six endohedral fullerenes containing one labile electron, the consistency between assigned NACs and ASMs was quantified for the Hirshfeld, CM5, Bader, and DDEC6 methods. Among these four methods, the DDEC6 method gave the most consistent agreement between assigned NACs and ASMs.

In closing, we note the DDEC6 atomic electron and spin distributions, $\left\{\rho_A\left(\vec{r}_A\right), \vec{m}_A\left(\vec{r}_A\right)\right\}$, can be used as the basis for computing additional AIM descriptors besides NACs and ASMs. A convenient method for computing bond orders based on DDEC6 partitioning has been developed by one of us (TAM) and will be described in a forthcoming publication. We have efficiently parallelized the computation of DDEC6 NACs, ASMs, and bond orders using OpenMP parallelization directives for Fortran. This parallelization will be the subject of a latter publication by us. Our parallelized code for computing DDEC6 NACs, ASMs, and bond orders is currently available in the CHARGEMOL program distributed via ddec.sourceforge.net.[183]

**Acknowledgments:** Supercomputing resources were provided by the Extreme Science and Engineering Discovery Environment (XSEDE). XSEDE is funded by NSF grant OCI-1053575. XSEDE project grant TG-CTS100027 provided allocations on the Trestles and Comet clusters at the San Diego Supercomputing Center (SDSC) and the Stampede cluster at the Texas Advanced Computing Center (TACC).

† **Electronic supplementary information (ESI) available:** Summary of alternative charge partitioning algorithms considered; summary of integration routines; iterative algorigthm for computing the convex functional NACs; and .xyz files (which can be read using any text editor or the free Jmol visualization program downloadable from jmol.sourceforge.net) containing geometries, net atomic charges, atomic dipoles and quadrupoles, fitted tail decay exponents, and atomic spin moments.

# Electronic Supplementary Information for

# DDEC6: A Method for Computing Even-Tempered Net Atomic Charges in Periodic and Nonperiodic Materials

Thomas A. Manz* and Nidia Gabaldon Limas

Department of Chemical & Materials Engineering, New Mexico State University, Las Cruces, New Mexico, 88003-8001.

*E-mail: tmanz@nmsu.edu

**Contents**

## 1. Additional Tables

Table S1. Calculated and experimental ground spin state multiplicities of the neutral atoms. Elements with different calculated and experimental values are shown in boldface type.

| Atomic number | Atomic symbol | Calc. | Exp.[a] | Atomic number | Atomic symbol | Calc. | Exp.[a] | Atomic number | Atomic symbol | Calc. | Exp.[a] |
|---|---|---|---|---|---|---|---|---|---|---|---|
| 1 | H | 2 | 2 | 38 | Sr | 1 | 1 | 75 | Re | 6 | 6 |
| 2 | He | 1 | 1 | 39 | Y | 2 | 2 | 76 | Os | 5 | 5 |
| 3 | Li | 2 | 2 | **40** | **Zr** | **5** | **3** | 77 | Ir | 4 | 4 |
| 4 | Be | 1 | 1 | 41 | Nb | 6 | 6 | **78** | **Pt** | **1** | **3** |
| 5 | B | 2 | 2 | 42 | Mo | 7 | 7 | 79 | Au | 2 | 2 |
| 6 | C | 3 | 3 | 43 | Tc | 6 | 6 | 80 | Hg | 1 | 1 |
| 7 | N | 4 | 4 | 44 | Ru | 5 | 5 | 81 | Tl | 2 | 2 |
| 8 | O | 3 | 3 | 45 | Rh | 4 | 4 | 82 | Pb | 3 | 3 |
| 9 | F | 2 | 2 | 46 | Pd | 1 | 1 | 83 | Bi | 4 | 4 |
| 10 | Ne | 1 | 1 | 47 | Ag | 2 | 2 | 84 | Po | 3 | 3 |
| 11 | Na | 2 | 2 | 48 | Cd | 1 | 1 | 85 | At | 2 | 2 |
| 12 | Mg | 1 | 1 | 49 | In | 2 | 2 | 86 | Rn | 1 | 1 |
| 13 | Al | 2 | 2 | 50 | Sn | 3 | 3 | 87 | Fr | 2 | 2 |
| 14 | Si | 3 | 3 | 51 | Sb | 4 | 4 | 88 | Ra | 1 | 1 |
| 15 | P | 4 | 4 | 52 | Te | 3 | 3 | 89 | Ac | 2 | 2 |
| 16 | S | 3 | 3 | 53 | I | 2 | 2 | 90 | Th | 3 | 3 |
| 17 | Cl | 2 | 2 | 54 | Xe | 1 | 1 | 91 | Pa | 4 | 4 |
| 18 | Ar | 1 | 1 | 55 | Cs | 2 | 2 | 92 | U | 5 | 5 |
| 19 | K | 2 | 2 | 56 | Ba | 1 | 1 | 93 | Np | 6 | 6 |
| 20 | Ca | 1 | 1 | 57 | La | 2 | 2 | 94 | Pu | 7 | 7 |
| 21 | Sc | 2 | 2 | **58** | **Ce** | **3** | **1** | 95 | Am | 8 | 8 |
| **22** | **Ti** | **5** | **3** | 59 | Pr | 4 | 4 | **96** | **Cm** | **7** | **9** |
| **23** | **V** | **6** | **4** | 60 | Nd | 5 | 5 | 97 | Bk | 6 | 6 |
| 24 | Cr | 7 | 7 | 61 | Pm | 6 | 6 | 98 | Cf | 5 | 5 |
| 25 | Mn | 6 | 6 | 62 | Sm | 7 | 7 | 99 | Es | 4 | 4 |
| 26 | Fe | 5 | 5 | 63 | Eu | 8 | 8 | 100 | Fm | 3 | 3 |
| 27 | Co | 4 | 4 | 64 | Gd | 7 | 9 | 101 | Md | 2 | 2 |
| 28 | Ni | 3 | 3 | 65 | Tb | 6 | 6 | 102 | No | 1 | 1 |
| 29 | Cu | 2 | 2 | 66 | Dy | 5 | 5 | 103 | Lr | 2 | [b] |
| 30 | Zn | 1 | 1 | 67 | Ho | 4 | 4 | 104 | Rf | 3 | [b] |
| 31 | Ga | 2 | 2 | 68 | Er | 3 | 3 | 105 | Db | 4 | [b] |
| 32 | Ge | 3 | 3 | 69 | Tm | 2 | 2 | 106 | Sg | 7 | [b] |
| 33 | As | 4 | 4 | 70 | Yb | 1 | 1 | 107 | Bh | 6 | [b] |
| 34 | Se | 3 | 3 | 71 | Lu | 2 | 2 | 108 | Hs | 5 | [b] |
| 35 | Br | 2 | 2 | 72 | Hf | 3 | 3 | 109 | Mt | 4 | [b] |
| 36 | Kr | 1 | 1 | 73 | Ta | 4 | 4 | | | | |
| 37 | Rb | 2 | 2 | **74** | **W** | **7** | **5** | | | | |

[a]Experimental data from references [1] and [2]. [b]Experimental data not available or not conclusive.

Table S2. Electronic spatial extent (square bohr) of selected reference ions in the +1, 0, and -1 charge states. These are for the seven elements where the PW91 and experimental low energy spin states differ for the neutral atoms. The +1 and -1 charge states used the charge compensation scheme described in the main text. Reference ion calculations performed with the PW91 functional near the complete basis set limit as described in the main text .

| | **+1 cation** | **neutral (exp. low energy spin)** | **neutral (PW91 low energy spin)** | **-1 anion** |
|---|---|---|---|---|
| **Ti** | 25.4 | 46.1 | 42.4 | 70.5 |
| **V** | 24.1 | 43.5 | 39.8 | 65.3 |
| **Zr** | 43.5 | 64.1 | 60.8 | 90.4 |
| **Ce** | 68.9 | 91.9 | 92.0 | 126.6 |
| **W** | 52.9 | 68.7 | 66.0 | 88.6 |
| **Pt** | 49.1 | 61.6 | 59.9 | 76.9 |
| **Cm** | 72.3 | 97.3 | 97.9 | 145.4 |

## 2. Summary of Charge Partitioning Alternatives Evaluated

We now briefly summarize some of the alternatives investigated during the course of developing the DDEC6 method. This section's purpose is to point out strategies we tried that did not perform well or that did not result in systematic improvements. This information is important to avoid duplicative efforts that could result if investigators retried these strategies without realizing they have already been tried.

Achieving good general purpose charge assignment is a balancing act of competing demands: (a) core electrons assigned to the host atom, (b) NACs with good conformational transferability, (c) chemically meaningful NACs that describe electron transfer trends (and core electron binding energy shifts in some materials), and (d) an efficiently converging polyatomic multipole expansion that reproduces the material's electrostatic potential. Hence, the term 'even-tempered' in this article's title. To approximately correlate with spectroscopic core electron binding energy shifts in transition metal compounds, the assigned atomic charge distributions should not be too delocalized. To give NACs that approximately reproduce molecular dipole moments and the electrostatic potential surrounding a material, the assigned $\{\rho_A(\vec{r}_A)\}$ should resemble their spherical averages, $\{\rho_A^{avg}(r_A)\}$. To achieve good conformational and chemical transferability among similar materials, it is preferable for the assigned $\{\rho_A(\vec{r}_A)\}$ to resemble real atoms. This can be achieved by optimizing $\{\rho_A(\vec{r}_A)\}$ to resemble reference ion densities. Therefore, we believe the most straightforward approach to achieving an even-tempered charge assignment method involves three components: (a) integrating the electron density in the local vicinity of each atomic nucleus (where 'vicinity' refers to positions close to the volume of space dominated by that atom), (b) optimizing $\{\rho_A(\vec{r}_A)\}$ to resemble their spherical averages, and (c) optimizing $\{\rho_A(\vec{r}_A)\}$ to resemble a set of reference ions.

Following these general principles, we tested a large number of new charge assignment algorithms. For various reasons, some of these charge assignment algorithms worked much better than others. Of all the algorithms we tested, the algorithm with the best overall performance was selected to be the DDEC6 method. We now briefly describe the other algorithms we tested.

In some schemes, we defined a localized net atomic charge as

$$q_A^{loc} = z_A - \oint \frac{\left(w_A(r_A)\right)^m}{\sum_{B,L}\left(w_B(r_B)\right)^m}\rho(\vec{r})d^3\vec{r}_A \qquad \text{(S1)}$$

where $w_A(r_A)$ is a DDEC-style atomic weighting factor combining a reference density and (optionally) spherical averaging with (optionally) exponential tail constraints. Different values of m > 1 were investigated. In each charge partitioning iteration, $q_A^{ref}$ was set equal to some linear combination of $q_A^{loc}$, $q_A$, $q_A^{HD}$, and (optionally) other factors. The $q_A^{loc}$ and $q_A$ were then updated in each iteration and iterated to convergence. This type of scheme does not work well, because in materials like boron nitride the cation is more diffuse than the anion leading to $|q_A^{loc}| > |q_A|$ resulting in increased NAC magnitude when any $q_A^{loc}$ is included in $q_A^{ref}$. (These larger NAC magnitudes degraded the quality of fitting the electrostatic potential in materials like the BN sheet and nanotube.) We tried to counterbalance this by mixing in a fraction of $q_A^{HD}$ (which usually has $|q_A^{HD}| \le |q_A|$), but this did not produce consistently good results across a wide range of materials. We also tried variations where $q_A^{ref}$ was set to whichever was smaller in magnitude, $q_A^{loc}$ or $q_A$. We also tried variations in which $q_A^{ref}$ was adjusted for each atom until $q_A^{loc} = q_A$. We also tried various schemes in which $q_A^{ref}$ was adjusted for each atom until $q_A^{loc}$ preferably lay between $q_A^{HD}$ and $q_A$ (or between 0 and $q_A$), subject to the condition that $q_A^{ref}$ should differ from $q_A$ by no more than a preset allowance ('trust radius'). In fact, we publically released one such scheme called DDEC4 in the CHARGEMOL 3.1 version released on ddec.sourceforge.net September 29, 2014. (This version computed the DDEC3 NACs by default, but contained the option to compute DDEC4 NACs instead.) Nevertheless, our further testing revealed more advantageous approaches, which resulted in the DDEC4 algorithm being abandoned.

After extensive trials, we finally realized that continuously updating $q_A^{loc}$ and $q_A^{ref}$ is inherently problematic. Specifically, some of the atoms get greedy and continuously take electrons from the other atoms. To our surprise, we found the difference $q_A^{loc} - q_A$ can remain nearly constant in some materials (e.g., TiO solid) over a large number of iterations in which $q_A$ changes by a total of >0.5 electrons. This means that a continuously updated $q_A^{loc}$ does not provide a reliable reference value to prevent atoms from becoming greedy. We briefly tried setting $q_A^{loc}$ equal to the Bader charge, but abandoned this strategy due to the presence of non-nuclear attractors in some materials.

We also tested more aggressive buried tail constraints in which $P(r_A)w_A(r_A)$ was constrained to decay exponentially with increasing $r_A$ in the atom's buried tail, where $P(r_A)$ was a polynomial of $r_A$. We tested a few different polynomials $P(r_A)$ chosen to reproduce the limits $P(r_A = 0) = 1$ and $P(r_A \to \infty) \propto (r_A)^p$ with p > 1. (The DDEC4 method briefly mentioned above contained this type of tail constraint.) After extensive testing, we concluded the extra complexity of $P(r_A)$ did not appreciably improve results across a wide range of materials, so we reverted to the simpler strategy of constraining just $w_A(r_A)$ to decay exponentially in the atom's buried tail. We also tried various schemes for computing the decay exponents applied to $w_A(r_A)$. Ultimately, we decided to use a strategy similar to that used in the DDEC3 method plus the addition of a constraint to prevent $w_A(r_A)$ from becoming too contracted.

We also tested strategies in which $q_A^{ref}$ was set equal to a linear combination of $q_A$, $q_A^{first_loc}$ (i.e., $q_A^{loc}$ computed in the first charge partitioning iteration), $q_A^{second_loc}$ (i.e., $q_A^{loc}$ computed in the second charge partitioning iteration), $q_A^{loc}$ (computed in the current charge cycle), $q_A^{HD}$, and the net atomic charge computed from the iterative-Hirshfeld like partitioning using the conditioned or unconditioned charged reference ion. After extensive testing, we could not attribute any tangible benefit to the continuous updating of $q_A^{ref}$ in each charge cycle, so we finally replaced this kind of strategy with a fixed $q_A^{ref}$ value.

Finally, after deciding to use a fixed $q_A^{ref}$ value for all charge cycles, we tested various schemes for computing the target $q_A^{ref}$ value. We tested a scheme similar to the DDEC5 method (see description next paragraph), except $q_A^{1,Stock}, q_A^{1,Loc}, q_A^{1,ref}, q_A^{2,Stock}, q_A^{2,Loc}, q_A^{2,ref}$ were computed based on the conditioned reference densities instead of the unconditioned reference densities. We also tried schemes that employed various combinations based on both the conditioned and unconditioned reference densities. Using conditioned reference densities to compute $q_A^{1,Stock}, q_A^{1,Loc}, q_A^{1,ref}, q_A^{2,Stock}, q_A^{2,Loc}, q_A^{2,ref}$ worsened the performance. We also tested schemes in which the ratio of $q_A^{1,Stock}$ to $q_A^{1,Loc}$ to form $q_A^{1,ref}$ was set different than the ratio of $q_A^{2,Stock}$ to $q_A^{2,Loc}$ to form $q_A^{2,ref}$. However, we did not notice any appreciable improvements and so decided on the simpler scheme of keeping these two ratios the same. We also investigated variations of this ratio before settling on the value used in the DDEC5 and DDEC6 methods.

One of the schemes we developed with fixed $q_A^{ref}$ was called the DDEC5 method. This method used $\sigma_A(r_A) = (\rho_A^{cond}(r_A))^{\chi}(\rho_A^{wavg}(r_A))^{1-\chi}$ with one conditioning step (i.e., c = 1) and $\chi = 1/3$ to yield $\chi_{equiv}^{DDEC5} = 1/4$. This method used the same formula for $\rho_A^{wavg}(r_A)$ and the same reference ion charges (and $\rho_A^{cond}(r_A)$) as the DDEC6 method. DDEC5 also applied the same exponential decay constraints on $w_A(r_A)$ in the fourth and later charge cycles as the DDEC6 method. DDEC5 also applied the constraint $N_A^{val} \geq 0$. We publically released DDEC5 in the CHARGEMOL 3.2.1 version released on ddec.sourceforge.net August 12, 2015. (This version computed the DDEC3 NACs by default, but contained the option to compute DDEC5 NACs instead.) While the DDEC5 method performed well, it did not have a provably convex optimization functional or provably unique solution. We also extensively tested one algorithm with the same form as DDEC5, except using two conditioning steps (i.e., c = 2) and $\chi = 1/2$ to yield $\chi_{equiv} = 1/4$. This algorithm did not converge for the ozone+1 B3LYP system. This led us to believe the DDEC5 method might not converge for some materials, due to its optimization functional not being provably convex. Desiring a proof of unique convergence, we then developed the Convex functional and later the DDEC6 method that have proven unique solutions.

We also tested schemes similar to those described above, but differing in parameter values such as the localization exponent m, the precise formulation of the $w_A(r_A)$ tail constraints, etc. We also tested a few schemes that are quite different from those described above. A few additional schemes computed $q_A^{loc}$ based on atom-atom overlap populations (computed via various schemes) instead of based on Eq. (S1) above. However, these were more computationally expensive than Eq. (S1), and we did not discern any performance improvements compared to Eq. (S1). We also tried schemes in which the AIM charge distribution $\rho_A(\vec{r}_A)$ was computed by using linear combinations of $\rho(\vec{r})w_A(r_A)/W(\vec{r})$ and $\rho(\vec{r})(w_A(r_A))^m / \sum_{B,L}(w_B(r_B))^m$ with m > 1 or other localization schemes instead of $(w_A(r_A))^m$.

While this list is not comprehensive of all of the charge partitioning algorithms we tested, it provides a general idea of the types of charge partitioning schemes we tested. In the end, we settled on the DDEC6 charge partitioning method, because it provided consistently good results across a wide range of material types.

In addition, we performed a large number of tests regarding optimization of the computational cost. In addition to computational tests on various materials, we developed iteration calculus with associated algebraic models (solved analytically) and finite difference numerical models (solved in spreadsheets) that accurately predicted and described the convergence performance of various computational algorithms. We used these mathematical models to derive the optimal parameters leading to fast and robust convergence. Using this iteration calculus, we designed efficient convergence accelerators (see Section 4.2.3 below) that optimize the convergence speed for self-consistent schemes. Our computational tests confirmed the theoretically derived optimal parameters and performance improvements associated with these convergence accelerators. These improvements ultimately led to the number of required charge cycles being reduced from <200 for the DDEC3 method to 7 for the DDEC6 method. We believe that 7 charge cycles is close to the minimum of what can be used to consistently obtain accurate results.

## 3. Integration Routines Employed

### 3.1 General Overview

In the limit of an arbitrarily fine grid spacing and sufficiently large cutoff radii, the converged DDEC6 properties should be independent of the specific choice of integration routine. The choice of integration routine primarily effects the computational efficiency and precision. The optimal integration routine depends on the type of input information available. A uniformly spaced grid is a convenient choice for quantum chemistry calculations using planewave basis sets, because this type of grid naturally lends itself to computing the electron and spin density grids via Fourier transform from the planewave coefficients. In general, using a uniformly spaced grid for charge partitioning is convenient when the quantum chemistry program (VASP, ONETEP, GPAW, etc.) used to generate the electron and spin distributions also uses this same grid type. A uniformly spaced grid is not the most computationally efficient choice for quantum chemistry calculations using Gaussian basis sets. For quantum chemistry calculations using Gaussian basis sets, computationally efficient atom-centered overlapping[3] and non-overlapping[4] grid types have been extensively described in the literature. Nevertheless, we used a uniformly spaced grid for all DDEC6 calculations described in this work. This was motivated by the fact that atom-centered overlapping and non-overlapping grids have not yet been programmed into the CHARGEMOL program used to compute the DDEC6 properties.

When using uniformly spaced grids, it is sometimes best to integrate core-like and valence-like electron distributions separately. Here, the term core-like electron distribution refers to an electron distribution concentrated near atomic nuclei. Core-like electron distributions can have extremely high density values near atomic nuclei. The term valence-like electron distribution refers to an electron distribution that has a significant fraction of its electrons in the atomic valence regions without extreme density spikes near the atomic nuclei. Unless special precautions are taken, the extreme density spikes near atomic nuclei in core-like electron distributions can lead to inaccurate integration of the number of core-like electrons. Using an extremely fine uniform grid to integrate the core-like electron distribution is one possible strategy, but this strategy would be too computationally expensive and impractical. Instead, we integrate the core-like and valence-like electron distributions separately. Then we correct the core-like density grid to force it to integrate to the correct number of core-like electrons. With this correction in place, accurate integrations over the core-like density grid can be performed. The following section describes details of this core grid correction.

During DDEC analysis of VASP PAW quantum chemistry calculations, the precision of integrating valence-like electron distributions was improved using the valence occupancy correction and all-electron spin density approximation described in the Supporting Information Section E (pages S9–S10) of Manz and Sholl.[5]

During DDEC analysis of GAUSSIAN 09 generated wfx files, the precision of integrating electron and spin distributions and multipole moments was improved using the valence occupancy corrections described in the Supporting Information Section F (pages S10–S11) of Manz and Sholl.[5] In the present work, we made two additional improvements in computational efficiency. First additional improvement in computational efficiency: Each Gaussian basis set product has the general form $(X-X_0)^{\ell_1}(Y-Y_0)^{\ell_2}(Z-Z_0)^{\ell_3}\exp\left(-\alpha|\vec{r}-\vec{r}_0|^2\right)$. Here, $\vec{r}_0=(X_0,Y_0,Z_0)$ represents the center of the Gaussian basis set product. The powers $(\ell_1,\ell_2,\ell_3)$ are non-negative integers. To improve computational efficiency, we sorted all Gaussian basis set products into blocks where Gaussian basis set products in each block shared the same $\alpha$ and $\vec{r}_0$. The $\exp\left(-\alpha|\vec{r}-\vec{r}_0|^2\right)$, $(X-X_0),(Y-Y_0)$, and $(Z-Z_0)$ terms were computed only once for each block at each grid point. This produced computational savings by avoiding recomputing these terms for every Gaussian basis set product within each block. Second additional improvement in computational efficiency: We added a grid interpolation scheme to increase the computational efficiency of generating valence, core, and spin density grids from Gaussian basis set coefficients. This grid interpolation decreases the computational cost by approximately a factor of five with negligible impact on the computational precision. Specifically, we used a set of grids explicitly including every $n^{th}$ grid point along each lattice direction, with n = 1 (finest), 2, 3, 4, 6, 8, or 12 (coarsest). The finest grid (i.e., n = 1) had a uniform spacing of ~0.14 bohr. Each Gaussian basis set product was assigned to one of these grids according to how diffuse it was. Specifically, a Gaussian basis set product proportional to $\exp\left(-\alpha|\vec{r}-\vec{r}_0|^2\right)$ was assigned to grids according to the following scheme:

(1) If $\alpha > 5$ (atomic units), the Gaussian basis set product was assigned to the core-like density grid with analytic integration to compute the occupancy corrections. This grid had a uniform spacing of ~0.14 bohr.

(2) If $5 \geq \alpha > 0.4$ (atomic units), the Gaussian basis set product was assigned to the n=1 valence density grid. This grid had a uniform spacing of ~0.14 bohr.

(3) If $\frac{0.16}{n_i^2} \geq \alpha > \frac{0.16}{(n_{i+1})^2}$ (atomic units), the Gaussian basis set product was assigned to the $n_i$ valence density grid, where $n_i = 2,3,4,6,8$. The Gaussian basis set product was assigned to the n=12 (coarsest) valence density grid if $(0.16/144) \geq \alpha$ (atomic units). This scheme scales the coarseness of the grid in the exact same manner that $\alpha$ scales. This means the 'relative coarseness' of the grid remains approximately constant independent of the $\alpha$ value.

The valence and spin density contributions for each Gaussian basis set product were computed on the corresponding assigned $n^{th}$ grid. For each Gaussian basis set product, a renormalization factor of up to ±5% was applied to ensure it integrated to the proper value over the grid.[5] After all Gaussian basis set products were computed over the corresponding grids, the coarser grids were interpolated back onto the finer grids in the following order: (a) the n = 12 grid was interpolated back onto the n = 6 grid, (b) the n = 8 grid was interpolated back onto the n = 4 grid, (c) the n = 6 grid was interpolated back onto the n = 3 grid, (d) the n = 4 grid was interpolated back onto the n = 2 grid, (e) the n = 3 grid was interpolated back onto the n = 1 grid, and (f) the n = 2 grid was interpolated back onto the n = 1 grid. This has the effect of interpolating the n = 12 grid onto the n = 1 grid by first interpolating the n = 12 grid onto the n = 6 grid, then interpolating the n = 6 grid onto the n = 3 grid, and finally interpolating the n = 3 grid onto the n = 1 grid. By the sequence of steps (a) to (f), all of the coarser grids were finally interpolated onto the n = 1 grid. A linear interpolation was used in each of steps (a) to (f). Such a linear interpolation yields the same integral of each Gaussian basis product over the coarser and finer grids.

In this work, we used a 5 Å cutoff radius for $w_A(r_A)$, which means $\rho_A(\vec{r}_A) = 0$ for $r_A > 5$ Å. For all charge distributions depending only on $r_A$ (i.e., spherically symmetric distributions), we used 100 radial shells evenly spaced between 0 and 5 Å.

### 3.2 Core Electron Partitioning with Core Grid Correction

#### 3.2.1 Overview

Core electron partitioning with core grid correction assigns core-like electron distributions, $\{\rho_A^{core}(\vec{r}_A)\}$, that integrate to yield the exact analytic number of core-like electrons for each atom

$$\oint \rho_A^{core}(\vec{r}_A) d^3\vec{r}_A = N_A^{core}. \qquad \text{(S2)}$$

Here, $N_A^{core}$ refers to the exact analytic number of core-like electrons that have been included in the core-like electron distribution, $\rho^{core}(\vec{r})$. In general, this depends upon the specific density grid setups not the chemical states of atoms. For example, a Mg atom could have $N_A^{core}$ set to 0, 2, 10, or other values, depending on how many core-like electrons were written to the core density grid.

Core grid correction is not necessary to compute accurate NACs. (When comparing NACs computed with to without this core grid correction, the NACs typically change by up to ~0.002 e due to integration artifacts arising from the finite grid spacing.) For example, the paper introducing the DDEC3 method did not use core grid correction.[5] The primary reason for including core grid correction is that it allows $\{\rho_A(\vec{r}_A)\}$ to be integrated to yield $N_A$ without worrying about errors in the integrated number of core-like electrons. This is critical for evaluating quantities that are nonlinear functionals of $\{\rho_A(\vec{r}_A)\}$. Bond orders quantify the number of electrons exchanged between two atoms. Because the exchange interaction is a nonlinear functional of the electron density, integrals for computing bond orders must be based on $\rho_A(\vec{r}_A)$ not just the atomic valence density $\rho_A^{val}(\vec{r}_A)$. This requires that $\rho_A(\vec{r}_A)$ integrate to the correct number of electrons—hence the need for a core grid correction.

If an atomic nucleus falls directly on a grid point, the density at that grid point may be very high. During integration, the number of electrons contributed by a pixel is calculated as the electron density at that pixel times the pixel volume. For a pixel centered on an atomic nucleus, the average electron density in the volume occupied by the pixel is less than the electron density exactly at the nuclear position. Therefore, grid points centered directly at atomic nuclei will produce integration errors if the density is taken to be that at the nuclear position. This error can be removed by estimating and using the average density for each pixel volume in place of the point density at the nuclear position.

**3.2.2 Design Criteria**

a) The core-like density assigned to each atom should integrate to the correct number of core-like electrons within a specified convergence tolerance (e.g., $10^{-5}$ e). For example, if a calculation is performed with 2 core electrons in Mg, the assigned core density for this atom should integrate to between 1.99999 and 2.00001 e. This is done by correcting the core density for pixels with the highest core density (i.e., the nuclear cusps).

b) The core grid correction should never produce a negative core-like electron density for any grid point.

c) For a particular atom, the core grid correction should not change the relative ordering of grid point core-like densities. Specifically, if grid point 1 contains a higher core-like density assigned to atom A than grid point 2, then after the correction is applied this should still be the case.

d) Because nuclear cusps contribute most of the integration error, the core grid correction should be localized to those grid points with the highest core-like densities (i.e., those closest to atomic nuclei).

e) The core grid correction should not interfere with the exponential decay constraint applied to the $\{w_A^{core}(r_A)\}$. Recall that atomic core densities decay at least as fast as exp(-2$r_A$) where $r_A$ is in bohr. This constraint is applied during the core partitioning. Consider two grid points near nucleus A such that grid point 1 is closer to nucleus A than grid point 2. Then,

$$w_A^{core}(r_2) \le w_A^{core}(r_1)\exp(-2(r_2 - r_1)) \qquad \text{(S3)}$$

where $r_1$ and $r_2$ are the distances from grid points 1 and 2, respectively, to nucleus A.

**3.2.3 Iterative Algorithm**

Two separate sets of iterations are performed: (i) a first set of iterations to determine a pre-corrected $w_A^{core}(r_A)$ and (ii) a second set of iterations to correct the core-like electron density grid to yield the correct number of core-like electrons for each atom.

**3.2.3.1 Iterations to determine a pre-corrected $w_A^{core}(r_A)$**

For atoms having no core-like electrons assigned to the core grid (e.g., if $N_A^{core} < 10^{-10}$), the assigned $w_A^{core}(r_A)$ and $\rho_A^{core,avg}(r_A)$ are set to zero. For all remaining atoms, the following sequence of steps is performed starting with the initial estimate $w_A^{core}(r_A) = \rho_A^{ref}(r_A, q_A = 0)$.

a) For each grid point:

$$W^{core}(\vec{r}) = \sum_{A,L} w_A^{core}(r_A) \quad \text{(S4)}$$

b) The spherical average core density is computed for each atom,

$$\rho_A^{core,avg}(r_A) = \left\langle \frac{w_A^{core}(r_A)}{W^{core}(\vec{r})} \rho^{core}(\vec{r}) \right\rangle_{r_A} \quad \text{(S5)}$$

c) If $\left| N_A^{core} - \oint \rho_A^{core,avg}(r_A) d^3\vec{r}_A \right| < 10^{-5}$ for every atom and at least five prior core iterations have been performed, the calculation is considered converged and exits. Otherwise, the calculation continues.

d) Starting with $w_A^{core}(r_A) = \rho_A^{core,avg}(r_A)$ as the initial guess, $w_A^{core}(r_A)$ is updated to satisfy constraint (S3) by recursively setting

$$w_A^{core}(r_A) = \min\left( w_A^{core}(r_A), w_A^{core}(r_A - \Delta r_A) \exp(-2\Delta r_A) \right) \quad \text{(S6)}$$

beginning with the second radial shell and continuing outward to the last radial shell. The calculation then repeats the sequence of steps b) to d) until it converges and exits in step c).

**3.2.3.2 Iterations to correct the core density grid**

For atoms having no core-like electrons assigned to the core grid (e.g., if $N_A^{core} < 10^{-10}$), the assigned $w_A^{core}(r_A)$ and $\rho_A^{core,avg}(r_A)$ are set to zero. For all remaining atoms, the following sequence of steps is performed to correct the core grid.

a) In each correction iteration i, a real variable $K_A$ is computed for each atom using the following equations

$$\rho_A^{core}(\vec{r}_A)\Big|_i = \frac{w_A^{core}(r_A)\Big|_i}{W^{core}(\vec{r})\Big|_i} \rho^{core}(\vec{r})\Big|_i \quad \text{(S7)}$$

$$K_A\Big|_i = \min\left( \left( \frac{N_A^{core} - \oint \rho_A^{core}(\vec{r}_A)\Big|_i d^3\vec{r}_A}{\oint \left( \rho_A^{core}(\vec{r}_A)\Big|_i \right)^3 d^3\vec{r}_A} \right), \frac{0.25}{\left( \rho_A^{core}\Big|_i (\max) \right)^2} \right) \quad \text{(S8)}$$

where $\rho_A^{core}\Big|_i(\max)$ is the largest value of $\rho_A^{core}(\vec{r}_A)\Big|_i$ over the set of all grid points.

b) $\rho^{core}(\vec{r})$ is updated by

$$\tilde{\rho}_A^{core}(\vec{r}_A)\Big|_{i+1} = \frac{\rho_A^{core}(\vec{r}_A)\Big|_i}{\sqrt{1-2K_A\Big|_i\left(\rho_A^{core}(\vec{r}_A)\Big|_i\right)^2}} \quad \text{(S9)}$$

$$\rho^{core}(\vec{r})\Big|_{i+1} = \sum_{A,L}\tilde{\rho}_A^{core}(\vec{r}_A)\Big|_{i+1}. \quad \text{(S10)}$$

c) $\rho_A^{core,avg}(r_A)$ is updated by

$$\rho_A^{core,avg}(r_A)\Big|_{i+1} = \left\langle \frac{w_A^{core}(r_A)\Big|_i}{W^{core}(\vec{r})\Big|_i}\rho^{core}(\vec{r})\Big|_{i+1}\right\rangle_{r_A} \quad \text{(S11)}$$

d) If $\left|N_A^{core} - \oint \rho_A^{core,avg}(r_A)\Big|_{i+1} d^3\vec{r}_A\right| < 10^{-5}$ for every atom, the calculation is considered converged and exits. Otherwise, the calculation continues.

e) Starting with $w_A^{core}(r_A)\Big|_{i+1} = \rho_A^{core,avg}(r_A)\Big|_{i+1}$ as the initial guess, $w_A^{core}(r_A)\Big|_{i+1}$ is updated to satisfy constraint (S3) by recursively setting

$$w_A^{core}(r_A)\Big|_{i+1} = \min\left(w_A^{core}(r_A)\Big|_{i+1}, w_A^{core}(r_A)\Big|_{i+1}\exp(-2\Delta r_A)\right) \quad \text{(S12)}$$

beginning with the second radial shell and continuing outward to the last radial shell.

f) $W^{core}(\vec{r})$ is updated:

$$W^{core}(\vec{r})\Big|_{i+1} = \sum_{A,L} w_A^{core}(r_A)\Big|_{i+1} \quad \text{(S13)}$$

The calculation then repeats the sequence of steps a) to f) until it converges and exits in step d).

### 3.2.4 Proof this Iterative Algorithm Satisfies the Design Criteria

a) The iterative scheme converges to the desired solution. *Proof*: Near the solution, we have

$$K_A\Big|_i = \frac{N_A^{core} - \oint \rho_A^{core}(\vec{r}_A)\Big|_i d^3\vec{r}_A}{\oint\left(\rho_A^{core}(\vec{r}_A)\Big|_i\right)^3 d^3\vec{r}_A} \text{ and } \left|K_A\Big|_i\left(\rho_A^{core}(\vec{r}_A)\Big|_i\right)^2\right| \ll 1 \quad \text{(S14)}$$

Therefore, we can expand Eq. (S9) as a Taylor series to give

$$\tilde{\rho}_A^{core}(\vec{r}_A)\Big|_{i+1} = \left(\rho_A^{core}(\vec{r}_A)\Big|_i + \left(\frac{N_A^{core} - \oint \rho_A^{core}(\vec{r}_A)\Big|_i d^3\vec{r}_A}{\oint\left(\rho_A^{core}(\vec{r}_A)\Big|_i\right)^3 d^3\vec{r}_A}\right)\left(\rho_A^{core}(\vec{r}_A)\Big|_i\right)^3\right) + \text{residual} \quad \text{(S15)}$$

Integrating Eq. (S15) yields

$$\oint \tilde{\rho}_A^{core}(\vec{r}_A)\Big|_{i+1} d^3\vec{r}_A = N_A^{core} + \text{residual} \quad \text{(S16)}$$

which shows $\oint \tilde{\rho}_A^{core}(\vec{r}_A)\Big|_{i+1} d^3\vec{r}_A$ converges to $N_A^{core}$. Combining Eqs. (S9) and (S16) gives

$$\oint\left(\frac{\rho_A^{core}\left(\vec{r}_A\right)\Big|_i}{\sqrt{1-2K_A\Big|_i\left(\rho_A^{core}\left(\vec{r}_A\right)\Big|_i\right)^2}}\right)d^3\vec{r}_A = N_A^{core} + \text{residual} \qquad (S17)$$

which shows the left side of Eq. (S17) converges to $N_A^{core}$. This can only be true if $\lim_{i\to\infty} K_A = 0$, because otherwise $\rho_A^{core}\left(\vec{r}_A\right)\Big|_i$ would increase ($\lim_{i\to\infty} K_A > 0$) or decrease ($\lim_{i\to\infty} K_A < 0$) without bound. With Eq. (S8) this means

$$\lim_{i\to\infty}\oint \rho_A^{core}\left(\vec{r}_A\right)\Big|_i d^3\vec{r}_A = N_A^{core} \qquad (S18)$$

$$\lim_{K_A\to 0}\tilde{\rho}_A^{core}\left(\vec{r}_A\right)\Big|_i = \rho_A^{core}\left(\vec{r}_A\right)\Big|_i \qquad (S19)$$

so the iterative process converges to the desired solution.

b) The corrected core density is nonnegative at every grid point. *Proof*: The minimum of the factor $\sqrt{1-2K_A\Big|_i\left(\rho_A^{core}\left(\vec{r}_A\right)\Big|_i\right)^2}$ occurs when $K_A\Big|_i > 0$ and for the grid point $\rho_A^{core}\left(\vec{r}_A\right)\Big|_i(\max)$. From Eq. (S8), it follows $K_A\Big|_i\left(\rho_A^{core}\left(\vec{r}_A\right)\Big|_i\right)^2 \le 0.25$. Therefore, $\sqrt{1-2K_A\Big|_i\left(\rho_A^{core}\left(\vec{r}_A\right)\Big|_i\right)^2} \ge \sqrt{1/2}$. Examining Eq. (S9), this means $\tilde{\rho}_A^{core}\left(\vec{r}_A\right)\Big|_{i+1} \ge 0$.

c) For a particular atom, the correction does not change the relative ordering of grid point core densities. *Proof*: Consider the function

$$\Xi(s) = \frac{s}{\sqrt{1-2Ks^2}} \qquad (S20)$$

which has the derivative

$$\frac{d\Xi}{ds} = \frac{1}{\left(1-2Ks^2\right)^{3/2}} \qquad (S21)$$

$\Xi(s)$ is a monotonically increasing function of s over the range $Ks^2 < 0.5$. Because Eq. (S9) has the functional form $\Xi(s)$, the relative ordering of grid point core densities is preserved for each atom.

d) The correction is localized to those grid points with the highest core densities (i.e., those closest to atomic nuclei). *Proof*: Combining

$$\rho^{core}\left(\vec{r}\right)\Big|_i = \sum_{A,L}\rho_A^{core}\left(\vec{r}_A\right)\Big|_i \qquad (S22)$$

with Eq. (S10) gives

$$\rho^{core}\left(\vec{r}\right)\Big|_{i+1} - \rho^{core}\left(\vec{r}\right)\Big|_i = \sum_{A,L}\rho_A^{core}\left(\vec{r}_A\right)\Big|_i\left(\frac{1}{\sqrt{1-2K_A\Big|_i\left(\rho_A^{core}\left(\vec{r}_A\right)\Big|_i\right)^2}}-1\right) \qquad (S23)$$

Consider the function

$$g(s) = s\left(\frac{1}{\sqrt{1-2Ks^2}} - 1\right) = \Xi(s) - s \qquad \text{(S24)}$$

which has the derivative

$$\frac{dg}{ds} = \frac{1}{\left(1-2Ks^2\right)^{3/2}} - 1. \qquad \text{(S25)}$$

$\frac{dg}{ds} = 0$ if and only if K = 0 or s=0. Moreover, $g(s)$ is a monotonically increasing (when K > 0) or decreasing (when K < 0) function of $s$ over the range $Ks^2 < 0.5$. For small s, $g(s)$ expands a Taylor series to

$$g(s) = Ks^3 + \text{residual} \leq \frac{0.25}{\left(s_{max}\right)^2} s^3 + \text{residual} \qquad \text{(S26)}$$

The inequality on the right-most side of Eq. (S26) arises from Eq. (S8). Due to the cubic dependence of $g(s)$ on s for small s, the points with largest $\rho_A^{core}\left(\vec{r}_A\right)\big|_i$ dominate the core correction for atom A. These points are typically located close to nucleus A.

e) <u>The correction does not interfere with the exponential decay constraint applied to</u> $\left\{w_A^{core}\left(r_A\right)\right\}$. Consider two grid points near nucleus A such that grid point 1 is closer to nucleus A than grid point 2. From Eq. (S9),

$$\ln\left(\frac{\tilde{\rho}_A^{core}\left(\vec{r}_1\right)\big|_{i+1}}{\tilde{\rho}_A^{core}\left(\vec{r}_2\right)\big|_{i+1}}\right) = \ln\left(\frac{\rho_A^{core}\left(\vec{r}_1\right)\big|_i}{\rho_A^{core}\left(\vec{r}_2\right)\big|_i}\right) + \frac{1}{2}\ln\left(\frac{1-2K_A\big|_i\left(\rho_A^{core}\left(\vec{r}_2\right)\big|_i\right)^2}{1-2K_A\big|_i\left(\rho_A^{core}\left(\vec{r}_1\right)\big|_i\right)^2}\right) \qquad \text{(S27)}$$

***case 1*:** $K_A > 0$. In this case, the last term in Eq. (S27) increases $\ln\left(\frac{\tilde{\rho}_A^{core}\left(\vec{r}_1\right)\big|_{i+1}}{\tilde{\rho}_A^{core}\left(\vec{r}_2\right)\big|_{i+1}}\right)$ so the core density increase performed during the core grid correction does not cause $\ln\left(\frac{\tilde{\rho}_A^{core}\left(\vec{r}_1\right)\big|_{i+1}}{\tilde{\rho}_A^{core}\left(\vec{r}_2\right)\big|_{i+1}}\right)$ to be $\leq \exp\left(2\left(r_2 - r_1\right)\right)$.

***case 2*:** $K_A < 0$. This occurs when the core density grid assigns too much core density to atom A. This can be due to a nucleus falling directly on a grid point, in which case $\ln\left(\frac{\tilde{\rho}_A^{core}\left(\vec{r}_1 = 0\right)\big|_{i+1}}{\tilde{\rho}_A^{core}\left(\vec{r}_2\right)\big|_{i+1}}\right)$ is too high and Eq. (S27) appropriately decreases it. Alternatively, this case can arise if the assigned core density is too diffuse for any reason. The lowering of $\tilde{\rho}_A^{core}\left(\vec{r}_1 = 0\right)\big|_{i+1}$ together with constraint (S3) corrects the problem of too much core density being assigned to this atom.

**3.2.5 What features cause this scheme to converge rapidly and robustly?**

a) The correction is localized to regions with highest $\rho_A^{core}(\vec{r}_A)$ where typically $w_A(r_A)/W(\vec{r}) \approx 1$. This makes corrections for different atoms almost independent of each other.

b) The relative ordering of core density values for an atom is preserved. This ensures a smooth behavior.

c) Convergence is rapid near the solution, as evidenced by the Taylor series expansion in Eqs. (S15) and (S16).

d) Examining Eqs. (S8) and (S9), the density changes are bounded by

$$\frac{1}{\sqrt{3}} \leq \frac{\tilde{\rho}_A^{core}(\vec{r}_A)\big|_{i+1}}{\rho_A^{core}(\vec{r}_A)\big|_i} \leq \sqrt{2} \qquad \text{(S28)}$$

The extreme values occur for the grid point corresponding to $\rho_A^{core}\big|_i(\max)$ when $\oint\left(\rho_A^{core}(\vec{r}_A)\big|_i\right)^3 d^3\vec{r}_A$ is dominated by $\rho_A^{core}\big|_i(\max)$ such that $\oint\left(\rho_A^{core}(\vec{r}_A)\big|_i\right)^3 d^3\vec{r}_A \approx \left(\rho_A^{core}\big|_i(\max)\right)^3 \times V_{pixel}$ (where $V_{pixel}$ is the pixel volume) and under the condition that $\left|N_A^{core} - \oint \rho_A^{core}(\vec{r}_A)\big|_i d^3\vec{r}_A\right|$ is large. Under these conditions, $K_A > 0$ gives the limiting behavior

$$\tilde{\rho}_A^{core}(\vec{r}_A)\big|_{i+1}(\max) = \frac{\rho_A^{core}(\vec{r}_A)\big|_i(\max)}{\sqrt{1 - 2\frac{0.25}{\left(\rho_A^{core}(\vec{r}_A)\big|_i(\max)\right)^2}\left(\rho_A^{core}(\vec{r}_A)\big|_i(\max)\right)^2}} \qquad \text{(S29)}$$
$$= \sqrt{2}\,\rho_A^{core}(\vec{r}_A)\big|_i(\max)$$

Under these conditions, $K_A < 0$ gives the limiting behavior

$$\tilde{\rho}_A^{core}(\vec{r}_A)\big|_{i+1}(\max) = \frac{\rho_A^{core}(\vec{r}_A)\big|_i(\max)}{\sqrt{1 - 2\frac{\left(N_A^{core} - \left(\rho_A^{core}(\vec{r}_A)\big|_i(\max)\right)V_{pixel}\right)}{\left(\rho_A^{core}(\vec{r}_A)\big|_i(\max)\right)^3 V_{pixel}}\left(\rho_A^{core}(\vec{r}_A)\big|_i(\max)\right)^2}} \qquad \text{(S30)}$$
$$\approx \frac{\rho_A^{core}(\vec{r}_A)\big|_i(\max)}{\sqrt{3}}$$

e) Noting that $\left(\sqrt{2}\right)^{20} = 1024$, this means about 20 iterations are required to increase a grid point density by a factor of $10^3$. Noting that $\left(\sqrt{3}\right)^{13} = 1262.665$, this means about 13 iterations are required to decrease a grid point density by a factor of $10^3$. Because the approach to convergence is smooth and the core density assigned to each grid point is never off by more than a factor of $10^6$, this means convergence is always achieved in fewer than 40 iterations. In practice, convergence is nearly always achieved in fewer than 20 iterations.

## 4. Computational Algorithm for Convex Functional

### 4.1 Iterative Algorithm

The complex functional was optimized using the following procedure. First, $q_A^{ref}$ was computed in the first two charge cycles as described in Section 2.5 of the main text. Second, the conditioned reference ion density, $\rho_A^{cond}(r_A)$, was computed in the third charge cycle as described in Section 2.6 of the main text.

The fourth charge cycle used the following procedure to compute the fixed reference density

$$\rho_A^{fixed_ref}(r_A) = H_A(r_A) \qquad \text{(S31)}$$

in the Convex functional. First, we computed

$$\sigma_A(r_A) = \rho_A^{cond}(r_A)\left\langle \rho(\vec{r}) / \rho^{cond}(\vec{r}) \right\rangle_{r_A} . \qquad \text{(S32)}$$

Then, $\sigma_A(r_A)$, is reshaped to form $G_A(r_A)$ exactly as described in item i) in Section 2.7 of the main text. Then, $G_A(r_A)$ is reshaped to form $H_A(r_A)$ exactly as described in item ii) of Section 2.7 of the main text.

The fifth (i.e., $i = 5$) and latter charge cycles form an iterative process to achieve a self-consistent solution. The fifth charge cycle starts with the following initial estimates: $w_A^5(r_A) = \rho_A^{fixed_ref}(r_A)$, $\kappa_A^5 = 0$, and $C_A^5 = 1$. These are refined to self-consistency using the following sequence of steps in the fifth and latter charge cycles:

1. In the first loop over grid points and atoms, the following sum is computed at each grid point:

$$W(\vec{r}) = \sum_{B,L} w_B(r_B) \qquad \text{(S33)}$$

2. In the second loop over grid points and atoms, the following quantities are computed:

$$\rho_A(\vec{r}_A) = w_A(r_A)\rho(\vec{r})/W(\vec{r}) \qquad \text{(S34)}$$

$$\rho_A^{avg}(r_A) = \left\langle \rho_A(\vec{r}_A) \right\rangle_{r_A} \qquad \text{(S35)}$$

$$N_A = \oint \rho_A(\vec{r}_A) d^3\vec{r} \qquad \text{(S36)}$$

$$u_A = 2\oint \left(1 - \frac{w_A(r_A)}{W(\vec{r})}\right)\left(1 + \frac{w_A(r_A)}{W(\vec{r})}\right)\rho_A(\vec{r}_A) d^3\vec{r}_A \qquad \text{(S37)}$$

All of these quantities except $\{\rho_A(\vec{r}_A)\}$ are stored.

3. If the $N_A$ and $\{\rho_A^{avg}(r_A)\}$ changes between successive charge cycles were less than $10^{-5}$e and $10^{-5}$e/bohr$^3$, respectively, for each atom two consecutive times in a row then the calculation is considered converged. Starting with the 10$^{th}$ charge cycle, the calculation breaks at this point if it is considered converged. If it is not converged or the charge cycle is less than 10, the calculation proceeds to # 4 below.
4. At the end of the i$^{th}$ charge cycle $(i \geq 5)$, the updated atomic weighting factors are given by

$$w_A^{i+1}(r_A) = C_A^i e^{\kappa_A^{i+1}} \sqrt{\rho_A^{fixed_ref}(r_A)\rho_A^{avg}(r_A)} \quad \text{(S38)}$$

where

$$\kappa_A^{i+1} = \text{const}\cdot\left(\max\left(0, \kappa_A^i - \left(N_A^{val}/u_A\right)\right)\right) \quad \text{(S39)}$$

if $u_A \geq 10^{-7}$, and $\kappa_A^{i+1} = 0$ if $u_A < 10^{-7}$. The constant appearing in Eq. (S39) affects only the convergence speed and robustness without affecting the converged solution. The approximately optimal value of $2-\sqrt[3]{4}$ is derived in SI Section 4.2 below. The convergence accelerator, $C_A^i$, minimizes the number of required charge cycles without changing the converged solution. For $i \geq 6$,

$$C^i = \frac{e^{\kappa_A^{i+1}} \sqrt{\rho_A^{fixed_ref}(r_A)\rho_A^{avg}(r_A)}}{\left(2-\sqrt{2}\right)e^{\kappa_A^{i+1}} \sqrt{\rho_A^{fixed_ref}(r_A)\rho_A^{avg}(r_A)} + \left(\sqrt{2}-1\right)w_A^i(r_A)}. \quad \text{(S40)}$$

As explained in SI Section 4.2 below, this form of the convergence accelerator maximizes the convergence speed. The convergence accelerator equals one when the calculation converges.

After generating $w_A^{i+1}(r_A)$ for each atom, the calculation returns to step #1 above and starts the next (i.e., $(i+1)$) charge cycle using $w_A^{i+1}(r_A)$ as the new estimate for $w_A(r_A)$ in Eq. (S33). This iterative process is continued until the calculation satisfies the convergence criteria and breaks in step #3 above.

We now show that convergence of $\{N_A\}$ and $\{\rho_A^{avg}(r_A)\}$ can occur only if $\{\kappa_A\}$ are also converged. Following Manz and Sholl,[5] we define

$$J_{AB} = \frac{\partial N_A}{\partial \kappa_B} = \oint \left( \frac{w_A(r_A)}{W(\vec{r})}\delta_{AB} - \sum_L \frac{w_A(r_A)w_B(r_B)}{\left(W(\vec{r})\right)^2} \right)\rho(\vec{r})d^3\vec{r}. \quad \text{(S41)}$$

The $\sum_{L}$ in Eq. (S41) accounts for periodic images (if any) of atom B. For a process in which $\{\rho_A^{avg}(r_A)\}$ are converged, changes in $\{w_A(r_A)\}$ can arise only from changes in $\{\kappa_A\}$. For such a process,

$$\sum_{A} d\kappa_A dN_A = \sum_{A}\sum_{B} d\kappa_A J_{AB} d\kappa_B \qquad \text{(S42)}$$

Inserting Eq. (S41) into (S42) and rearranging gives

$$\sum_{A} d\kappa_A dN_A = \frac{1}{2}\sum_{A}\sum_{B,L}\left[\left(d\kappa_A - d\kappa_B\right)^2 \oint\left(\frac{w_A(r_A)w_B(r_B)}{\left(W(\vec{r})\right)^2}\right)\rho(\vec{r})d^3\vec{r}\right] \geq 0. \qquad \text{(S43)}$$

If $\{N_A\}$ are converged, then both sides of Eq. (S43) are identically zero. For every pair of atoms A and B, this requires either $\left(d\kappa_A - d\kappa_B\right)^2 = 0$ or else the AB overlap integral in Eq. (S43) is zero. Thus, for any sets of atoms with non-zero overlaps, $\{d\kappa_A\} = 0$ when $\{dN_A\} = 0$. For an atom without any overlaps (i.e., isolated atomic ion limit), $u_A = 0$ and the converged $\rho_A(\vec{r}_A)$ is independent of $\kappa_A$. Therefore, when $u_A$ is negligible (e.g., $u_A < 10^{-7}$) we set $\kappa_A = 0$. For all other atoms, $\left(d\kappa_A - d\kappa_B\right)^2 = 0$ when $\{dN_A\} = 0$. Therefore, $\{N_A\}$ cannot be converged between successive charge cycles unless $\{\kappa_A\}$ are converged between successive charge cycles. Therefore, $\{w_A(r_A)\}$ are converged in the DDEC6 method if and only if $\{N_A\}$ and $\{\rho_A^{avg}(r_A)\}$ are converged.

## 4.2 Derivation of Optimal Convergence Parameters

### 4.2.1 Derivation of the form of $u_A$ in Eq. (S37)

The careful reader will observe the quantity $u_A$ appearing in Eq. (S37) has a different form than the quantity $u_A$ appearing in Eq. (84) of the main text. During DDEC6 charge cycles, $e^{\kappa_A(r_A)}$ is multiplied by a fixed term (i.e., $w_A^{fixed}(r_A)$ ). In contrast, when optimizing the Convex functional, $\kappa_A(r_A)$ affects $\rho_A^{avg}(r_A)$ which in turn affects $w_A(r_A)$. For generality, we consider a an atomic weighting factor of the form

$$w_A^{convex}(r_A) = e^{\kappa_A}\left(\rho_A^{fixed_ref}(r_A)\right)^{\chi}\left(\rho_A^{avg}(r_A)\right)^{1-\chi} \qquad \text{(S44)}$$

where $0 < \chi \leq 1$. Note that $\chi = 1/2$ for the Convex functional described in Section 4.1 above. Inserting Eq. (S44) into Eq. (12) of the main text and rearranging yields

$$\rho_A^{avg}(r_A) = e^{\kappa_A(r_A)/\chi}\rho_A^{fixed_ref}(r_A)\left(\left\langle \rho(\vec{r})/W(\vec{r})\right\rangle_{r_A}\right)^{1/\chi}. \qquad \text{(S45)}$$

<u>Case 1</u>: When $W(\vec{r}) \approx \rho(\vec{r})$, changes in $w_A^{convex}(r_A)$ tend to be absorbed by the other atoms. In this case, taking the partial derivative of Eq. (S45) yields

$$\left(\frac{\partial \rho_A^{avg}(r_A)}{\partial \kappa_A(r_A)}\right)_{\kappa_B(r_B)} \approx \frac{\rho_A^{avg}(r_A)}{\chi}. \quad \text{(S46)}$$

$$\left(\frac{\partial w_A^{convex}(r_A)}{\partial \kappa_A(r_A)}\right)_{\kappa_B(r_B)} \approx \frac{w_A^{convex}(r_A)}{\chi} \qquad \text{(S47)}$$

Case 2: At the other extreme, where changes in $w_A^{convex}(r_A)$ are not absorbed by the other atoms, then

$$\begin{aligned}\left(\frac{\partial N_A}{\partial \kappa_A(r_A)}\right)_{\kappa_B(r_B)} &= \frac{\partial}{\partial \kappa_A(r_A)} \oint w_A(r_A)\rho(\vec{r})/W(\vec{r}) d^3\vec{r}_A \\ &= \oint \frac{\partial w_A(r_A)}{\partial \kappa_A(r_A)} \frac{\rho(\vec{r})}{W(\vec{r})}\left(1 - \frac{w_A(r_A)}{W(\vec{r})}\right) d^3\vec{r}_A\end{aligned} . \qquad \text{(S48)}$$

Taking the partial derivative of Eq. (S44) with the help of Eq. (S45) yields

$$\left(\frac{\partial w_A^{convex}(r_A)}{\partial \kappa_A(r_A)}\right)_{\kappa_B(r_B)} \lesssim \frac{w_A^{convex}(r_A)}{\chi}. \qquad \text{(S49)}$$

Substituting Eq. (S49) into (S48) gives

$$\left(\frac{\partial N_A}{\partial \kappa_A(r_A)}\right)_{\kappa_B(r_B)} \lesssim \frac{1}{\chi} \oint w_A^{convex}(r_A) \frac{\rho(\vec{r})}{W(\vec{r})}\left(1 - \frac{w_A^{convex}(r_A)}{W(\vec{r})}\right) d^3\vec{r}_A . \qquad \text{(S50)}$$

Cases 1 and 2 can be combined by noting that case 1 dominates when $w_A^{convex}(r_A) \ll W(\vec{r})$, and case 2 dominates when $w_A^{convex}(r_A) \approx W(\vec{r})$. Assuming a linear interpolation where $\left(1 - \frac{w_A^{convex}(r_A)}{W(\vec{r})}\right)$ is the fraction assigned to case 1 and $\frac{w_A^{convex}(r_A)}{W(\vec{r})}$ is the fraction assigned to case 2 yields:

$$u_A = \left(\frac{\partial N_A}{\partial \kappa_A(r_A)}\right)_{\kappa_B(r_B)} \lesssim \frac{1}{\chi} \oint w_A^{convex}(r_A) \frac{\rho(\vec{r})}{W(\vec{r})}\left(1 - \frac{w_A^{convex}(r_A)}{W(\vec{r})}\right)\left(1 + \frac{w_A^{convex}(r_A)}{W(\vec{r})}\right) d^3\vec{r}_A . \qquad \text{(S51)}$$

The safest approach corresponds to setting $u_A$ to its approximate upper bound. (This corresponds to estimating $\kappa_A(r_A)$ changes as conservatively as possible to minimize overshoot.) This explains the basis for the form of $u_A$ appearing in Eq. (S37).

### 4.2.2 Derivation of the constant value in Eq. (S39)

To derive this constant, we construct a convergence model in which the error at charge cycle i is characterized by

$$N_A^{val}\Big|_i = -\varepsilon_i \qquad \text{(S52)}$$

where $\varepsilon_i$ is some positive number. Substituting Eq. (S52) into (S39) yields

$$\kappa_A^{i+1} - \kappa_A^{i} = \mathrm{const} \cdot \varepsilon_i / u_A . \qquad \text{(S53)}$$

During charge cycle i+1, this change in $\kappa_A$ impacts $w_A(r_A)$ only through the $e^{\kappa_A(r_A)}$ prefactor and not through the $\rho_A^{avg}(r_A)$ term, because the $\rho_A^{avg}(r_A)$ term will be impacted only during the i+2 and subsequent charge cycles. From Eq. (S44), the change in $w_A^{convex}(r_A)$ at constant $\rho_A^{avg}(r_A)$ is

$$\left( \frac{\partial w_A^{convex}(r_A)}{\partial \kappa_A(r_A)} \right)_{\rho_A^{avg}(r_A), \kappa_B(r_B)} = w_A^{convex}(r_A) . \qquad \text{(S54)}$$

Comparing Eqs. (S54) and (S47) gives

$$\left( \frac{\partial w_A^{convex}(r_A)}{\partial \kappa_A(r_A)} \right)_{\rho_A^{avg}(r_A), \kappa_B(r_B)} \approx \chi \left( \frac{\partial w_A^{convex}(r_A)}{\partial \kappa_A(r_A)} \right)_{\kappa_B(r_B)} . \qquad \text{(S55)}$$

Combining Eqs. (S39) and (S55), the impact of the $\kappa_A$ change felt during the i+1 charge cycle is

$$\left( \frac{\partial N_A}{\partial \kappa_A(r_A)} \right)_{\rho_A^{avg}(r_A), \kappa_B(r_B)} \left( \kappa_A^{i+1}(r_A) - \kappa_A^{i}(r_A) \right) = \mathrm{const} \cdot \varepsilon_i \cdot \chi . \qquad \text{(S56)}$$

Thus, neglecting the $\rho_A^{avg}(r_A)$ changes, the error during charge cycle i+1 will be

$$\varepsilon_{i+1} = \varepsilon_i \left( 1 - \mathrm{const} \cdot \chi \right) . \qquad \text{(S57)}$$

Applying once more,

$$\varepsilon_{i+2} = \varepsilon_{i+1} \left( 1 - \mathrm{const} \cdot \chi \right) = \varepsilon_i \left( 1 - \mathrm{const} \cdot \chi \right)^2 . \qquad \text{(S58)}$$

Combining Eqs. (S53) with (S57) and (S58) yields

$$\kappa_A^{i+2} - \kappa_A^{i+1} = \mathrm{const} \cdot \varepsilon_i \left( 1 - \mathrm{const} \cdot \chi \right) / u_A \qquad \text{(S59)}$$

$$\kappa_A^{i+3} - \kappa_A^{i+2} = \mathrm{const} \cdot \varepsilon_i \left( 1 - \mathrm{const} \cdot \chi \right)^2 / u_A . \qquad \text{(S60)}$$

Therefore, the sum of $\kappa_A$ changes over three successive charge cycles is

$$\kappa_A^{i+3} - \kappa_A^{i} = \mathrm{const} \cdot \varepsilon_i \left[ \left( 1 - \mathrm{const} \cdot \chi \right)^2 + \left( 1 - \mathrm{const} \cdot \chi \right) + 1 \right] / u_A . \qquad \text{(S61)}$$

The key to deriving an appropriate value for the constant is to note that eventually the $\rho_A^{avg}(r_A)$ changes will kick in and effect $w_A(r_A)$. A reasonable value of the constant will correspond to the sum of $\kappa_A$ changes over three successive charge cycles not overshooting the eventual $N_A^{val}$ changes even when including the eventual effects of changing $\rho_A^{avg}(r_A)$ on $w_A(r_A)$. Since the eventual $N_A^{val}$ change is $u_A$ times the $\kappa_A$ change, the non-overshoot condition derived from Eq. (S61) is

$$\mathrm{const} \left[ \left( 1 - \mathrm{const} \cdot \chi \right)^2 + \left( 1 - \mathrm{const} \cdot \chi \right) + 1 \right] \leq 1 . \qquad \text{(S62)}$$

The approximately optimal constant corresponds to the equality condition. Solving Eq. (S62) yields

$$\text{const} = \frac{1}{\chi} - \sqrt[3]{\frac{1}{\chi^3} - \frac{1}{\chi^2}} \; . \qquad \text{(S63)}$$

For $\chi = 1/2$ , we have $\text{const} = 2 - \sqrt[3]{4} = 0.41259...$

We performed computational tests (using model systems in spreadsheet) with different $\chi$ values and different constant values, which verified Eq. (S63) yields nearly optimal convergence performance. With the constant value from Eq. (S63), the system converges rapidly with highly damped overshoot if the $N_A^{val} \geq 0$ constraint is binding. Using $\chi = 1/2$, $\text{const} = 2 - \sqrt[3]{4} = 0.41259...$, and the convergence accelerator described in Section 4.2.3, convergence to $10^{-5}$ electrons is achieved in approximately 20 total charge cycles for the Convex functional.

**4.2.3 Derivation of the optimal convergence accelerator**

When using an atomic weighting factor of the form shown in Eq. (S44), the spherical average atomic density, $\rho_A^{avg}(r_A)$, is computed based on the $w_A(r_A)$ that used the prior $\rho_A^{avg}(r_A)$. This causes the situation that if the estimate for $\rho_A^{avg}(r_A)$ in charge cycle i is too large (small), the new estimate for $\rho_A^{avg}(r_A)$ in charge cycle i+1 will also be too large (small). In general, therefore, it will take many charge cycles to work off errors in the estimated $\{\rho_A^{avg}(r_A)\}$. The higher the proportion of spherical averaging (i.e., the smaller the $\chi$ value) in $w_A(r_A)$, the more pronounced this problem will be.

This problem can be solved using a convergence accelerator. A convergence accelerator causes changes in the estimated $\{\rho_A^{avg}(r_A)\}$ to take effect in fewer charge cycles, thereby allowing the calculation to be converged in fewer charge cycles. All feasible functional forms of a convergence accelerator become linear as the change to $\{\rho_A^{avg}(r_A)\}$ becomes relatively small. Therefore, we choose the linear form

$$w_A^{i+1}(r_A) = \psi_A^i(r_A) + m\left(\psi_A^i(r_A) - \psi_A^{i-1}(r_A)\right) \quad \text{(S64)}$$

where m is a constant and

$$\psi_A^i(r_A) = e^{\kappa_A^{i+1}} \left(\rho_A^{fixed_ref}(r_A)\right)^{\chi} \left(\rho_A^{avg}(r_A)\Big|_i\right)^{(1-\chi)} . \qquad \text{(S65)}$$

We begin by defining a set of variables that quantify the approach to convergence:

$$y1_A^i(r_A) = \ln\left(\rho_A^{avg}(r_A)\Big|_i \Big/ \rho_A^{avg}(r_A)\Big|_{converged}\right) \qquad \text{(S66)}$$

$$y2_A^i(r_A) = \ln\left(w_A^i(r_A) \big/ w_A^{converged}(r_A)\right) \qquad \text{(S67)}$$

where $\rho_A^{avg}(r_A)\big|_i$ is the value of $\rho_A^{avg}(r_A)$ computed during charge cycle i, and $\rho_A^{avg}(r_A)\big|_{converged}$ and $w_A^{converged}(r_A)$ are the final converged values of $\rho_A^{avg}(r_A)$ and $w_A(r_A)$, respectively. Combining Eqs. (S65) and (S66) yields

$$\ln\left(\psi_A^i(r_A)/\psi_A^{converged}(r_A)\right) = (1-\chi)\,y1_A^i(r_A). \quad \text{(S68)}$$

According to Eq. (S44),

$$\psi_A^{converged}(r_A) = w_A^{converged}(r_A) \quad \text{(S69)}$$

In regions where atom-atom overlaps are significant (i.e., $w_A(r_A) \ll W(\vec{r})$), the spherical average computed during charge cycle i (i.e., $\rho_A^{avg}(r_A)\big|_i$) is proportional to $w_A^i(r_A)$:

$$\frac{\partial \ln\left(\rho_A^{avg}(r_A)\big|_i\right)}{\partial \ln\left(w_A^i(r_A)\right)} \approx 1\ . \quad \text{(S70)}$$

From Eq. (S70), it directly follows that

$$y1_A^i(r_A) \approx y2_A^i(r_A) \approx \varepsilon_A^i(r_A) \quad \text{(S71)}$$

where $\varepsilon_A^i(r_A)$ quantifies the approach to convergence.

Computational tests on real systems studied with the CHARGEMOL program, as well as numerical model systems studied in spreadsheet, showed that convergence in the fewest number of charge cycles is achieved when the constant m is set to the largest value giving non-oscillatory convergence. For steady non-oscillatory convergence, the errors are reduced to a nearly constant fraction between successive charge cycles:

$$f \approx \frac{\varepsilon_A^i(r_A)}{\varepsilon_A^{i-1}(r_A)}. \quad \text{(S72)}$$

Dividing Eq. (S64) by $\psi_A^{converged}(r_A)$ and taking the log of both sides in the limit of small $\varepsilon(r_A)$ yields:

$$\varepsilon_A^{i+1}(r_A) = (1-\chi)\varepsilon_A^i(r_A) + m(1-\chi)\left(\varepsilon_A^i(r_A) - \varepsilon_A^{i-1}(r_A)\right). \quad \text{(S73)}$$

Substituting Eq. (S72) into (S73) gives

$$f^2\varepsilon_A^{i-1}(r_A) = (1-\chi)f\varepsilon_A^{i-1}(r_A) + m(1-\chi)\left(f\varepsilon_A^{i-1}(r_A) - \varepsilon_A^{i-1}(r_A)\right) \quad \text{(S74)}$$

which simplifies to the characteristic equation

$$f^2 - (1+m)(1-\chi)f + m(1-\chi) = 0. \quad \text{(S75)}$$

Solving Eq. (S75) gives

$$f = \frac{(1+m)(1-\chi) \pm \sqrt{(1+m)^2(1-\chi)^2 - 4m(1-\chi)}}{2}. \quad \text{(S76)}$$

The limiting value of f occurs when the discriminant (i.e., quantity under square root) is zero:

$$(1+m)^2(1-\chi)^2 - 4m(1-\chi) = 0 \quad \text{(S77)}$$

which yields

$$m = \frac{\left(1-\sqrt{\chi}\right)^2}{1-\chi}. \qquad \text{(S78)}$$

Substituting Eq. (S78) into (S76) yields the limiting value of f:

$$f = 1-\sqrt{\chi}. \qquad \text{(S79)}$$

We chose the following form for our convergence accelerator:

$$w_A^{i+1}\left(r_A\right) = \frac{e^{2\kappa_A^{i+1}}\left(\rho_A^{fixed_ref}\left(r_A\right)\right)^{2\chi}\left(\rho_A^{avg}\left(r_A\right)\right)^{2(1-\chi)}}{\left(1-a\right)e^{\kappa_A^{i+1}}\left(\rho_A^{fixed_ref}\left(r_A\right)\right)^{\chi}\left(\rho_A^{avg}\left(r_A\right)\right)^{(1-\chi)} + \left(a\right)w_A^{i}\left(r_A\right)} \qquad \text{(S80)}$$

where $0 \le a < 1$ is a constant. We chose this form, because it has the key advantage of guaranteeing $w_A^{i+1}\left(r_A\right) \ge 0$. For small $\varepsilon_A^{i}\left(r_A\right)$, dividing both sides of Eq. (S80) by $w_A^{converged}\left(r_A\right)$ and simplifying as a perturbation series in $\varepsilon_A^{i}\left(r_A\right)$ gives the leading order result

$$f\varepsilon_A^{i}\left(r_A\right) \approx \varepsilon_A^{i}\left(r_A\right)\left(\left(1-\chi\right)-a\chi\right). \qquad \text{(S81)}$$

Substituting Eq. (S79) into (S81) and solving gives the optimized convergence accelerator parameter

$$a = \sqrt{\frac{1}{\chi}} - 1. \qquad \text{(S82)}$$

Notably, the rate of convergence is practically independent of the material. The largest number of charge cycles required to converge the NACs and $\left\{\rho_A^{avg}\left(r_A\right)\right\}$ within convergence_threshold e and e/bohr$^3$, respectively, is

$$charge_cycles \le \frac{\ln\left(convergence_threshold\right) - \ln\left(\Delta q_A\right)}{\ln(f)} + 4 + 2 \qquad \text{(S83)}$$

where $\Delta q_A$ is the maximum NAC error on the fourth charge cycle. convergence_threshold is the error on the (last -2)$^{th}$ charge cycle. Two final charge cycles are required to demonstrate the NAC and $\left\{\rho_A^{avg}\left(r_A\right)\right\}$ changes between successive charge cycles are below the convergence_threshold two times in a row. The first four and last two charge cycles are thus added in Eq. (S83). We used convergence_threshold = $10^{-5}$ e and e/bohr$^3$ on the NACs and $\left\{\rho_A^{avg}\left(r_A\right)\right\}$, respectively. A reasonable approximation is that the NACs on the fourth charge cycle are within ~±0.2 e of the final NACs. Substituting into Eq. (S83) with $f = 1-\sqrt{1/2}$ yields charge_cycles ≤ 14. Indeed, more than 99% of the materials studied in this paper converged within 14 charge cycles when using the Convex functional with the convergence accelerator.

We performed an extensive set of computational tests confirming all aspects of the theory described above. These computational tests included both tests on real materials using the CHARGEMOL code as well as numerical finite difference models in spreadsheet. All aspects of the above theory were doubly confirmed (i.e., both for the real materials and for the finite difference models), including:

1. The errors between successive charge cycles follows a nearly constant ratio f.
2. We compared f values for m = 0 and the optimal m value (i.e., also a = 0 and the optimal a value) for both $\chi$ = 1/2 and 1/3. All of the computational results were in precise agreement with Eq. (S76). In these cases, the number of charge cycles required for convergence closely followed Eq. (S83).
3. As m and a are decreased below their optimal values, the calculation takes more charge cycles to converge. As m and a are increased above their optimal values, the calculations do not converge in fewer charge cycles. As m and a are increased to unreasonably large values (i.e., many times larger than their optimal values) the calculations begin to oscillate notably.

#### 4.2.4 Convergence speed of spin partitioning

**Finally, we note that the DDEC spin partitioning method[6] follows a similar convergence law as that noted above for the Convex functional.** Specifically, the spin partition method uses $\chi_{spin} = 1/2$ which utilizes a geometric average between $\vec{m}_A^{avg}(r_A)$ and $\vec{m}_A^{proportional}(\vec{r}_A)$.[6] The DDEC spin partitioning method uses an optimized convergence algorithm that achieves convergence as rapidly as feasible.[6] For the same reasons as described above, the DDEC spin partitioning method converges at the same rate for all materials with a constant error fraction between successive spin cycles. Theoretical analysis shows the optimal f value depends only on $\chi$ independent of the particulars of the optimization scheme. Specifically, the analog of Eq. (S79) is

$$f_{spin} = 1 - \sqrt{\chi_{spin}} \quad . \qquad \text{(S84)}$$

Using $\chi_{spin} = 1/2$, this means the ASM errors on spin cycle i+1 are only about 29% as large as the errors on spin cycle i. We confirmed this prediction using numerous computational tests on real collinear and non-collinear magnetic materials. For both collinear and non-collinear magnetism, the required number of spin cycles follows this analog of Eq. (S83):

$$spin_cycles \le \frac{\ln(spin_threshold) - \ln(\Delta M_A)}{\ln(f_{spin})} + 1 + 1 \qquad \text{(S85)}$$

where $\Delta M_A$ is the maximum ASM error on the first spin cycle. In Eq. (S85), the first +1 accounts for the first spin cycle. spin_threshold is the error on the (last -1)$^{th}$ spin cycle. A final spin cycle is required to demonstrate the ASM changes are below the spin_threshold; this accounts for the second +1 appearing in Eq. (S85). For spin partitioning, we used a spin_threshold of $5\times10^{-5}$ e with $\chi_{spin} = 1/2$. Substituting these values into Eq. (S84) yields

$$spin_cycles \le 9 \, . \qquad \text{(S86)}$$

Indeed, all of the collinear and non-collinear magnetic materials we have examined to date followed Eq. (S86). Although in the end we decided to go with the DDEC6 (seven fixed charge partitioning steps) rather than the self-consistent Convex functional for the charge partitioning, this convergence theory still fully applies to the self-consistent spin partitioning used in the DDEC6 method. Thus, a key advantage of our methodology is that both the charge and spin partitioning converge within a small number of cycles for all materials.